\documentclass{./egPublStyle-cgf/egpubl}
\usepackage{./egPublStyle-cgf/eg2026}
\CGFStandardLicense

\ConferencePaper
\usepackage[T1]{fontenc}
\usepackage{egPublStyle-cgf/dfadobe}
\usepackage{cite}
\BibtexOrBiblatex
\electronicVersion
\PrintedOrElectronic

\usepackage[export]{adjustbox}

\usepackage{colortbl}
\usepackage{amsmath}
\usepackage{amssymb}

\usepackage[dvipsnames, table]{xcolor}
\usepackage{array}
\usepackage{stfloats}
\usepackage{multirow}
\usepackage{url}
\usepackage{wrapfig}
\usepackage{algorithm}
\usepackage[noend]{algpseudocode}
\usepackage{tabularx}
\usepackage[labelfont=bf, font=it]{caption}
\usepackage{ccicons}
\usepackage{siunitx}
\usepackage{orcidlink}
\newcommand{\greencheck}{{\textcolor[RGB]{112, 200, 89}{\checkmark}}}

\newcommand{\redX}{{\color{red}$\times$}}

\newcommand{\vect}[1]{\textbf{#1}}

\usepackage{./egPublStyle-cgf/egweblnk}
\usepackage{authblk}
\usepackage{sidecap}

\title[Convex Primitive Decomposition for Collision Detection]%
    {Convex Primitive Decomposition for Collision Detection}

\author{{\parbox{\textwidth}{\centering Julian Knodt\orcidlink{0000-0003-4461-2036} and Xifeng Gao\orcidlink{0000-0003-0829-707510}}}\newline\newline{\parbox{\textwidth}{\centering Lightspeed Studios, Bellevue, Washington, USA}}}

\begin{document}

\maketitle
\begin{abstract}
Creation of collision objects for 3D models is a time-consuming task, requiring modelers to manually place primitives such as bounding boxes, capsules, spheres, and other convex primitives to approximate complex meshes. While there has been work in automatic approximate convex decompositions of meshes using convex hulls, they are not practical for applications with tight performance budgets such as games due to slower collision detection and inability to manually modify the output while maintaining convexity as compared to manually placed primitives. Rather than convex decomposition with convex hulls, we devise an approach for bottom-up decomposition of an input mesh into convex primitives specifically for rigid body simulation inspired by quadric mesh simplification. This approach fits primitives to complex, real-world meshes that provide plausible simulation performance and are guaranteed to enclose the input surface. We test convex primitive decomposition on over 60 models from Sketchfab, showing the algorithm's effectiveness. On this dataset, convex primitive decomposition has lower one-way mean and median Hausdorff and Chamfer distance from the collider to the input compared to V-HACD and CoACD, with less than one-third of the complexity as measured by total bytes for each collider. On top of that, rigid-body simulation performance measured by wall-clock time is consistently improved across 24 tested models.
\end{abstract}

    

%\keywords{Shape Approximation, Convex Decomposition, Collision Detection, Primitive Abstraction}

\begin{figure*}
\centering
\setlength{\tabcolsep}{1pt}
\renewcommand{\arraystretch}{0.81}
\begin{tabular}{c c c c}
    Input & Ours & CoACD{\tiny \cite{coacd}} & V-HACD{\tiny \cite{vhacd}} \\
    \includegraphics[width=0.24\linewidth]{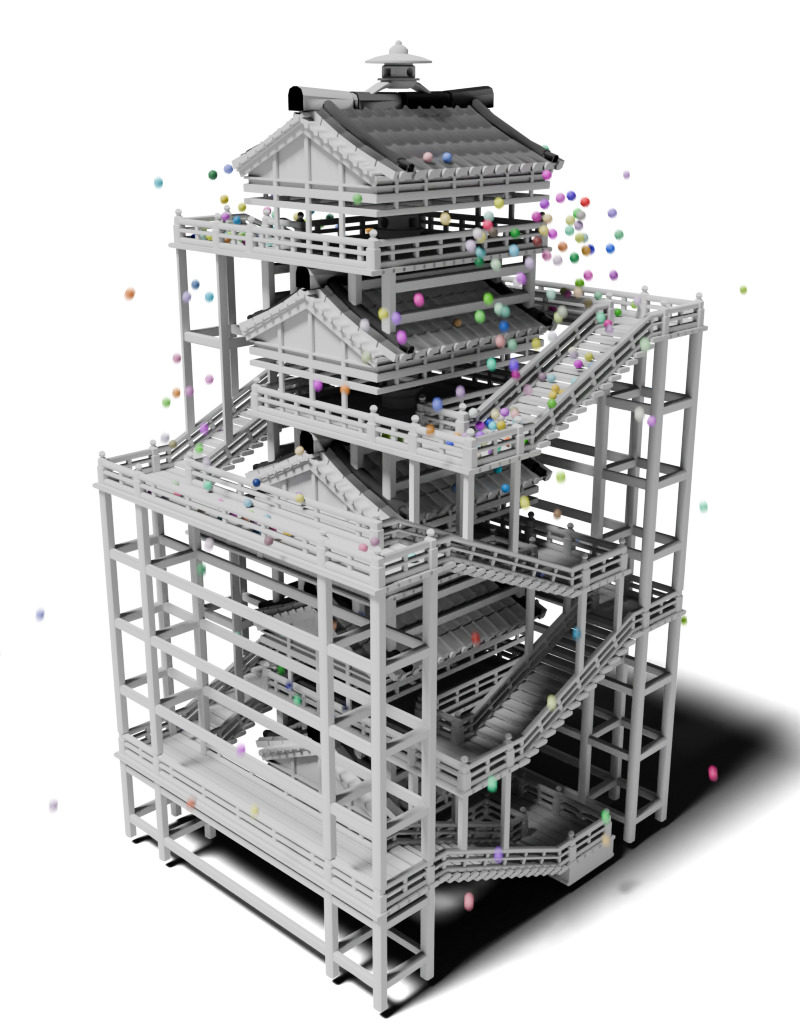}
    & \includegraphics[width=0.24\linewidth]{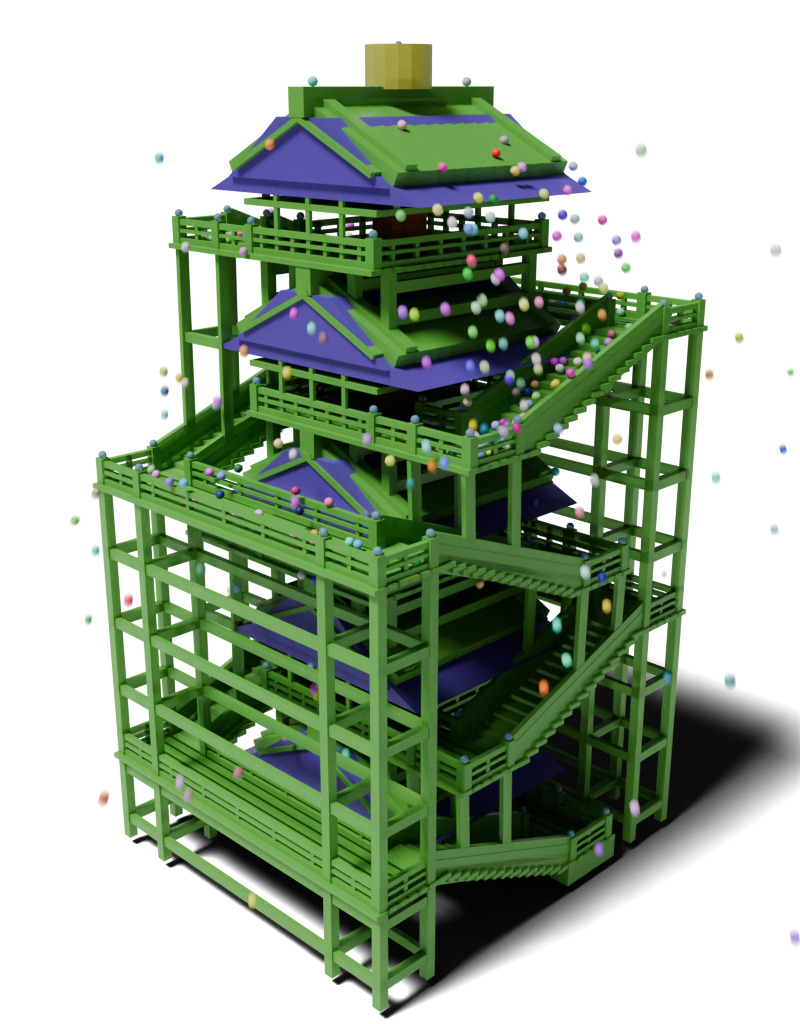}
    & \includegraphics[width=0.24\linewidth]{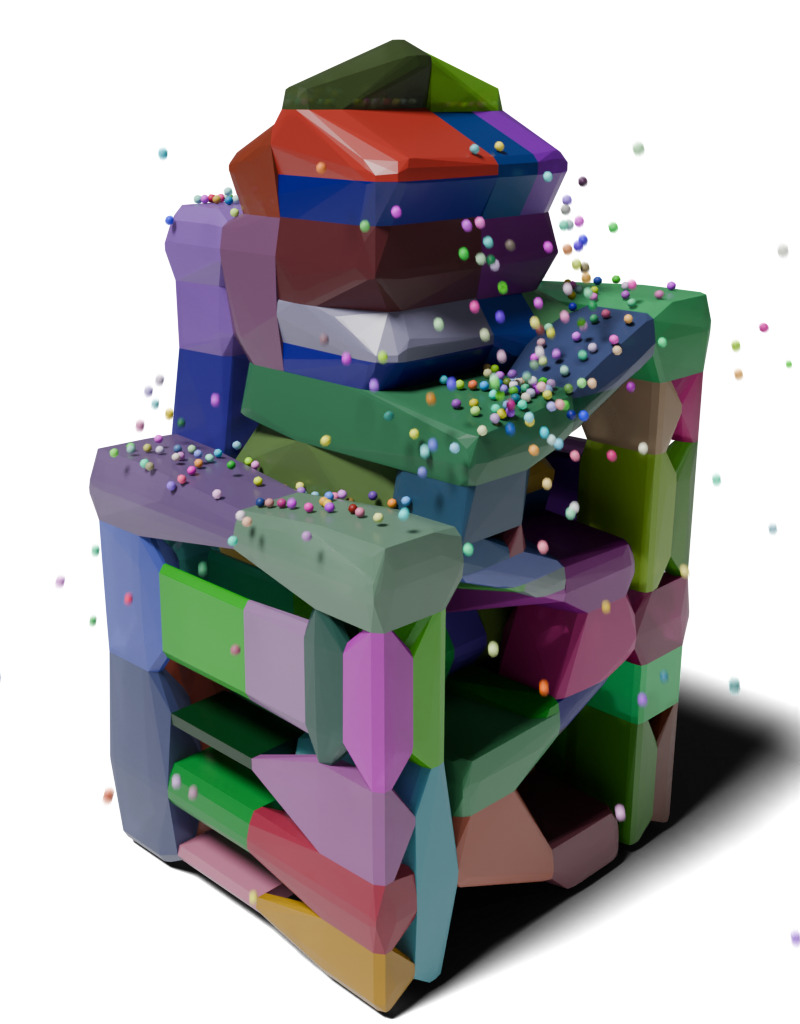}
    & \includegraphics[width=0.24\linewidth]{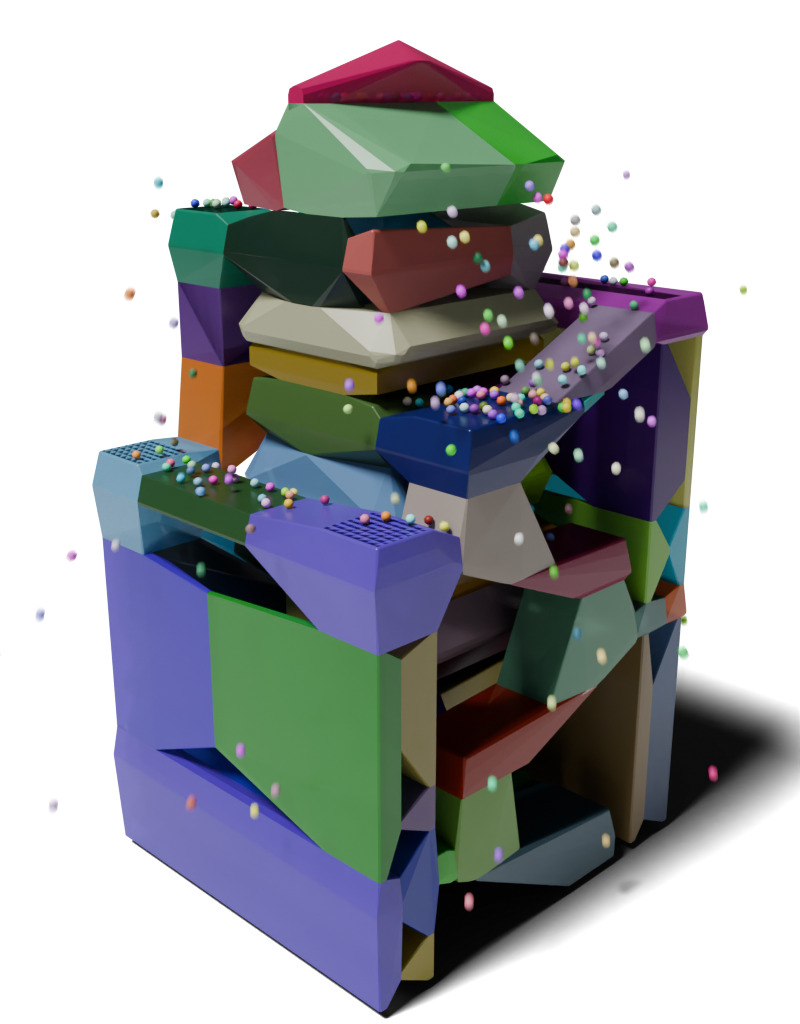} \\
    
    \frame{\includegraphics[width=0.24\linewidth]{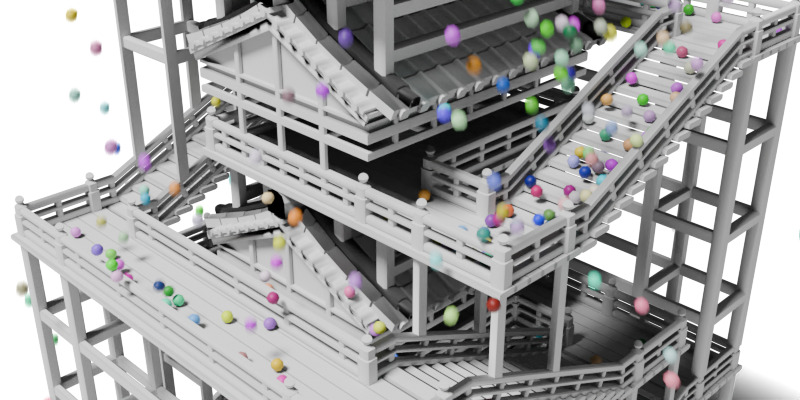}}
    & \frame{\includegraphics[width=0.24\linewidth]{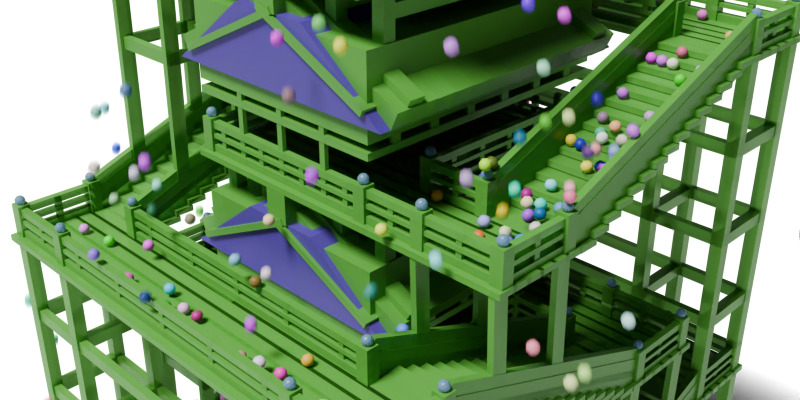}}
    & \frame{\includegraphics[width=0.24\linewidth]{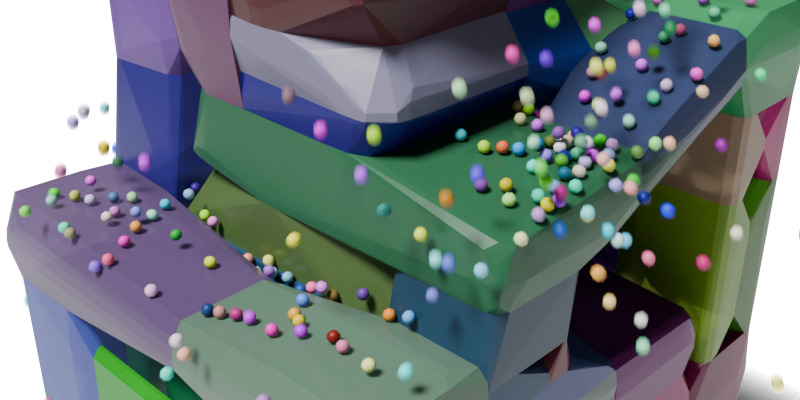}}
    & \frame{\includegraphics[width=0.24\linewidth]{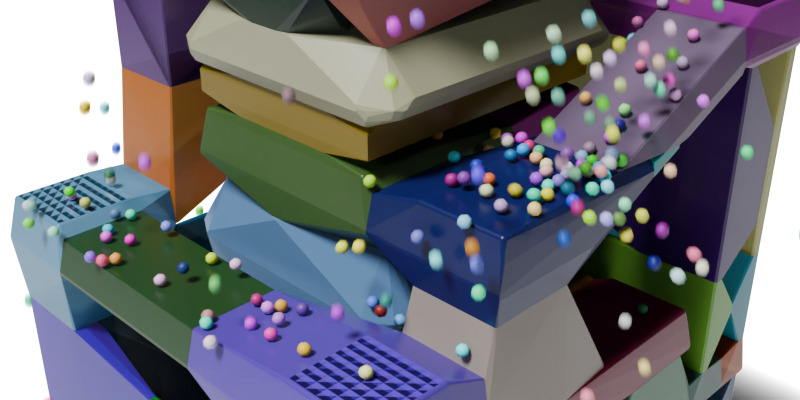}} \\
    
    $\frac{\text{Hausdorff/Chamfer New to Input}}{\lVert\text{Bounding Box Diag}\rVert_2}^\downarrow$ & $0.0384/\num{9.58e-3}$ & $0.0391/\num{9.42e-3}$ & $0.0507/\num{8.51e-3}$ \\
    |F| = 264328 & \scriptsize 1387 Boxes, 80 Spheres, 1 Cap, 1 Cyl, 11 Prism  &  107 Hulls (|F| = 11812) & 75 Hulls (|F| = 12408) \\
    \multicolumn{4}{c}{\includegraphics[width=0.8\linewidth]{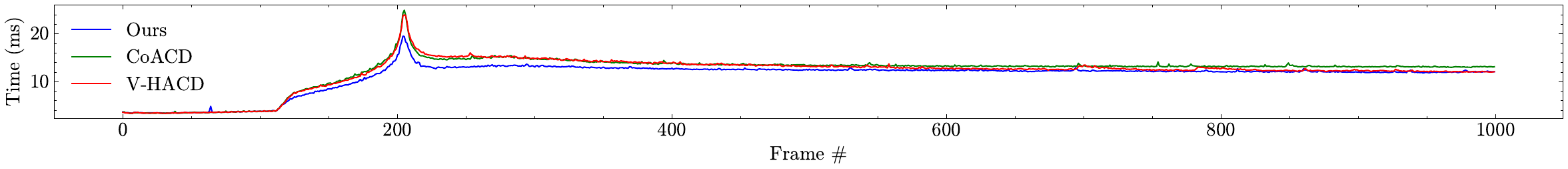}}
    \vspace{-0.8em} %\vspace{-1.5em}
\end{tabular}
\caption{\label{fig:teaser} Convex primitive decomposition on a complex non-manifold mesh with boundaries. Output primitives tightly adhere to the input, and is more efficient in simulation than convex hulls. To test performance, we drop spheres on each collider and measure frame times$^\downarrow$. Our approach has the closest simulation to the original mesh (i.e. balls in the stairwell), but with higher efficiency than CoACD or V-HACD shown in the plot, where x-axis indicates the frame number, y-axis indicates time taken. Green in our approach indicates a bounding box, yellow a cylinder, dark blue a trapezoidal prism, light blue a sphere, and red a capsule. For CoACD and V-HACD, colors are randomly assigned per convex component. Rendering artifacts on V-HACD are due to flipped faces. \ccby sin\_nass.}
%\Description{A watchtower stylized as a Shinto temple, decomposed into convex primitives, compared against CoACD and V-HACD. The watchtower has a number of floors and has a staircase wrapping around it.}
\vspace{-1.5em}
\end{figure*}

\section{Introduction}

Collision objects are an important component of 3D games, defining the interactions between the player and the world. The current flow for game developers and artists for constructing collider meshes can be slow and time-consuming, requiring manual placement of primitives such as capsules, boxes, and spheres, or construction of convex hulls and back-and-forth tuning of performance. Prior work in this area focuses on convex decomposition of input meshes~\cite{coacd, acd_polyhedra, vhacd,animated_decomp} into convex hulls, and there has not been much emphasis on fitting primitives such as boxes, capsules, and other convex primitives shapes to approximate an input mesh. On the other hand, support within physics engines such as PhysX~\cite{physx} for collision with primitives (boxes, capsules, spheres) is heavily optimized, and is considered faster than convex hulls. Because of this belief, artists will manually build colliders despite the effort required to do so, especially compared to the minimal effort required to use prior fully-automatic collider mesh generation approaches.

Previous approaches for automatic collision object construction focus mostly on convex hull decomposition~\cite{coacd, vhacd, animated_decomp, nav_approx_acd} for static and animated meshes. These approaches generate approximately convex components for an input mesh based on concavity metrics which measure how concave each component is. Prior work generates a small number of convex components which look good at first glance, but suffer from a few practical problems. First, the decomposition may hide complexity that slows simulation, as each convex component can have an arbitrary number of faces and vertices. Second, decomposition is difficult to control, sometimes producing poor approximations in concave regions, filling holes in empty space, and cutting up planar surfaces. Finally, the output primitives are not similar to artist created colliders, due to reliance on plane cutting, asymmetry of output shapes, and overcomplexity of each component. While this final reason may not appear a problem, it prevents artists from easily modifying colliders using tools such as Blender~\cite{Blender} or Unreal Engine~\cite{UE5}. For these reasons, we diverge from the prior approach of approximate convex decomposition and focus instead on fitting a subset of parametric convex primitives to represent an input mesh, which can be easily manipulated with existing tools and closely matches artist's manual workflow.

Our approach takes inspiration from Quadric Mesh Reduction~\cite{qem, qem_hoppe,qem_uv_texture} and Spherical Quadric Error Metrics~\cite{sqem, anim_sqem}, with each mesh face representing a primitive. These primitives are then greedily combined together bottom-up, producing a simplified representation. This simplified representation, consisting of a set of primitives that cover the surface of an input mesh, is similar to Oriented Bounding Box Trees~\cite{obbtree}, while allowing for more diversity in primitives.

Our approach is efficient and can compute an arbitrary number of primitives for meshes that have millions of faces. The meshes produced are suitable for accurate and efficient collision detection, which we verify using an off-the-shelf collision simulation.

We collect and test our approach on a dataset of models from Sketchfab~\cite{sketchfab}, showing our approach is efficient and robust in practice. We compare our approach to prior approximate convex decomposition algorithms V-HACD~\cite{vhacd} and CoACD~\cite{coacd} on distance from the input mesh, complexity of output collider, and performance in downstream collision detection simulation.

In summary, we hope to demonstrate that convex primitive decomposition serves to fill the gap between research in creating efficient colliders and actual artistic modeling of collider meshes for rigid body simulation. Our tests validate that our approach is suitable for collision detection and more closely adheres to the input mesh than prior work while at the same time has reduced complexity and better performance.
\section{Related Work}

\subsection{Quadric Mesh Reduction}

Our approach is loosely based on Quadric Mesh Reduction~\cite{qem, qem_uv_texture, qem_hoppe, probabilistic_quadrics}, which reduces the number of triangles through edge collapse~\cite{fast_and_memory_efficient_polygonal_simp, mesh_optimization} by representing vertices of triangles as linear operators of the form $Q_v = \Sigma pp^\top$, with $p\in\mathbb{R}^4$ defined as the equation of the plane of each face containing vertex $v\in\mathbb{R}^3$. These linear operators can be efficiently combined as $Q^* = Q_1 + Q_2$, and the optimal position of a combination of vertices can be computed by solving: \begin{equation}
\begin{bmatrix}
    Q_{00} & Q_{01} & Q_{02} & Q_{03} \\
    Q_{10} & Q_{11} & Q_{12} & Q_{13} \\
    Q_{20} & Q_{21} & Q_{22} & Q_{23} \\
    0 & 0 & 0 & 1
\end{bmatrix}
\begin{bmatrix}
\vect{v} \\ 1
\end{bmatrix}= \begin{bmatrix}
    0 \\ 0 \\ 0 \\ 1
\end{bmatrix}
\end{equation}
Because of the simplicity of combining $Q$ due to its linearity, this approach scales well as the size of the input mesh increases. Edges are merged greedily, based on the error $\vect{v}^\top (Q_{v_0} + Q_{v_1}) \vect{v}$, and are used to build progressively coarser representations of the input mesh, similar in spirit to ~\cite{qem, silhouette_clipping, sqem}.
Quadric mesh reduction is used in industry and research as a way to reduce the number of elements in complex meshes. It has also been extended to handle other attributes on meshes, including vertex colors, normals, and texture coordinates. Quadrics have been applied to many problems, including handling topology computations that rely on mass~\cite{QuadricTopology}, shape reconstruction~\cite{shape_recon_quadric}, spectral simplification~\cite{spectral_mesh_simp}, and vertex positions in dual contouring~\cite{dual_contouring}.

Quadrics have also been adapted to handle approximation of meshes as a linear interpolation of spheres~\cite{sqem, anim_sqem}. Instead of the standard quadric metric, a newly proposed quadric metric is based on the signed distance to a sphere: $d = \vect{n}^\top (\vect{p} - \vect{c}_\text{sphere}) - r_\text{sphere}$. From this formulation, an approach that solves for the center $\vect{c}_\text{sphere}$ and radius $r_\text{sphere}$ is derived, and elements can be merged together in a similar way as standard quadrics. Unlike standard quadrics though, the cost function is replaced with the sum of squared oriented distances of each sphere to the faces of the vertices that it subsumes. Spherical Quadrics (SQEM) lead to a coarse approximation of the input mesh as an interconnected series of spheres, which can then be used for animation. Because of its simplicity and similarity to our goal of abstracting an input with coarse shapes, we draw inspiration for our approach from Spherical Quadric Error Metrics. Note that SQEM is not comparable to our method in rigid body simulation, since the parameters of spheres are linearly interpolated along mesh faces and edges and thus it cannot be represented as discrete components for collision.

Our approach builds on the edge collapse operator, replacing the vertices usually merged together for mesh simplification with primitives for each face that are progressively merged together to form a larger primitive. Our formulation is also simpler than normal quadrics, as we only need to maintain a single matrix in $\mathbb{R}^{3\times 3}$.

A few mesh simplification approaches in theory could be applied to collision detection~\cite{bounding_proxies_for_shape_approximation, silhouette_clipping}, but in practice triangle meshes used directly in rigid body simulation often lead to objects clipping through faces and poor performance.

\subsection{Shape Abstraction}

Another approach to simplifying meshes is abstraction using simpler geometric primitives. Geometric primitives, often represented as parametric models such as signed distance functions, quadrics, or explicit primitives, are used in constructive solid geometry for computer aided design~\cite{realtime_csg} along with parametric curves~\cite{abstraction_of_shapes}. There are a number of different approaches to recovering shapes from input meshes such as random sampling (RANSAC)~\cite{efficient_ransac, generic_primitive_detection_in_point_clouds} and data-driven approaches \cite{cuboid_abstraction, superquadrics, marching_primitives, surface_edge_detect_point_cloud}. These approaches primarily work on simple inputs and cannot preserve high frequency details.
One work related to our approach is ~\cite{quadric_surface_extraction}, which fits quadrics to a surface with Lloyd's algorithm and a Euclidean distance metric. Their fitting also requires the computation of eigenvectors to fit quadrics, distinct from our approach which uses them for alignment only. A number of works stemmed from that work~\cite{variational_3d_shape_segmentation}. Another similar work is ~\cite{hierarchical_mesh_segmentation_based_on_fitting} which seeks to perform shape approximation by merging faces, similar to our approach. In contrast though, this prior work only shows 3 shapes with custom cost functions per shape (whereas this work shows 6 with the same cost function), and is not guaranteed to enclose the input shape. Furthermore, that work does not have a specific target application and tested on a much smaller dataset, whereas this work has a large emphasis on validation specifically for collision detection.

There are also many works using clustering for shape abstraction. For such approaches, there are a few design choices. First is choice of clustering algorithm, usually Lloyd's algorithm/iterative region growing~\cite{quadric_surface_extraction, variational_mesh_segmentation, d_charts, variational_3d_shape_segmentation, quasi_optimal_region_growing, variational_hierarchical_directed_bounding_box_construction}, a greedy approach which merges elements together bottom up (similar to our approach)~\cite{hierarchical_face_clustering_on_polygonal_surfaces, qem, simple_primitive_recognition_via_face_clustering, hierarchical_mesh_segmentation_based_on_fitting}, or splitting approaches ~\cite{split_merge_refine, hierarchical_mesh_segmentation_based_on_quadric}. Prior work may target different outputs, such as clustering faces into primitives represented by quadrics~\cite{quadric_surface_extraction, variational_mesh_segmentation, generic_primitive_detection_in_point_clouds}, a small non-general set of primitives~\cite{ellipsoid_decomposition_of_3d_models, simple_primitive_recognition_via_face_clustering, primitive_based_3d_segmentation_cad_models, hierarchical_mesh_segmentation_based_on_quadric, variational_3d_shape_segmentation}, oriented bounding boxes~\cite{obbtree, split_merge_refine, variational_hierarchical_directed_bounding_box_construction, variational_3d_shape_segmentation, variational_obb_tree}, and ellipsoids~\cite{variational_3d_shape_segmentation, ellipsoid_decomposition_of_3d_models}. Our approach is close to prior work that produces primitives, but our approach is more general as any parametric shape that satisfies a simple interface can be used. This includes, spheres, capsules, boxes, frustums, trapezoidal prisms and cylinders, which no previous approach can produce together. Furthermore, prior work relies on primitive specific features for fitting. Our work introduces the use of the eigendecomposition for the primitive's orientation, which can be used to fit other parameters using the enclosed points, with less reliance on primitive specific features. Our work also focuses on the specific problem of rigid body simulation. For a comprehensive analysis on prior work, see a recent survey on geometric primitive fitting~\cite{survey_of_simple_geometric_primitives}.

\subsection{Approximate Convex Decomposition}

Prior work on convex decomposition for collision detection has revolved around approximate convex decomposition~\cite{model_composition_from_interchangeable_components, vhacd, nav_approx_acd, acd_polyhedra, animated_decomp, shape_segmentation_by_approx_convex, hierarchical_convex_approximation_of_3d_shapes}. Approximate convex decomposition relies on splitting an input mesh into convex hulls, approximating the shape of the input mesh using a concavity metric to determine whether to a split a shape into subparts. Many of these approaches remesh or voxelize the input to make it manifold, then partition the manifold mesh top-down along cutting planes, tightly matching the input shape. What we found when investigating whether this approach is suitable for game development is that convex hulls are more complex than the count of hulls indicates, can occasionally be imprecise, and are not easy to modify within existing DCC tools or engines while maintaining convexity. This motivates our divergence from convex hull decomposition.

One component of prior work (and specifically CoACD) is their use of plane-cutting of hulls into separate components, relying on Monte Carlo search to find a good cutting plane. Due to the infinite number of cutting plane candidates, prior approaches cut on a limited set of axes, leading to excessive and poor cuts. We evade this problem entirely by performing bottom-up merging of primitives. This allows us to take into account local coordinate frames, ridding ourselves of suboptimal axis-aligned constraints.

We also note that our approach's error metric using volume differences is built on prior works such as ~\cite{animated_decomp} which uses volume to determine when to split a hull. One key difference is that prior work often requires watertight meshes to compute the volume of the input mesh. Our approach does not require a watertight input, since we use the volume of the already computed primitives rather than the mesh itself.
\section{Method}
Given an input mesh $\mathbb{M} = (V, F), \textbf{V$_i$} \in\mathrm{R}^3, F_i \subseteq V, |F_i| \geq 3$, with each face a polygon of three or more vertices, we output a set of primitives $P, |P| \leq N$  where $N\in\mathbb{Z}_+$ is a user-defined target positive number of primitives. In our implementation, $P$ can be a capped cylinder, a capsule, a sphere, an isosceles trapezoidal prism, a frustum, or an oriented bounding box. Our approach follows traditional quadric mesh reduction~\cite{qem} and constructs a linear operator, in our case a $3\times 3$ matrix, corresponding to each primitive. Linear operators support addition, $(f+g)(\vect{x}) = f(\vect{x}) + g(\vect{x})$, and scalar multiplication, $kf(\vect{x}) = f(k\vect{x}), k \in\mathbb{R}$. \vect{Bold text} indicates a vector in $\mathbb{R}^3$.

We outline our approach, starting from initializing linear operators (Sec. \ref{subsec:linear_operator}) for each face of the input mesh, corresponding to one primitive per face, and describe how to convert linear operators into primitives in Sec. \ref{subsec:primitive}. We then describe our greedy approach to combine topologically adjacent linear operators, using the input mesh's face adjacency as topology. Akin to QEM's edge-collapse, we use a minimal excess volume cost function to select primitives to merge (Sec. \ref{subsec:merge}), and terminate once there are either no more elements to simplify or a user-defined criteria is met (Sec. \ref{subsec:terminate}). We discuss the implementation details in Sec. \ref{sec:impl_details}, and the full algorithm for our approach is given in Alg.~\ref{alg:prim_mesh_reduction}.
\begin{algorithm}
\caption{Convex Primitive Decomposition\label{alg:prim_mesh_reduction}}
\begin{algorithmic}[1]
    \Statex \textbf{Input: } Mesh = $V\in\mathbb{R}^3, F_i\subseteq V$, \# Prims N$\in\mathbb{Z}_+$
    \Statex Volume Threshold $M\in\mathbb{R}_+ \stackrel{\text{default}}{=}\inf$
    \Statex \textbf{Output: } Primitives $P \text{ s.t. } |P| \leq $ N
    \Statex Preprocessing \Comment{Remove overlapped vertices}
    \State $\vect{n}_i = \text{normal}(F_i)$, $\vect{t}_i = \text{tangent}(F_i)$
    \State $Q_i$ = \text{area}($F_i$)($\vect{n}_i \vect{n}_i^\top + \epsilon\vect{t}_i \vect{t}_i^\top)$, $P_i$ = \text{Prim}($Q_i$), $\text{Vol}(f_i) = \text{Vol}(P_i)$
    \State pq : Priority Queue
    \For{adjacent faces $f_0, f_1$ $\in$ F} \Comment{Initialize Priority Queue}
        \State $P^* = \text{Prim}(f_0 + f_1)$
        \If{Vol($P^*$) - (Vol($f_0$) + Vol($f_1$)) > M}
            \textbf{continue}
        \EndIf
        \State pq.push(priority =  Vol($P^*$) - (Vol($f_0$) + Vol($f_1$)), ($P^*$, $f_0$, $f_1$) )
    \EndFor
    \While{!pq.empty() \textbf{and} $|P| > \text{N}$} \Comment{Greedily Merge}
        \State ($P^*$, $f_0$, $f_1$) = pq.pop()
        \State Vol($f_0$) = Vol($f_1$) = Vol($P^*$)
        \State $P_{f_0} = P_{f_1} = $ $P^*$ \Comment{Set both faces to new primitive}
        \State Update Costs of Adjacent Primitives
    \EndWhile
    \Statex Postprocessing \Comment{{\small Remove primitives enclosed by other primitives}} \\
    \Return Unique Primitives $P_i$
\end{algorithmic}
\end{algorithm}

\subsection{Linear Operator}\label{subsec:linear_operator}
Our linear operator corresponding to a primitive is a $3 \times 3$ area-weighted matrix per face of the input mesh, $Q$. We define $Q = \vect{n}\vect{n}^\top$, $\vect{n}^\top \vect{n} = 1, \vect{n}\in\mathbb{R}^3$. $\vect{n}\vect{n}^\top\in\mathbb{R}^{3\times3}$ is the outer product of the normal of a face with itself, which characterizes each face's plane. $Q$ is a subset of QEM's metric, $\text{QEM}(\vect{x}) = \vect{x}^\top(\vect{n}\vect{n}^\top)\vect{x} - 2 \vect{n}^\top(\vect{p}^\top\vect{x})\vect{x} + (\vect{p} - \vect{x})^\top(\vect{p} - \vect{x})$, with the linear and constant terms  dropped since they identify position. We do not need positions since we use the vertices within each primitive to compute positions directly. To derive a primitive from $Q$, we use $Q$'s eigendecomposition $Q = W\Lambda W^{-1}, \Lambda = \text{diag}(\begin{bmatrix}\lambda_2 & \lambda_1 & \lambda_0 \end{bmatrix}),
W = \begin{bmatrix} \vect{w}_2 & \vect{w}_1 & \vect{w}_0 \end{bmatrix}^\top, 
|\lambda_2| \geq |\lambda_1| \geq |\lambda_0|$: for a single face, $Q$'s largest eigenvector
$Q\vect{w}_2 = \lambda_2 \vect{w}_2$ is the face's area-weighted normal. When adding another face $Q_a + Q_b$, the largest eigenvector $\vect{w}_{2,a+b}$ corresponds to the largest shared direction of both faces' normals, and the second $\vect{w}_{1,a+b}$ will be the shared component orthogonal to the first. When adding faces, the largest eigenvector corresponds to the largest area-weighted direction that the faces are oriented towards, and the others will be orthogonal. The orthogonal basis of eigenvectors $\vect{w}_2,\vect{w}_1,\vect{w}_0$ defines an oriented bounding box which bounds the corresponding set of faces. We provide a visualization of $Q$ defined per input face and the eigendecomposition of output primitives in the inset figure. An analysis of this operators eigendecomposition and its relation to principal curvature is provided in ~\cite{quadric_based_polygonal_surface_simplification}.

\begin{SCfigure}
    \caption*{A visualization of the area weighted $Q$ of an input mesh (shown in blue), and our approach's output primitives. The output primitives from our approach are aligned with the faces they enclose. In the output, cylinders are yellow, oriented boxes are green, and eigenvectors of the input $Q$ for each face are red arrows.}
    \label{fig:q_vis}
    \setlength{\tabcolsep}{-0.1em}
    \begin{tabular}{m{0.23\linewidth} m{0.22\linewidth}} 
        \includegraphics[width=0.9\linewidth]{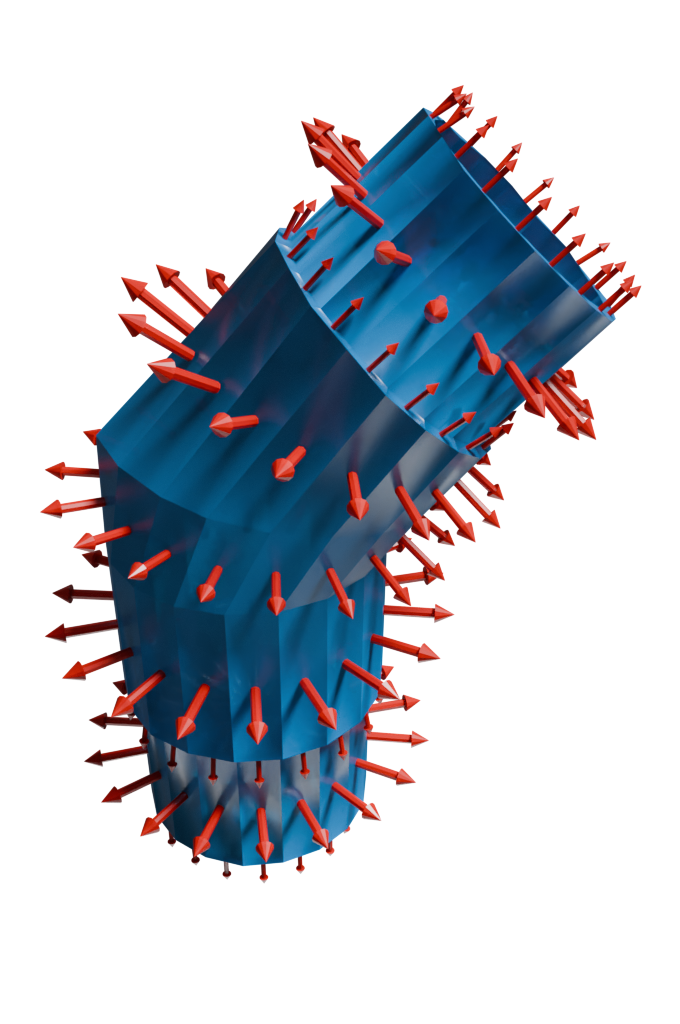} &
        \includegraphics[width=0.9\linewidth]{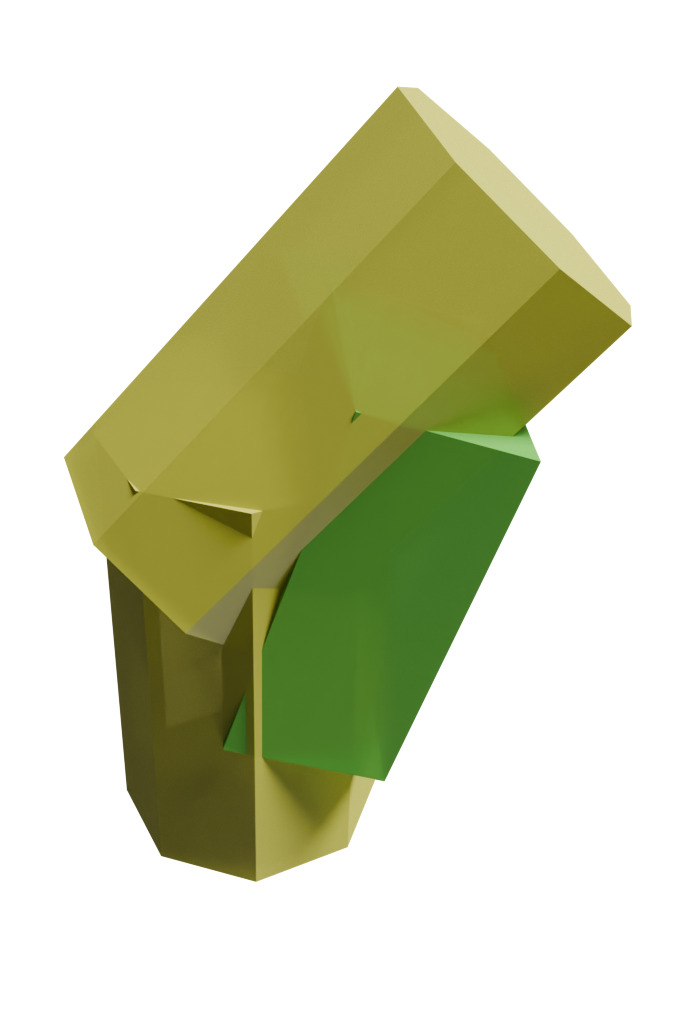} \\
        \multicolumn{1}{c}{Input} & \multicolumn{1}{c}{Output} \\
        \multicolumn{1}{c}{|F| = 216} & \multicolumn{1}{c}{1 Box, 2 Cyl.}
    \end{tabular}
\end{SCfigure}

We arrived upon this generalization while crafting an operator for cylinders. Instead of more parameters, the smallest eigenvalue's eigenvector closely aligns with the cylinder's axis. We then observed other eigenvectors align with reasonable bounding boxes for shapes and can serve as orientations. This operator identifies an orientation that aligns closely with the largest area-weighted tangent and normal of the enclosed faces. This operator is not globally optimal, but gives good approximations. An illustration of the eigenvectors that correspond to primitives is shown in the inset below. 

\setlength{\intextsep}{0pt}
\begin{wrapfigure}{o}{0.4\linewidth}
    \centering
    \renewcommand{\arraystretch}{0}
    \begin{tabular}{c c}
        \includegraphics[width=0.09\textwidth]{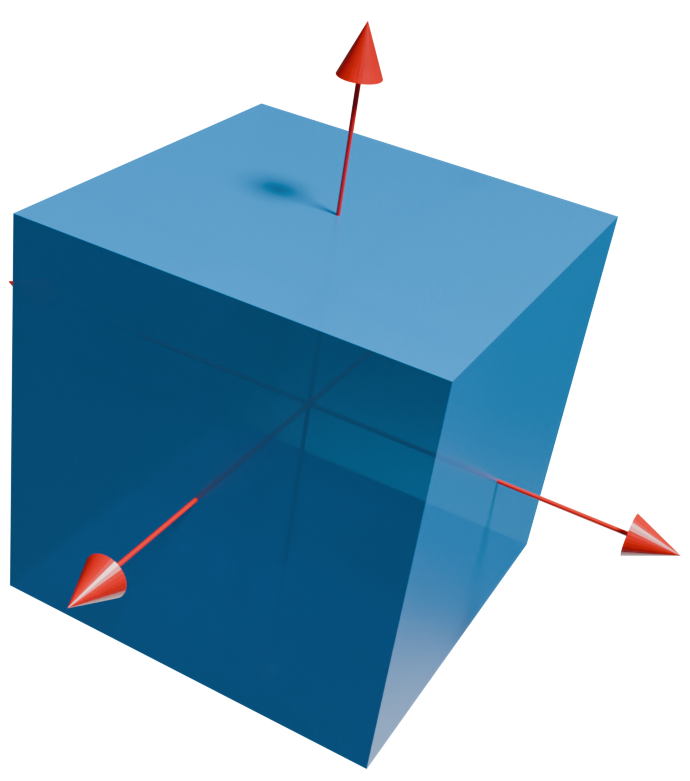} &
        \includegraphics[width=0.09\textwidth]{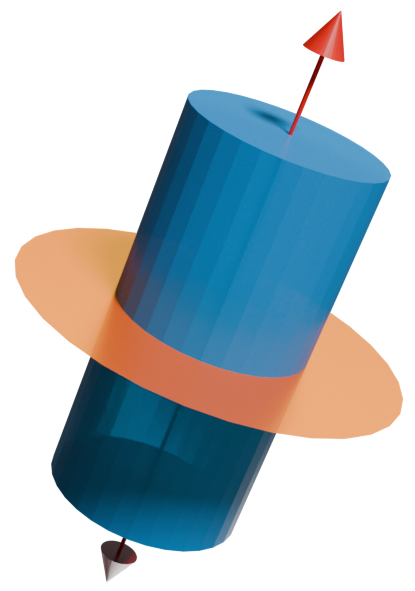}
        \\
        OBB/Prism &  Cyl/Cap \\
    \end{tabular}
    \label{fig:prim_eigen}
    %\Description{A blue cylinder with a red arrow through it indicating an eigenvector, and an orange circle indicating the space which the other eigenvectors lie in. The same for a cube, except now there is an arrow through each face.}
\end{wrapfigure}

For OBBs and isosceles trapezoidal prisms eigenvectors align with faces. For capsules and cylinders, one eigenvector is aligned with the cylinder's direction, and other eigenvectors will lie orthogonal to this axis.

% \subsection{Understanding Linear Operator Q}
% While the operator $Q$ was derived from observation, we make an informal argument why the eigenvectors of this operator approximate ``good'' axes for bounding shapes, and how a ``good'' set of axes can be used to fit a primitive.
% We first consider what a ``good'' approximation would be. From prior work on fitting minimal bounding boxes~\cite{o_rourke_minimal_enclosing}, for a convex polyhedron a tight bounding box will be flush with at least two adjacent faces. We then claim that a ``good'' enclosing primitive is nearly flush with the subset of faces that are nearly coplanar and have the largest area. 
% We then note that the eigenvectors of $Q$ are the principal components of the set of area-weighted normals for the mesh faces subsumed by  $Q$. Since the eigenvectors reflect the largest set of area weighted normals, the principal component of $Q$ exactly captures the largest set of nearly coplanar faces for any subsumed convex shape. Furthermore, as we align primitives with the eigenvectors, the constructed primitive can be constructed to be nearly flush with with these large faces, making it ``good'' enough.

\paragraph*{Face Normal Computation} For a triangle $\vect{v}_0, \vect{v}_1, \vect{v}_2$, the normal is given by the cross-product of the edges: $\vect{n}_\text{tri} = (\vect{v}_2 - \vect{v}_0) \times (\vect{v}_1 - \vect{v}_0)$. For quads (faces with four vertices) the face's normal is not well-defined if the quad is non-planar. To compute an approximate normal for both planar and non-planar quads, we define the normal of a quad $\vect{v}_0, \vect{v}_1, \vect{v}_2, \vect{v}_3$ as the cross-product of the diagonals:
$
    \vect{n}_\text{quad} = \frac{
        (\vect{v}_0 - \vect{v}_2)\times(\vect{v}_1 - \vect{v}_3)
        }{
    \lVert (\vect{v}_0 - \vect{v}_2)\times(\vect{v}_1 - \vect{v}_3) \rVert_2
    }
$.
For planar quads, this is exactly correct.
For polygons with more than four vertices, we create a triangle fan rooted at the 0th vertex with separate normals per triangle.

\subsection{Primitive Construction from Linear Operators}\label{subsec:primitive}

To convert $Q$ into a primitive, we first decompose it into its eigenvectors: $\vect{w}_0, \vect{w}_1, \vect{w}_2 \in\mathbb{R}^3, \vect{w}_i^\top \vect{w}_j = 0 \text{ iff } i \neq j, \vect{w}_i^\top \vect{w}_i = 1$, $Q\vect{w}_i = \lambda_i\vect{w}_i, |\lambda_0|\leq|\lambda_1|\leq|\lambda_2|$, implemented using the eigendecomposition described in ~\cite{SingularValue3x3}. With these eigenvectors as the axes, we fit all additional parameters for each primitive by expanding them to fully enclose the corresponding mesh faces.

\begin{figure}
    \centering
    \includegraphics[width=0.8\linewidth]{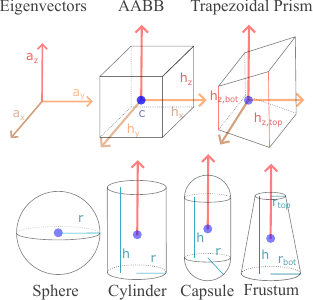}
    \caption{Primitives supported in our approach. First, axes are computed as the eigendecomposition of a quadric, $a_z$ corresponds to the minimum eigenvalue's eigenvector, $a_x$ the max. These axes are then used to compute an oriented bounding box (OBB). The OBB's center $c$ and the axes are then used to fit all other primitives.}
    \label{fig:primitives}
\end{figure}

Our approach supports arbitrary primitives which satisfy an interface. Primitives must have efficient volume computation, be constructible as zero-volume elements with a given orientation and position, and support strictly increasing their parameters to enclose an unordered set of points. For our implementation, we use oriented bounding boxes, spheres, capped cylinders, capsules, frustums, and isosceles trapezoidal prisms, shown in Fig.~\ref{fig:primitives}. We outline our implementation of the interface for each primitive below. For each primitive, the points it subsumes are denoted as $\vect{p}_i\in\mathbb{R}^{n\times3}$. We informally define \textit{subsuming} as when a primitive \textit{must} enclose a point. This does not include points primitives enclose but are not required to.

\paragraph*{Oriented Bounding Boxes}
Oriented bounding boxes are the foundation of all other primitives since they do not require an initial position and their position can be used for other primitives.
The bounds ($u_x,u_y,u_z,l_x,l_y,l_z\in\mathbb{R}$) of OBBs are computed per axis as $u_i = \max_{\vect{p}\in P}(\vect{v}_i^\top \vect{p})$ and $l_i =\min_{\vect{p}\in P}(\vect{v}_i^\top\vect{p}), i\in \{x , y, z\}$. The half-extent along each axis is $h_i = \max(\frac{1}{2}(u_i - l_i), \num{1e-3})$, where $\num{1e-3}$ is included to prevent degenerate volumes. We apply the same clamping to parameters for other shapes. The center of the OBB is $c_i = \frac{1}{2} (u_i + l_i)$, the volume is $8\Pi_{i=1}^3 h_i$, with a factor of 8 since $h_i$ are half-extents. OBBs are fit first, as other primitives initialize their position to the center of the OBB.

\paragraph*{Spheres}
Spheres are the simplest primitive, starting from a fixed center $\vect{c}$ (from the previously computed OBB), the radius is $r = \max_{\vect{p}\in P}(\lVert \vect{c} - \vect{p}\rVert_2)$. The volume of a sphere is $\frac{4}{3}\pi r^3$. Spheres are the cheapest primitive for collision detection, and are used despite their coarse shape.

\paragraph*{Capped Cylinders}
For capped cylinders with a fixed axis $\vect{a}, \vect{a}^\top\vect{a} = 1$ and a point on the axis, $\vect{p}_\text{cyl}$, we derive the height and radius:
\begin{align}
\text{r} &= \max\limits_{\vect{p}\in P}(\lVert(\mathrm{I} - \vect{a}\vect{a}^\top) (\vect{p} - \vect{p}_\text{cyl}) \rVert_2) \nonumber\\
\text{h} &= \max\limits_{\vect{p}\in P}(\vect{a}^\top (\vect{p} - \vect{p}_\text{cyl})) - \min\limits_{\vect{p}\in P}(\vect{a}^\top (\vect{p} - \vect{p}_\text{cyl}))
\end{align}
The cylinder's volume is $\pi r^2 h$. We compute 1 cylinder per axis, and choose the cylinder with minimal cost.

\paragraph*{Capsules}
Capsules are similar to cylinders, with a fixed axis $\vect{a} \text{ such that } \vect{a}^\top\vect{a} = 1$ and point on $\vect{a}$, $\vect{p}_\text{cap}$, the radius computation is the same as a capped cylinder. There is a slight difference for the height, taking into consideration that the ends of the capsules are spheres, and the equation for the height is adjusted as follows:
\begin{align}
    r(\vect{p}) &= \lVert(\mathrm{I} - \vect{a}\vect{a}^\top)(\vect{p} - \vect{p}_\text{cap}) \rVert_2 \\\nonumber
    \text{h}(\vect{p}) &= \vect{a}^\top (\vect{p} - \vect{p}_\text{cap}) - \sqrt{r^2 - r(\vect{p})^2} \\\nonumber
    \text{height} &= \max\limits_{\vect{p}\in P}(\text{h}(\vect{p})) - \min\limits_{\vect{p}\in P}(\text{h}(\vect{p})) \\\nonumber
\end{align}
Capsules are supported in downstream physics applications due to their ease of computing distance, which is why we include them. Similar to cylinders, we compute one capsule per axis.

\paragraph*{Frustum}
To capture conical structures, we also implement frustums. Frustums are initialized from the axis of the minimum cost cylinder, with a position on the base $\vect{p}$, and axis $\vect{a}$. The height is set identically to cylinders. The radius of the top and bottom faces are set using the algorithm shown in. Alg.~\ref{alg:frustum}. The volume of the output frustum is given by $\frac{\pi h}{3}(r_\text{top}^2 + r_\text{top} r_\text{bot} + r_\text{bot}^2)$. 

\paragraph*{Isosceles Trapezoidal Prisms}
Finally, we implement isosceles trapezoidal prisms. No prior work uses explicit trapezoidal prisms, but we found buildings often have gable roofs that resemble a triangular prism and are poorly modeled by other primitives. We found isosceles triangular prisms to be numerically unstable because of the singularity (zero-width side) for subsuming sets of points. Instead, isosceles trapezoidal prisms can closely approximate a triangular prism and are numerically stable.

Given the orthogonal basis $\vect{a}_x, \vect{a}_y, \vect{a}_z$ and center $\vect{c}$, we check all 6 possible orderings of axes and compute the half-extents $(h_x, h_y, h_{zt}, h_{zb}\in\mathbb{R})$, where $h_{zt}$, $h_{zb}$ are along the axis $\vect{a}_z$ on $+\vect{a}_y, -\vect{a}_y$ respectively. $h_x$ and $h_y$ are computed identically to OBBs as $\max(\vect{a}_x^\top(\vect{p}-\vect{c})) \text{ and } \max(\vect{a}_y^\top(\vect{p}-\vect{c}))$.
% To compute $h_{zt}, h_{zb}$ we apply the following equations on each point $\vect{p}$, and update running estimates for $h_{zt}, h_{zb}$:
% \begin{align}\label{eq:iso_trap_eqs}
%     y(\vect{p}) &= \frac{\vect{a}_y^\top(\vect{p}-\vect{c}) + 1}{2h_y}, z(\vect{p}) = |\vect{a}_z^\top (\vect{p}-\vect{c})|, h_{zt}^0 = h_{bt}^0 = 0 \\\nonumber
%     h_{zt}(z, y, h_{zb}) &= \frac{|z| - h_{zb}(1 - y)}{y}, h_{zb}(z, y, h_{zt}) = \frac{|z| - h_{zt} y}{1 - y} \\\nonumber
% \end{align}
% We then guarantee all points are contained within the prism by fixing the larger half-extent and computing the optimal opposite half-extent, and then doing the same for the other half-extent.
Pseudocode for computing $h_{zt}, h_{zb}$ is given in the Appendix, Alg.~\ref{alg:isotrap}. The volume of the resulting trapezoidal prism is $4h_x h_y (h_{zt} + h_{zb})$, with the factor of 4 as $h_\bullet$ are half-extents.

Our construction of isosceles trapezoidal prisms is fast because it uses an unordered stream of points without allocation, but is not optimal. We chose this tradeoff as it is recomputed every collapse.

An implementation can be extended with additional shapes depending on what is accepted in the downstream application. Our approach does not require that primitives be optimal, since only relative costs between primitives matters. Furthermore, for specific kinds of shapes it is possible that more optimal algorithms can be used, such as exact OBBs and spheres~\cite{o_rourke_minimal_enclosing}, but for the other primitives exact algorithms are unknown. Since only a few primitives have exact algorithm our approach uses a more general fit.

\paragraph*{Face-Based Mesh Reduction}
\cite{qem} and \cite{sqem} rely on merging vertices to approximate positions on the input mesh, and the surface is defined by the edges and faces connecting these vertices, but our approach uses discrete primitives to represent the input's surface. Allowing each vertex to be covered by only one primitive does not cover the surface, leaving holes in the output. For example, consider the output primitives if no edges are contracted: the output would be small primitives around each vertex, a poor approximation of the input shape. To remedy this, primitives subsume \textit{faces} of the mesh, similar to \cite{hierarchical_face_clustering_on_polygonal_surfaces}. Each primitive covers all vertices of each face it subsumes, allowing vertices to be covered by multiple primitives and the union of all primitives covers the mesh's surface. We show the poor output of using one primitive per vertex in Sec.~\ref{sec:ablate-vertex-face-merging}, Fig.~\ref{fig:ablate-vertex-merging}.

\subsection{Optimal Primitive Selection}
\label{subsec:merge}
Given possible primitives that enclose a point set, we must decide which primitive is best. We measure best in two ways: the geometric tightness of the primitive, as measured by the additional volume introduced when merging two shapes and an abstract cost function for downstream applications, such as the cost of collision detection. For our implementation, we select parameters that balance geometric similarity and support in physics engines.

\paragraph*{Collapse Cost Function} Our end-goal is to approximate the input mesh as closely as possible for collisions; we want collision primitives to be as tight-fitting to the input surface as possible. To measure this, we minimize the excess volume introduced by each primitive merge. This leads to the following cost function for merging two primitives $p_0, p_1$:
\begin{equation}\label{eq:exact_cost}
    C(p_0, p_1) = V(\text{merge}(p_0, p_1)) - (V(p_0) + V(p_1))
\end{equation}
where $V(p)$ is the volume defined by the primitive $p$. This cost function penalizes \textit{excess} volume introduced by merging two primitives, and promotes removing primitives which overlap. When this cost-function is 0, there is no penalty to merge two primitives together, and when it is negative it will reduce the total volume of the shape. Usually, this cost-function will be positive, indicating an increase in volume due to merging. Like QEM, we collect all edges between mesh elements in one priority queue, and iteratively take the minimum cost contraction. For each mesh element, we store the volume of its primitive and update the volume after each collapse.

\paragraph*{Overlapping Primitives} Eq.~\ref{eq:exact_cost} does not account for double-counted volume of intersecting primitives, which would instead lead to the following cost function:
\begin{equation}\label{eq:cost_with_isect}
    C(p_0, p_1) = V(\text{merge}(p_0, p_1)) - (V(p_0) + V(p_1) - V(p_0 \cap p_1))
\end{equation}
This primary reason we do not include the volume of the intersection is because such a computation is expensive, especially if computed exactly using mesh booleans, and we experimentally find little quality increase when adding in this term. Even an approximate approach, such as rejection sampling, incurs a high penalty. In practice, primitives do not overlap much, especially for clean meshes designed for games. Despite Eq.~\ref{eq:exact_cost}'s simplicity, it performs well in practice, and we leave exploration of alternatives to future work. We ablate Eq.~\ref{eq:exact_cost} against Eq.~\ref{eq:cost_with_isect} using rejection sampling in Sec.~\ref{sec:ablate-isect}, Fig.~\ref{fig:ablate-isect}, and find minuscule improvement but high computational cost, at least 30$\times$ the wall-clock time.

\paragraph*{Handling Different Primitive Costs.} In downstream applications, geometric fit is not the only consideration, as primitive variants incur different costs. For example, physics engines have primitives for boxes, capsules, and spheres giving them better performance. When merging primitives based only on volume though, prisms often have lower volume than OBBs, a subset of trapezoidal prisms with $h_{zt} = h_{zb}$, which may lead to lower simulation performance. To incorporate preferences over shape variants, we combine our cost function with a user-provided weighting: $V'(p) = k(p) V(p)$. $k(p)$ depends on the type of primitive, and can be set per target application. We use a weight of 1.05 for cylinders, 1.4 for trapezoidal prisms, 1.0 for capsules, spheres and bounding boxes, and 2.1 for frustums. Our weighting heavily prefers bounding boxes, capsules and spheres due to their support in physics engines. We penalize cylinders, as when they have small radius and large height there is little visible difference between cylinders and capsules, but capsules are more performant. Depending on use case, this weighting can be tuned, but these values strike a good balance between performance and geometric accuracy. We ablate our choice of weights as compared to a uniform and alternative weighting in Sec.~\ref{sec:ablate-primitive-costs}, Fig.~\ref{fig:ablate-primitive-cost}.

The choice of primitives in downstream applications is dependent on use case. For example, in a platformer or first-person shooter which requires precise control it may be better to use expensive colliders to keep accuracy. For cases such as a puzzle game, it may be fine to use coarser boxes. We believe that currently there is no one correct answer for which primitives to use in all cases, and that consideration must be taken case by case.

\paragraph*{Termination}
\label{subsec:terminate}
Termination happens when there are no more collapsible edges or the target number of primitives is reached. Since our approach always decreases the number of primitives, our approach is guaranteed to terminate. After termination, we output parameters for all primitives or quantize primitives to a mesh.

Like QEM~\cite{qem}, our approach relies on the user providing the target number of primitives, which affects the ability to represent the input mesh.

\paragraph*{Excess Volume Thresholding}\label{sec:excess_volume_thresh}
Manually tuning the number of primitives gives good results, but may provide insufficient control over order of merges. Furthermore, some merges introduce more volume than merging two disconnected components. To allow better control of order, we add an optional user-defined \textit{excess volume threshold}, relative to the volume of the axis-aligned bounding box of the input. If combining two primitives increases the volume above this threshold, that merge is prevented. This can be used to prune undesirable merges with less manual tuning.

\subsection{Implementation Details}\label{sec:impl_details}
\paragraph*{Removing Redundant Primitives}
After termination, we observe that sometimes the points subsumed by one primitive are fully enclosed by a larger primitive, and there is no point to maintain nested primitives since they incur computation without functionality. These primitives can be culled after our approach, by checking if all points subsumed for each primitive are entirely within another primitive. We perform a concurrent pairwise check after simplification and cull internal primitives, with negligible impact to the total execution cost.

\paragraph*{Coplanar Vertices}\label{sec:coplanar-vertices}
For coplanar faces, the quadric may have degenerate eigenvectors, a problem inherited from the original QEM.
To fix this, we add a quadric in the tangent space. For quads, we define a tangent basis that follows the directions of the quad. To handle general quads including non-planar ones, we use the following as the tangent direction $\vect{t}$ given vertices $\vect{v}_0, \vect{v}_1, \vect{v}_2, \vect{v}_3$:
$\vect{t} = \vect{v}_0 - \vect{v}_2 + \vect{v}_1 - \vect{v}_3$.
This is the halfway vector between the quad's two diagonals, which follow the direction of the quad's edges if it is regular and planar, and gives a reasonable value otherwise. Quads are specially handled due to their prevalence in game assets.

We treat triangles $t$ with counterclockwise edges $\vect{e}_0, \vect{e}_1,\vect{e}_2 \in \mathbb{R}^3, \lVert \vect{e}_0 \rVert_2 < \lVert \vect{e}_1 \rVert_2 < \lVert \vect{e}_2 \rVert_2$ as half of a regular quad, where $\vect{e}_2$ corresponds to one of the quads diagonals, by computing the equivalent of the above equation using the triangle's edges: $\vect{t}_0 = \frac{1}{2} (\vect{e}_0 - \vect{e}_1 + \vect{e}_2)$.
We flip the sign of the tangent based on the orientation of the vertex $\vect{v}\notin e_2$: $\vect{t} = (-\vect{t}_0) \text{ if } (\vect{v} - \vect{v}_{\in e_2})\cdot(\vect{e}_2\times \vect{n}) < 0 \text{ else } \vect{t}_0$,
where $n$ is the normal of the corresponding face and $\vect{v}_{\in e_2}$ is one of the vertices of $e_2$. This is designed for quad meshes which were triangulated, such as when a mesh is imported into UE~\cite{UE5}.

We add the weighted outer product of $\vect{t}$ to $Q$, to help improve stability: $Q' = Q + \epsilon  \vect{t}\vect{t}^\top$. This reduces error in ambiguous cases, but may reduce adherence to the input in other cases. We show an ablation of one example where it improves the output quality in Sec.~\ref{sec:ablate-tangent-quadric}, Fig.~\ref{fig:ablate-tangent-quadric}. We decide per mesh whether to include this factor.

\paragraph*{Vertex Deduplication}
A common pitfall we find when running our approach on uncleaned meshes is that vertices may be overlapped. This can happen due to kitbashing or modeling of parts which are later combined. We show deduplicating such vertices can change the output result in Sec.~\ref{sec:merge-vertices}, Fig.~\ref{fig:ablate-vertex-merging}. Merging these duplicate vertices by distance is similar to non-edge contractions in ~\cite{qem} (also known as virtual edge collapses). A recent preprint~\cite{simplifying_wild_tris} explored virtual edge collapses between distinct components, and we leave exploration of alternative virtual edge collapses to future work.

\paragraph*{Pairwise Component Merging}
After performing reduction based on the connectivity of primitives, there may be no remaining edges even if the target number of primitives is not met due to disconnected components. To meet the target number of primitives, edges are added to create a fully-connected graph and decimation is continued. Theoretically, this is costly since the number of edges is $\frac{1}{2} |C|(|C| - 1) \approx O(C^2)$, where $|C|$ is the number of components in the input. In practice, we observe the number of components for many meshes is small, and all pairs can be computed quickly. In cases where it is intractable, edges can be culled based on excess volume thresholding from Sec.~\ref{sec:excess_volume_thresh}.

We find if the input mesh has separate components, this may help our approach, as separate components often delineate distinct convex objects which should not be merged, and note that constraining collapses to existing topology has not hindered prior work \cite{qem_hoppe, sqem, silhouette_clipping, hierarchical_face_clustering_on_polygonal_surfaces}. Vertex deduplication and pairwise merges also mitigate effects of input topology.

\paragraph*{Collapsible Primitive Data Structure}
Triangle mesh reduction requires bookkeeping for each faces' indices, but our implementation does not need to retain faces. Instead, we must manage which faces are subsumed by which primitive, for which we use a forest of cyclic linked-lists. We initialize one list per primitive, with each primitive subsuming a single face. When merging two primitives, we select an arbitrary node from each list, and swap pointers to their next element, merging the two lists together. This allows for $O(1)$ merging with efficient iteration of groups at the cost of pointer chasing. We also use Disjoint Set Unions~\cite{UnionFind} for quickly identifying which faces have been merged together, allowing for fast pruning of deleted edges in the priority queue.

\section{Results}

To evaluate our approach, we focus on two primary metrics: similarity of simulation of the new mesh to the original (\textbf{correctness}), and \textbf{efficiency} measured first-and-foremost by timing in simulation, then secondarily the complexity of colliders measured in bytes used, and the number and kind of each component. For correctness, our primary metric is qualitative appearance in simulation. If the simulation plausibly resembles the original, our approach is good enough for downstream usage, since there are no good measures for similarity of simulation. We also measure one-directional distance from points on the surface of the new mesh to the original, which has been used in prior work such as ~\cite{animated_decomp}, but we note that other prior works have ignored geometric distance~\cite{coacd, nav_approx_acd}. We choose to measure the one-directional distance from the new collider mesh to the original, as it penalizes falsely filling in holes and concave regions.

To measure rigid body simulation efficiency, we test collision performance directly in simulation by dropping 5000 spheres on each mesh and measure the frame duration in the first 1000 frames of simulation. Frame duration measurements reflect two things: different outcomes of collision detection and different performance of colliders. We include our rigid body simulator in the supplemental as an executable, with convex decompositions from our approach. We do not use frame-rate (frames per second), because it changes non-linearly with performance, and frame time is easier to reason about when budgeting time while containing equivalent information to frame-rate.

We test our approach on meshes from Sketchfab, which have a mix of triangle, quad, and polygonal faces, and some other data shown in the appendix. We test on props, buildings, characters, plants, and levels/environments. Unlike prior work, our approach handles non-manifold, non-watertight meshes directly without preprocessing. Our focus is primarily on meshes for games which have clean topology and a mix of tris and quads, but we also test on a few noisy scanned models. For each model, we manually decide how many primitives to keep.

\subsection{Efficiency of Collision Objects}

To demonstrate the usability of our approach in rigid body simulation, our primary metric is in-simulation wall-time benchmarking to validate collider performance and correctness. We caveat that performance of any simulation is heavily dependent on hardware and physics engine, but trends should be similar across systems. To determine if a collision is similar to the input, we manually inspect the behavior, similar to prior work on simulation.
\paragraph*{Simulation details} We run all methods on a 8-core AMD Ryzen 7800X3D processor with all processing done on the CPU, using \href{https://rapier.rs/}{\color{blue} Rapier}. While most physics engines support boxes, capsules, and spheres, support for cylinders is mixed, and no engines known to the authors support trapezoidal prisms or frustums. Rapier supports cylinders, otherwise they can be discretized and represented as convex hulls, which is the approach we took for trapezoidal prisms and frustums.

\begin{figure}[ht]
    \centering
    \setlength{\tabcolsep}{1pt}
    \begin{tabular}{c c c}
        Input & Ours & CoACD \\
        \includegraphics[width=0.3\linewidth]{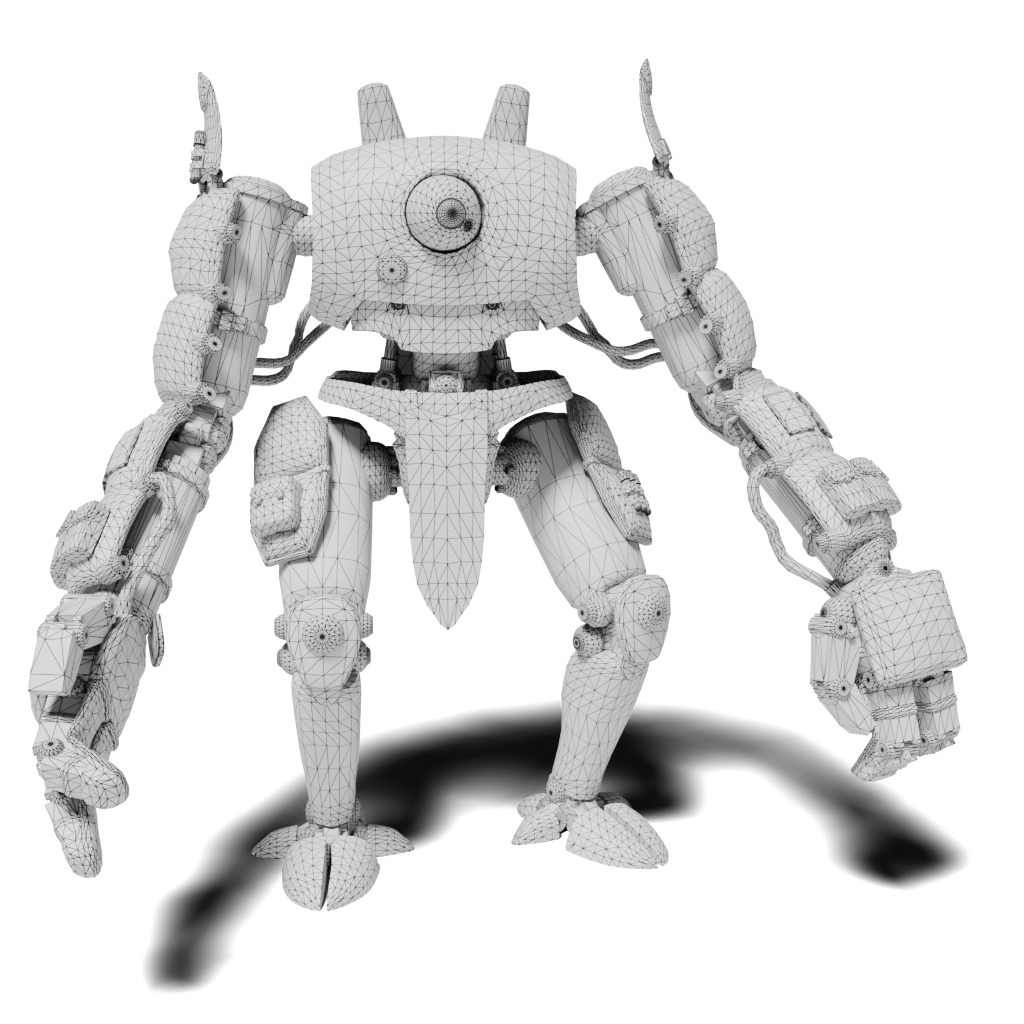} & 
        \includegraphics[width=0.3\linewidth]{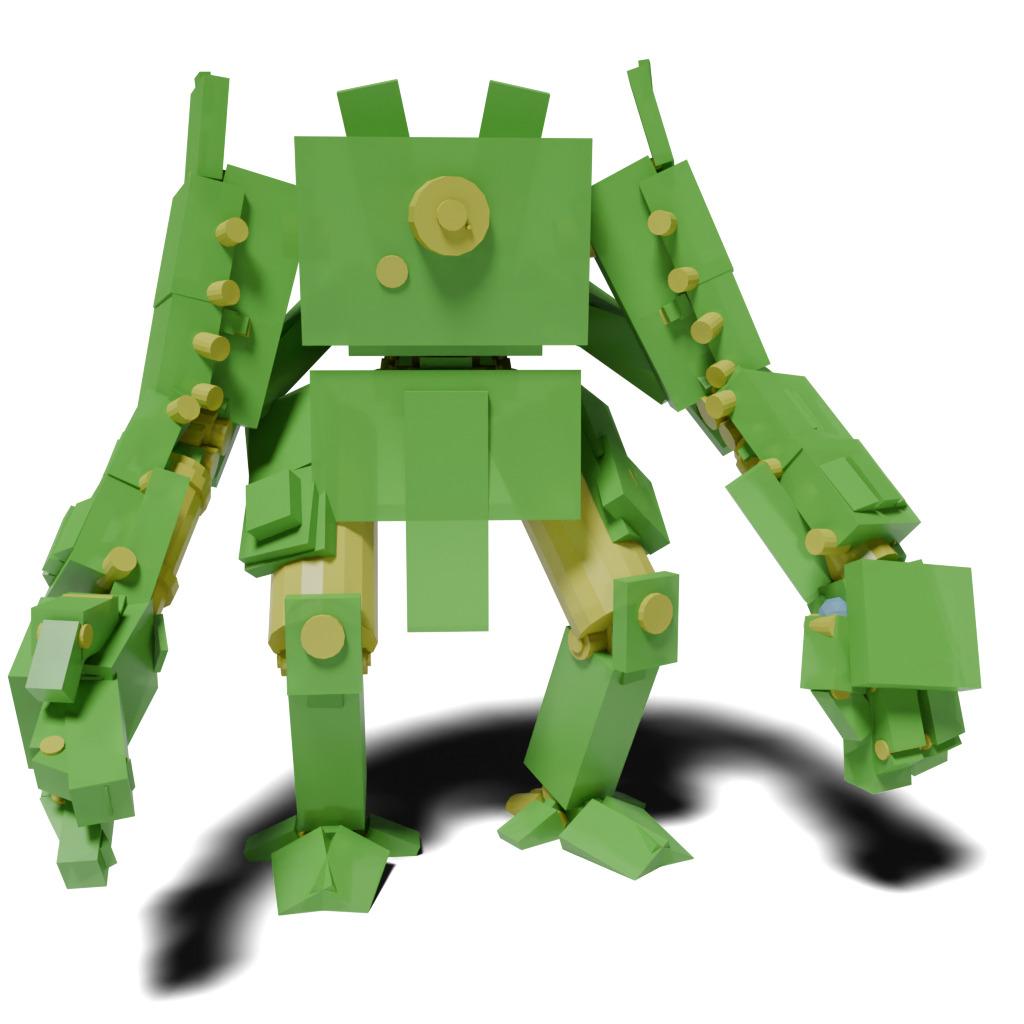} & 
        \includegraphics[width=0.3\linewidth]{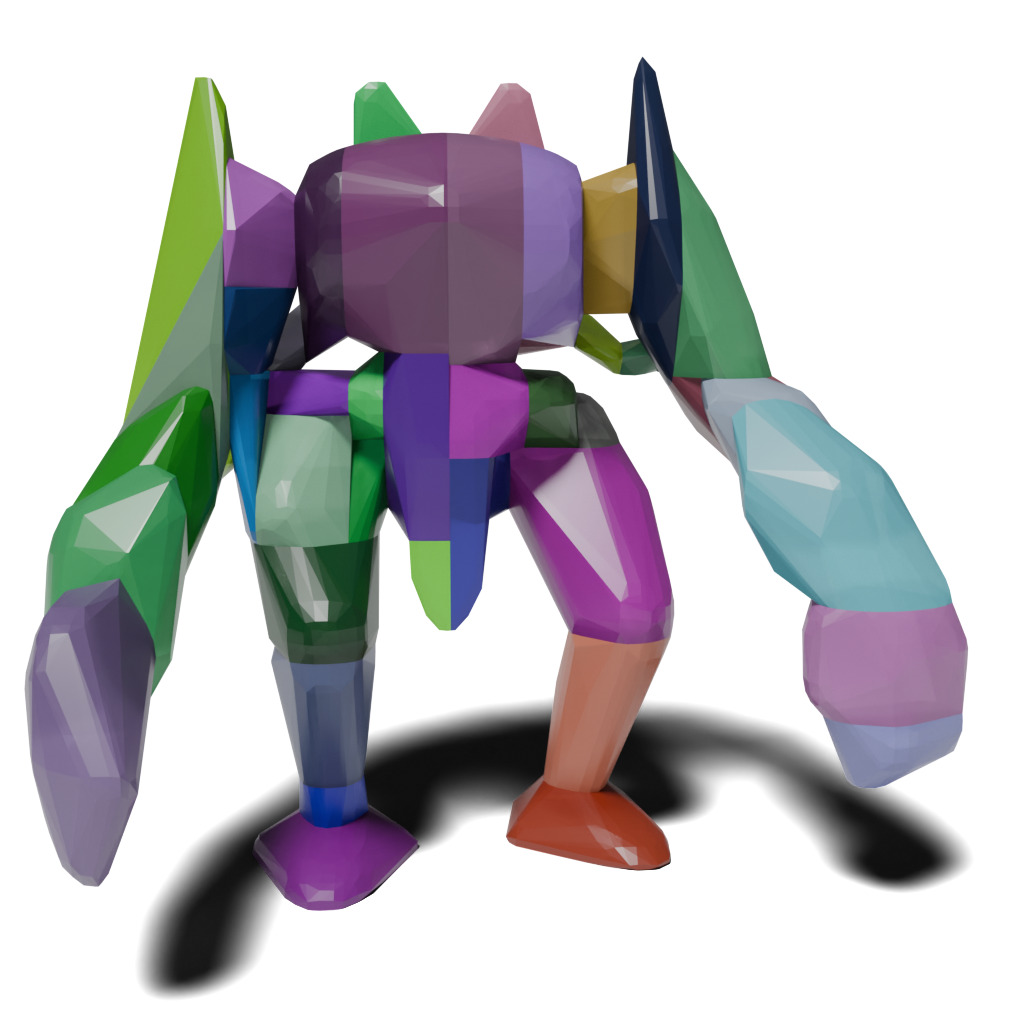} \\
        %& \includegraphics[width=0.24\linewidth]{assets/simulation_vera} \\
        |F| = 47790 & 298 Prim. & \footnotesize 46 Hulls, |F| = 9774 \\ %& (Ours shown) \\
        \multicolumn{3}{c}{\tiny 138 Boxes, 146 Cyl., 12 Sph.} \\

        {\footnotesize $\frac{\text{Haus/Cham New to Input}}{\lVert\text{Bounding Box Diag}\rVert_2}^\downarrow$}
        & {\footnotesize $0.0112/\num{1.15e-3}$} & {\footnotesize $0.0340/\num{7.10e-3}$} \\
        \multicolumn{3}{c}{\includegraphics[width=0.9\linewidth]{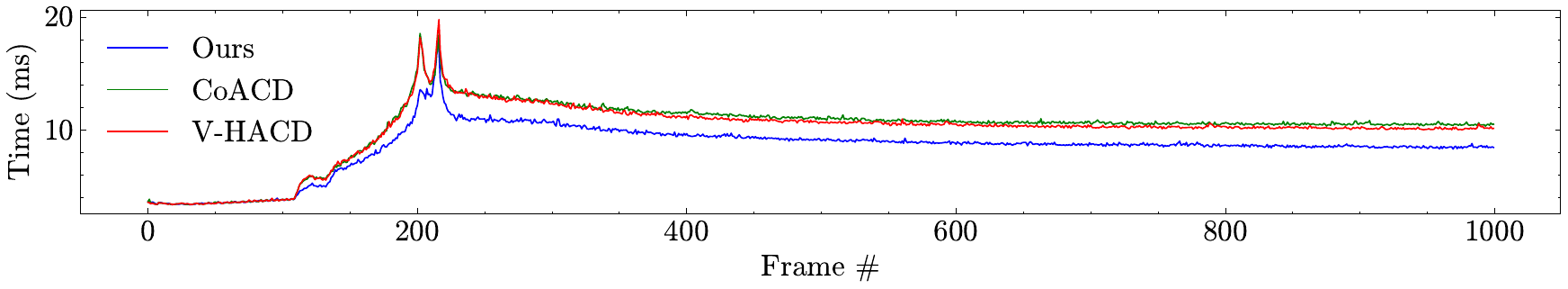}} \\
    \end{tabular}
    \caption{The Robot Vera model, tested in collision simulation by dropping 5000 spheres on top of it. A snapshot of our test simulation is shown on the right, with the model from our approach. Our approach is closer to the input model than CoACD, while having better performance as shown in the plot above. More model comparisons are in Fig.~\ref{fig:additional-sim-times}. \ccby JohnMesplay.}
    \label{fig:sim-time}
    %\Description{A chart comparing the duration of each frame for collision detection, for our approach, CoACD and V-HACD. The three approaches are quite similar, our approach is lower than the alternates, except at the tail-end where our approach is higher.}
    \vspace{0.1em}
\end{figure}

We evaluated collision time on 24 meshes such as the robot shown in Fig.~\ref{fig:sim-time}, with more results in Fig.~\ref{fig:additional-sim-times}. When comparing our approach with CoACD and V-HACD on the two key factors of simulation similarity and wall-clock time, our approach has more similar collisions as compared to the input with better performance. For example, for the mesh from Fig.~\ref{fig:teaser} our approach holds spheres inside the stairwell and on platforms. On the other hand, CoACD and V-HACD produce coarse outputs, causing spheres to roll off the mesh. This boosts their performance since there are fewer collisions but with inaccurate behavior. Despite our approach being more similar to the input than CoACD and V-HACD, our approach has better performance, often by at least one to two milliseconds per frame. When dealing with a performance budget of 16 milliseconds per 
frame (60 fps), that saved time is meaningful, giving more time for other computations.

\subsection{Comparing Complexity/Costs of Colliders}
It is difficult to provide an apples-to-apples comparison of complexity of our approach and CoACD/V-HACD, since their fundamental components are convex hulls with vertices and faces, versus our parametric primitives. Even worse, since capsules, spheres and cylinders do not have a finite number of faces, it is not possible to compare face counts. To demonstrate that we are using fewer resources than CoACD/V-HACD, even though the primitive count is higher than the hull count, we use the lowest common denominator, the number of bytes of each approach.
We compute the number of bytes for all methods, showing our approach is less complex, while at the same time is geometrically closer to the input with better simulation wall-clock times. To measure complexity, we count the number of bytes per component as per Tab.~\ref{tab:memory-costs}, with total costs for all models shown in the Appendix, Tab.~\ref{tab:raw-memory-cost}. Note we underestimate the cost of convex hulls by assuming that integers fit in two bytes (\texttt{uint\_16t}), when in some rarer cases it is necessary to use four bytes (\texttt{uint\_32t}).
When comparing aggregate statistics on our dataset, convex primitive decomposition, CoACD, and V-HACD uses on average \textbf{22523.7}, 93809.3, and 68934.1 bytes, and a median of \textbf{6362}, 76572, and 44592 bytes respectively. Clearly, convex primitive decomposition creates colliders with fewer resources than approximate convex hull decomposition, with better simulation frame durations than convex hulls. Since each primitive is cheaper than each convex hull, even though our approach has a larger number of primitives compared to the number of hulls, overall it is cheaper.
Furthermore, even with an equal number of primitives and hulls, we show primitives have lower wall-clock simulation time in Sec.~\ref{sec:ablate-hull-count}, Fig.~\ref{fig:ident_num_prim_hull}.

\begin{table}[!hb]
    \setlength{\tabcolsep}{1pt}
    \centering
    \begin{tabular}{|c|c|c|c|}
        \hline
        Primitive Kind & Minimum Floats Required & Total & \small Engine \\\hline
        Oriented Box & 3 position, 3 length, 4 orientation & 10 & \textbf{\textcolor{PineGreen}{Yes}} \\\hline
        Capsule & 3 start-point, 3 end-point, 1 radius & 7 & \textbf{\textcolor{PineGreen}{Yes}} \\\hline
        Sphere & 3 center, 1 radius & 4 & \textbf{\textcolor{PineGreen}{Yes}} \\\hline
        Cylinder & 3 start-point, 3 end-point, 1 radius & 7 & \textbf{\textcolor{Goldenrod}{Some}} \\\hline
        Frustum & 3 start-point 3 end-point, 2 radii & 8 & \textbf{\textcolor{BurntOrange}{Quantized}} \\\hline
        Prism & 3 position, 4 length, 4 orientation & 11 &\textbf{\textcolor{Goldenrod}{As Hull}} \\\hline\hline
        Convex Hull & 3 floats per vertex, 3 ints per tri & - & \textbf{\textcolor{PineGreen}{Yes}} \\\hline
    \end{tabular}
    \caption{Memory costs required for each primitive and for convex hulls. Orientations are stored as quaternions. For cylinders, some physics engines have direct support. Frustums are not supported, but can be quantized for usage, and prisms can be represented as convex hulls.}
    \label{tab:memory-costs}
    %\Description{See caption.}
    \vspace{0em}
\end{table}

\paragraph*{Implementations of Rigid-Body Collision} To understand the performance of rigid body simulation with many primitives/convex hulls, it is useful to understand collision detection's implementation. Collision detection happens in two phases, a broad phase followed by a narrow phase. In the broad phase, many comparisons are culled using coarse checks such as axis-aligned bounding boxes and axis overlap checks. The broad phase is independent of the complexity of each collider, and thus is more agnostic to primitives versus hulls. In the narrow phase, precise algorithms such as GJK~\cite{gjk}, or primitive-to-primitive checks are used. Due to broad phase culling, it is difficult to predict performance, as usually collision detection skips most checks, and there are only a few expensive comparisons. In the worst case though, objects may collide with all primitives in the scene, nullifying the broad phase. In our case, since collision with primitives is cheaper than collision with convex hulls and only a small set of expensive checks are used, convex primitive decomposition is faster than approximate convex hull decomposition.

\subsection{Measuring Similarity of Collision Objects}
To demonstrate that our approach is robust at producing faithful decompositions, we show results on a variety of cases, including inputs with holes, multiple components, and environments.

\begin{figure}[!h]
    \centering
        \setlength{\tabcolsep}{0em}
    \begin{tabular}{c c c}
        Input & Convex Prim. Decomp. & CoACD \\
        \includegraphics[width=0.32\linewidth]{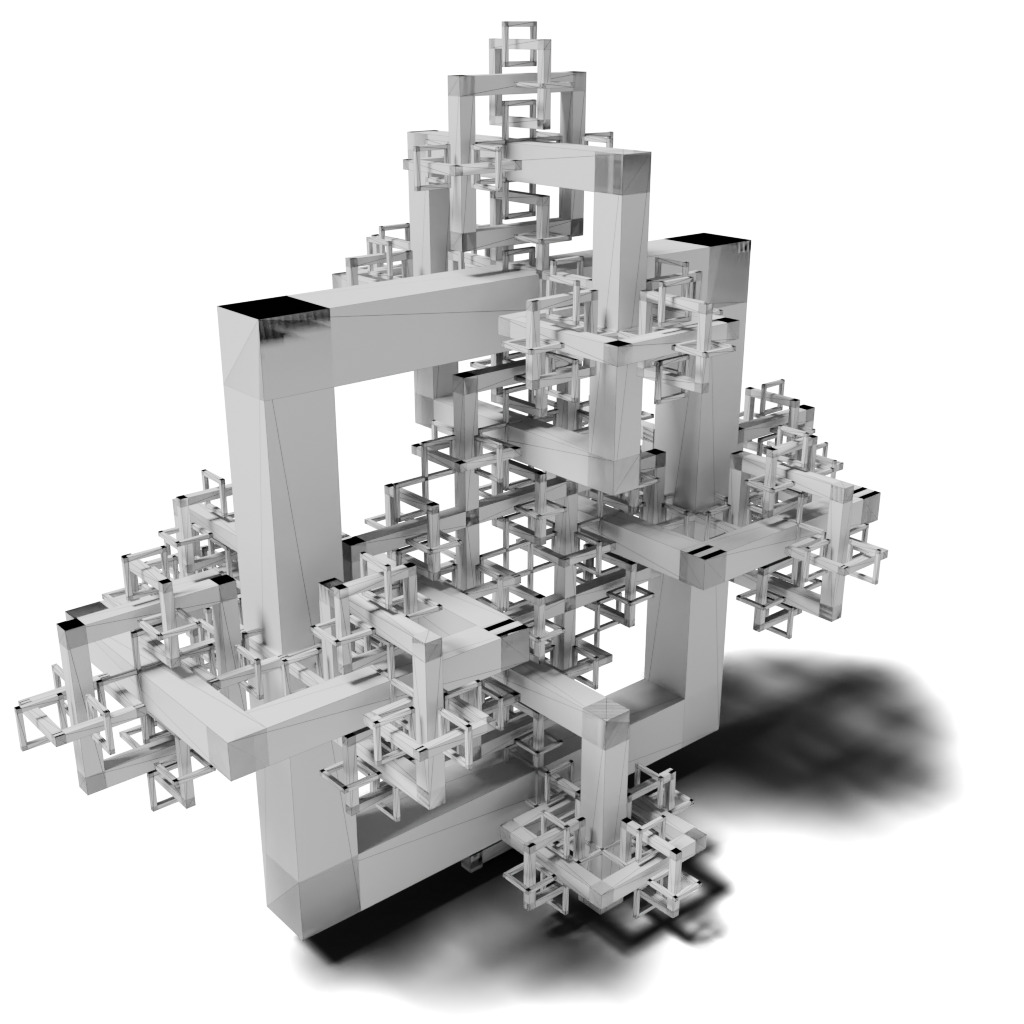} &
        \includegraphics[width=0.32\linewidth]{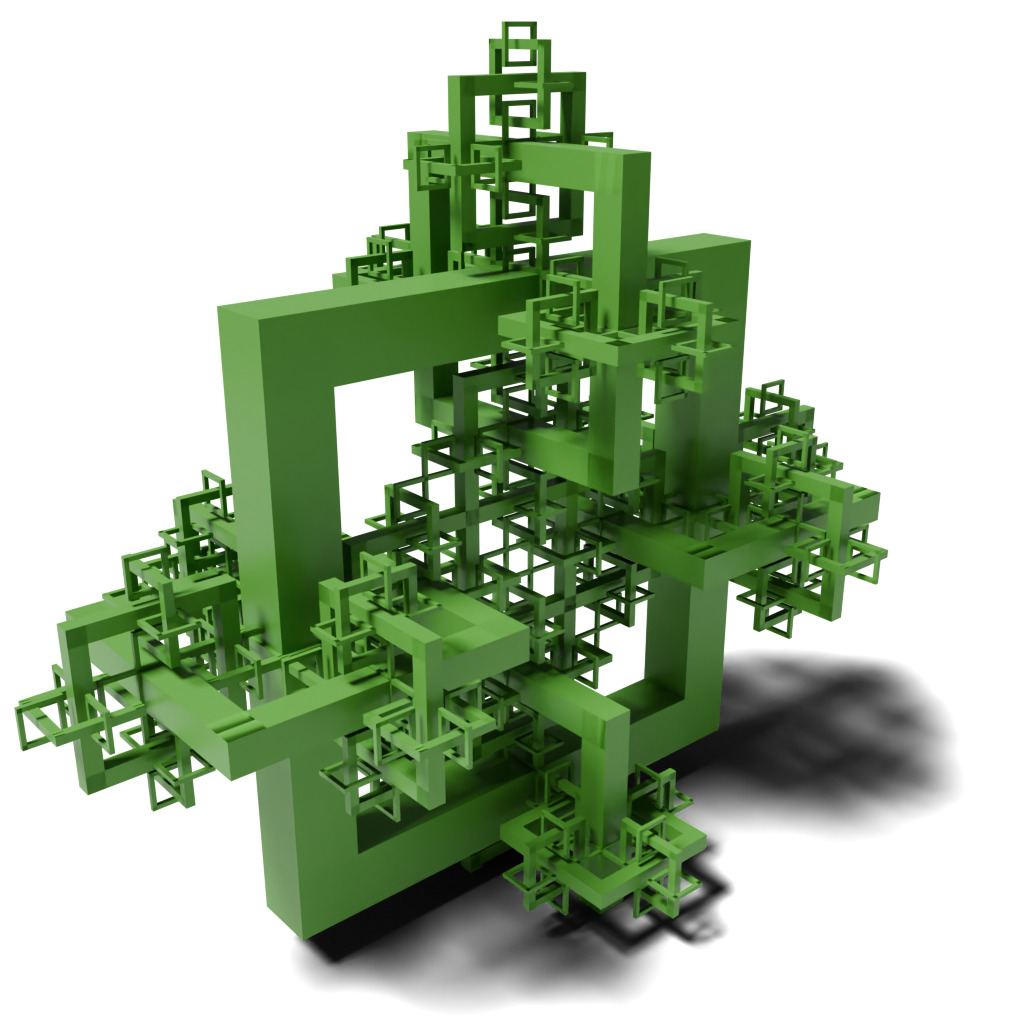} &
        \includegraphics[width=0.32\linewidth]{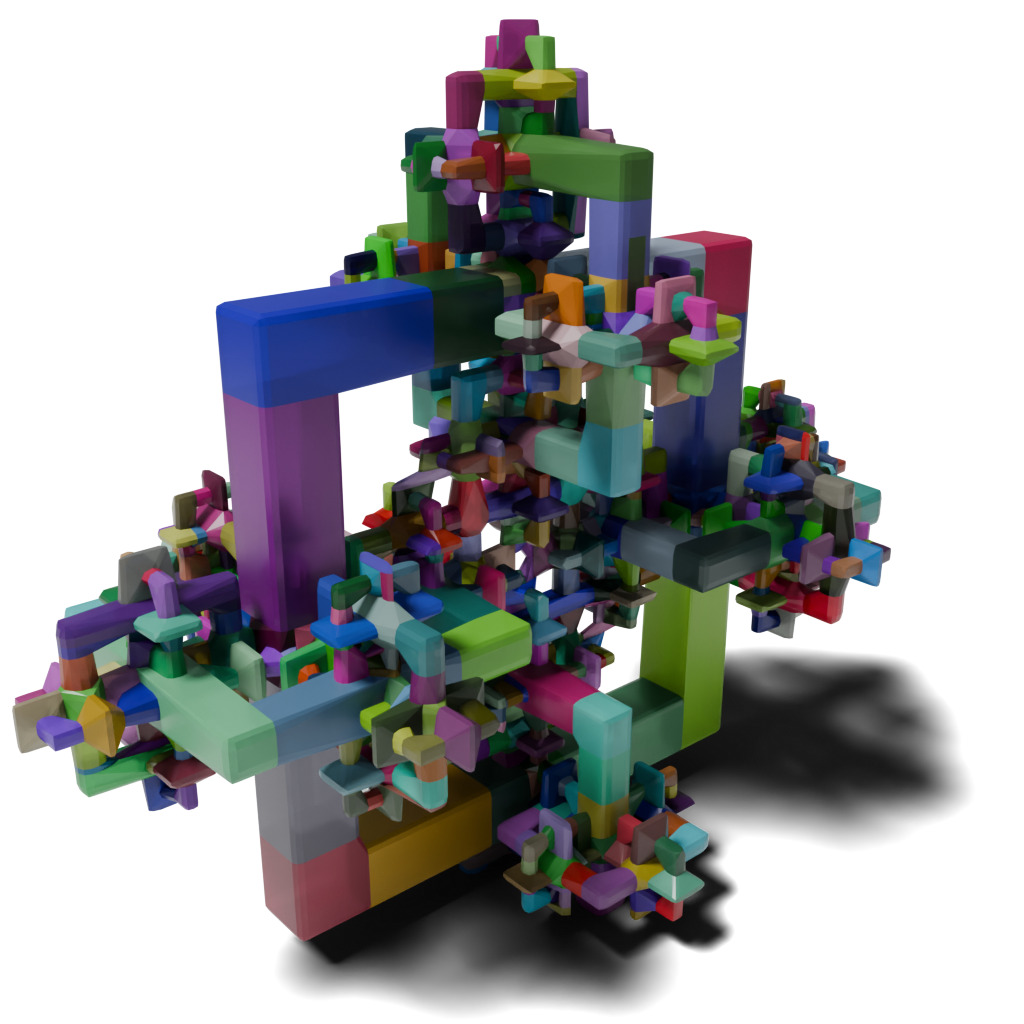} \\
        |F| = 8184 & 5456 Boxes & 1126 Hulls {\small (|F| = 53148)} \\
        {\small $\frac{\text{Haus./Cham. New to Input}}{\lVert\text{Bounding Box Diag}\rVert_2}^\downarrow$} & \scriptsize $\num{3.34e-5}/\num{3.25e-5}$ &
        \scriptsize $\num{3.07e-2}/\num{8.02e-3}$ \\
        \multicolumn{3}{c}{\includegraphics[width=0.9\linewidth]{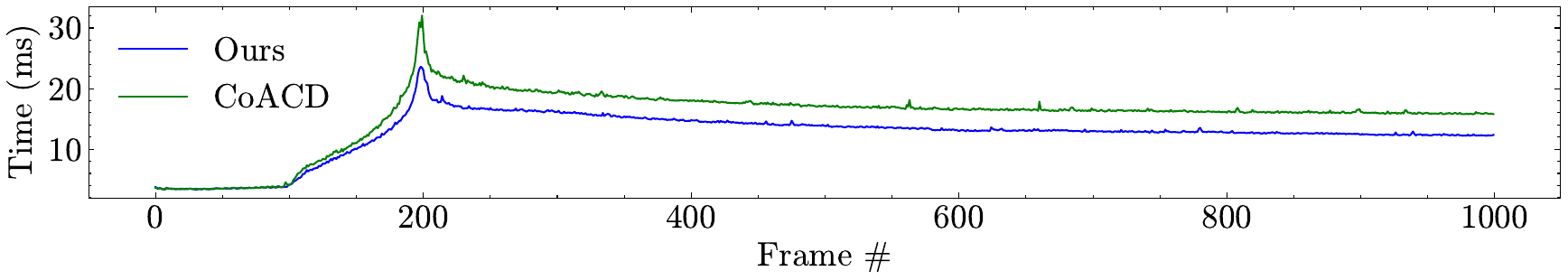}} \\
    \end{tabular}
    \vspace{-1em}
    \caption{Our approach decomposes the input mesh into a number of primitives which more closely adhere to the input mesh than CoACD, while allowing faster collision detection. Dark regions on the input mesh are due to overlapping non-manifold faces. Frame duration comparisons for simulation of collision with 5000 dropping balls are shown below; our approach has faster simulation than CoACD on all frames. \ccbync Dixbit.}
    \label{fig:fractal}
    %\Description{A fractal shape decomposed using our method and CoACD. CoACD's output is more coarse and is not able to as precisely preserve holes in the input compared to our approach.}
\end{figure}

We compare our approach's preservation of sharp edges on a fractal shape as compared to CoACD in Fig.~\ref{fig:fractal}. On this model, our approach is two orders of magnitude closer to the input than CoACD, since CoACD rounds the edges of the input shape. We also compare the simulation frame duration, showing that even with a more precise decomposition, our approach is faster.

\begin{figure*}
    \centering
    \renewcommand{\arraystretch}{0.0}
    \begin{tabular}{c c c c}
        Input & Convex Prim. Decomp. & CoACD & V-HACD \\
        \includegraphics[width=0.22\linewidth]{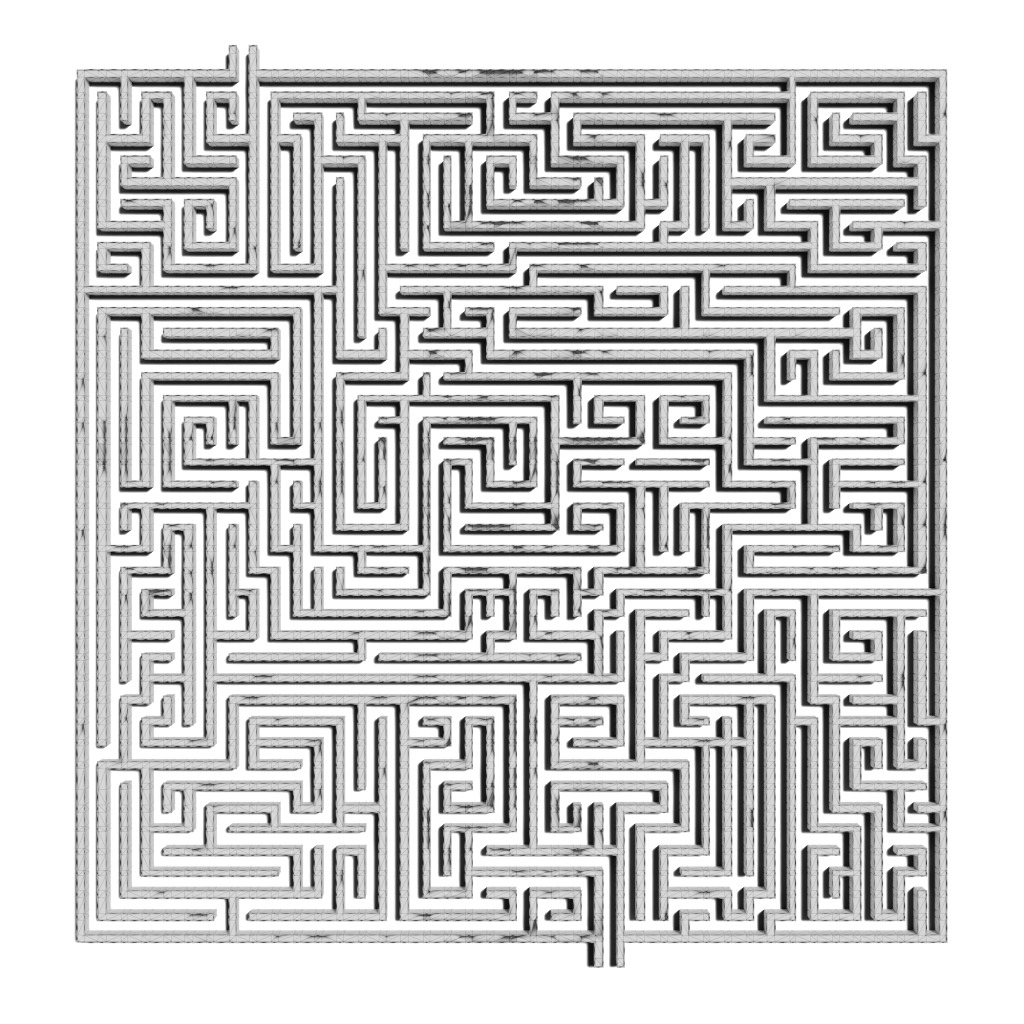} &
        \includegraphics[width=0.22\linewidth]{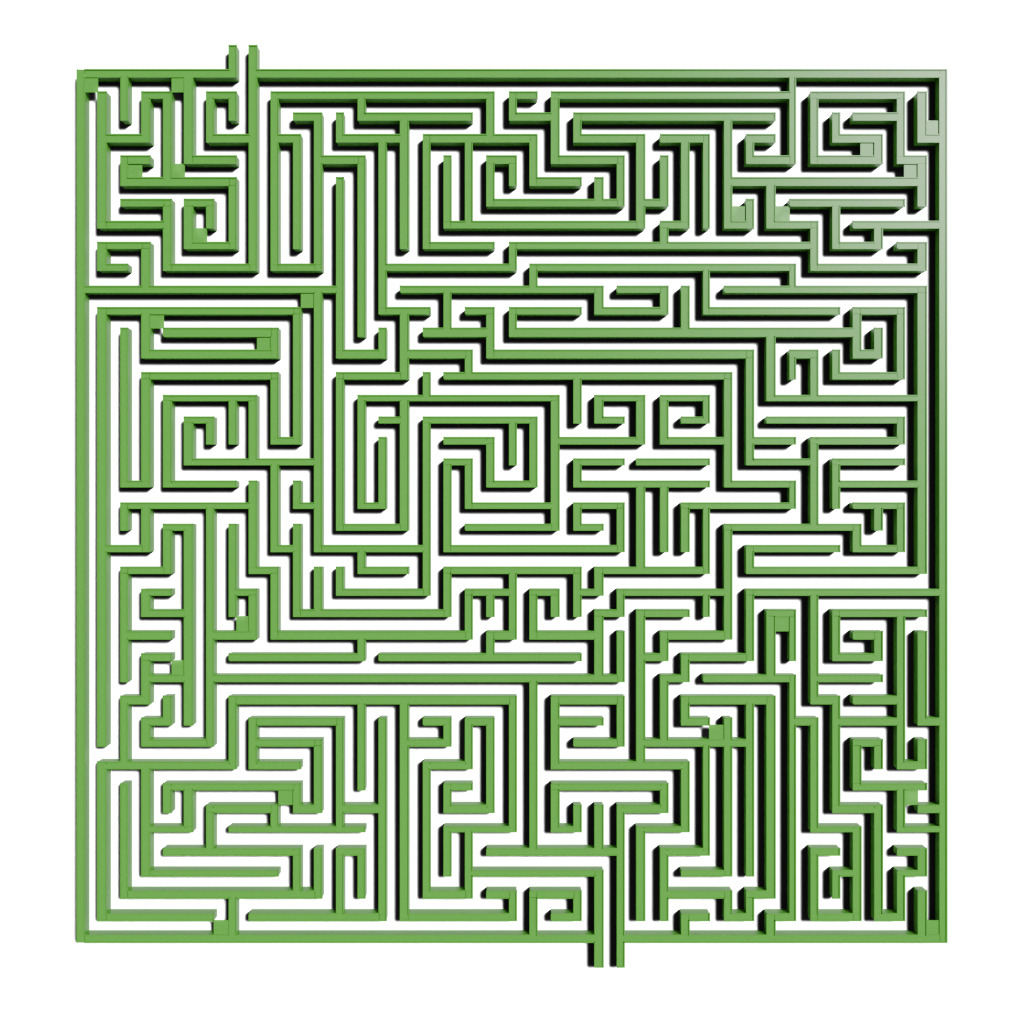} &
        \includegraphics[width=0.22\linewidth]{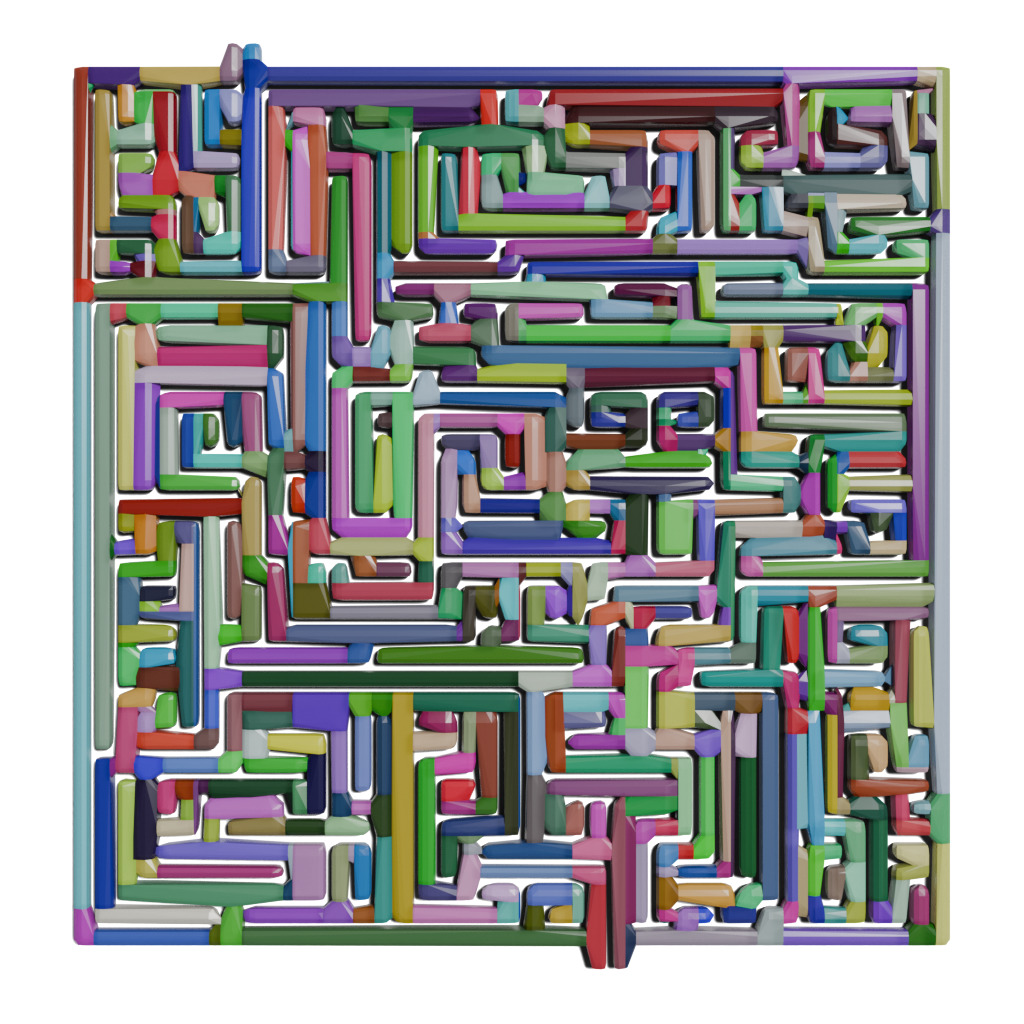} &
        \includegraphics[width=0.22\linewidth]{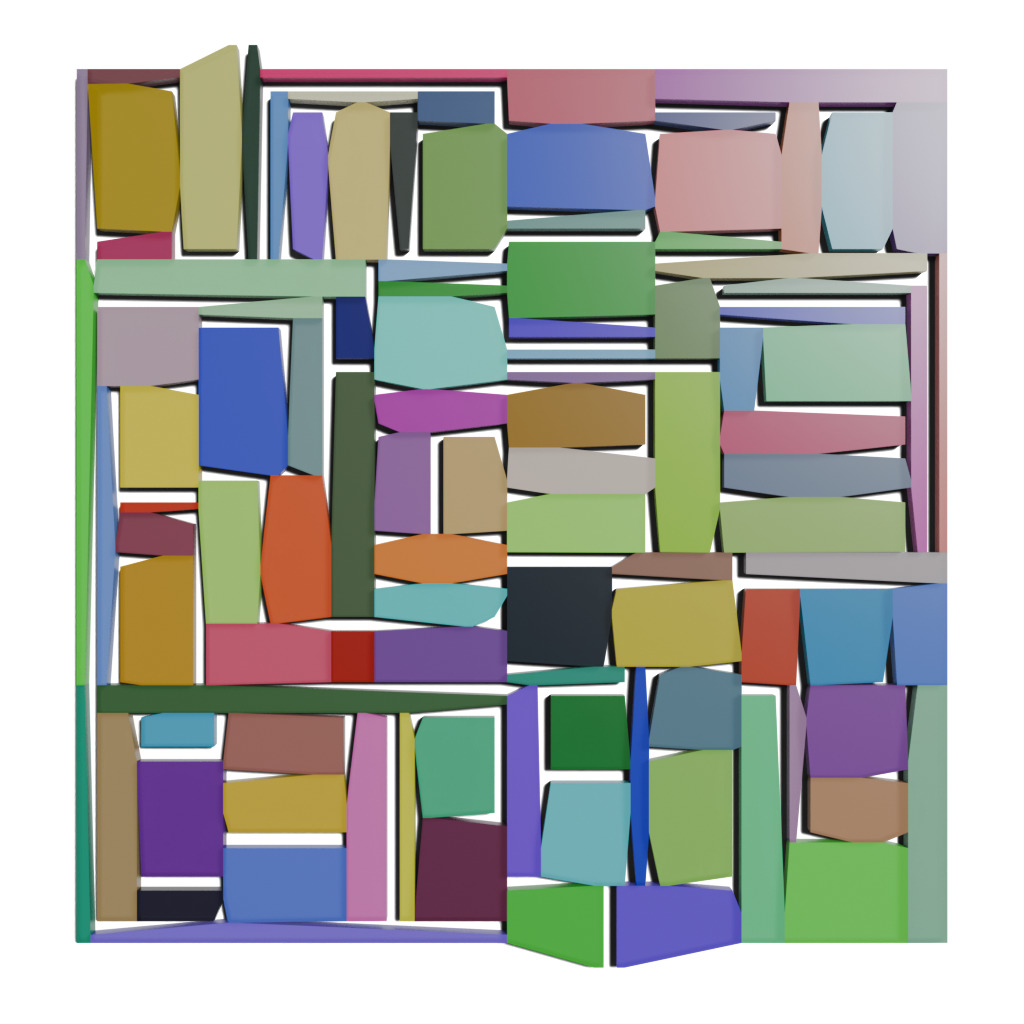} \\
        
        \includegraphics[width=0.22\linewidth]{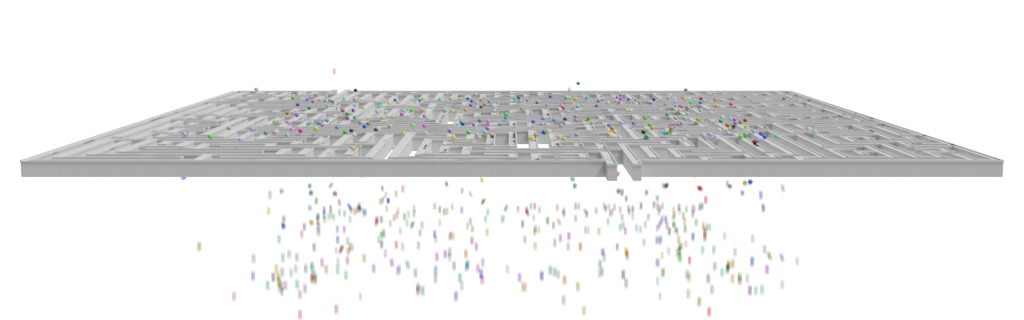} &
        \includegraphics[width=0.22\linewidth]{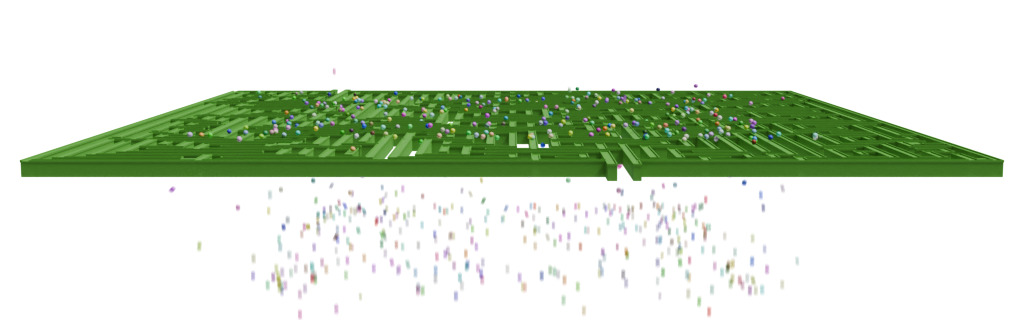} &
        \includegraphics[width=0.22\linewidth]{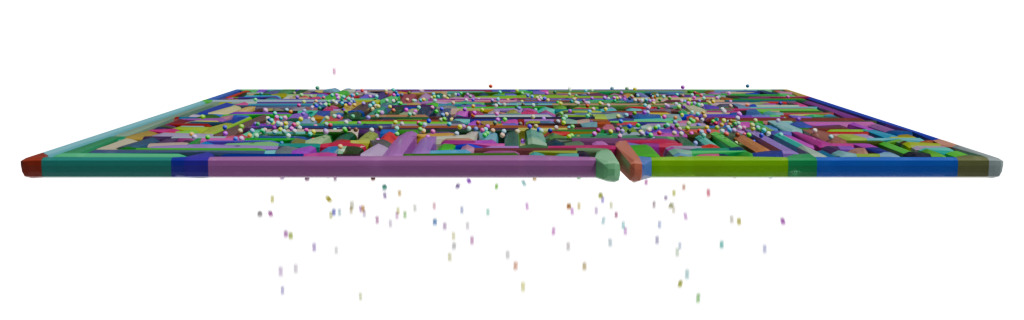} &
        \includegraphics[width=0.22\linewidth]{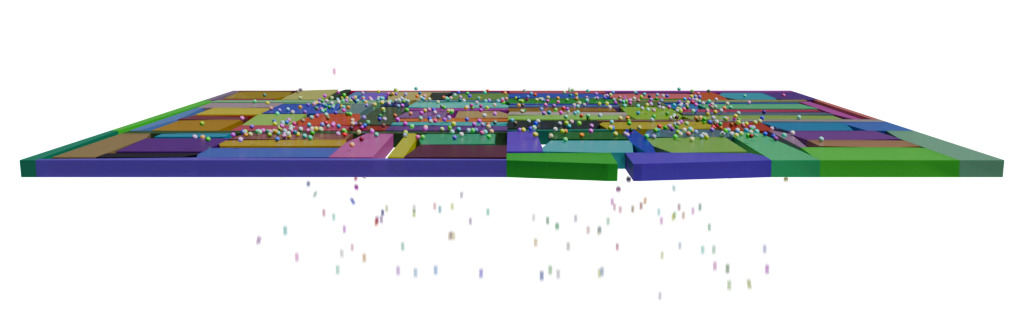} \\
        
        |F| = 37191 & 6982 Boxes & 669 Hulls (|F| = 69962) & 105 Hulls (|F| = 3628) \\ 
        $\frac{\text{Haus./Cham. New to Input}}{\lVert\text{Bounding Box Diag}\rVert_2}^\downarrow$ &
        $\num{5.72e-4}/\num{4.35e-4}$ & $0.0111/\num{5.524e-3}$ & $0.0163/\num{1.94e-3}$ \\
    \end{tabular}
    \caption{Our approach can cleanly preserve holes in the non-manifold, non-watertight input mesh, maintaining its traversability as shown by the number of balls that can pass through in the second row. Prior convex decomposition approaches reduce the size of holes due to voxelization and preprocessing needed to make the mesh manifold and watertight, changing the collision behavior of the output collider. \ccby Talaei-dev.}
    \label{fig:maze}
    %\Description{A flat maze with comparisons of our approach to CoACD and V-HACD. Our approach mostly preserves holes in the mesh, whereas CoACD and V-HACD round edges and have large hulls that block passage.}
\end{figure*}

We also demonstrate our approach's generation of colliders with holes by running on a maze while preserving traversability in Fig.~\ref{fig:maze}. To compare fairly to alternatives, we increase CoACD's resolution from 30 units to 80 units, and set the concavity to its minimum value of $0.01$. For V-HACD, we use the off-the-shelf settings. While CoACD preserves some holes, the gaps are thinned and edges rounded, whereas our approach preserves most holes. V-HACD closes most holes, making it unsuitable for replacing the original mesh. We visualize differences in simulation, by dropping 500 balls in Blender~\cite{Blender} (5000 is used in simulation, 500 is only for visualization). The balls pass through our output and the original mostly unfettered, whereas CoACD and V-HACD prevent many from going through.

\begin{figure*}
    \centering
    \setlength{\tabcolsep}{1pt}
    \begin{tabular}{c c c c c}
        Input, $|F| = 17986$  & \tiny Ours (Coarse), 54 Boxes, 5 Cylinders & \tiny Ours (Precise), 536 Boxes, 5 Cylinders & \tiny CoACD, 90 Hulls (|F| = 6984) & \tiny V-HACD, 17 Hulls (|F| = 1298) \\
        \includegraphics[width=0.19\linewidth]{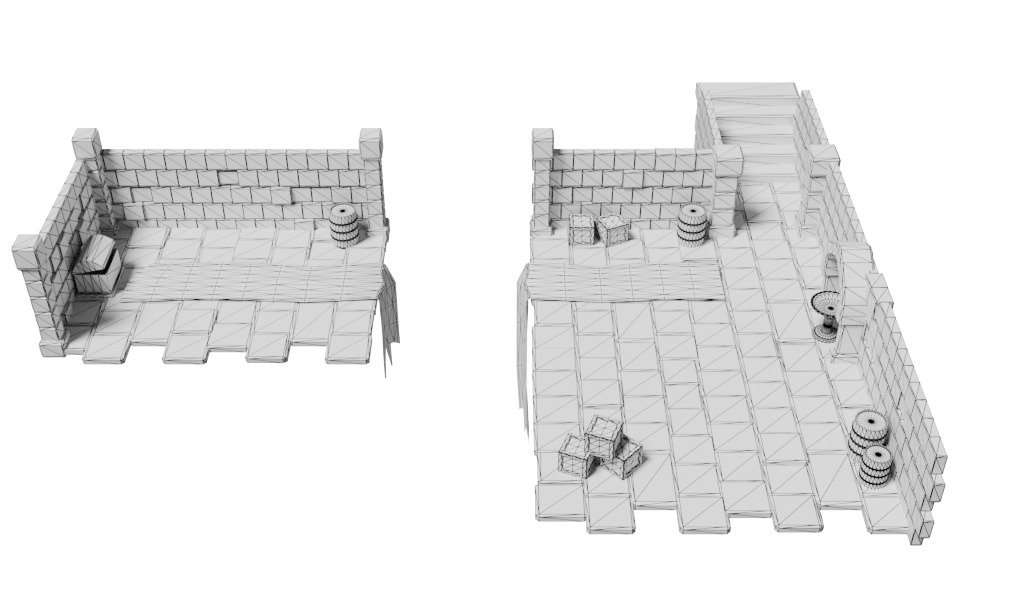} &
        \includegraphics[width=0.19\linewidth]{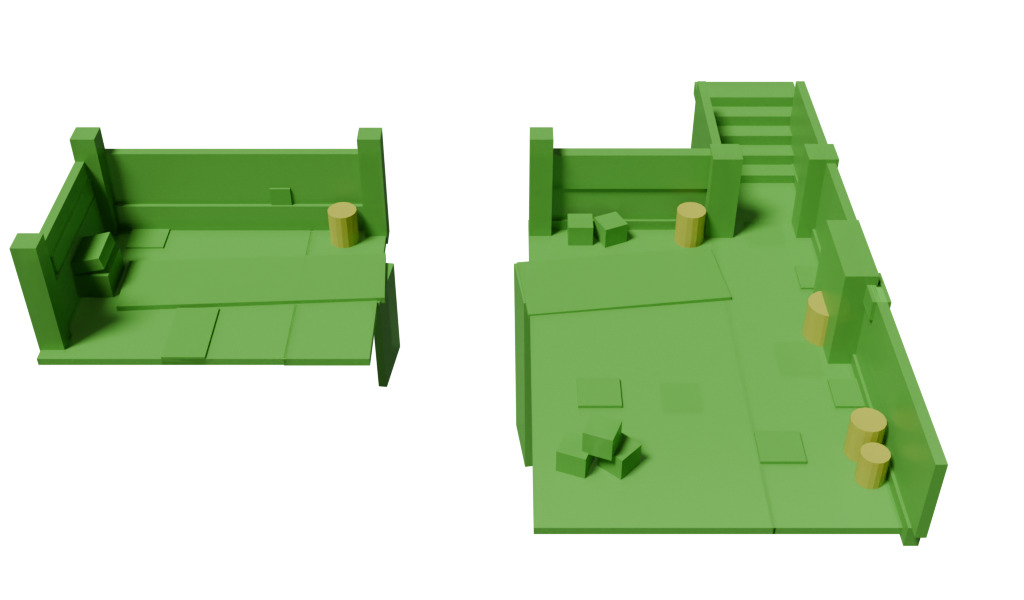} & \includegraphics[width=0.19\linewidth]{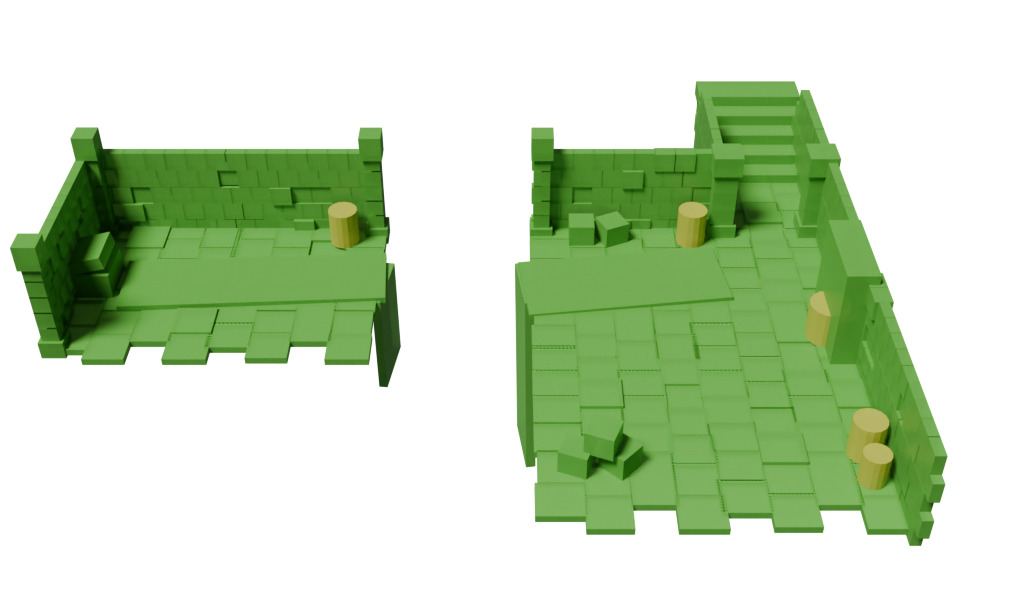} &
        \includegraphics[width=0.19\linewidth]{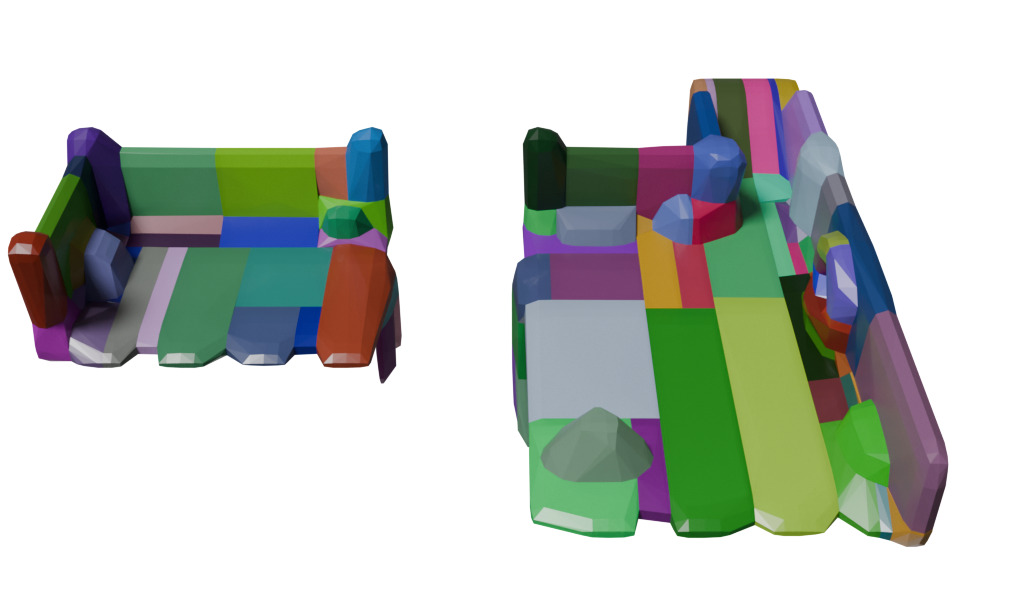} &
        \includegraphics[width=0.19\linewidth]{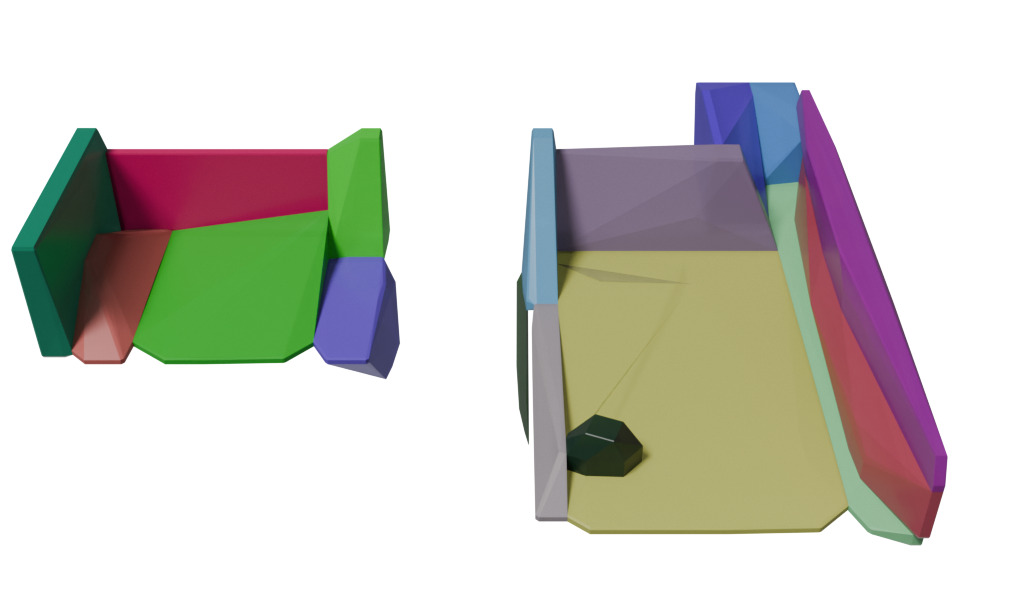} \\
        \small $\frac{\text{Hausdorff/Chamfer New to Input}}{\lVert\text{Bounding Box Diag}\rVert_2}^\downarrow$ &
        $\num{3.09e-2}/\num{5.47e-3}$ & $\num{3.06e-2}/\num{1.55e-3}$ & $\num{1.67e-2}/\num{5.48e-3}$ & $\num{5.48e-2}/\num{4.83e-3}$ \\
    \end{tabular}
    \caption{Our approach can automatically fit colliders for environments at multiple resolutions, while more closely adhering to the input as compared to CoACD and V-HACD, maintaining sharp features such as the boxes, barrels, and chest, and can be edited easily. \ccby Karthik Naidu.}
    \label{fig:dungeon}
    %\Description{A level which is decomposed twice by our approach compared to CoACD and V-HACD. Our approach is cleaner than CoACD and V-HACD, which round edges and introduce a number of components which could be merged together.}
\end{figure*}

We then demonstrate our approach on environments. For game levels, convex decomposition is overkill since floors and walls are replaced with box colliders. Yet, current tools cannot automatically fit boxes, requiring artists to place them and balance trade-offs in precision versus runtime, as some applications need details such as height changes in the floor and others use one box for performance. To reflect the ability to control this trade-off, we show two granularity of output on an environment in Fig.~\ref{fig:dungeon}, and one output on the Bistro~\cite{ORCAAmazonBistro} scene in the Appendix, Fig.~\ref{fig:bistro}. Our outputs are visually and quantitatively faithful to the original mesh at both resolutions, and preserve items such as the barrels and boxes in Fig.~\ref{fig:dungeon}, and wine glasses and utensils in Fig.~\ref{fig:bistro}. The variation in the floors and walls show how our approach can be used to tune precision. By comparison, convex decomposition smooths over the props and floor with arbitrary cuts, giving implausible simulation.

\setlength{\intextsep}{0pt}
\setlength{\columnsep}{0pt}
\begin{wrapfigure}{r}{0.26\textwidth}
    \centering
    \includegraphics[width=0.26\textwidth]{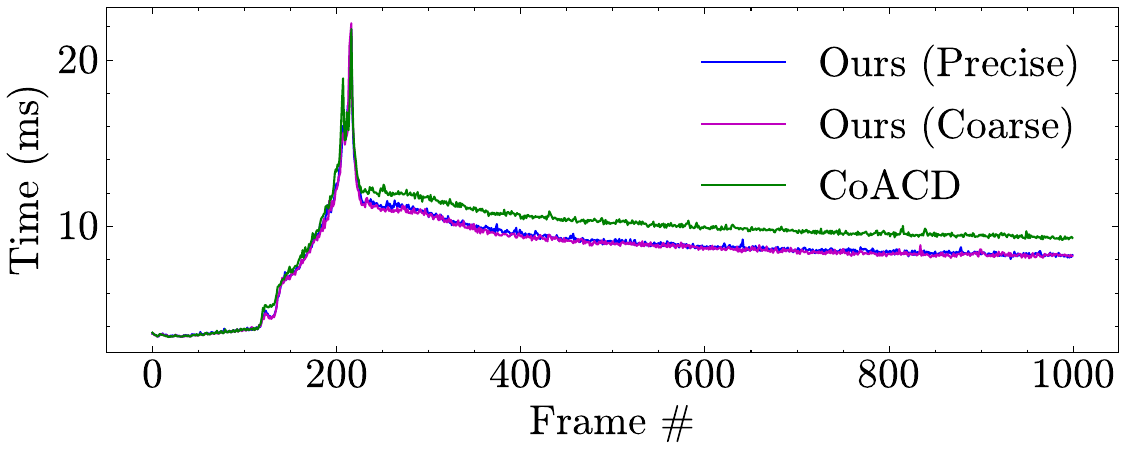}
\end{wrapfigure}

We measure frame duration of simulation of CoACD and our approach from Fig.~\ref{fig:dungeon} shown in the inset. Our coarse and precise meshes both have better performance than CoACD, with better geometric similarity.

\subsection{Distance to Original Mesh}

We measure geometric similarity by the one-way Hausdorff and Chamfer distances from the collider to the input mesh, like ~\cite{animated_decomp}. The key motivation for choosing this metric is that each point on the collider should be as close to the input as possible. Complete results are provided in Tab.~\ref{tab:distance_comparison}. Across our tested meshes, our approach has the lowest Hausdorff distance for 44 meshes, and lowest Chamfer distance for 43, shown in Tab.~\ref{tab:count1way}. The average one-way hausdorff distance on our dataset for convex primitive decomposition, CoACD, and V-HACD are \textbf{0.0445}, 0.0514, and 0.0710, and median \textbf{0.0346}, 0.0383, and 0.0593 respectively. The average chamfer distances are $\textbf{\num{6.95e-3}}, \num{9.91e-3}, \num{8.82e-3}$, and median $\textbf{\num{5.40e-3}}, \num{9.41e-3}, \num{7.60e-3}$ respectively. This shows that convex primitive decomposition consistently has closer geometry to the original, with less total complexity, measured by byte count. This also shows how CoACD's optimization for max distance does so at the cost of average distance, whereas our approach is closer on both metrics, even though it optimizes for volume.

\begin{table}
\centering
\small
\setlength{\tabcolsep}{0.3pt}
\begin{tabular}{|c|c|c|c|}
    \hline
    Summary Statistics & Ours & CoACD & V-HACD \\\hline
    \scriptsize \#Models w/ Lowest 1-way Hausdorff$^\uparrow$ & \cellcolor{orange!5} 44 & 20 & 4 \\\hline
    \scriptsize \#Models w/ Lowest 1-way Chamfer$^\uparrow$ & \cellcolor{orange!5} 43 & 3 & 22 \\\hline
    Mean 1-way Chamfer$^\downarrow$ & \cellcolor{orange!5} $\num{6.95e-3}$ & $\num{9.91e-3}$ & $\num{8.82e-3}$ \\\hline
    Median 1-way Chamfer$^\downarrow$ & \cellcolor{orange!5} $\num{5.40e-3}$ & $\num{9.41e-3}$ & $\num{7.60e-3}$ \\\hline
    Mean 1-way Hausdorff$^\downarrow$ & \cellcolor{orange!5} 0.0445 & 0.0514 & 0.0710 \\\hline
    Median 1-way Hausdorff$^\downarrow$ & \cellcolor{orange!5} 0.0346 & 0.0383 & 0.0593 \\\hline
\end{tabular}
\caption{
    \label{tab:count1way}Our approach has more lower 1-way distances on our dataset from the generated mesh to the input mesh as compared to CoACD~\cite{coacd} and V-HACD~\cite{vhacd}.
}
\vspace{-1em}
\end{table}

% To show primitive decompositions have reduced total complexity compared to prior work, we compare the number of parts, shown in the Appendix, Tab.~\ref{tab:complete_results}, and the number of bytes used in Tab.~\ref{tab:raw-memory-cost}. While our approach uses more primitives than hulls produced by prior work, when comparing the complexity of hulls versus primitives by their components with the costs from Tab.~\ref{tab:memory-costs}, primitives are simpler.
% In Tab.~\ref{tab:metric_comparison}, we show the count of models which outperforms the alternatives on models in our dataset. Out of the models tested, ours has the lowest byte cost for 44 models, CoACD for 6, and V-HACD for 1. Our approach also has lower Chamfer and Hausdorff distance across the majority of models in our dataset. Furthermore, when considering both bytes used and distance our approach has significantly more than alternatives.

% \begin{table}
% \begin{tabular}{|c|c|c|c|}
%     \hline
%     \# Models & Ours & CoACD & V-HACD \\\hline
%     Lowest Byte Usage & 54 & 8 & 5 \\\hline
%     Lowest Haus. Dist. (New $\to$ Input) & 36 & 26 & 5 \\\hline
%     Lowest Haus. \& Bytes$^*$ & 26 & 2 & 0 \\\hline
%     Lowest Cham. Dist. (New $\to$ Input) & 35 & 5 & 27 \\\hline
%     Lowest Cham. \& Bytes$^*$ & 24 & 0 & 0 \\\hline
% \end{tabular}
% \caption{Our approach outperforms CoACD and V-HACD for both distance and number of bytes used across the majority of models in our dataset. $^*$This is only models which are lower on both metrics.}
% \label{tab:metric_comparison}
% \end{table}

\begin{figure*}
    \centering
    \setlength{\tabcolsep}{0pt}
    \begin{tabular}{c c c c c}
        \multicolumn{5}{c}{Level Of Detail Comparison of Our Approach For Simulation} \\
        Input & \multicolumn{4}{c}{Decreasing Resolution $\rightarrow$} \\
        \put(0.1\linewidth,0.1\linewidth){\includegraphics[width=0.76\linewidth]{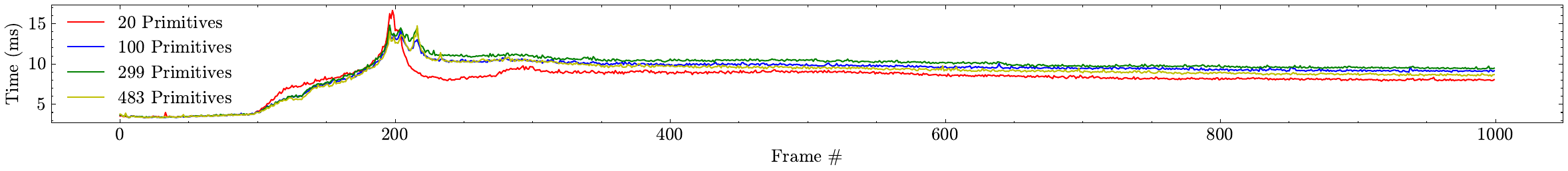}} 
        \includegraphics[width=0.19\linewidth]{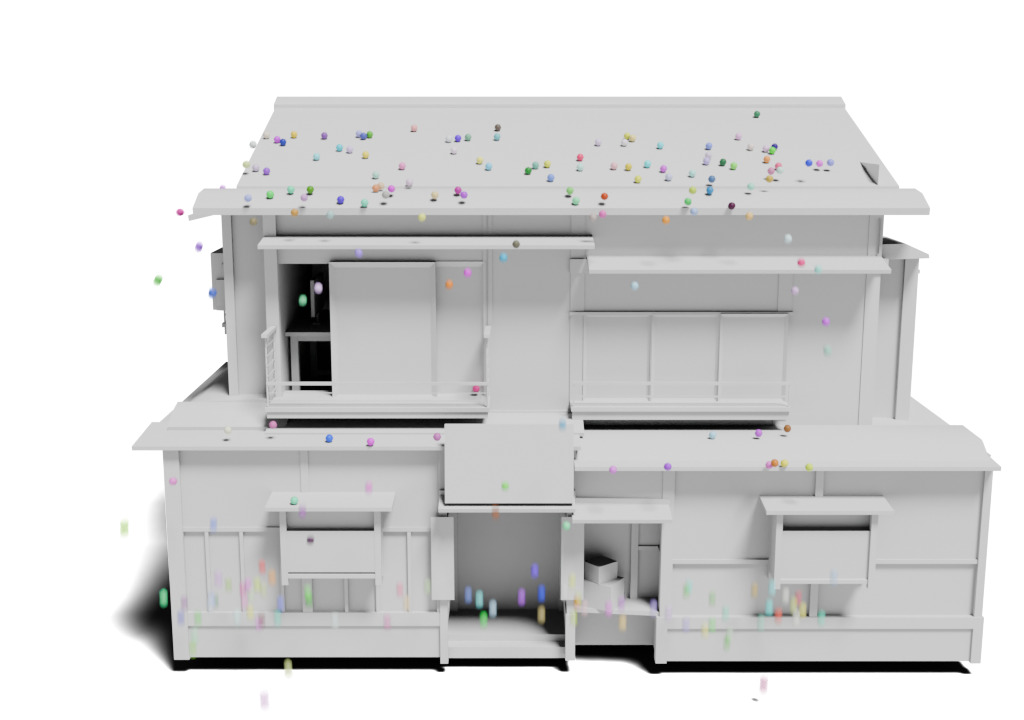} &
        \includegraphics[width=0.19\linewidth]{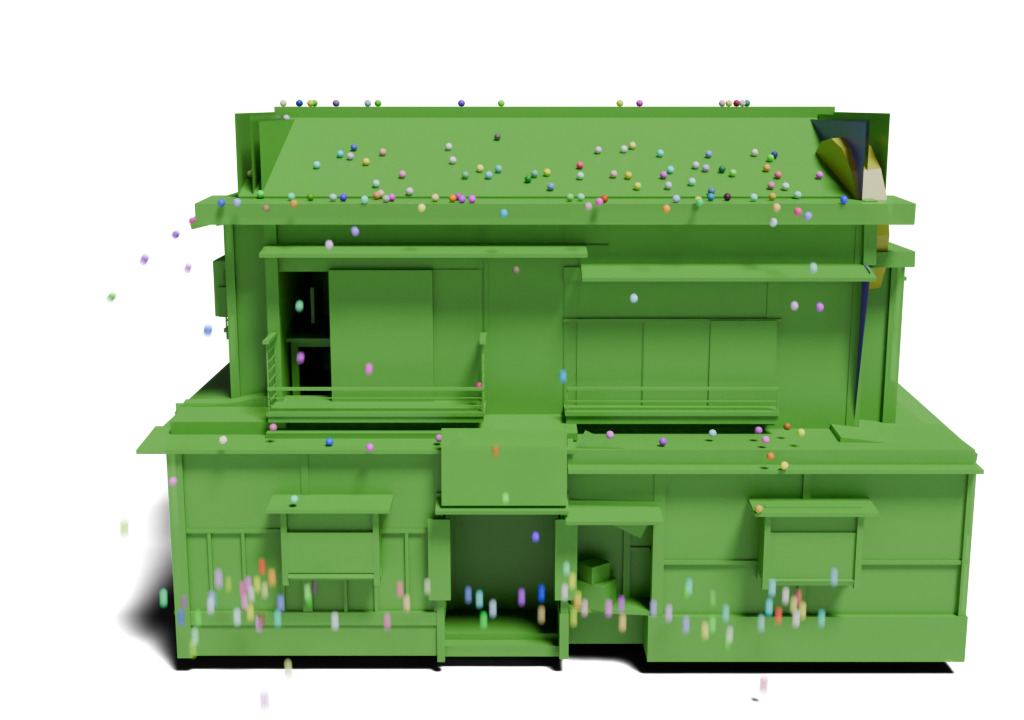} &
        \includegraphics[width=0.19\linewidth]{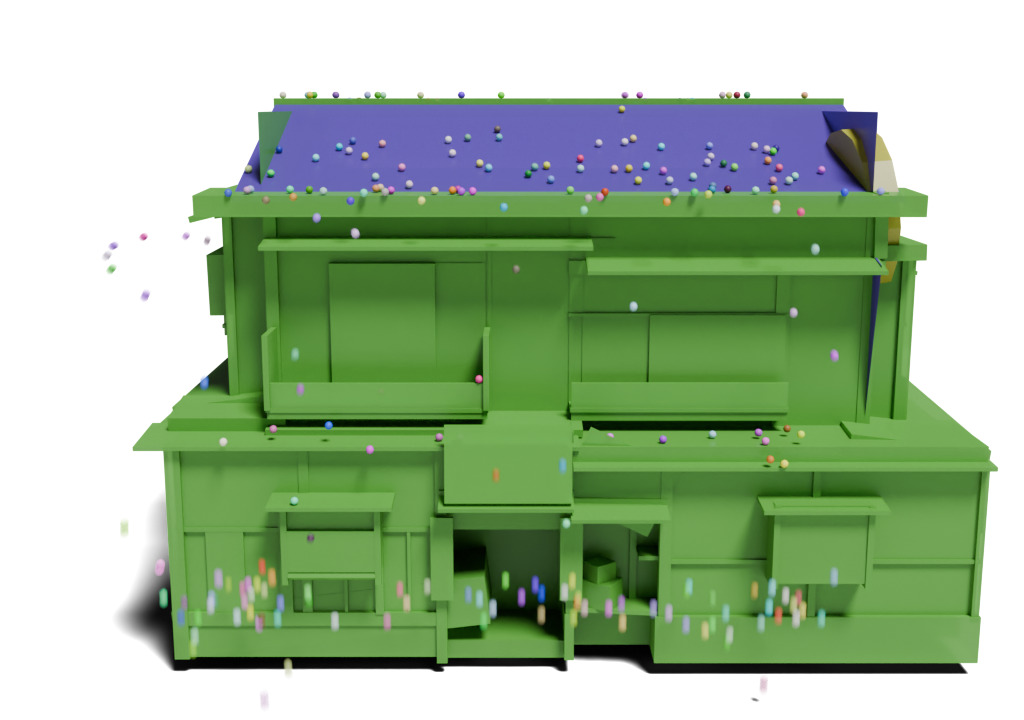} &
        \includegraphics[width=0.19\linewidth]{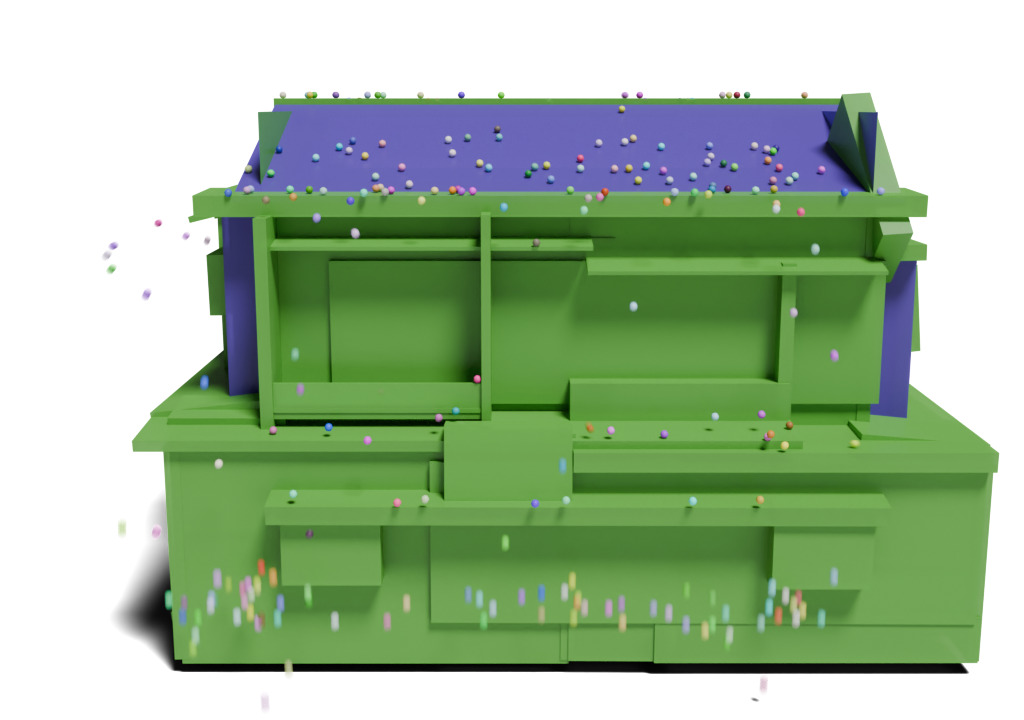} &
        \includegraphics[width=0.19\linewidth]{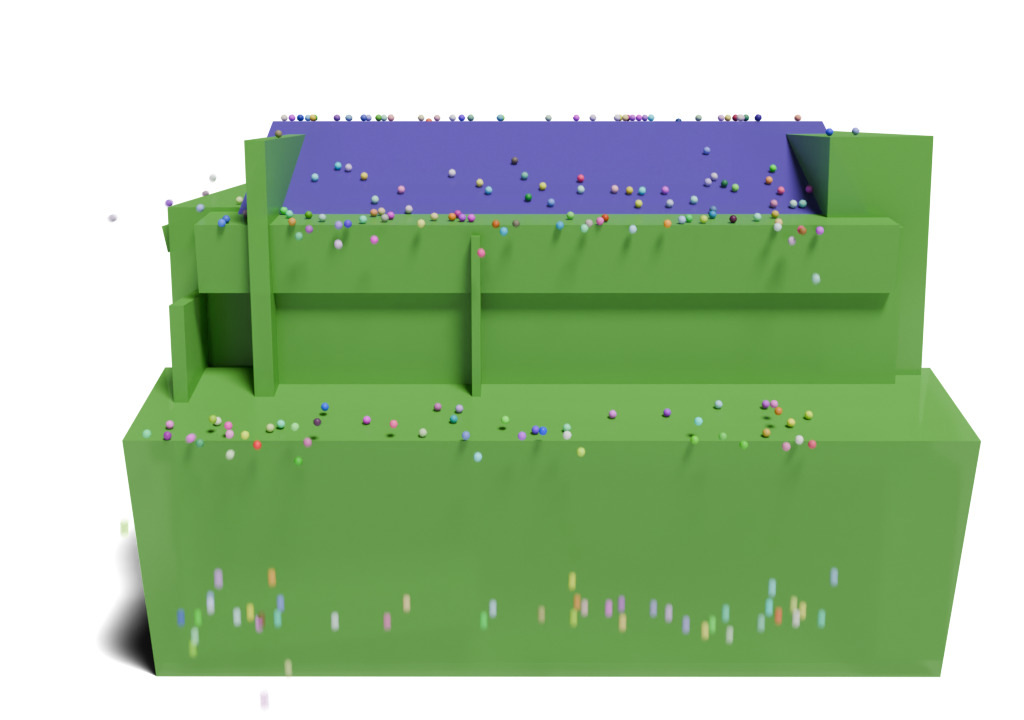} \\
        $|F| = 4569$ &
        \scriptsize 470 Boxes, 1 Cap, 8 Cyl, 4 Prism &
        292 Boxes, 3 Cyl, 4 Prism &
        94 Boxes, 6 Prism &
        19 Boxes, 1 Prism \\
        Generation Time (sec): & 0.208 & 0.977 & 1.408 & 1.640 \\ 
    \end{tabular}
    \caption{We compare our approach at multiple levels of detail for a single mesh, varying only the target number of output primitives. Our approach gracefully degrades from a high-resolution approximation of the input to a lower resolution approximation. At different resolutions, the simulation using our collider still resembles the input mesh. Rigid-body collision time for each resolution is shown in the plot. The more precise meshes are generated faster than coarse meshes, as our approach is bottom up. \ccby Mikail Karaca.}
    \label{fig:level-of-detail-collision}
    %\Description{Multiple levels of details for a single japanese dojo, with balls visualized falling on top of the mesh. Between each level of detail, collision looks relatively similar.}
\end{figure*}

\paragraph*{Level of Detail Collision Objects}
To demonstrate our approach for generating different levels of detail for colliders, we create multiple resolutions for one mesh in Fig.~\ref{fig:level-of-detail-collision}, all tested in simulation. At each level our approach maintains the similarity of balls rolling off the roof, demonstrating our approach's versatility. The simulation run-time at each level also shows that varying the resolution provides a way to tune performance versus geometric similarity.

\begin{figure}
    \centering
    \begin{tabular}{c c}
        \scriptsize Input & \scriptsize No Max Added Volume \\
        \small \#Primitives, $\frac{\text{Chamfer New to Input}}{\lVert\text{Bounding Box Diag}\rVert_2}^\downarrow$  & 13 Boxes,  $\num{2.37e-3}$ \\
        \includegraphics[width=0.45\linewidth]{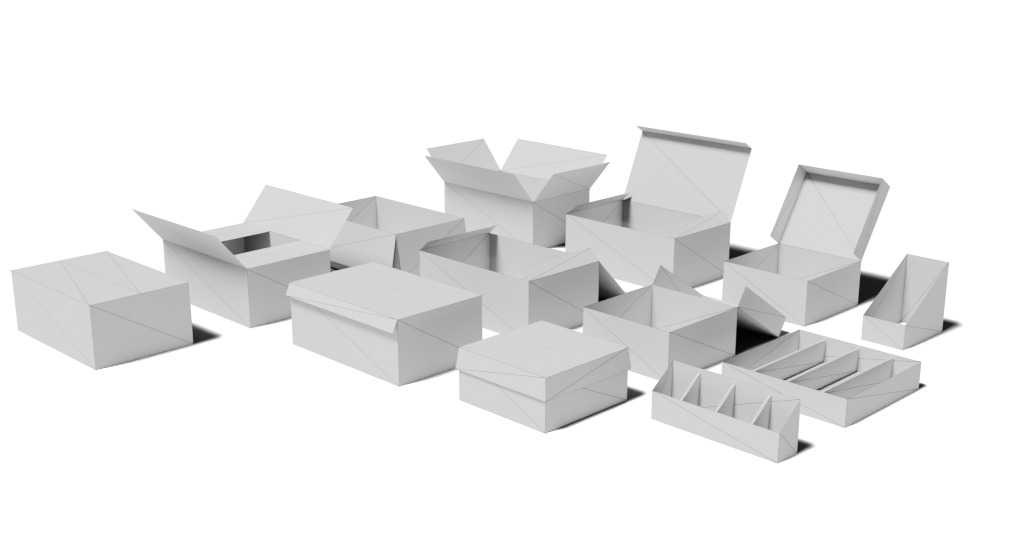} 
        \put(-0.45\linewidth, -0.15\linewidth){\includegraphics[width=0.45\linewidth]{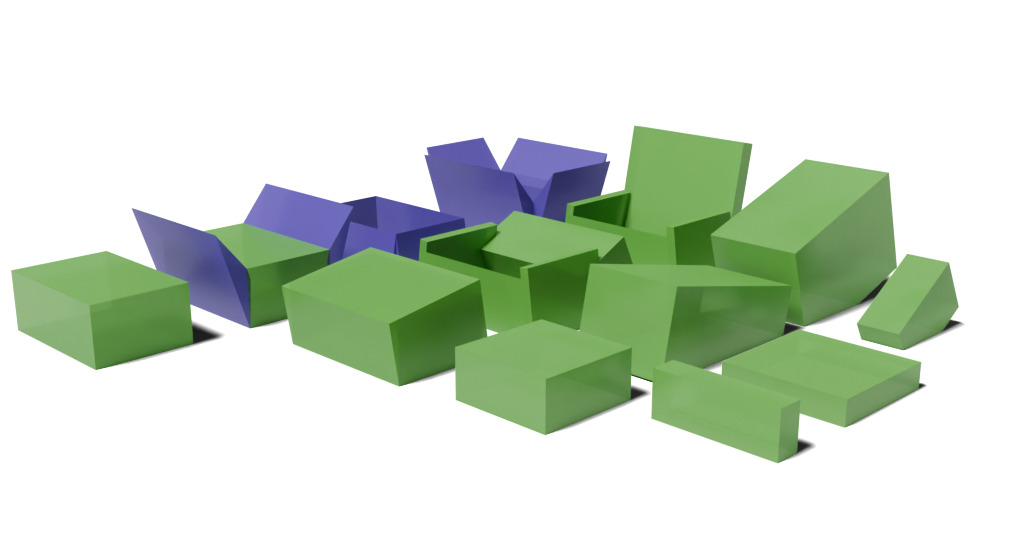}} &
        \includegraphics[width=0.45\linewidth]{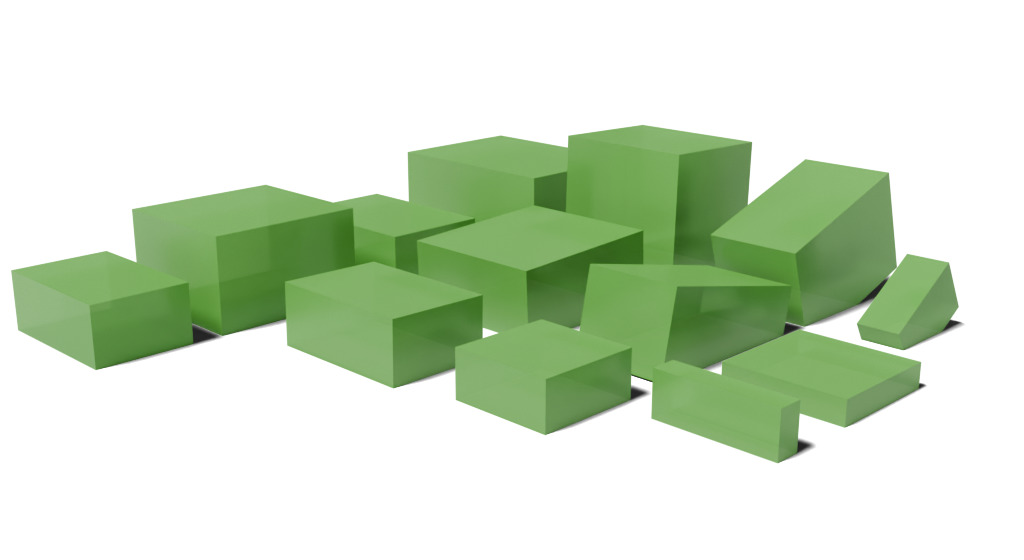}
        \put(-0.45\linewidth, -0.15\linewidth){\includegraphics[width=0.45\linewidth]{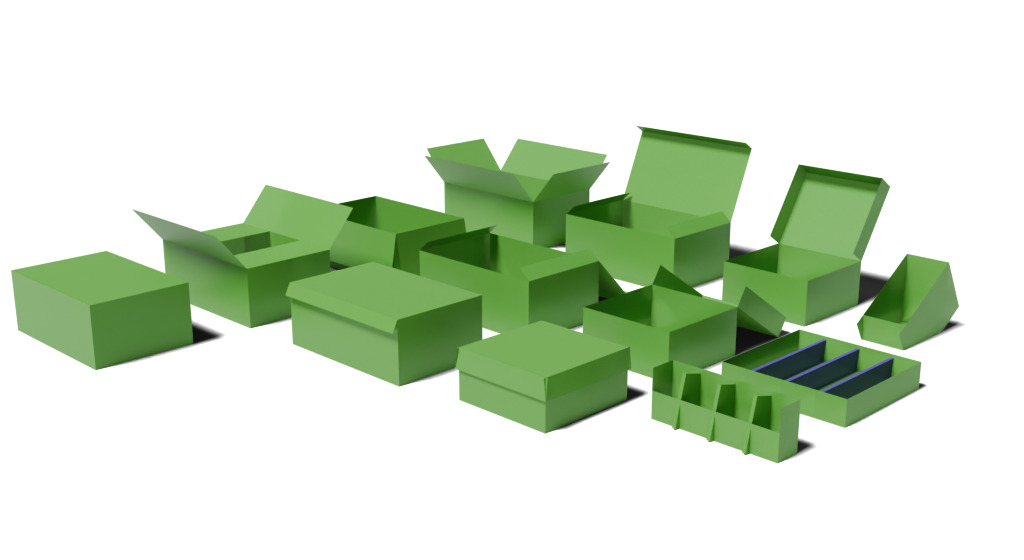}}\\
        \scriptsize $\num{1e-2}$ Bounding Volume & \scriptsize $\num{1e-4}$ Bounding Volume \\
        25 Boxes, 10 Prisms, $\num{2.35e-3}$ & 107 Boxes, $\num{1.21e-7}$ \\
    \end{tabular}
    \caption{Our approach can be controlled using the maximum allowed volume added when merging two primitives. If merging two primitives together exceeds some (See label above images for this example's values) fraction of the input mesh's axis-aligned bounding box's volume, merging is forbidden. This allows closer adherence to the original mesh. \ccby jellystuff.}
    \label{fig:variable_excess_volume}
    %\Description{A collection of a boxes at various states of opening, decomposed by our approach. The leftmost image is the input, the 2nd from left is a coarse set of boxes which do not preserve any of the openings. The 3rd from the left preserves some larger openings. The final image is with all openings preserved, but with many smaller primitives.}
    \vspace{-1em}
\end{figure}

\paragraph*{Limiting Excess Volume}
To control how close the collection of primitives should adhere to the the input mesh, users can control the output mesh's adherence to the original using a maximal excess volume.
In Fig.~\ref{fig:variable_excess_volume}, the input mesh is a collection of cardboard boxes with flaps in various positions. By limiting the excess volume introduced per collapse, the output has higher geometric similarity, but more primitives. Depending on the user's goal, excess volume can be tuned to balance the trade-off between precision and efficiency.

\begin{SCfigure}
    \centering
    \includegraphics[width=0.5\linewidth]{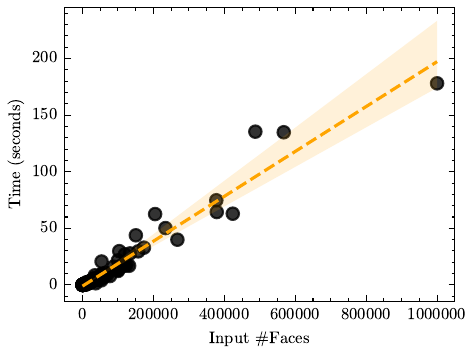}
    \caption{The wall-clock runtime of our implementation scales linearly with the input mesh's number of faces, which allows our approach to scale, processing meshes with millions of faces. In the worst case, performance may be $O(|V|^2)$ if one primitive contains nearly all vertices, but we never observe this in practice.}
    \label{fig:decomposition_time}
    %\Description{A scatter plot of our approach comparing the number of faces to time in seconds, with a regression line. Our approach appears linear.}
\end{SCfigure}

\paragraph*{Timing}
Our implementation efficiently decomposes meshes and even though the theoretical time complexity of collapse is at least O($n\log n$) due to reordering the priority queue the wall-clock time scales linearly with the size of the input mesh. We visualize the time to decompose an input mesh in Fig.~\ref{fig:decomposition_time}, with exact time per mesh in Tab.~\ref{tab:complete_results}. For small meshes, our approach completes in under a second. This is acceptable as the approach is designed to process meshes offline. In the worst case, our algorithm may be slow if one primitive repeatedly merges and no others do, as it will have many neighbors and enclose a large number of points with $O(|V|^2)$ complexity. In practice this case does not appear, since it would lead to a high volume error.

\section{Ablations}

In the following sections, we ablate some of our design choices.

\begin{figure}[!htb]
    \centering
    \setlength{\tabcolsep}{0em}
    \begin{tabular}{c c c}
        % \includegraphics[width=0.33\linewidth]{ablations/input_melon_cubes} &
        % \includegraphics[width=0.33\linewidth]{ablations/edge_weight_0} &
        % \includegraphics[width=0.33\linewidth]{ablations/edge_weight_1e-2} \\
        %  Input & Weight = 0 & Weight = $\num{1e-2}$ \\
        %  $|F| = 45016$ & 35 Boxes, 3 Cylinders & 38 Boxes \\
        Input & Weight = 0 & Weight = $\num{1e-2}$ \\
        \includegraphics[width=0.33\linewidth]{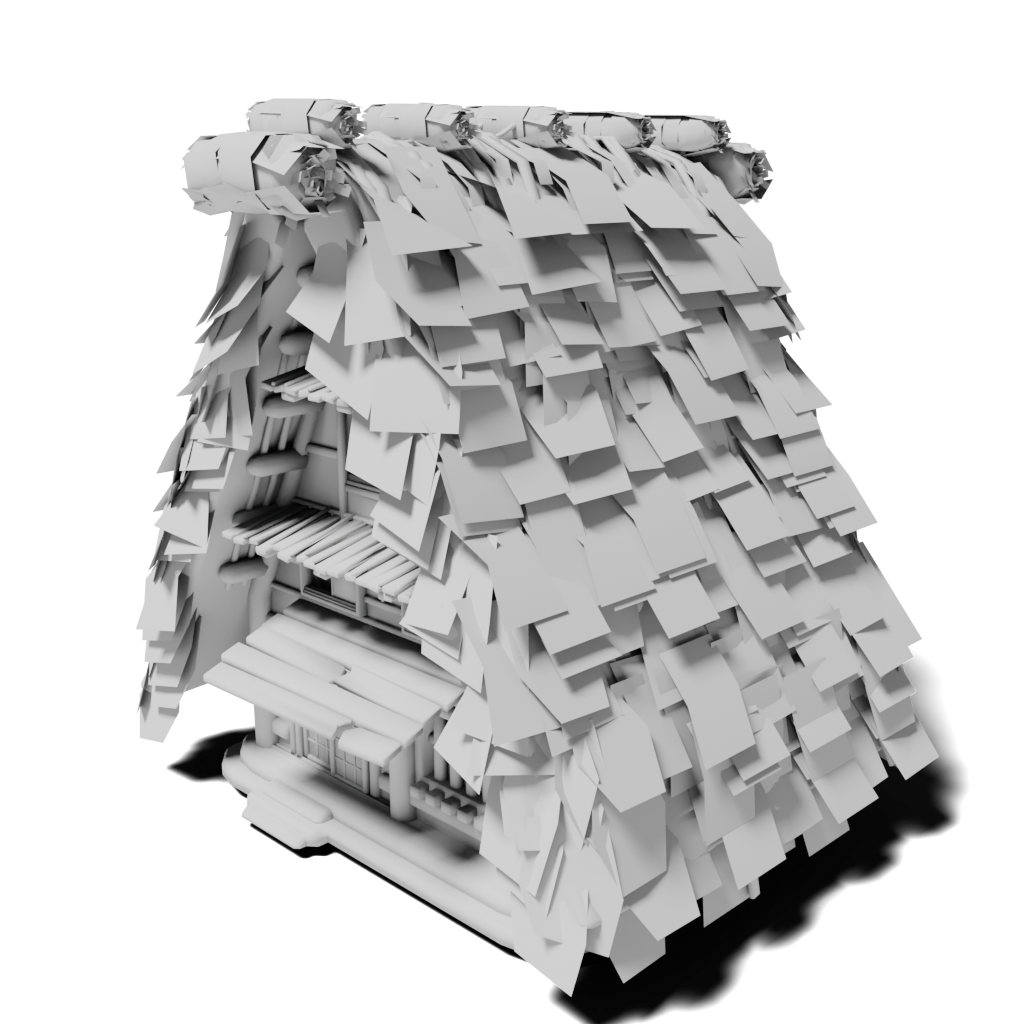} &
        \includegraphics[width=0.33\linewidth]{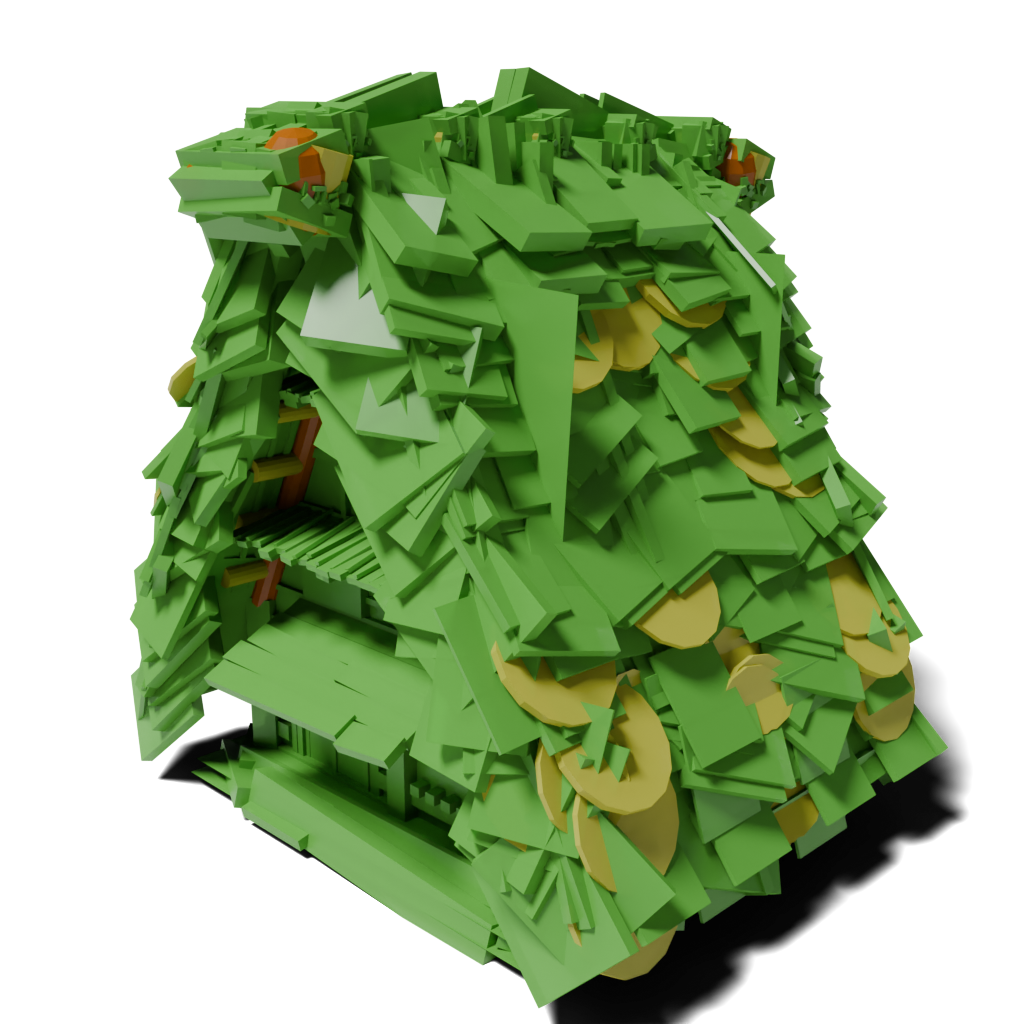} &
        \includegraphics[width=0.33\linewidth]{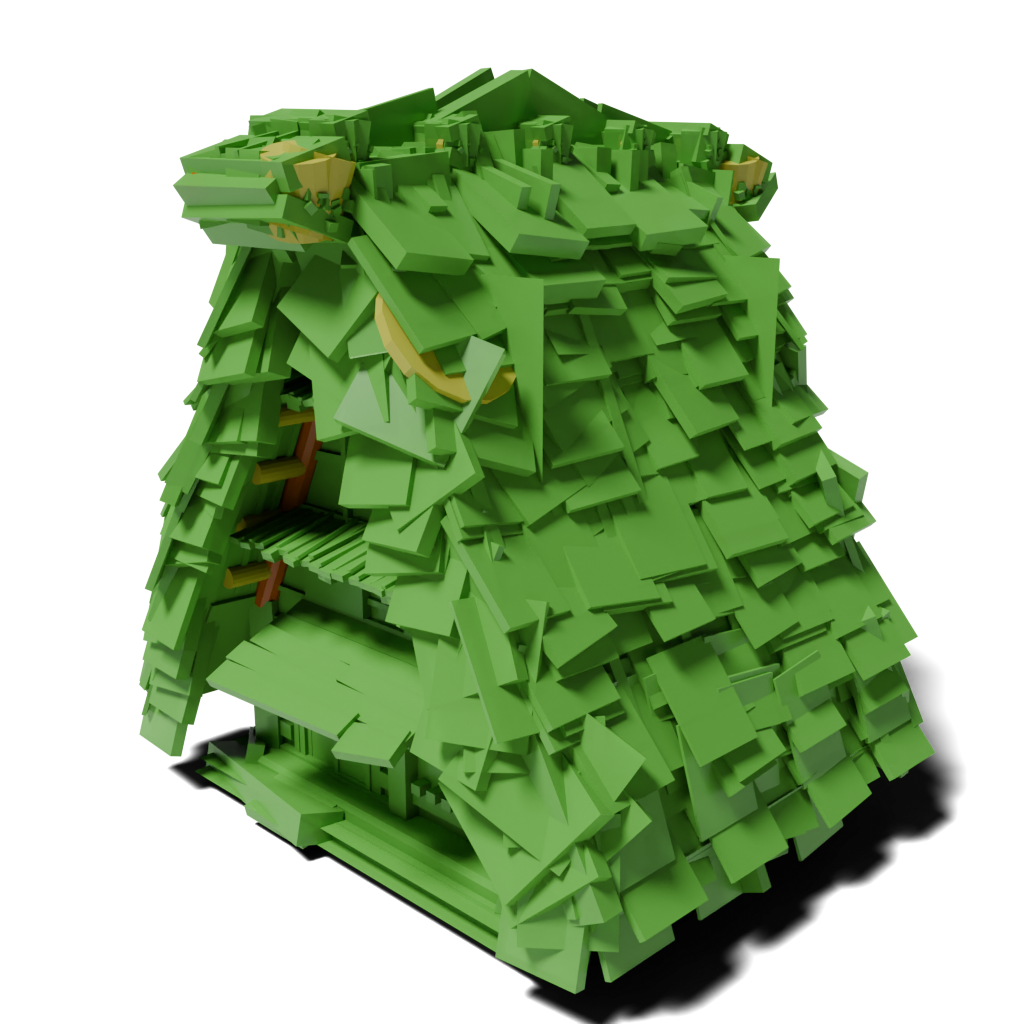} \\
        $|F| = 15965$ & \tiny 1272 Boxes, 12 Cap, 116 Cyl, 2 Sph & \tiny 1327 Boxes, 12 Cap, 53 Cyl, 2 Sph \\
        \scriptsize $\frac{\text{Haus./Chamfer New to Input}}{\lVert\text{Bounding Box Diag}\rVert_2}^\downarrow$ & \scriptsize $\num{6.33e-2}/\num{5.95e-3}$ & \scriptsize $\num{5.54e-2}/\num{4.38e-3}$ \\
    \end{tabular}
    \caption{Comparison of our approach with and without tangent weights for each input face's quadric. Without tangents, primitives may not properly align with planar faces, resulting in worse approximations as shown by the cylinders. With an added quadric in the longest edge's direction, primitives align more closely with each face. \ccby Lokomoto.
    }
    \label{fig:ablate-tangent-quadric}
    %\Description{A triangular house with a large number of quad faces floating on the side of it, representing the hay that is used as a roof. The middle column has some of these represented as a cylinder, and the right column most are represented by boxes.}
\end{figure}

\paragraph*{Coplanar Vertices Tangent Quadric}\label{sec:ablate-tangent-quadric}
To demonstrate adding a tangent weight from Sec.~\ref{sec:coplanar-vertices}, we visualize outputs with and without tangent edge weights in Fig.~\ref{fig:ablate-tangent-quadric}. Many coplanar faces in the input are converted to cylinders due to ambiguity with no weight, but with edge-weights these primitives more closely align with the input mesh's faces, allowing them to be tighter OBBs.

\begin{figure}[!htb]
    \centering
    \setlength{\tabcolsep}{-1em}
    \begin{tabular}{c c c}
        Input & Vertex Dedup. & No Dedup. \\
        \includegraphics[width=0.35\linewidth]{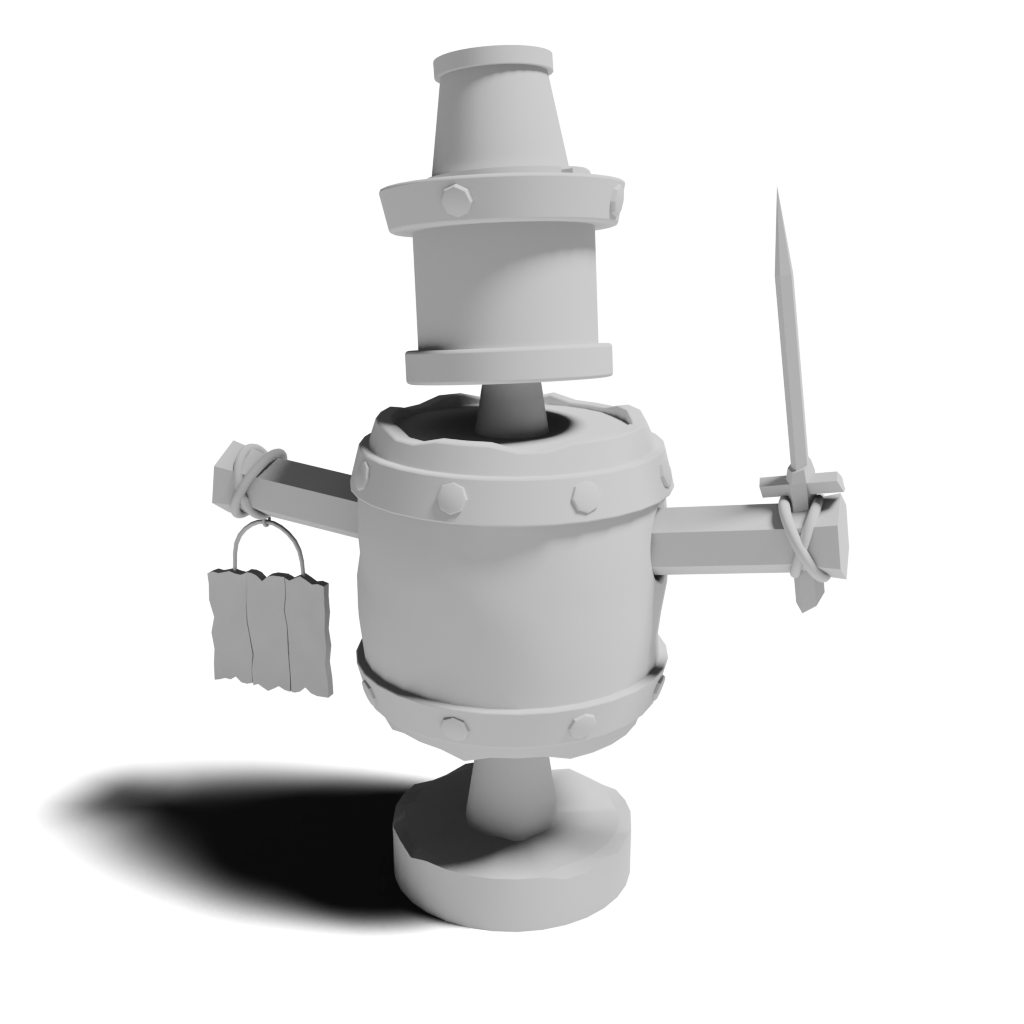} &
        \includegraphics[width=0.35\linewidth]{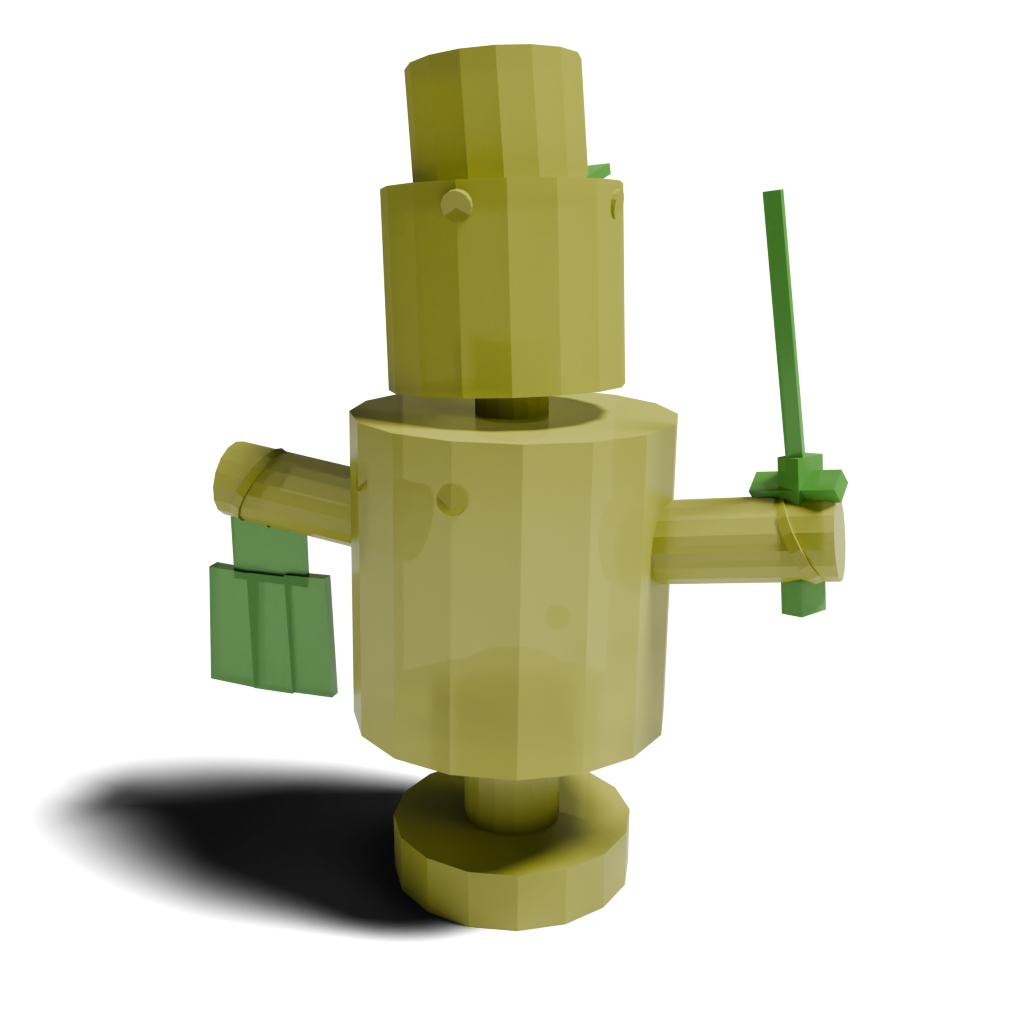} &
        \includegraphics[width=0.35\linewidth]{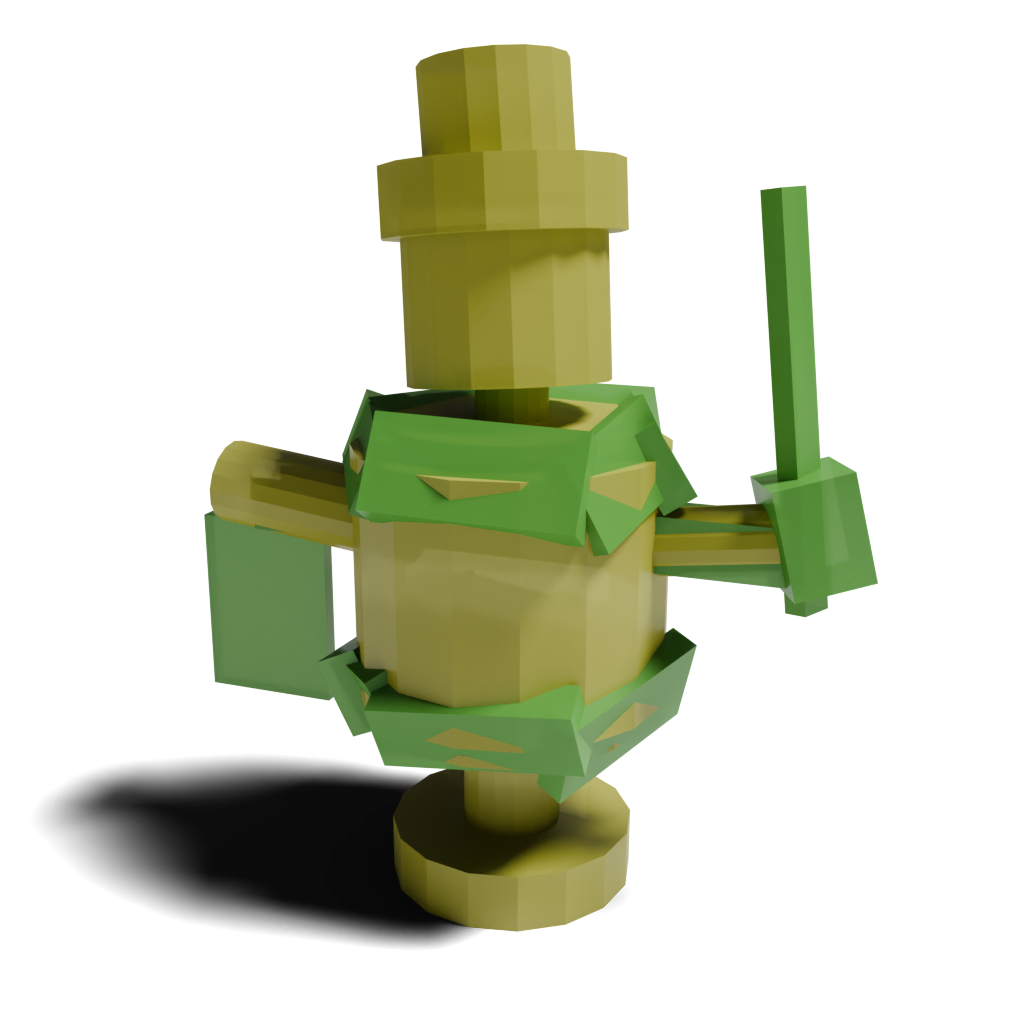} \\
        $|F| = 20816$ & 8 Boxes, 19 Cyl & 18 Boxes, 12 Cyl \\
    \end{tabular}
    \caption{Comparison of our approach with and without vertex deduplication and the same number of target primitives. The deduplicated model has additional primitives enclosed, leading to a lower final number of primitives. Deduplicating vertices changes the topology of the input mesh, leading to different results. \ccby ``FinBeenWhere?''.}
    \label{fig:ablate-vertex-merging}
    %\Description{A training dummy with a sword in one hand a sign in the other. 2nd from the left is our approach with vertices deduplicated, which is a coarse approximation of the input. In the final column is our approach without vertices deduplicated, which has many more primitives and in some regions more accurately resembles the input.}
\end{figure}

\paragraph*{Vertex Deduplication}\label{sec:merge-vertices}
We test our approach with and without vertex deduplication in Fig.~\ref{fig:ablate-vertex-merging}. Deduplicating vertices can change results since it increases possible edge collapses. Without deduplicating, components such as barrel hoops are not merged with the body, but other regions such as the hat are preserved differently. For most cases, we find exactly overlapped vertex merging improves the output quality of the result.

\begin{figure}[!htb]
    \centering
    \setlength{\tabcolsep}{0.125em}
    \begin{tabular}{c c c}
        Input & Vertex Merging & Face Merging \\
        \includegraphics[width=0.28\linewidth]{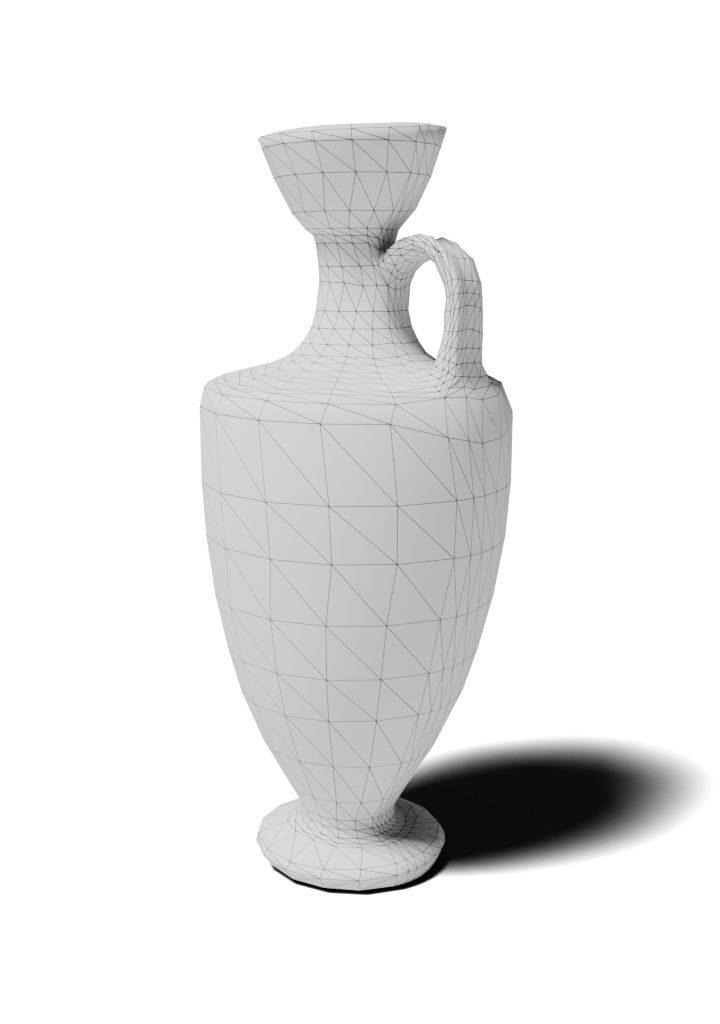} &
        \includegraphics[width=0.28\linewidth]{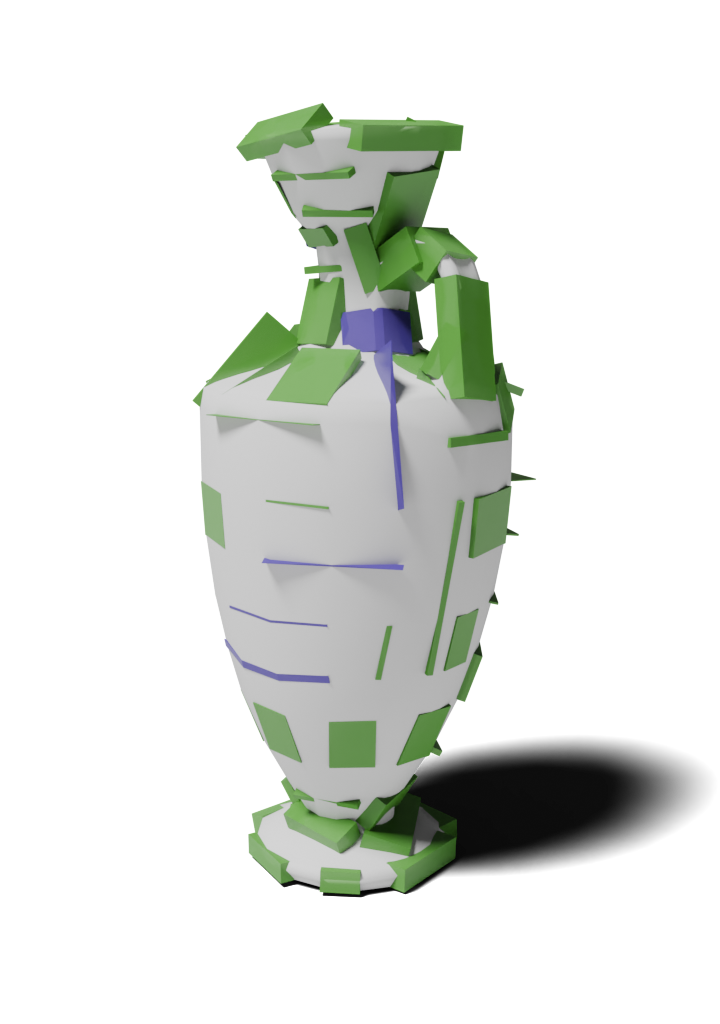} &
        \includegraphics[width=0.28\linewidth]{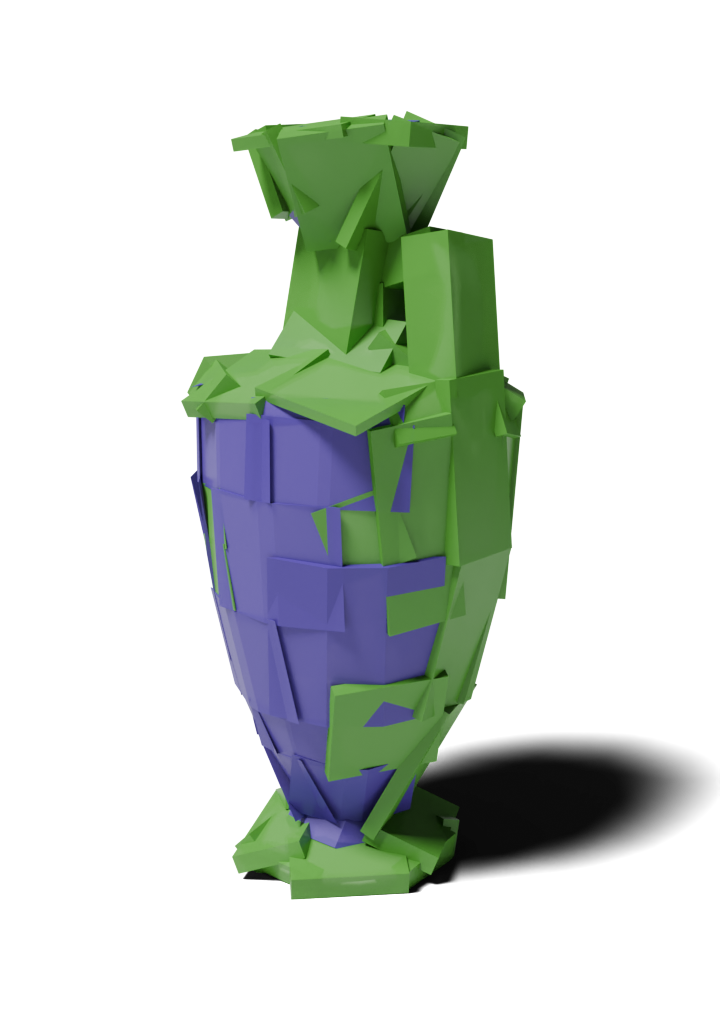} \\
        |F| = 1075 & {\scriptsize 86 Boxes, 1 Cyl, 13 Prisms} & {\scriptsize 62 Boxes, 3 Cyl, 35 Prisms} \\
    \end{tabular}
    \caption{Our approach using per-vertex merging and per-face merging. Per-vertex merging leaves large portions of the input (shown in white) uncovered, whereas merging faces covers the whole surface. \ccbync Global Digital Heritage and GDH-Afrika.}
    \label{fig:ablate-vertex-face-merging}
    %\Description{Ablation of our approach for merging vertices together versus merging faces together. Left is the input mesh. 2nd from left is vertex merging which leaves large regions of the input mesh peaking through. Right is face-merging, which looks like it has some regions peaking through but this is due to quantization of cylinders.}
    \vspace{0em}
\end{figure}

\paragraph*{Vertex vs. Face Merging}\label{sec:ablate-vertex-face-merging}
To show each vertex must correspond to multiple primitives, we show our approach on a lekythos with primitives per vertex or face in Fig.~\ref{fig:ablate-vertex-face-merging}. Vertex merging leaves much of the input surface uncovered, unsuitable for use as a collider, showing the importance of using faces to represent primitives.

\begin{figure}[!htb]
    \setlength{\tabcolsep}{2pt}
    \renewcommand{\arraystretch}{0.2}
    \centering
    \begin{tabular}{c c c}
        Input & Ours & CoACD \\
        \includegraphics[width=0.3\linewidth]{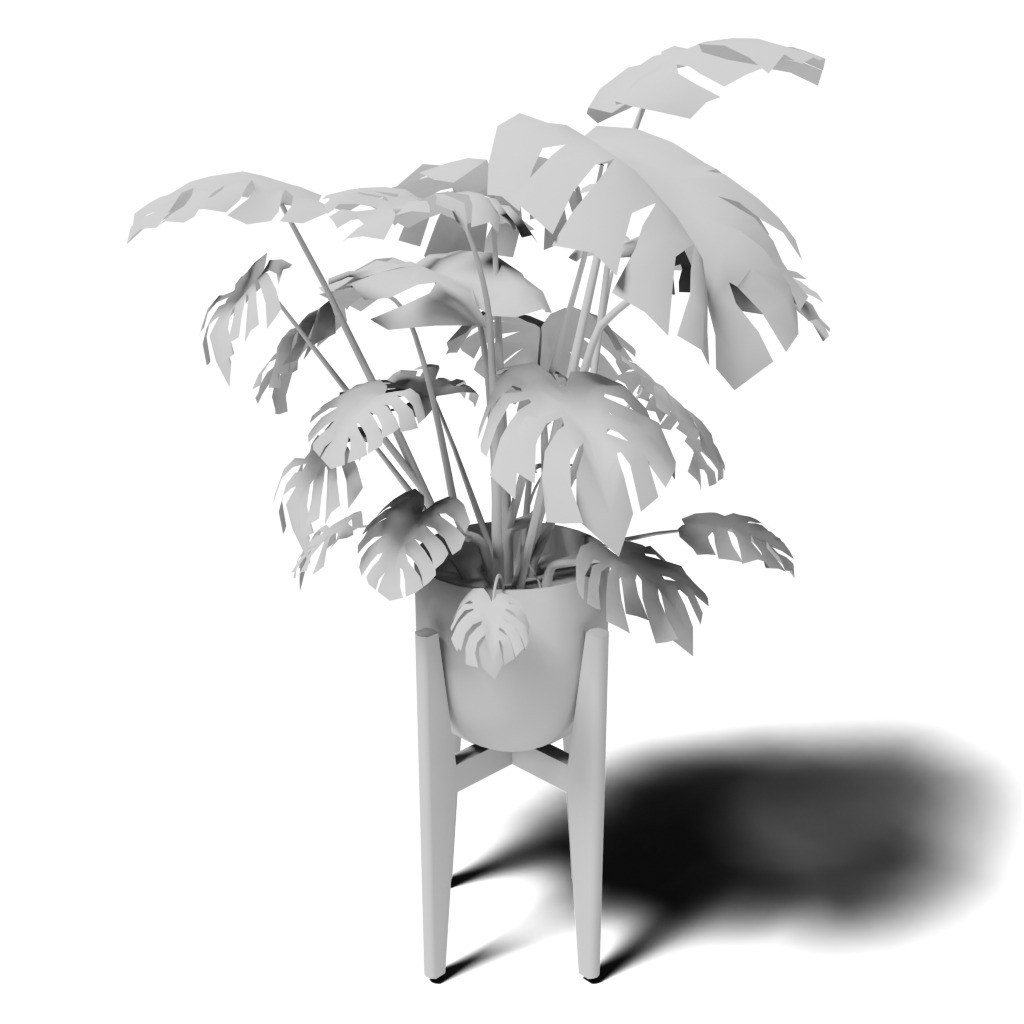} &
        \includegraphics[width=0.3\linewidth]{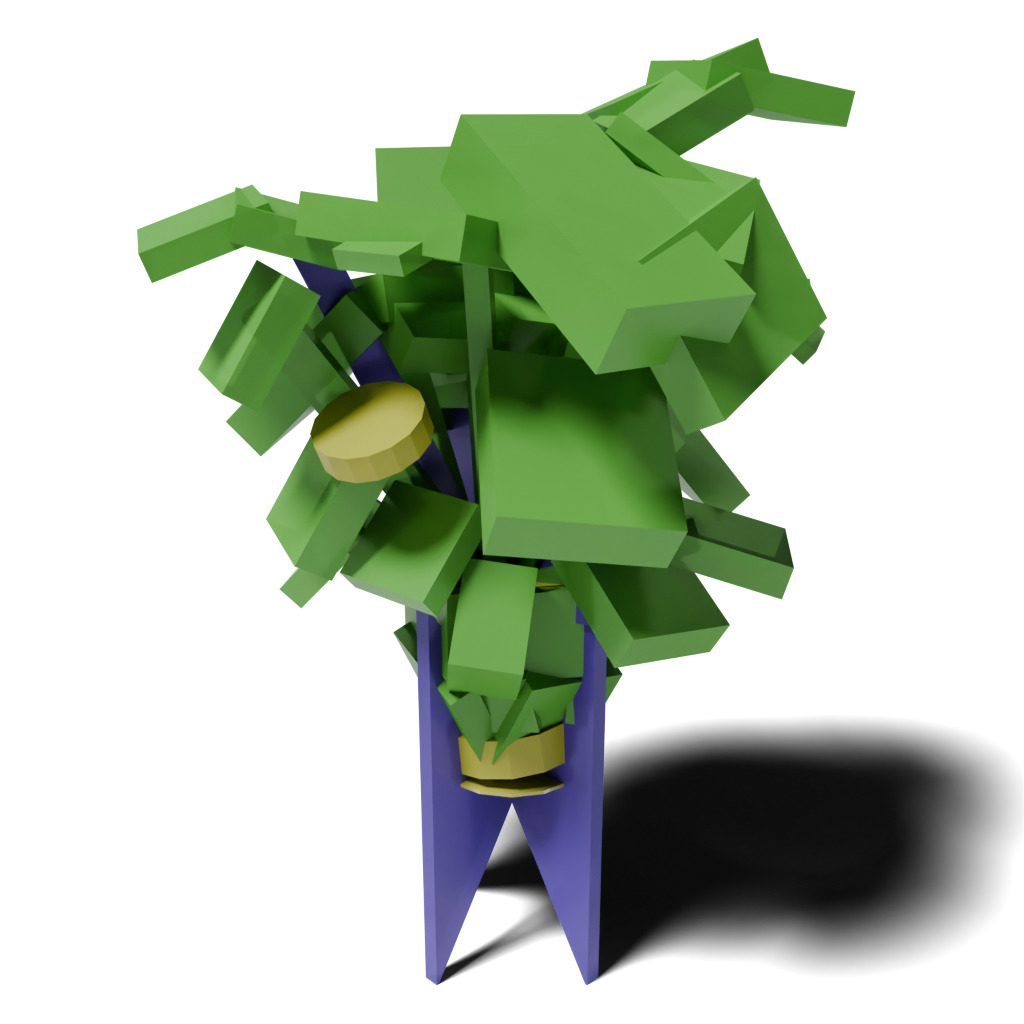} &
        \includegraphics[width=0.3\linewidth]{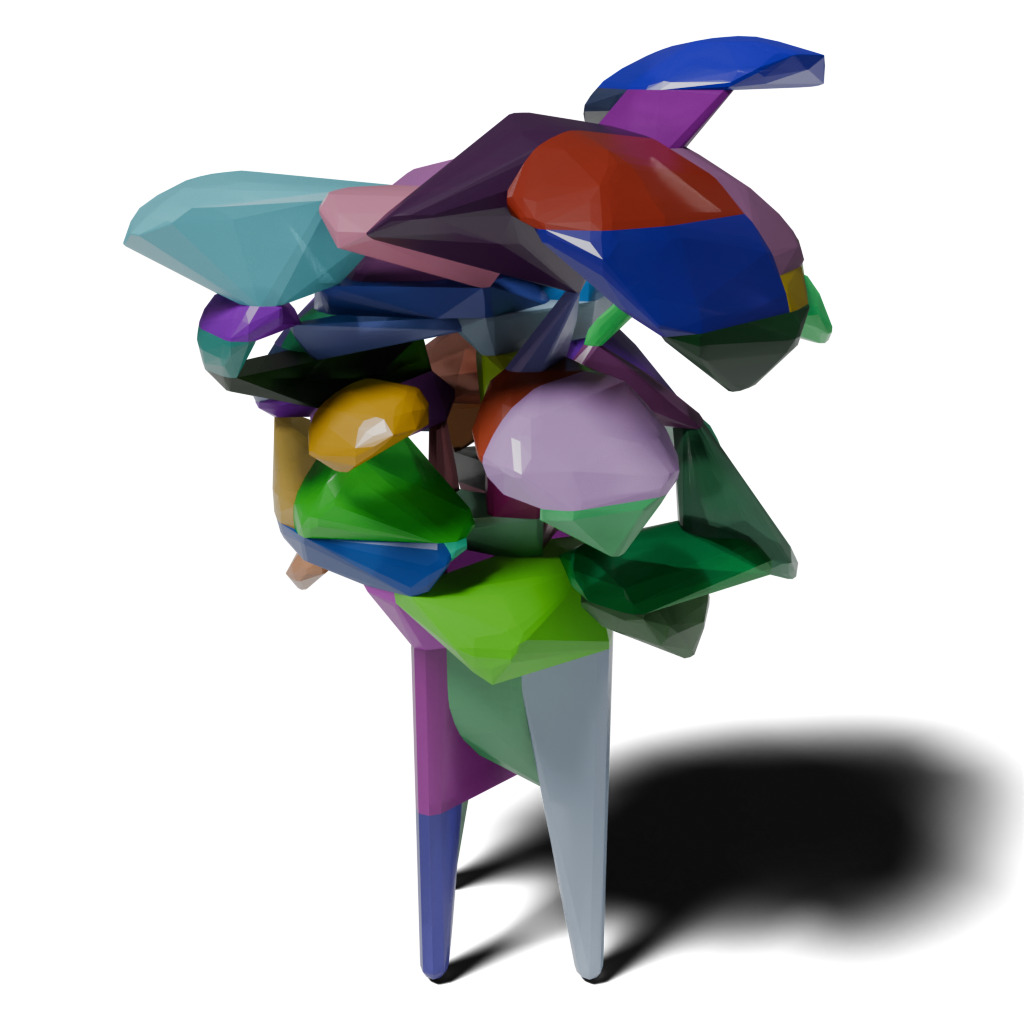} \\
        |F| = 4861 & 70 Prim. {\tiny (57 OBB, 7 Cyl, 6 Prism)} & 70 Hull (|F| = 10462) \\
        \scriptsize $\frac{\text{Haus./Chamfer New to Input}}{\lVert\text{Bounding Box Diag}\rVert_2}^\downarrow$ & \scriptsize $\num{3.57e-2}/\num{1.26e-2}$ & \scriptsize $\num{5.53e-2}/\num{1.06e-2}$ \\
        \multicolumn{3}{c}{\includegraphics[width=0.9\linewidth]{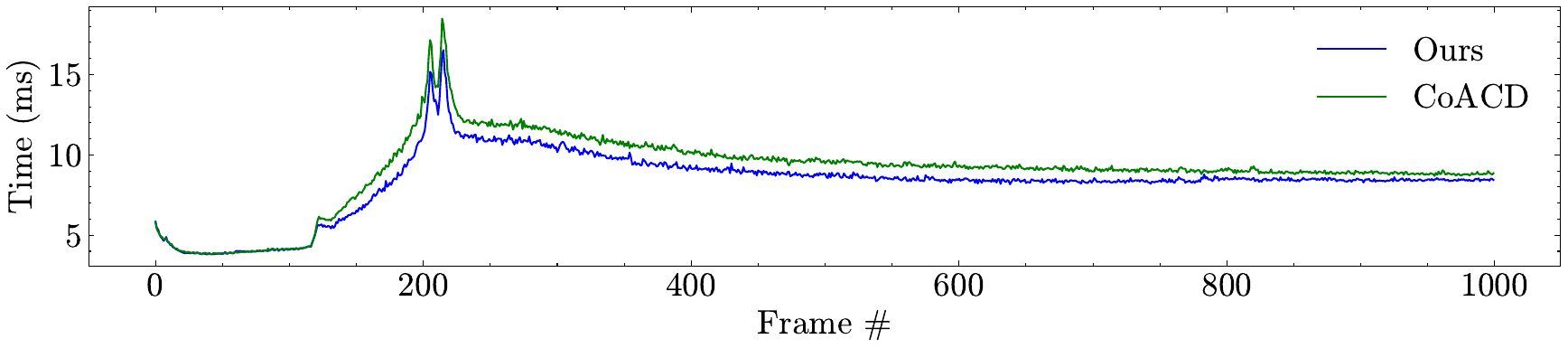}} \\
    \end{tabular}
    \caption{Our approach compared to CoACD with an identical number of primitives and convex hulls, and similar geometric similarity. Our approach has faster simulation than CoACD, as the collision with hulls depends on the number of vertices and faces. \ccby Giora.}
    \label{fig:ident_num_prim_hull}
    %\Description{3 potted plants, left is a gray mesh, center is boxes, cylinders, and prisms representing the mesh, and right is convex hulls of varying colors. Below, a plot where the blue line representing time of collision detection for primitives is below a blue line indicating the duration of CoACD.}
\end{figure}

\paragraph*{Comparing Primitive \& Hull Counts}\label{sec:ablate-hull-count}
We show hull and primitive counts do not reflect cost by comparing simulation with an equal number of primitives and hulls in Fig.~\ref{fig:ident_num_prim_hull}. Despite matching counts, primitives are faster in simulation, as the number of hulls masks the number of faces and vertices used in GJK.

\begin{figure}[!htb]
    \centering
    \renewcommand{\arraystretch}{0.5}
    \begin{tabular}{c c}
        \multicolumn{2}{c}{Comparison of Weighting Per Primitive Kind} \\
        Input & Our Weighting \\
        $|F| =  106464$ & {\scriptsize 221 Boxes, 11 Cap, 126 Cyl, 9 Prism} \\
        \scriptsize $\frac{\text{Haus./Chamfer New to Input}}{\lVert\text{Bounding Box Diag}\rVert_2}^\downarrow$ & $\num{9.62e-3}/\num{1.92e-3}$ \\
        \includegraphics[width=0.48\linewidth]{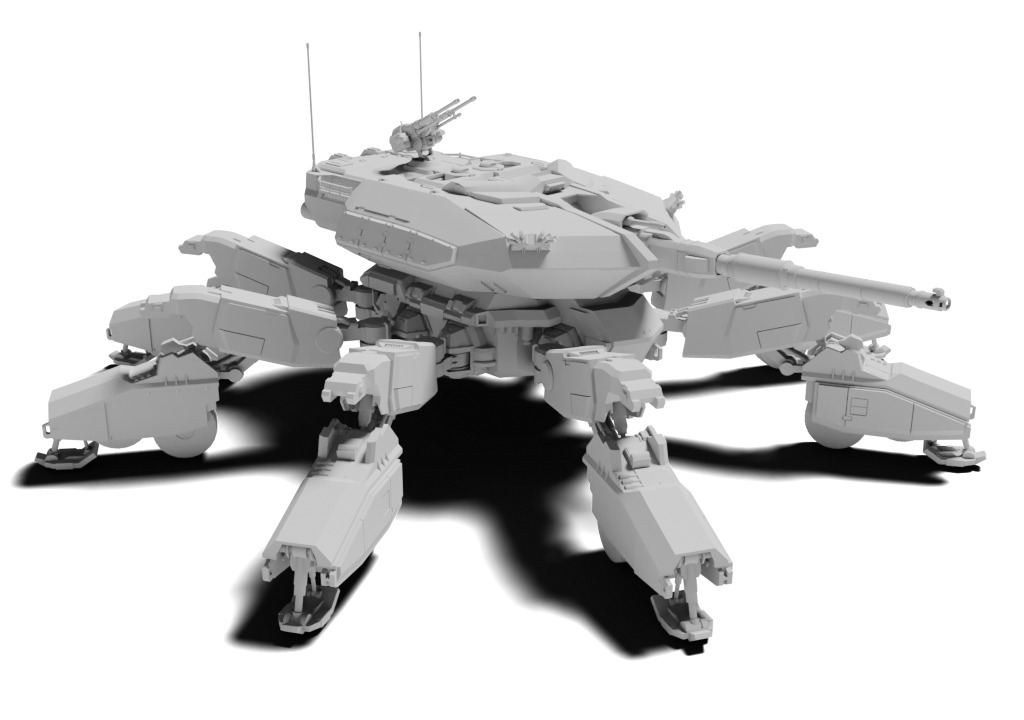}
        \put(-0.485\linewidth,-0.29\linewidth){\includegraphics[width=0.48\linewidth]{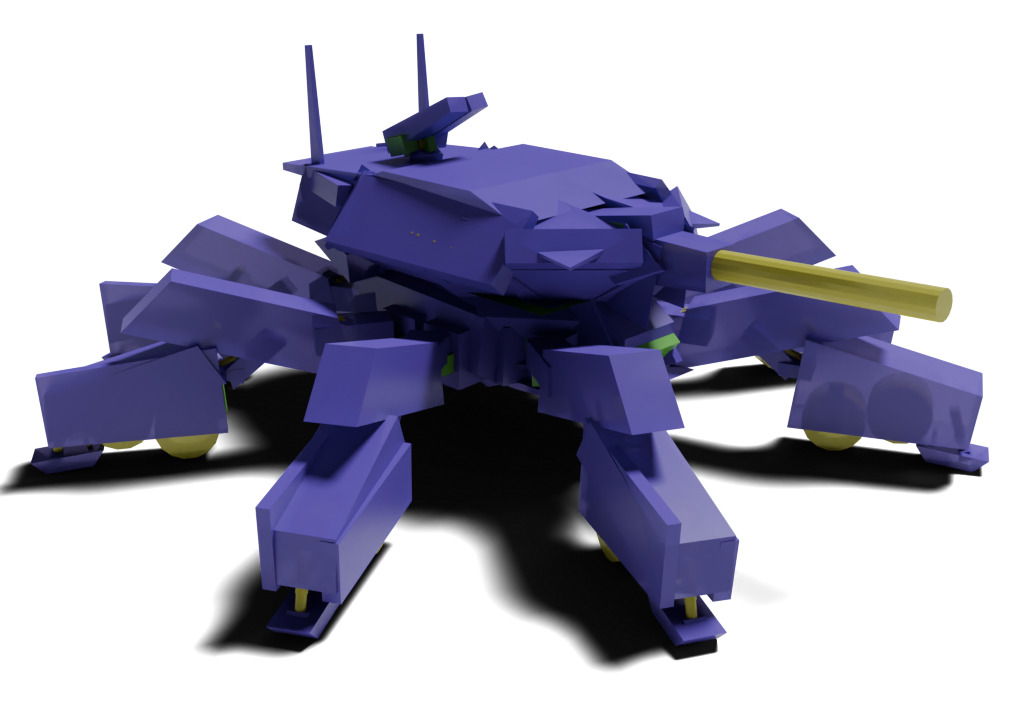}}
        &
        \includegraphics[width=0.48\linewidth]{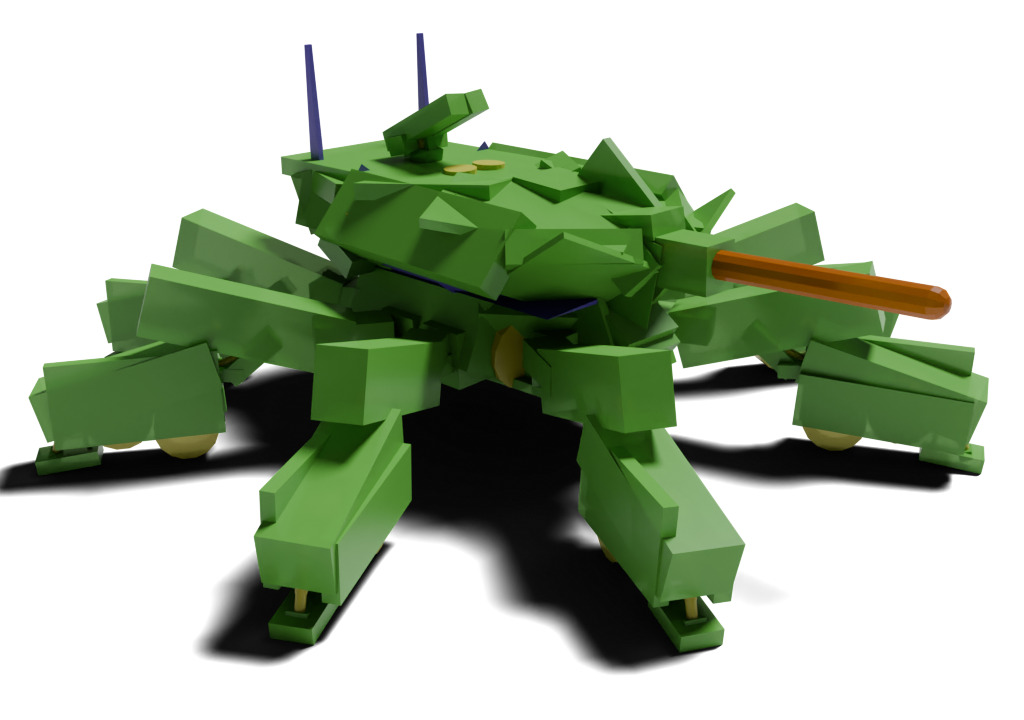} 
        \put(-0.485\linewidth, -0.29\linewidth){
        \includegraphics[width=0.48\linewidth]{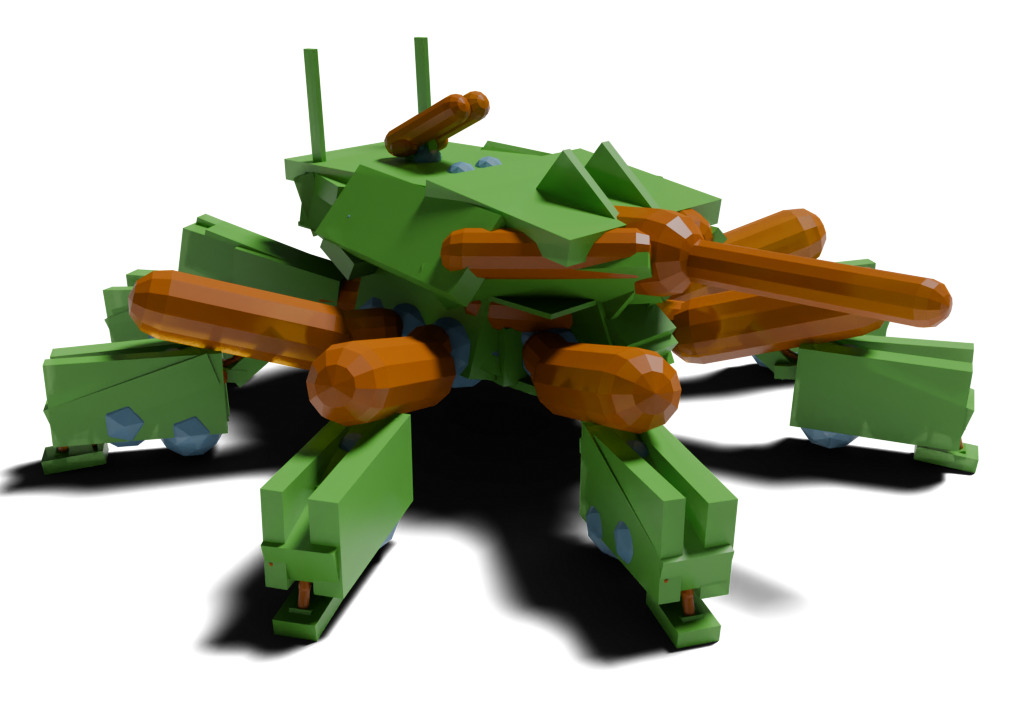}}
        \\
        Uniform(All = 1) & Sph(0.5)/Cap(0.75)/OBB(1) \\
        {\scriptsize 25 Boxes, 139 Cyl, 201 Prism} & {\scriptsize 147 Boxes, 142 Cap, 66 Spheres} \\
        $\num{7.33e-3}/\num{2.87e-3}$ & $\num{1.17e-2}/\num{4.01e-3}$ \\
        \multicolumn{2}{c}{\includegraphics[width=0.9\linewidth]{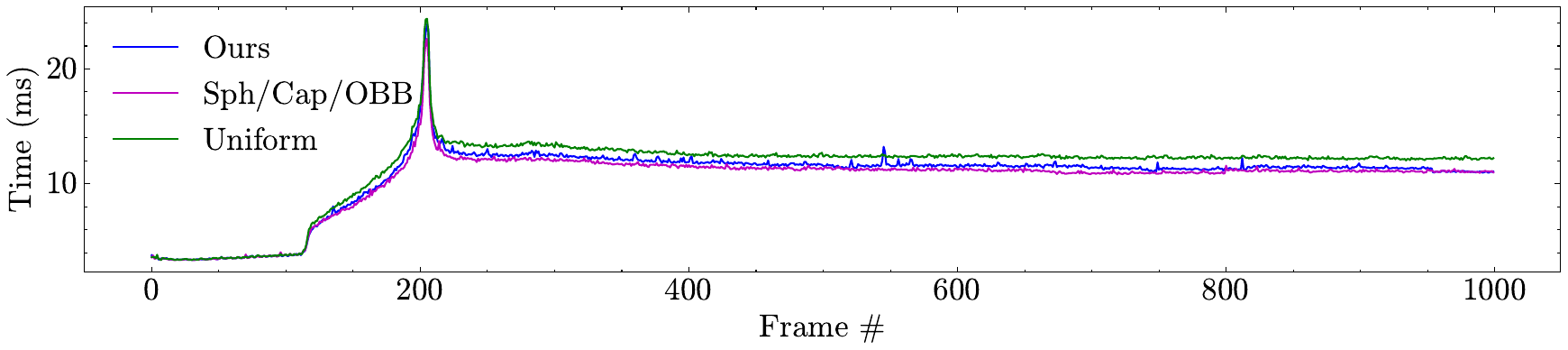}}
    \end{tabular}
    \caption{Our approach with different sets of weights. Uniform weighting outputs more isosceles trapezoidal prisms, reducing performance in simulation as shown in the plot. We also compare a weighting which heavily favors spheres (0.5), capsules (0.75), and OBBs (1) with everything else disabled, and observe that this can improve simulation performance but with less geometric similarity. OBBs are shown in green, capsules in orange, cylinders in yellow, prisms in blue, and spheres in light blue. \ccby Adoni.}
    \label{fig:ablate-primitive-cost}
    %\Description{Different approximations of an input spider tank.}
    \vspace{0em}
\end{figure}

\paragraph*{Primitive Variant Cost Ablation}\label{sec:ablate-primitive-costs}
We ablate our set of costs for each variant of primitive, by comparing our weighting to uniform weighting (all weights set to 1) and a weighting with only spheres, capsules, and OBBs in Fig.~\ref{fig:ablate-primitive-cost}. When using uniform weighting, many prisms are output. We also test a weighting that only uses the common primitives, where spheres, capsules and OBBs have weights of 0.5, 0.75, and 1.5 respectively. The output of this weighting is also reasonable, but is coarser in some regions like the legs, compared to our weighting. We plot the simulation time for each weighting, to highlight the trade-off of geometric similarity and simulation performance. As the uniform weighting has prisms that are converted to hulls, it is slower than our weighting. On the other hand, using only directly supported primitives leads to faster simulation time, but with less geometric similarity. Our weighting balances the two approaches.

\begin{figure}[!h]
    \setlength{\tabcolsep}{0em}
    \renewcommand{\arraystretch}{0.6}
    \begin{tabular}{c c}
        Original & Ours \\
        \includegraphics[width=0.42\linewidth]{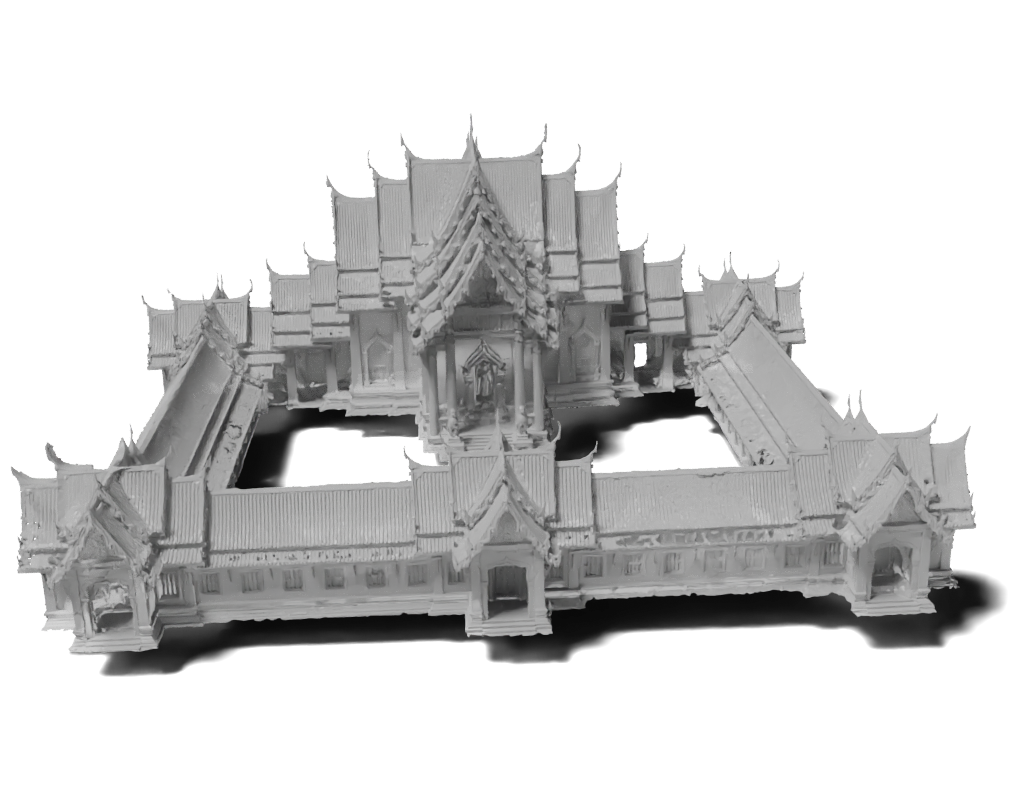}
        \put(-0.42\linewidth, -0.2\linewidth){\includegraphics[width=0.42\linewidth]{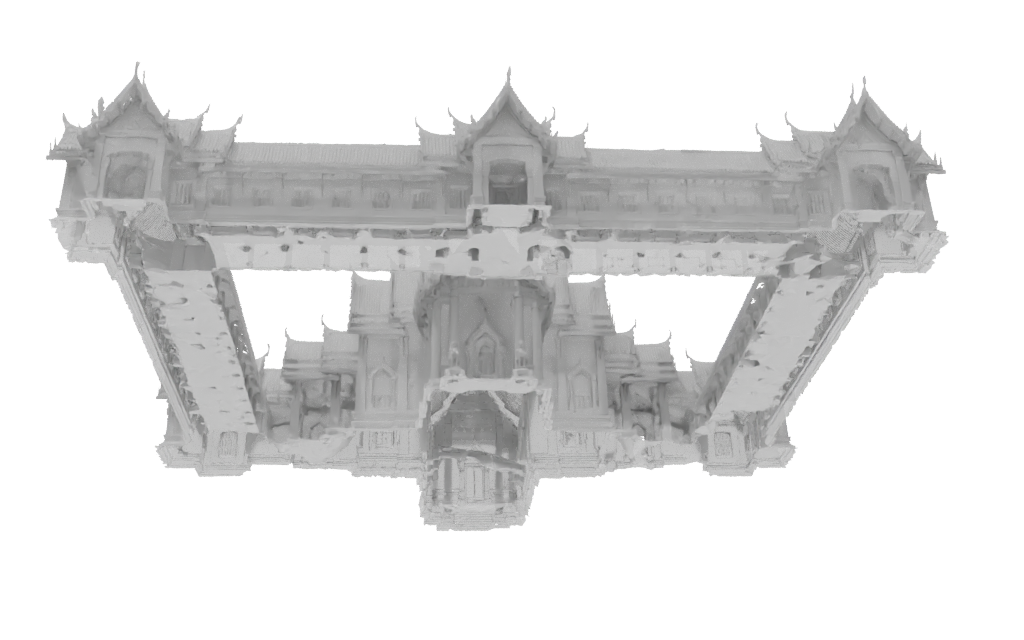}} &
        \includegraphics[width=0.42\linewidth]{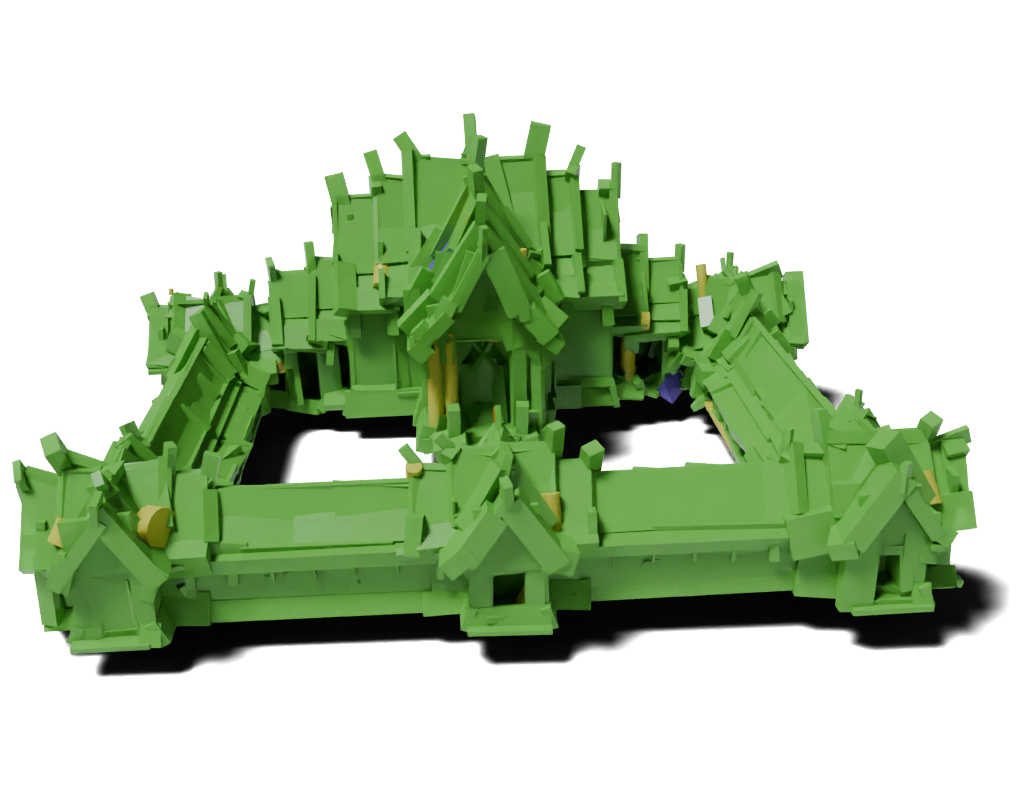}
        \put(-0.42\linewidth, -0.2\linewidth){\includegraphics[width=0.42\linewidth]{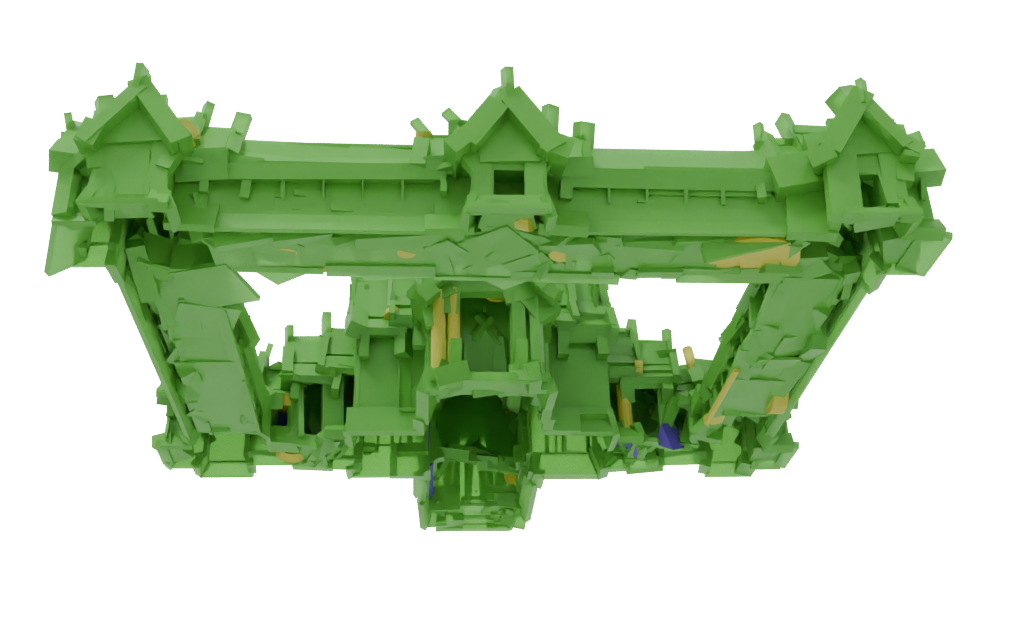}} \\
        $|F| = 999956$ & \tiny 1620 Boxes, 55 Cyl, 17 Cap, 8 Prisms, 6 Sph \\
        $\frac{\text{Haus./Cham. New to Input}}{\lVert\text{Bounding Box Diag}\rVert_2}^\downarrow$ & $\num{1.64e-2}/\num{1.95e-3}$ \\
    \end{tabular}
    \caption{Our approach on a noisy scanned 3D model with an open bottom. Despite the input's density, our approach adheres to its structure including small details such as columns and chofa. \ccbyncsa GoodScan 3D.}
    \label{fig:ours_scanned_data}
    %\Description{A scan of a Buddhist temple decomposed into a number of primitives, with a top and bottom view. The input mesh on the left is dense, and consists of a walled area with one wall containing a large temple building. Our approach preserves the coarse shape, and the bottom also contains holes which the original mesh had.}
\end{figure}

\paragraph*{Scanned Data}
To demonstrate our approach on noisy data, we benchmark our approach on a 3D scan of the Wat Benchamabophit in Fig.~\ref{fig:ours_scanned_data}. This scan contains 999956 faces, with holes on the bottom. Our approach still accurately decomposes the input mesh, and preserves holes even with the high density, while maintaining small features such as columns of the inner gate and chofa on the roof.

\begin{figure}
    \centering
    \setlength{\tabcolsep}{0.1pt}
    \begin{tabular}{c c c c}
        Input & Eq.~\ref{eq:exact_cost} & Eq.~\ref{eq:cost_with_isect} (Dense) & Eq.~\ref{eq:cost_with_isect} (Sparse) \\
        
        \includegraphics[width=0.24\linewidth]{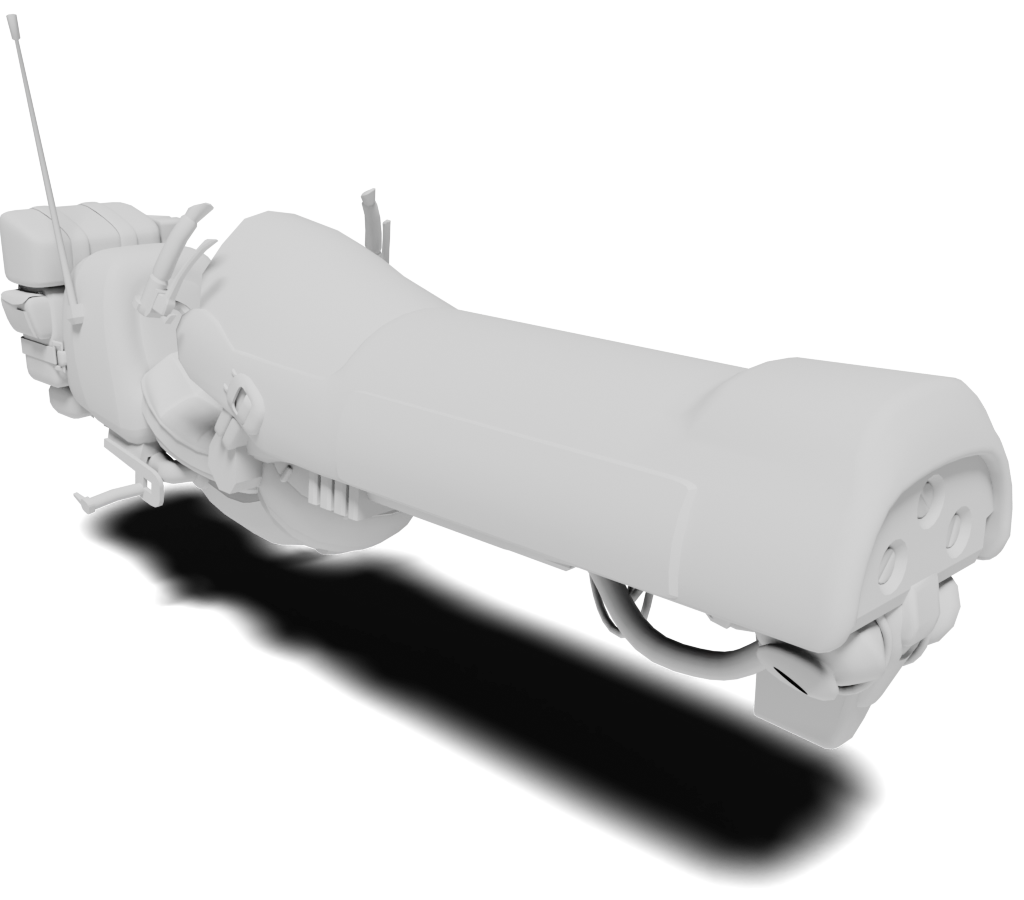} &
        \includegraphics[width=0.24\linewidth]{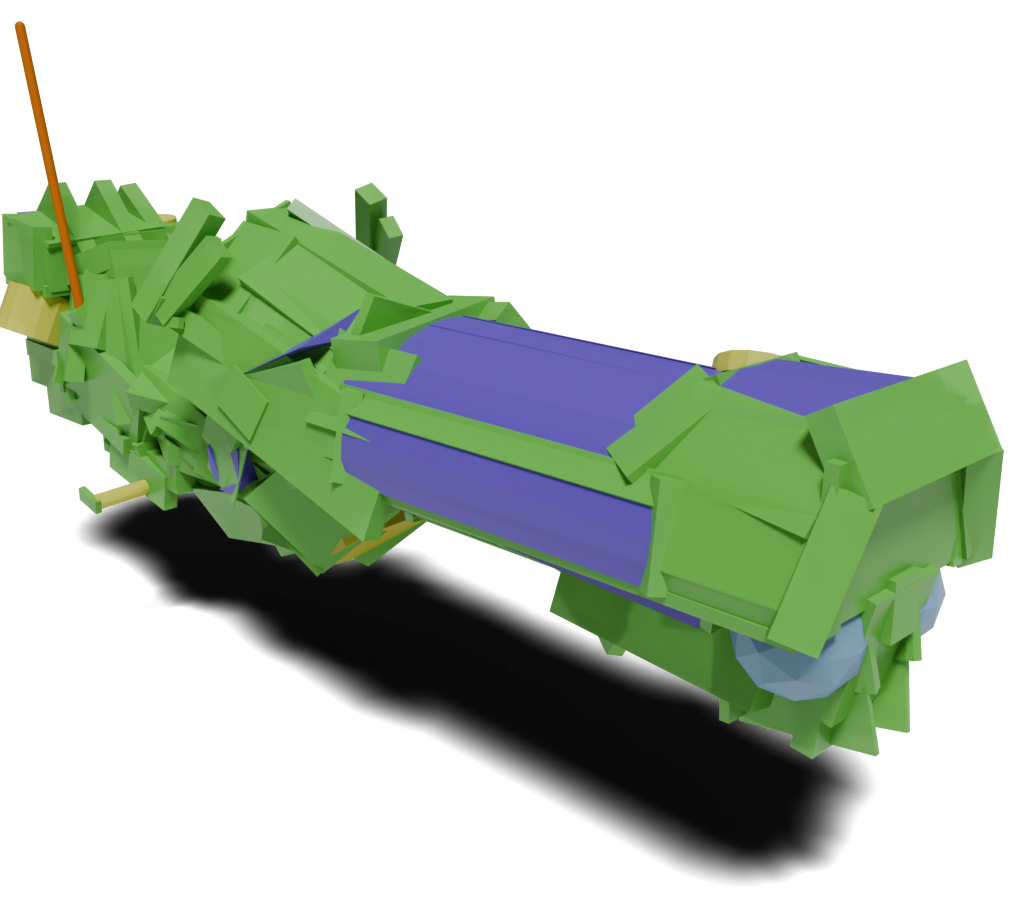} &
        \includegraphics[width=0.24\linewidth]{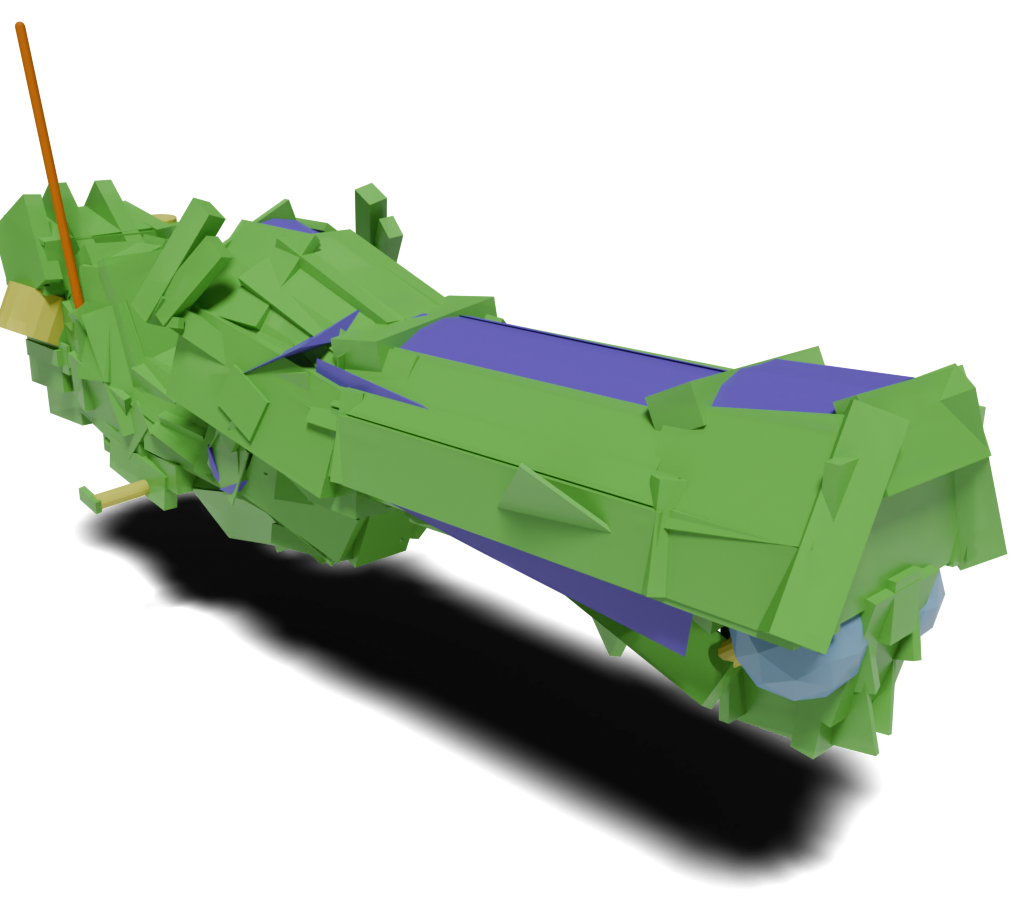} &
        \includegraphics[width=0.24\linewidth]{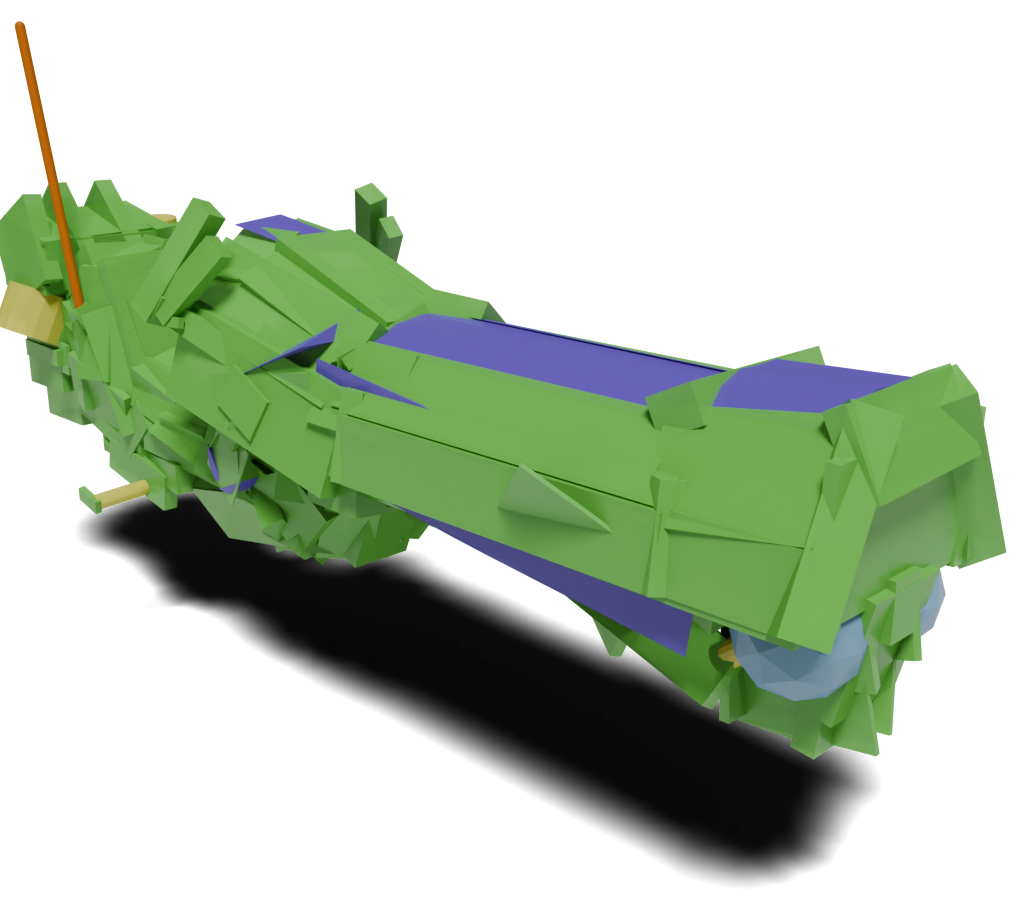} \\
        {\scriptsize $\frac{\text{$\frac{\text{Haus}}{\text{Cham}}$ New to Input}}{\lVert\text{Bounding Box Diag}\rVert_2}^\downarrow$} & \tiny $\frac{\num{2.43e-2}}{\num{5.031e-3}}$ & \tiny $\frac{\num{2.39e-2}}{\num{5.019e-3}}$ &
        \tiny $\frac{\num{2.44e-2}}{\num{5.028e-3}}$ \\
        \\
        |F| = 11926 & \tiny \shortstack{235 Boxes, 21 Cyl \\ 1 Cap, 26 Prism, 3 Sph} & \tiny \shortstack{238 Boxes, 20 Cyl \\ 1 Cap, 21 Prism, 3 Sph} & \tiny \shortstack{238 Boxes, 20 Cyl \\ 1 Cap, 22 Prism, 3 Sph} \\
        Time (sec) & 1.29 & 523.53 & 34.045 \\
    \end{tabular}
    \caption{We compare the cost function from Eq.~\ref{eq:exact_cost} with an approximation of Eq.~\ref{eq:cost_with_isect} using rejection sampling. For dense sampling, we sample 1,000,000 points for every pair of primitives, and for sparse sampling we sample 50,000 points. In practice there is little benefit using Eq.~\ref{eq:cost_with_isect}, with no visible improvement, and the cost of computation makes it intractable for real usage. \ccbync Tasha.Lime.}
    \label{fig:ablate-isect}
    %\Description{4 speeder bikes, left is the input mesh, which has a seat and a half-cylindrical tank in front of it. In the right 3 columns there are approximations of the speeder bike using primitives.}
    \vspace{-1em}
\end{figure}

\paragraph*{Intersecting Primitive Ablation}\label{sec:ablate-isect}
We ablate the cost function from Eq.~\ref{eq:exact_cost} versus Eq.~\ref{eq:cost_with_isect} in Fig.~\ref{fig:ablate-isect}, by approximating the intersection's volume using dense rejection sampling. We sample points from inside the smaller primitive, and estimate the intersection's volume as the fraction of points contained in the other primitive times the smaller primitive's volume. We do not see significant gains from Eq.~\ref{eq:cost_with_isect}, but sampling even with a small number of points is $30\times$ slower. Since there is little benefit with large cost, we do not use Eq.~\ref{eq:cost_with_isect}.

\section{Discussion}

Our approach decomposes an input mesh into a number of convex primitives, suitable for rigid-body collision, drastically and robustly simplifying complex real-world meshes. Convex primitive decomposition scales with mesh size and can handles millions of triangles.

While previous approaches focus on tight decompositions using convex hulls, they are impractical due to slower simulation and imprecision due to voxelization and planar cuts. We flip the top-down approach with bottom-up merging, removing the complexity of cutting while producing efficient decompositions.

Unlike prior approaches in primitive abstraction, we do not use deep learning, RANSAC, or differentiability, making our approach robust and fast for quick artistic iteration. Our approach can be built into existing tooling for artists to quickly generate collision objects which can then be directly manipulated.

\section{Limitations}

\begin{figure}
    \centering
    \setlength{\tabcolsep}{-8pt}
    \begin{tabular}{c c c}
        \multicolumn{3}{c}{Coarse Decomposition Failure Case} \\
        Input & Precise & Coarse \\
        \includegraphics[width=0.38\linewidth]{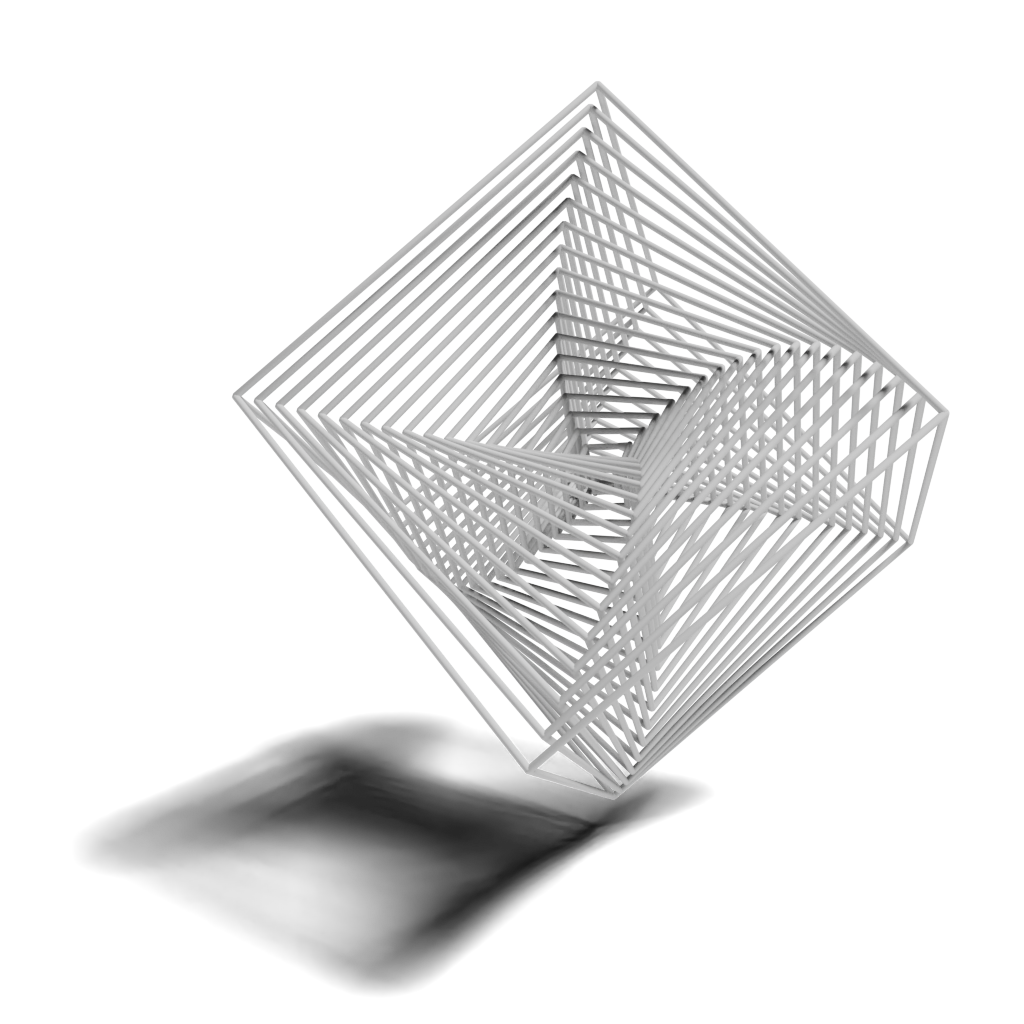} &
        \includegraphics[width=0.38\linewidth]{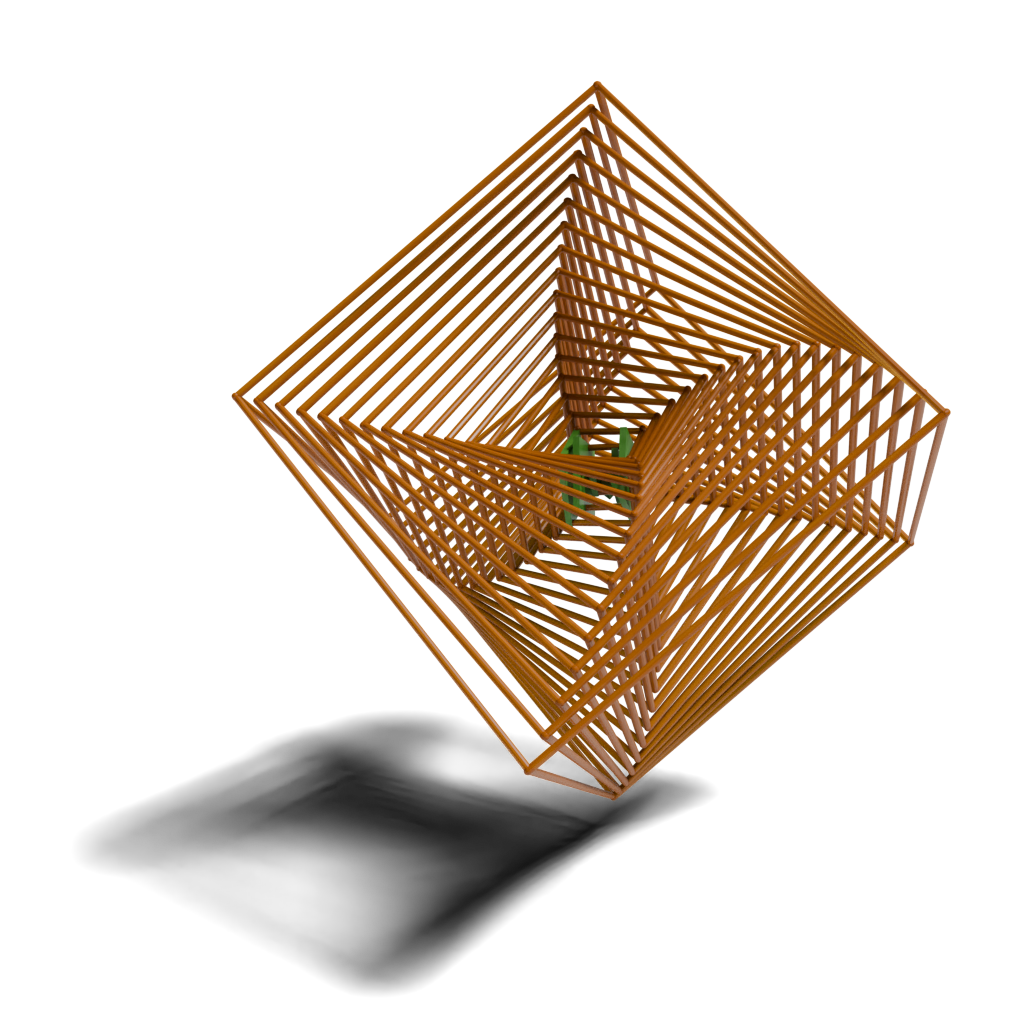} &
        \includegraphics[width=0.38\linewidth]{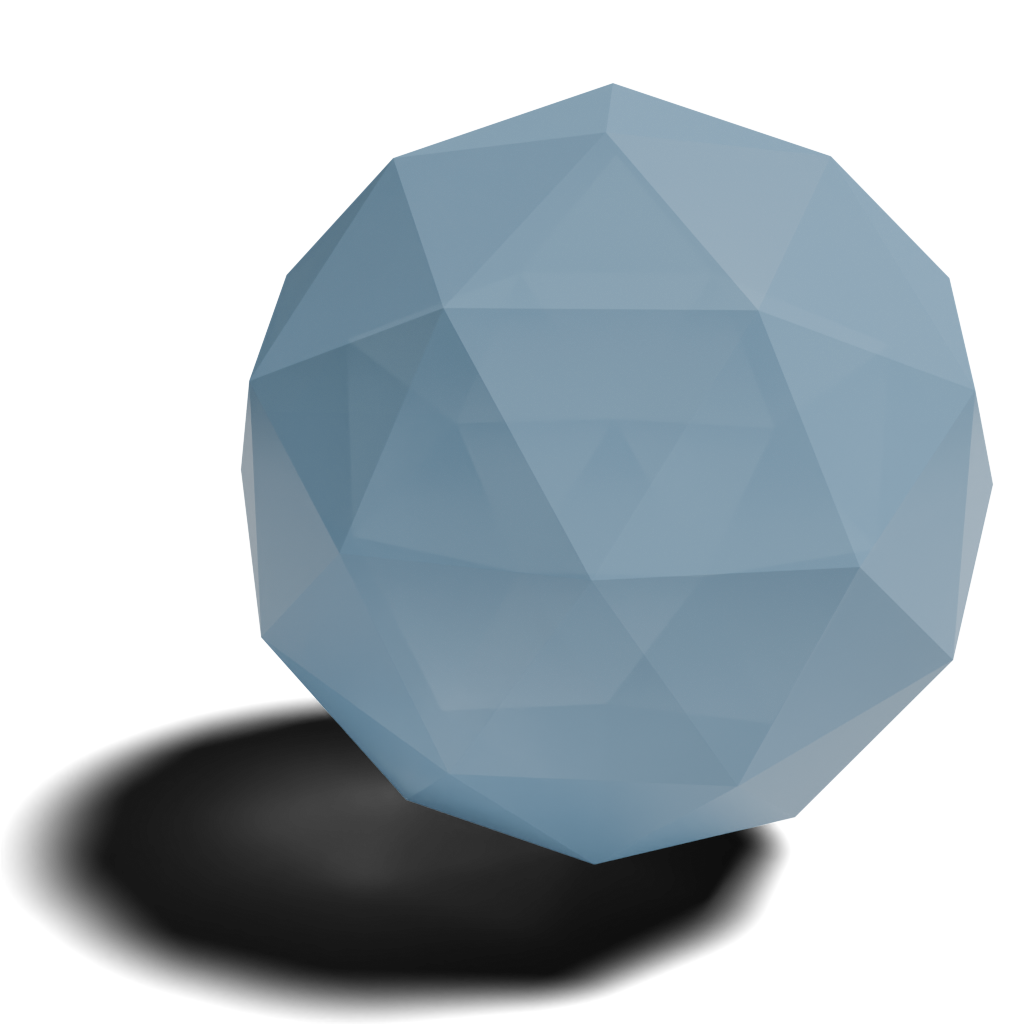} \\
        |F| = 5756 & 6 Boxes, 209 Cap & 1 Sphere \\
    \end{tabular}
    \caption{While our approach can decompose each component of the input well separately (middle), when combining them it does not recover a tight bounding box (right). This is because internal components unnecessarily affect the orientation of the output primitive. \ccby deslancer.}
    \label{fig:cube_failure}
    %\Description{Our approach on a wireframe cube with twisted wireframe cubes inside the original wireframe cube. 2nd from left is our approach which closely matches the input, which uses a capsule on each wireframe. Finally, our approach with a single primitive is only able to parameterize it as a large sphere which does not tightly bound the input.}
\end{figure}

One limitation of our approach is meshes with internal components  will have coarser bounding primitives, because normals of enclosed faces affect the eigendecomposition of each primitive. We show an example of this failure case in Fig.~\ref{fig:cube_failure} on a cube with twisted cubes inside of it. It is clear that the tightest bounding box only needs to fit the largest cube, but our approach does not recover that due to the influence of the other cubes.

Another limitation is that primitives that subsume coplanar faces may be degenerate and depend on the input mesh's orientation. This can be mitigated by adding the tangent direction (Sec.~\ref{sec:ablate-tangent-quadric}), but does not always remove the influence of the input orientation. Furthermore, perfectly coplanar faces may be overeagerly merged together. This is because these faces have no width along their shared plane, introducing near-zero volume when they are merged.

One limitation due to the coarseness of primitives is that they may be less suitable for high frequency details. For organic objects which often have more detail and curvature, primitive decomposition must output more primitives to maintain geometric similarity. As one future direction, this approach could be extended with curved primitives, or for those portions of the mesh convex hulls can be used to preserve detail.

Finally, it may be considered a limitation that convex primitive decomposition relies on the topology of the input. Small modifications in the topology of the input mesh can change the output mesh as can be seen by vertex deduplication in Fig.~\ref{fig:ablate-vertex-merging}. We do not find this to be a problem, and perform vertex deduplication and pairwise collapse to mitigate the effects of topology. Often, topology also serves as a good indicator as to where to split the mesh into different components, so it can be considered a way for artists to indicate where to separate colliders.

\paragraph*{Future Work}
We are able to handle cases where disconnected components of the mesh are fused together, using something similar to non-edge contraction from Garland and Heckbert ~\shortcite{qem}. In collision generation though, disconnected components may be more amenable to merging as compared to mesh simplification. We leave it to future work to explore edge collapses not defined by the input topology, using techniques such as the one from Liu et al.~\cite{simplifying_wild_tris}. 

\section{Conclusion}

In summary, we develop a concise and robust algorithm for decomposing an arbitrary mesh into a number of convex primitives usable in downstream applications such as collision detection. Our approach produces a small number of tight-fitting primitives designed for efficiency at runtime. We hope that our approach motivates further work into simple algorithms that closely align with the processes artists take, and that this work is adapted into game engines and physics/collider generation tooling.
\bibliographystyle{./egPublStyle-cgf/eg-alpha-doi}
\bibliography{citations}

@misc{
    sketchfab,
    author="Sketchfab",
    title={The best 3D viewer on the web},
    url={https://sketchfab.com/},
    year={2022}
}

@Manual{Blender,
   title = {Blender - a 3D modelling and rendering package},
   author = {{Blender Online Community}},
   organization = {Blender Foundation},
   address = {Stichting Blender Foundation, Amsterdam},
   year = {2018},
   url = {http://www.blender.org},
}

@misc{UE5,
  title = {Unreal Engine 5},
  author = {{Epic Games}},
  publisher = {Epic Games},
  year = {2022},
  url = {https://www.unrealengine.com/en-US/unreal-engine-5},
  urldate = {2022-11-01}
}

@inproceedings{silhouette_clipping,
author = {Sander, Pedro V. and Gu, Xianfeng and Gortler, Steven J. and Hoppe, Hugues and Snyder, John},
title = {Silhouette Clipping},
year = {2000},
isbn = {1581132085},
publisher = {ACM Press/Addison-Wesley Publishing Co.},
address = {USA},
booktitle = {Proceedings of the 27th Annual Conference on Computer Graphics and Interactive Techniques},
pages = {327–334},
numpages = {8},
series = {SIGGRAPH '00},
url = {https://doi.org/10.1145/344779.344935},
doi = {10.1145/344779.344935},
}

@inproceedings{qem,
author = {Garland, Michael and Heckbert, Paul S.},
title = {Surface Simplification Using Quadric Error Metrics},
year = {1997},
isbn = {0897918967},
publisher = {ACM Press/Addison-Wesley Publishing Co.},
address = {USA},
url = {https://doi.org/10.1145/258734.258849},
doi = {10.1145/258734.258849},
booktitle = {Proceedings of the 24th Annual Conference on Computer Graphics and Interactive Techniques},
pages = {209–216},
numpages = {8},
series = {SIGGRAPH '97}
}

@article{ransac,
author = {Fischler, Martin A. and Bolles, Robert C.},
title = {Random Sample Consensus: A Paradigm for Model Fitting with Applications to Image Analysis and Automated Cartography},
year = {1981},
issue_date = {June 1981},
publisher = {Association for Computing Machinery},
address = {New York, NY, USA},
volume = {24},
number = {6},
issn = {0001-0782},
url = {https://doi.org/10.1145/358669.358692},
doi = {10.1145/358669.358692},
journal = {Commun. ACM},
month = {jun},
pages = {381–395},
numpages = {15}
}

@inproceedings{SingularValue3x3,
  title={Computing the Singular Value Decomposition of 3x3 matrices with minimal branching and elementary floating point operations},
  author={Aleka McAdams and Andrew Selle and Rasmus Tamstorf and Joseph Teran and Eftychios Sifakis},
  year={2011},
  url={https://api.semanticscholar.org/CorpusID:18183079}
}

@article{sqem,
		author = {Jean-Marc Thiery and Emilie Guy and Tamy Boubekeur},
		title = {Sphere-Meshes: Shape Approximation using Spherical Quadric Error Metrics},
		journal ={ACM Transaction on Graphics (Proc. SIGGRAPH Asia 2013)},
		year = {2013},
		volume = {32},
		number = {6},
		pages = {Art. No. 178},
        url = {https://doi.org/10.1145/2508363.2508384},
        doi = {10.1145/2508363.2508384},
}

@article{coacd,
  title={Approximate convex decomposition for 3d meshes with collision-aware concavity and tree search},
  author={Wei, Xinyue and Liu, Minghua and Ling, Zhan and Su, Hao},
  journal={ACM Transactions on Graphics (TOG)},
  volume={41},
  number={4},
  pages={1--18},
  year={2022},
  publisher={ACM New York, NY, USA},
  url = {https://doi.org/10.1145/3528223.3530103},
  doi = {10.1145/3528223.3530103},
}

@article{UnionFind,
author = {Tarjan, Robert E. and van Leeuwen, Jan},
title = {Worst-case Analysis of Set Union Algorithms},
year = {1984},
issue_date = {April 1984},
publisher = {Association for Computing Machinery},
address = {New York, NY, USA},
volume = {31},
number = {2},
issn = {0004-5411},
url = {https://doi.org/10.1145/62.2160},
doi = {10.1145/62.2160},
journal = {J. ACM},
month = {mar},
pages = {245–281},
numpages = {37}
}

@article {probabilistic_quadrics,
    journal = {Computer Graphics Forum},
    title = {{Fast and Robust QEF Minimization using Probabilistic Quadrics}},
    author = {Trettner, Philip and Kobbelt, Leif},
    year = {2020},
    publisher = {The Eurographics Association and John Wiley & Sons Ltd.},
    ISSN = {1467-8659},
    DOI = {10.1111/cgf.13933}
}

@inproceedings{qem_uv_texture,
author = {Garland, Michael and Heckbert, Paul S.},
title = {Simplifying surfaces with color and texture using quadric error metrics},
year = {1998},
isbn = {1581131062},
publisher = {IEEE Computer Society Press},
address = {Washington, DC, USA},
booktitle = {Proceedings of the Conference on Visualization '98},
pages = {263–269},
numpages = {7},
location = {Research Triangle Park, North Carolina, USA},
series = {VIS '98},
doi={10.1109/VISUAL.1998.745312}
}

@inproceedings{qem_hoppe,
author = {Hoppe, Hugues},
title = {New quadric metric for simplifiying meshes with appearance attributes},
year = {1999},
isbn = {078035897X},
publisher = {IEEE Computer Society Press},
address = {Washington, DC, USA},
booktitle = {Proceedings of the Conference on Visualization '99: Celebrating Ten Years},
pages = {59–66},
numpages = {8},
location = {San Francisco, California, USA},
series = {VIS '99}
}

@INPROCEEDINGS{vhacd,
  author={Mamou, Khaled and Ghorbel, Faouzi},
  booktitle={2009 16th IEEE International Conference on Image Processing (ICIP)}, 
  title={A simple and efficient approach for 3D mesh approximate convex decomposition}, 
  year={2009},
  volume={},
  number={},
  pages={3501-3504},
  doi={10.1109/ICIP.2009.5414068}}

@article{animated_decomp,
author = {Thul, Daniel and Ladick\'{y}, L'ubor and Jeong, Sohyeon and Pollefeys, Marc},
title = {Approximate convex decomposition and transfer for animated meshes},
year = {2018},
issue_date = {December 2018},
publisher = {Association for Computing Machinery},
address = {New York, NY, USA},
volume = {37},
number = {6},
issn = {0730-0301},
url = {https://doi.org/10.1145/3272127.3275029},
doi = {10.1145/3272127.3275029},
journal = {ACM Trans. Graph.},
month = {dec},
articleno = {226},
numpages = {10}
}

@inproceedings{obbtree,
author = {Gottschalk, S. and Lin, M. C. and Manocha, D.},
title = {OBBTree: a hierarchical structure for rapid interference detection},
year = {1996},
isbn = {0897917464},
publisher = {Association for Computing Machinery},
address = {New York, NY, USA},
url = {https://doi.org/10.1145/237170.237244},
doi = {10.1145/237170.237244},
booktitle = {Proceedings of the 23rd Annual Conference on Computer Graphics and Interactive Techniques},
pages = {171–180},
numpages = {10},
series = {SIGGRAPH '96}
}

@article{QuadricTopology,
author = {Hafner, Christian and Ly, Micka\"{e}l and Wojtan, Chris},
title = {Spin-It Faster: Quadrics Solve All Topology Optimization Problems That Depend Only On Mass Moments},
year = {2024},
issue_date = {July 2024},
publisher = {Association for Computing Machinery},
address = {New York, NY, USA},
volume = {43},
number = {4},
issn = {0730-0301},
url = {https://doi.org/10.1145/3658194},
doi = {10.1145/3658194},
journal = {ACM Trans. Graph.},
month = {jul},
articleno = {78},
numpages = {13}
}

@inproceedings{shape_recon_quadric, 
        author = {Zhao, Tong and Bus\'{e}, Laurent and Cohen-Steiner, David and Boubekeur, Tamy and Thiery, Jean-Marc and Alliez, Pierre}, 
        title = {Variational Shape Reconstruction via Quadric Error Metrics}, 
        year = {2023}, 
        isbn = {9798400701597}, 
        publisher = {Association for Computing Machinery}, 
        address = {New York, NY, USA}, url = {https://doi.org/10.1145/3588432.3591529}, 
        doi = {10.1145/3588432.3591529}, 
        booktitle = {ACM SIGGRAPH 2023 Conference Proceedings}, 
        articleno = {45}, 
        numpages = {10}, 
        location = {Los Angeles, CA, USA}, 
        series = {SIGGRAPH '23} 
    }

@article{dual_contouring,
author = {Ju, Tao and Losasso, Frank and Schaefer, Scott and Warren, Joe},
title = {Dual contouring of hermite data},
year = {2002},
issue_date = {July 2002},
publisher = {Association for Computing Machinery},
address = {New York, NY, USA},
volume = {21},
number = {3},
issn = {0730-0301},
url = {https://doi.org/10.1145/566654.566586},
doi = {10.1145/566654.566586},
journal = {ACM Trans. Graph.},
month = {jul},
pages = {339–346},
numpages = {8}
}

@article{anim_sqem,
 author = {Thiery, Jean-Marc and Guy, \'{E}milie and Boubekeur, Tamy and Eisemann, Elmar},
 title = {Animated Mesh Approximation With Sphere-Meshes},
 journal = {ACM Trans. Graph.},
 issue_date = {May 2016},
 volume = {35},
 number = {3},
 month = may,
 year = {2016},
 issn = {0730-0301},
 pages = {30:1--30:13},
 articleno = {30},
 numpages = {13},
 url = {http://doi.acm.org/10.1145/2898350},
 doi = {10.1145/2898350},
 acmid = {2898350},
 publisher = {ACM},
 address = {New York, NY, USA},
}

@article{abstraction_of_shapes,
author = {Mehra, Ravish and Zhou, Qingnan and Long, Jeremy and Sheffer, Alla and Gooch, Amy and Mitra, Niloy J.},
title = {Abstraction of man-made shapes},
year = {2009},
issue_date = {December 2009},
publisher = {Association for Computing Machinery},
address = {New York, NY, USA},
volume = {28},
number = {5},
issn = {0730-0301},
url = {https://doi.org/10.1145/1618452.1618483},
doi = {10.1145/1618452.1618483},
journal = {ACM Trans. Graph.},
month = {dec},
pages = {1–10},
numpages = {10}
}

@inproceedings{realtime_csg,
author = {Lysenko, Mikola},
title = {Realtime constructive solid geometry},
year = {2007},
isbn = {9781450347266},
publisher = {Association for Computing Machinery},
address = {New York, NY, USA},
url = {https://doi.org/10.1145/1278780.1278789},
doi = {10.1145/1278780.1278789},
booktitle = {ACM SIGGRAPH 2007 Sketches},
pages = {7–es},
location = {San Diego, California},
series = {SIGGRAPH '07}
}

@inproceedings{nav_approx_acd,
author = {Andrews, James},
title = {Navigation-Driven Approximate Convex Decomposition},
year = {2024},
isbn = {9798400705250},
publisher = {Association for Computing Machinery},
address = {New York, NY, USA},
url = {https://doi.org/10.1145/3641519.3657479},
doi = {10.1145/3641519.3657479},
booktitle = {ACM SIGGRAPH 2024 Conference Papers},
articleno = {56},
numpages = {9},
location = {Denver, CO, USA},
series = {SIGGRAPH '24}
}

@inproceedings{acd_polyhedra,
author = {Lien, Jyh-Ming and Amato, Nancy M.},
title = {Approximate convex decomposition of polyhedra},
year = {2007},
isbn = {9781595936660},
publisher = {Association for Computing Machinery},
address = {New York, NY, USA},
url = {https://doi.org/10.1145/1236246.1236265},
doi = {10.1145/1236246.1236265},
booktitle = {Proceedings of the 2007 ACM Symposium on Solid and Physical Modeling},
pages = {121–131},
numpages = {11},
location = {Beijing, China},
series = {SPM '07}
}

@article{o_rourke_minimal_enclosing,
  author       = {Joseph O'Rourke},
  title        = {Finding minimal enclosing boxes},
  journal      = {Int. J. Parallel Program.},
  volume       = {14},
  number       = {3},
  pages        = {183--199},
  year         = {1985},
  url          = {https://doi.org/10.1007/BF00991005},
  doi          = {10.1007/BF00991005},
  bibsource    = {dblp computer science bibliography, https://dblp.org}
}

@article{efficient_ransac,
author = {Schnabel, R. and Wahl, R. and Klein, R.},
title = {Efficient RANSAC for Point-Cloud Shape Detection},
journal = {Computer Graphics Forum},
volume = {26},
number = {2},
pages = {214-226},
doi = {https://doi.org/10.1111/j.1467-8659.2007.01016.x},
url = {https://onlinelibrary.wiley.com/doi/abs/10.1111/j.1467-8659.2007.01016.x},
year = {2007}
}

@article{cuboid_abstraction,
  title     = {Learning Adaptive Hierarchical Cuboid Abstractions of 3D Shape Collections},
  author    = {Sun, Chunyu and Zou, Qianfang and Tong, Xin and Liu, Yang},
  journal   = {ACM Transactions on Graphics (SIGGRAPH Asia)},
  volume    = {38},
  number    = {6},
  year      = {2019},
  publisher = {ACM},
  url = {https://doi.org/10.1145/3355089.3356529},
  doi = {10.1145/3355089.3356529},
}

@inproceedings{superquadrics,
    title = {Superquadrics Revisited: Learning 3D Shape Parsing beyond Cuboids},
    author = {Paschalidou, Despoina and Ulusoy, Ali Osman and Geiger, Andreas},
    booktitle = {Proceedings IEEE Conf. on Computer Vision and Pattern Recognition (CVPR)},
    month = jun,
    year = {2019},
    doi={10.1109/CVPR.2019.01059}
}

@Inproceedings{marching_primitives,
    title = {Marching-Primitives: Shape Abstraction from Signed Distance Function},
    author = {Liu, Weixiao and Wu, Yuwei and Ruan, Sipu and Chirikjian, Gregory},
    booktitle = {Proceedings IEEE Conf. on Computer Vision and Pattern Recognition (CVPR)},
    year = {2023},
    doi={10.1109/CVPR52729.2023.00847}
}

@inproceedings{surface_edge_detect_point_cloud,
author = {Li, Yuanqi and Liu, Shun and Yang, Xinran and Guo, Jianwei and Guo, Jie and Guo, Yanwen},
title = {Surface and Edge Detection for Primitive Fitting of Point Clouds},
year = {2023},
isbn = {9798400701597},
publisher = {Association for Computing Machinery},
address = {New York, NY, USA},
url = {https://doi.org/10.1145/3588432.3591522},
doi = {10.1145/3588432.3591522},
booktitle = {ACM SIGGRAPH 2023 Conference Proceedings},
articleno = {44},
numpages = {10},
location = {Los Angeles, CA, USA},
series = {SIGGRAPH '23}
}

@misc{physx,
author = {{Nvidia Corporation}},
key = {Nvidia Corporation},
title = {PhysX},
year = {2017},
url = {https://developer.nvidia.com/physx-sdk},
language = {eng},
version = {3.3},
organization = {Nvidia Corporation},
date = {2017},
month = {04},
urldate = {2017-04-30},
}

@article{spectral_mesh_simp, 
  title = "Spectral Mesh Simplification", 
  author = {Thibault Lescoat and Hsueh-Ti Derek Liu and Jean-Marc Thiery and Alec Jacobson and Tamy Boubekeur and Maks Ovsjanikov},
  year = "2020", 
  journal = "Computer Graphics Forum (Proc. of EUROGRAPHICS 2020)",
  number = "2",
  volume = "39",
  pages  = "315--324",
  doi = {https://doi.org/10.1111/cgf.13932},
  url = {https://onlinelibrary.wiley.com/doi/abs/10.1111/cgf.13932},
}

@misc{ORCAAmazonBistro,
   title = {Amazon Lumberyard Bistro, Open Research Content Archive (ORCA)},
   author = {Amazon Lumberyard},
   year = {2017},
   month = {July},
   note = {\small \texttt{http://developer.nvidia.com/orca/amazon-lumberyard-bistro}},
   url = {http://developer.nvidia.com/orca/amazon-lumberyard-bistro}
}

@inproceedings{quadric_surface_extraction,
    author = {Yan, Dong-Ming and Liu, Yang and Wang, Wenping},
    title = {Quadric surface extraction by variational shape approximation},
    year = {2006},
    isbn = {354036711X},
    publisher = {Springer-Verlag},
    address = {Berlin, Heidelberg},
    url = {https://doi.org/10.1007/11802914_6},
    doi = {10.1007/11802914_6},
    booktitle = {Proceedings of the 4th International Conference on Geometric Modeling and Processing},
    pages = {73–86},
    numpages = {14},
    location = {Pittsburgh, PA},
    series = {GMP'06}
}

@article{variational_3d_shape_segmentation,
author = {Lu, Lin and Choi, Yi-King and Wang, Wenping and Kim, Myung-Soo},
title = {Variational 3D Shape Segmentation for Bounding Volume Computation},
journal = {Computer Graphics Forum},
volume = {26},
number = {3},
pages = {329-338},
doi = {https://doi.org/10.1111/j.1467-8659.2007.01055.x},
url = {https://onlinelibrary.wiley.com/doi/abs/10.1111/j.1467-8659.2007.01055.x},
year = {2007}
}

@inproceedings{fast_and_memory_efficient_polygonal_simp,
author = {Lindstrom, Peter and Turk, Greg},
title = {Fast and memory efficient polygonal simplification},
year = {1998},
isbn = {1581131062},
publisher = {IEEE Computer Society Press},
address = {Washington, DC, USA},
booktitle = {Proceedings of the Conference on Visualization '98},
pages = {279–286},
numpages = {8},
location = {Research Triangle Park, North Carolina, USA},
series = {VIS '98},
doi={10.1109/VISUAL.1998.745314}
}

@inproceedings{model_composition_from_interchangeable_components,
author = {Kreavoy, Vladislav and Julius, Dan and Sheffer, Alla},
title = {Model Composition from Interchangeable Components},
year = {2007},
isbn = {0769530095},
publisher = {IEEE Computer Society},
address = {USA},
url = {https://doi.org/10.1109/PG.2007.43},
doi = {10.1109/PG.2007.43},
booktitle = {Proceedings of the 15th Pacific Conference on Computer Graphics and Applications},
pages = {129–138},
numpages = {10},
series = {PG '07}
}

@ARTICLE{generic_primitive_detection_in_point_clouds,
  author={Birdal, Tolga and Busam, Benjamin and Navab, Nassir and Ilic, Slobodan and Sturm, Peter},
  journal={IEEE Transactions on Pattern Analysis and Machine Intelligence}, 
  title={Generic Primitive Detection in Point Clouds Using Novel Minimal Quadric Fits}, 
  year={2020},
  volume={42},
  number={6},
  pages={1333-1347},
  doi={10.1109/TPAMI.2019.2900309}}

@inproceedings{hierarchical_face_clustering_on_polygonal_surfaces,
author = {Garland, Michael and Willmott, Andrew and Heckbert, Paul S.},
title = {Hierarchical face clustering on polygonal surfaces},
year = {2001},
isbn = {1581132921},
publisher = {Association for Computing Machinery},
address = {New York, NY, USA},
url = {https://doi.org/10.1145/364338.364345},
doi = {10.1145/364338.364345},
booktitle = {Proceedings of the 2001 Symposium on Interactive 3D Graphics},
pages = {49–58},
numpages = {10},
series = {I3D '01}
}

@article{d_charts,
    journal = {Computer Graphics Forum},
    title = {{D-Charts: Quasi-Developable Mesh Segmentation}},
    author = {Julius, Dan and Kraevoy, Vladislav and Sheffer, Alla},
    year = {2005},
    publisher = {The Eurographics Association and Blackwell Publishing, Inc},
    ISSN = {1467-8659},
    DOI = {10.1111/j.1467-8659.2005.00883.x}
    }

@misc{simplifying_wild_tris,
      title={Simplifying Triangle Meshes in the Wild}, 
      author={Hsueh-Ti Derek Liu and Xiaoting Zhang and Cem Yuksel},
      year={2024},
      eprint={2409.15458},
      archivePrefix={arXiv},
      primaryClass={cs.GR},
      url={https://arxiv.org/abs/2409.15458}, 
}

@ARTICLE{gjk,
  author={Gilbert, E.G. and Johnson, D.W. and Keerthi, S.S.},
  journal={IEEE Journal on Robotics and Automation}, 
  title={A fast procedure for computing the distance between complex objects in three-dimensional space}, 
  year={1988},
  volume={4},
  number={2},
  pages={193-203},
  doi={10.1109/56.2083}
}

@article{variational_mesh_segmentation,
  title={Variational mesh segmentation via quadric surface fitting},
  author={Yan, Dong-Ming and Wang, Wenping and Liu, Yang and Yang, Zhouwang},
  journal={Computer-Aided Design},
  volume={44},
  number={11},
  pages={1072--1082},
  year={2012},
  publisher={Elsevier},
  doi = {https://doi.org/10.1016/j.cad.2012.04.005},
}

@inproceedings{quasi_optimal_region_growing,
  title={Quasi-optimal mesh segmentation via region growing/merging},
  author={Mizoguchi, Tomohiro and Date, Hiroaki and Kanai, Satoshi and Kishinami, Takeshi},
  booktitle={International Design Engineering Technical Conferences and Computers and Information in Engineering Conference},
  volume={48078},
  pages={547--556},
  year={2007},
  doi = {10.1115/DETC2007-35171}
}

@article{simple_primitive_recognition_via_face_clustering,
  title={Simple primitive recognition via hierarchical face clustering},
  author={Yang, Xiaolong and Jia, Xiaohong},
  journal={Computational Visual Media},
  volume={6},
  pages={431--443},
  year={2020},
  publisher={Springer},
  doi = {10.1007/s41095-020-0192-6}
}

@misc{variational_hierarchical_directed_bounding_box_construction,
  title={Variational Hierarchical Directed Bounding Box Construction for Solid Mesh Models}, 
  author={Rui Wang and Wei Hua and Gaofeng Xu and Yuchi Huo and Hujun Bao},
  year={2022},
  eprint={2203.10521},
  archivePrefix={arXiv},
  primaryClass={cs.GR},
  url={https://arxiv.org/abs/2203.10521}, 
}

@INPROCEEDINGS{split_merge_refine,
  author={Park, Chanhyeok and Sung, Minhyuk},
  booktitle={2024 International Conference on 3D Vision (3DV)}, 
  title={Split, Merge, and Refine: Fitting Tight Bounding Boxes via Over-Segmentation and Iterative Search}, 
  year={2024},
  volume={},
  number={},
  pages={1468-1477},
  doi={10.1109/3DV62453.2024.00146}
}

@inproceedings{hierarchical_mesh_segmentation_based_on_quadric,
  title={Hierarchical mesh segmentation based on quadric surface fitting},
  author={Zhang, Huijuan and Li, Chong and Gao, Leilei and Wang, Guoping},
  booktitle={2015 14th International Conference on Computer-Aided Design and Computer Graphics (CAD/Graphics)},
  pages={33--40},
  year={2015},
  organization={IEEE},
  doi={10.1109/CADGRAPHICS.2015.26}
}

@INPROCEEDINGS{ellipsoid_decomposition_of_3d_models,
  author={Bischoff, S. and Kobbelt, L.},
  booktitle={Proceedings. First International Symposium on 3D Data Processing Visualization and Transmission}, 
  title={Ellipsoid decomposition of 3D-models}, 
  year={2002},
  volume={},
  number={},
  pages={480-488},
  doi={10.1109/TDPVT.2002.1024103}
}

@article{primitive_based_3d_segmentation_cad_models,
    title={A primitive-based 3D segmentation algorithm for mechanical CAD models},
    author={Le, Truc and Duan, Ye},
    journal={Computer Aided Geometric Design},
    volume={52},
    pages={231--246},
    year={2017},
    publisher={Elsevier},
    url = {https://doi.org/10.1016/j.cagd.2017.02.009},
    doi = {10.1016/j.cagd.2017.02.009},
}

@Inbook{variational_obb_tree,
author="Bao, Hujun and Hua, Wei",
title="Variational OBB-Tree Approximation for Solid Objects",
bookTitle="Real-Time Graphics Rendering Engine",
year="2011",
publisher="Springer Berlin Heidelberg",
address="Berlin, Heidelberg",
pages="281--293",
isbn="978-3-642-18342-3",
doi="10.1007/978-3-642-18342-3_6",
url="https://doi.org/10.1007/978-3-642-18342-3_6"
}

@article{survey_of_simple_geometric_primitives,
    journal = {Computer Graphics Forum},
    title = {{A Survey of Simple Geometric Primitives Detection Methods for Captured 3D Data}},
    author = {Kaiser, Adrien and Ybanez Zepeda, Jose Alonso and Boubekeur, Tamy},
    year = {2019},
    publisher = {© 2019 The Eurographics Association and John Wiley & Sons Ltd.},
    ISSN = {1467-8659},
    DOI = {10.1111/cgf.13451}
}

@article{shape_segmentation_by_approx_convex,
author = {Kaick, Oliver Van and Fish, Noa and Kleiman, Yanir and Asafi, Shmuel and Cohen-OR, Daniel},
title = {Shape Segmentation by Approximate Convexity Analysis},
year = {2015},
issue_date = {November 2014},
publisher = {Association for Computing Machinery},
address = {New York, NY, USA},
volume = {34},
number = {1},
issn = {0730-0301},
url = {https://doi.org/10.1145/2611811},
doi = {10.1145/2611811},
journal = {ACM Trans. Graph.},
month = dec,
articleno = {4},
numpages = {11}
}

@article{hierarchical_convex_approximation_of_3d_shapes,
author = {Attene, Marco and Mortara, Michela and Spagnuolo, Michela and Falcidieno, Bianca},
title = {Hierarchical Convex Approximation of 3D Shapes for Fast Region Selection},
journal = {Computer Graphics Forum},
volume = {27},
number = {5},
pages = {1323-1332},
doi = {https://doi.org/10.1111/j.1467-8659.2008.01271.x},
url = {https://onlinelibrary.wiley.com/doi/abs/10.1111/j.1467-8659.2008.01271.x},
year = {2008}
}

@article{bounding_proxies_for_shape_approximation,
 author = {St\'{e}phane Calderon and Tamy Boubekeur},
 title = {Bounding Proxies for Shape Approximation},
 journal = {ACM Transactions on Graphics (Proc. SIGGRAPH 2017)},
 volume = {36},
 number = {5},
 month = {july},
 year = {2017},
 articleno = {57},
 issn = {0730-0301},
 url = {https://doi.org/10.1145/3072959.3073714},
 doi = {10.1145/3072959.3073714},
}

@phdthesis{quadric_based_polygonal_surface_simplification,
    author = {Garland, Michael and Heckbert, Paul},
    title = {Quadric-based polygonal surface simplification},
    year = {1999},
    isbn = {0599519908},
    publisher = {Carnegie Mellon University},
    address = {USA},
    note = {AAI9950005}
}

@inproceedings{mesh_optimization,
author = {Hoppe, Hugues and DeRose, Tony and Duchamp, Tom and McDonald, John and Stuetzle, Werner},
title = {Mesh optimization},
year = {1993},
isbn = {0897916018},
publisher = {Association for Computing Machinery},
address = {New York, NY, USA},
url = {https://doi.org/10.1145/166117.166119},
doi = {10.1145/166117.166119},
booktitle = {Proceedings of the 20th Annual Conference on Computer Graphics and Interactive Techniques},
pages = {19–26},
numpages = {8},
location = {Anaheim, CA},
series = {SIGGRAPH '93}
}

@article{hierarchical_mesh_segmentation_based_on_fitting,
author = {Attene, Marco and Falcidieno, Bianca and Spagnuolo, Michela},
issn = {0178-2789},
journal = {The Visual computer.},
lccn = {2004233402},
number = {3},
publisher = {Springer-Verlag},
title = {Hierarchical mesh segmentation based on fitting primitives},
volume = {22},
year = {2006},
url={https://doi.org/10.1007/s00371-006-0375-x},
doi={10.1007/s00371-006-0375-x}
}
\appendix

\paragraph{Meshing of Cylinders and Capsules}
Our approach outputs cylinders, capsules, spheres, and frustums modeled as parametric functions. For some applications though (such as rendering), all inputs must be given as a mesh, so we convert these primitives to quantized meshes. For cylinders and frustums, we output prisms capped by one face on each end. For spheres, we use a UV sphere approximation. For capsules, we output prisms capped with UV spheres on each end. This makes our approach suitable for any downstream applications.

\section{\label{sec:additional-results}Additional Results}
\begin{figure*}
    \centering
    \begin{tabular}{c c}
    \frame{\includegraphics[width=0.38\linewidth]{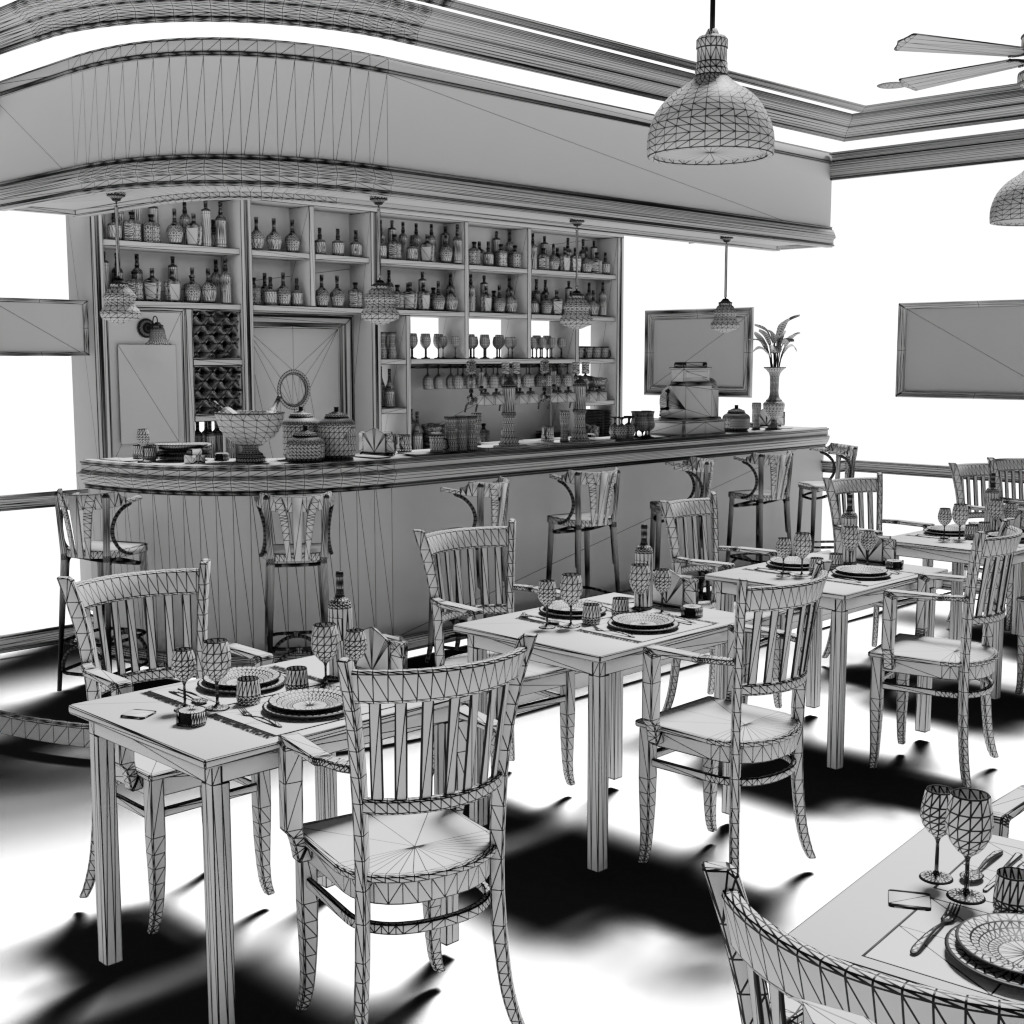}} &
    \frame{\includegraphics[width=0.38\linewidth]{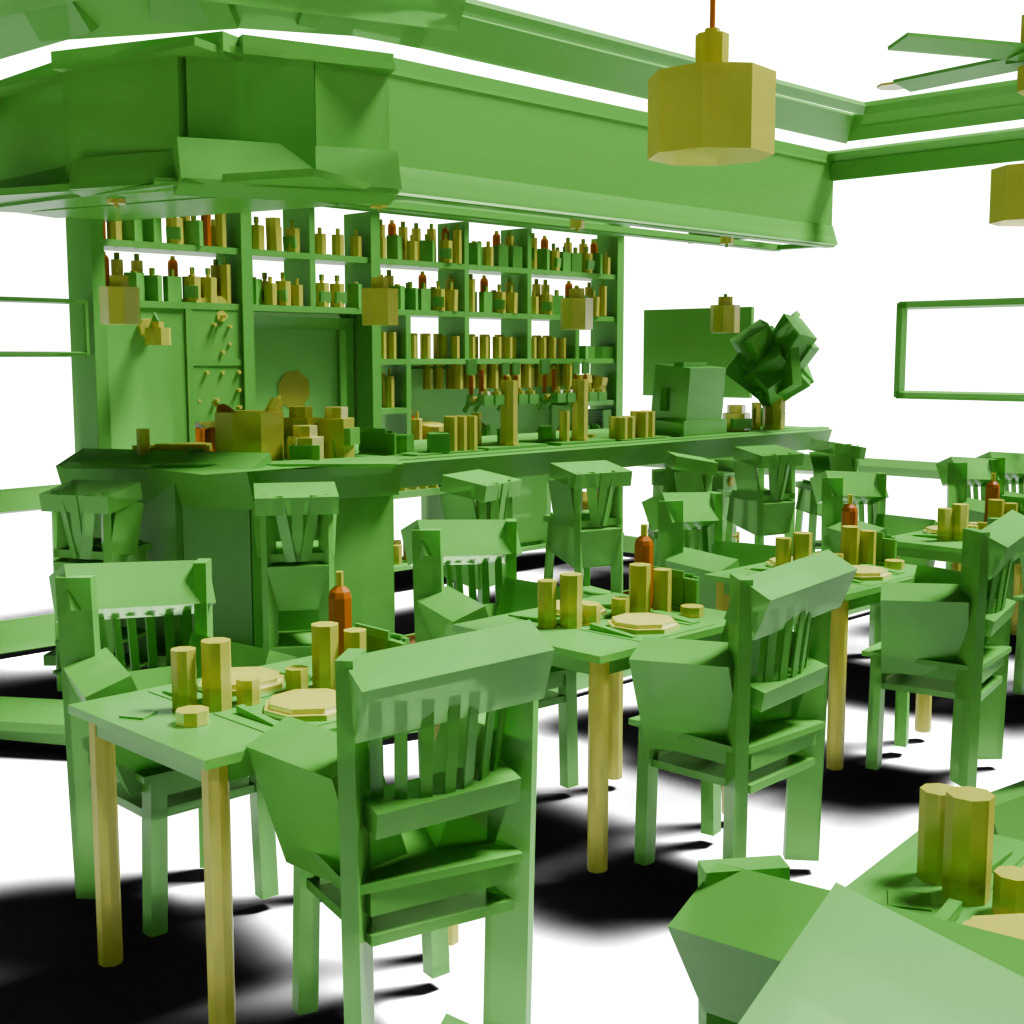}} \\
    |F| $= 570588$ & 2343 Boxes, 115 Capsules, 1343 Cylinders \\
    \end{tabular}
    \caption{An additional example of our approach on a dense, complex scene~\cite{ORCAAmazonBistro}. The entire scene is processed in one pass in under 2 minutes, and naturally decomposes objects into constituent parts. We use our approach to generate an initial decomposition, then automatically delete any extremely thin bounding boxes which have radius on any axis $\leq\num{1e-4}$. This postprocessing is included because around the placemats there are a number of frills which contribute a large number of primitives but have essentially 0 volume. Furthermore, many walls are entirely planar but may not be rectangular, leading to regions jutting out. Removing thin bounding boxes fixes both problems.
    Green indicates bounding boxes, yellow for cylinders, and red for capsules.
    \label{fig:bistro}}
    %\Description{The inside of a restaurant, with set tables and a wine bar. Behind the bar are a number of bottles in shelves. The scene is decomposed into cylinders and boxes.}
\end{figure*}

\begin{figure*}
    \centering
    \begin{minipage}[b]{0.5\textwidth}
    \raggedleft
    \begin{tabular}{c c}
        \includegraphics[width=0.57\textwidth]{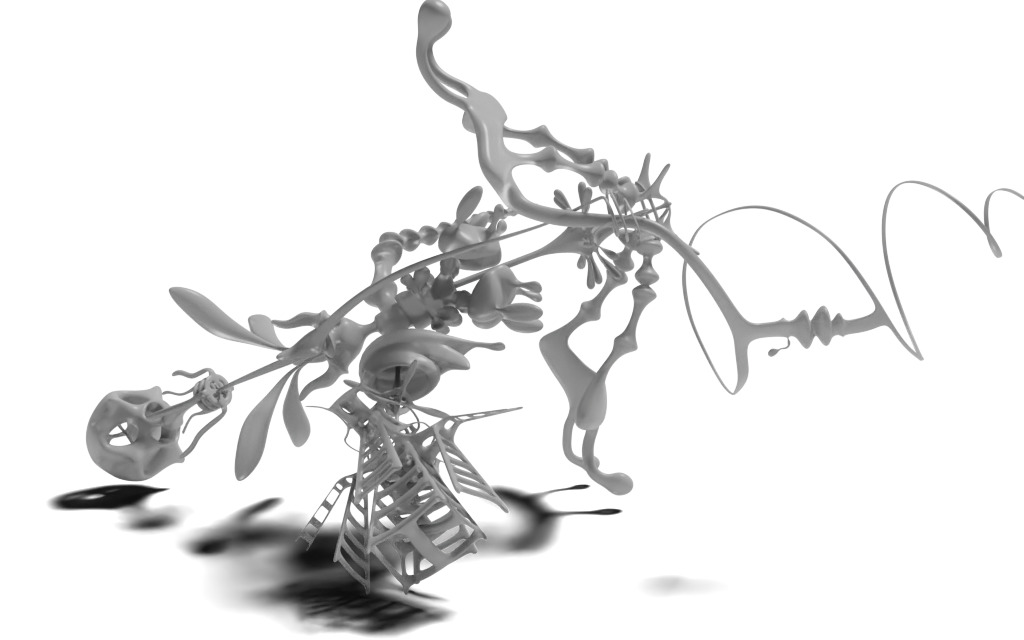} &
        \includegraphics[width=0.57\textwidth]{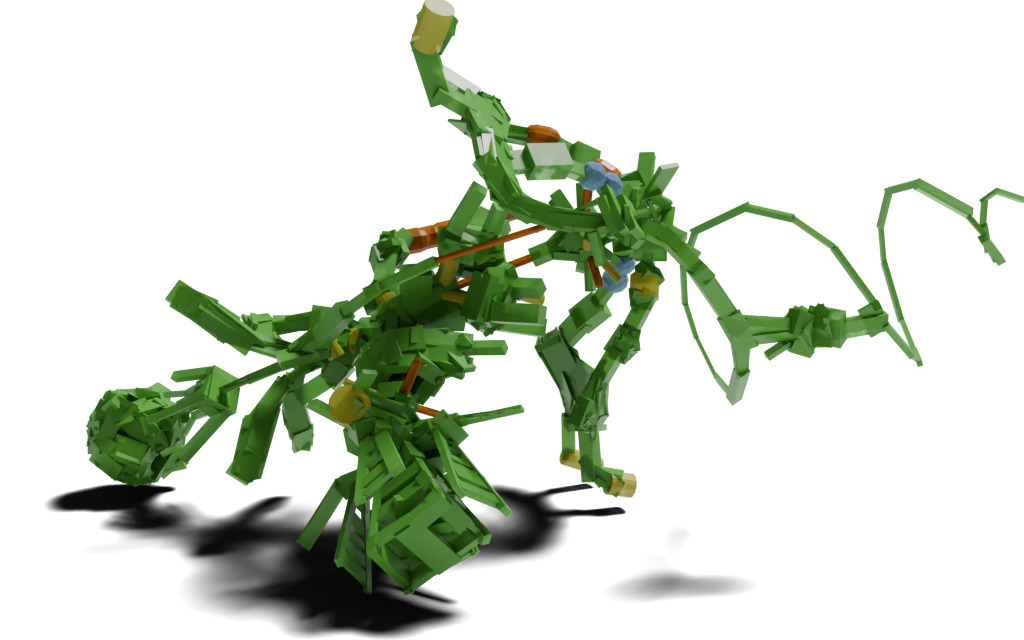} \\
        Input (|F| = 377344) & \small 761 Boxes, 13 Cap, 22 Cyl, 4 Spheres \\
    \end{tabular}

    \caption{Our approach on a particularly challenging mesh, ``Yeahright''. Our approach can preserve thin structures and holes. \cczero Keenan Crane.}
    \label{fig:yeahright}

    %\Description{Our approach on an extremely abnormal mesh. The input mesh is a bulbous wireframe with a number of trypophobia inducing components.}
    \end{minipage}
    \hspace{6em}
    \begin{minipage}[b]{0.3\textwidth}
        \raggedright
        \centering
        \frame{\includegraphics[width=0.666\textwidth]{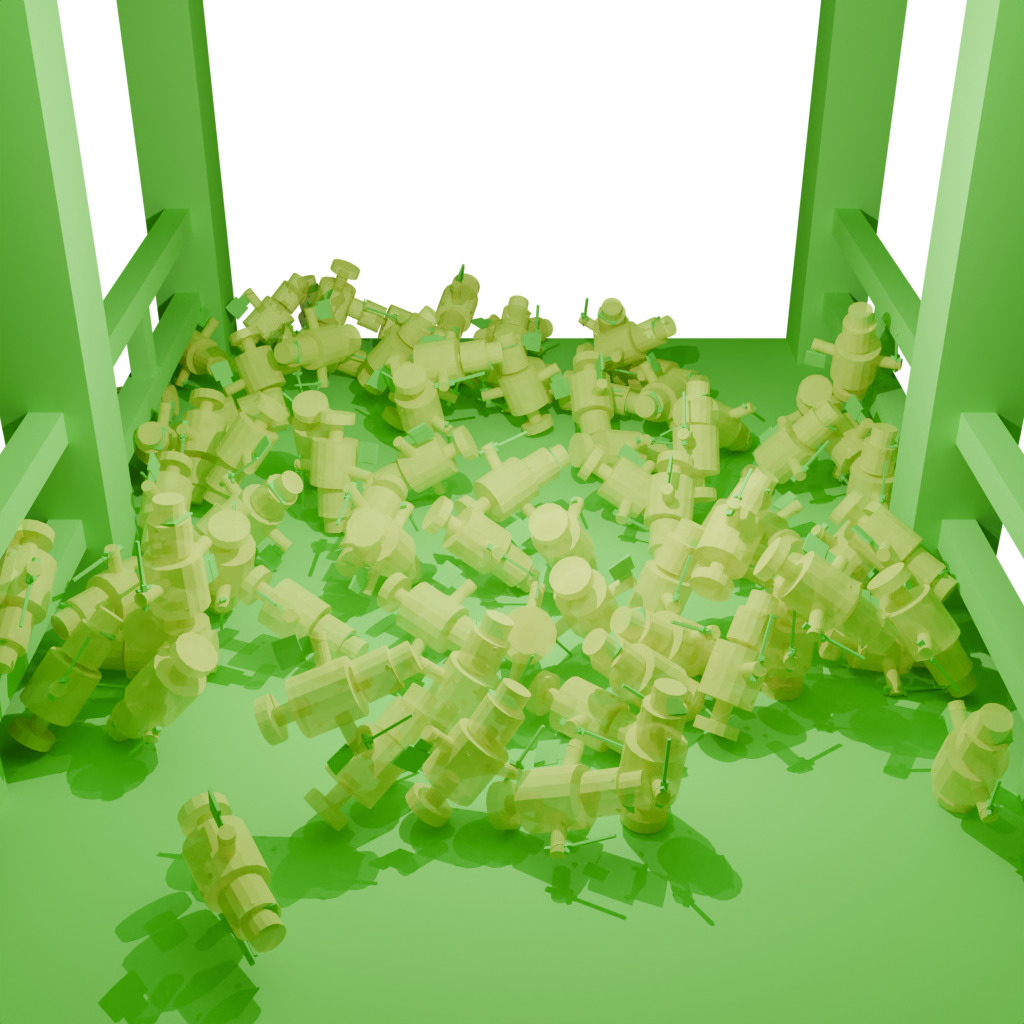}}
        \caption{Visualization of many training dummies from Fig.~\ref{fig:ablate-vertex-merging}, colliding inside of ``Jpn. Corridor Middle''.}
    \end{minipage}
\end{figure*}

We show additional results from our approach and provide pseudo-code for fitting frustums in Alg.~\ref{alg:frustum} and isosceles trapezoidal prisms in Alg.~\ref{alg:isotrap}. Then, we plot frame durations for collisions with 5000 spheres for many meshes in Fig.~\ref{fig:additional-sim-times}, and Fig.~\ref{fig:sim-times-component-comparison}. We show outputs on buildings, characters, props, and environments in Fig.~\ref{fig:additional_qualitative_results}. We include more comparisons to CoACD and V-HACD in Fig.~\ref{fig:additional_comparisons}. Complete statistics for the primitives generated on our dataset are in Tab.~\ref{tab:complete_results}, with all 1-way distances in Tab.~\ref{tab:distance_comparison}. We show byte counts for all models in Tab.~\ref{tab:raw-memory-cost}. We show our approach on a complex scene~\cite{ORCAAmazonBistro}, to generate an initial decomposition, then automatically delete thin bounding boxes to get the output shown in Fig.~\ref{fig:bistro}. We then show our approach on two organic objects in Fig.~\ref{fig:organic-meshes}, and one complex example in Fig.~\ref{fig:yeahright}. Finally, we show additional examples with collision detection comparisons from the V-HACD dataset~\cite{vhacd} in Fig.~\ref{fig:vhacd-data-comp}.

\newpage

\begin{algorithm}
\caption{Frustum Point Subsuming\label{alg:frustum}}
\begin{algorithmic}[1]
\Statex \textbf{Input: } Points $\in\mathbb{R}^{N\times3}$, Base Point $\vect{c}$, Axis $\vect{a}$
\Statex \textbf{Output: } height $h$, top and bottom radius $r_\text{top}, r_\text{bot}$
\State $h = 0, r_\text{bot} = 0, r_\text{top} = 0$
\For{p $\in$ Points} $h = \max(h, |(p-c)^\top a|)$
\EndFor
\State $r_\text{top}^* = 0, y_\text{top}^* = 1, r_\text{bot}^* = 0, y_\text{bot}^* = 0$ \Comment{Approx. constraints}
\Procedure{Y}{p}:
\Return $\frac{(p - c)^\top a}{h}$
\EndProcedure
\Procedure{$R_\text{side}$}{r, y, $r_\text{opp}$}:
\Return \textbf{if} side \textbf{is} bot \\ $|\frac{r - r_\text{opp}y}{1 - y}|$ \textbf{else} $|\frac{r - r_\text{opp}(1-y)}{y}|$
\EndProcedure
\For{p $\in$ Points}:
    \State side, opp $ = $ \textbf{if} $Y(p) \leq 0.5$ \{ bot, top \} \textbf{else} \{ top, bot \}
    \State $r = \lVert (p - c) - h(p - c)^\top a\rVert_2$
    \State $\text{next} = R_\text{side}(r, Y(p), r_\text{opp})$
    \If{$\text{next} > r_\text{side}^*$}
        \State $r_\text{side}^* = r, y_\text{side}^* = Y(p), r_\text{side} = \text{next}$
        \State $r_\text{opp}^* = R_\text{opp}(r_\text{opp}^*, y_\text{opp}^*, r_\text{side}^*)$
    \EndIf
\EndFor
\Statex Similar FixSide(...) Procedure as Alg.~\ref{alg:isotrap} \\
\Return $h, r_\text{top}, r_\text{bot}$
\end{algorithmic}
\end{algorithm}

\vfill\null

\begin{algorithm}
\caption{Isosceles Trapezoid Point Subsuming\label{alg:isotrap}}
\begin{algorithmic}[1]
\Statex \textbf{Input: } Points $\in\mathbb{R}^{N\times3}$, Center $c$, Axes $a_x, a_y, a_z$
\Statex \textbf{Output: } Half-radii $h_x, h_y, h_{zt}, h_{zb}$
\State $h_x = 0, h_y = 0, h_{zt} = 0, h_{zb} = 0$
\For{p $\in$ Points} \Comment{Compute $h_x$, $h_y$}
    \State $h_x = \max(h_x, |(p-c)^\top a_x|)$, $h_y = \max(h_y, |(p-c)^\top a_y|)$
\EndFor
\State $z_{zt}^* = 0, y_{zt}^* = 1, z_{zb}^* = 0, y_{zb}^* = 0$ \Comment{Current approx. constraints}
\For{p $\in$ Points}:
    \Procedure{UpdateConstraint}{side $\in \{ \text{zt, zb} \}$}
        \State \text{opp} = \textbf{if} side \textbf{is} zt \{ zb \} \textbf{else} \{ zt \}
        \State $\text{next} = h_\text{side}(z(p), y(p), h_\text{opp})$ \Comment{$h_{zt} \text{ or } h_{zb}$}
        \If{$\text{next} > h_\text{side}$}
            \State $z_\text{side}^* = z(p), y_\text{side}^* = y(p), h_\text{side} = \text{next}$
            \State $h_\text{opp} = |h_\text{opp}(z_\text{opp}^*, y_\text{opp}^*, h_\text{side})|$ \label{line:update_half_extent}
        \EndIf
    \EndProcedure
    \State\Call{UpdateConstraint}{\textbf{if} $y(p) \leq 0.5$ \{ zb \} \textbf{else} \{ zt \}}
\EndFor
\Procedure{FixSide}{top $\in$ bool} \\
    \Comment{Fix one side, find min. that subsumes all points for other.}\\
    \Comment{Guarantees that all input points are subsumed.}
    \State $v, o =$ \textbf{if} top \{ $zt, zb$ \} \textbf{else} \{ $zb, zt$ \}
    \State $h_{v} = 0$
    \For{p $\in$ Points}:
    $h_v = \max(h_v, h_v(z(p), y(p), h_o))$
    \EndFor
\EndProcedure
\State FixSide($h_{zt} < h_{zb}$) \Comment{Fix larger side and set smaller}
\State FixSide($h_{zt} \geq h_{zb}$) \Comment{Then vice-versa} \\
\Return $h_x, h_y, h_{zt}, h_{bt}$
\end{algorithmic}
\end{algorithm}

{
\setlength{\tabcolsep}{0em}
\begin{figure*}
    \centering
    \begin{tabular}{c c c c}
        \multicolumn{4}{c}{Additional Results} \\
        Input Mesh & Convex Prim. Decomp. & Input Mesh & Convex Prim. Decomp. \\
        \includegraphics[width=0.25\linewidth]{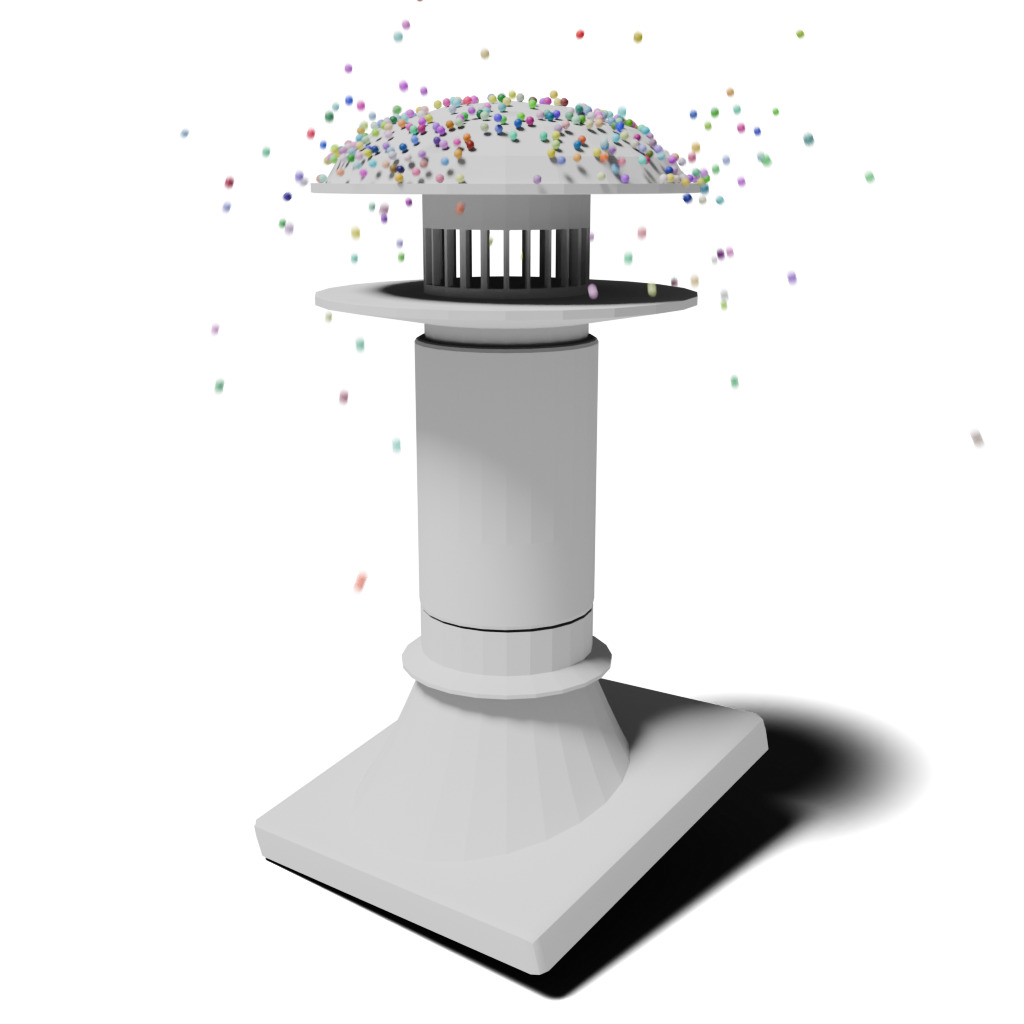} &
        \includegraphics[width=0.25\linewidth]{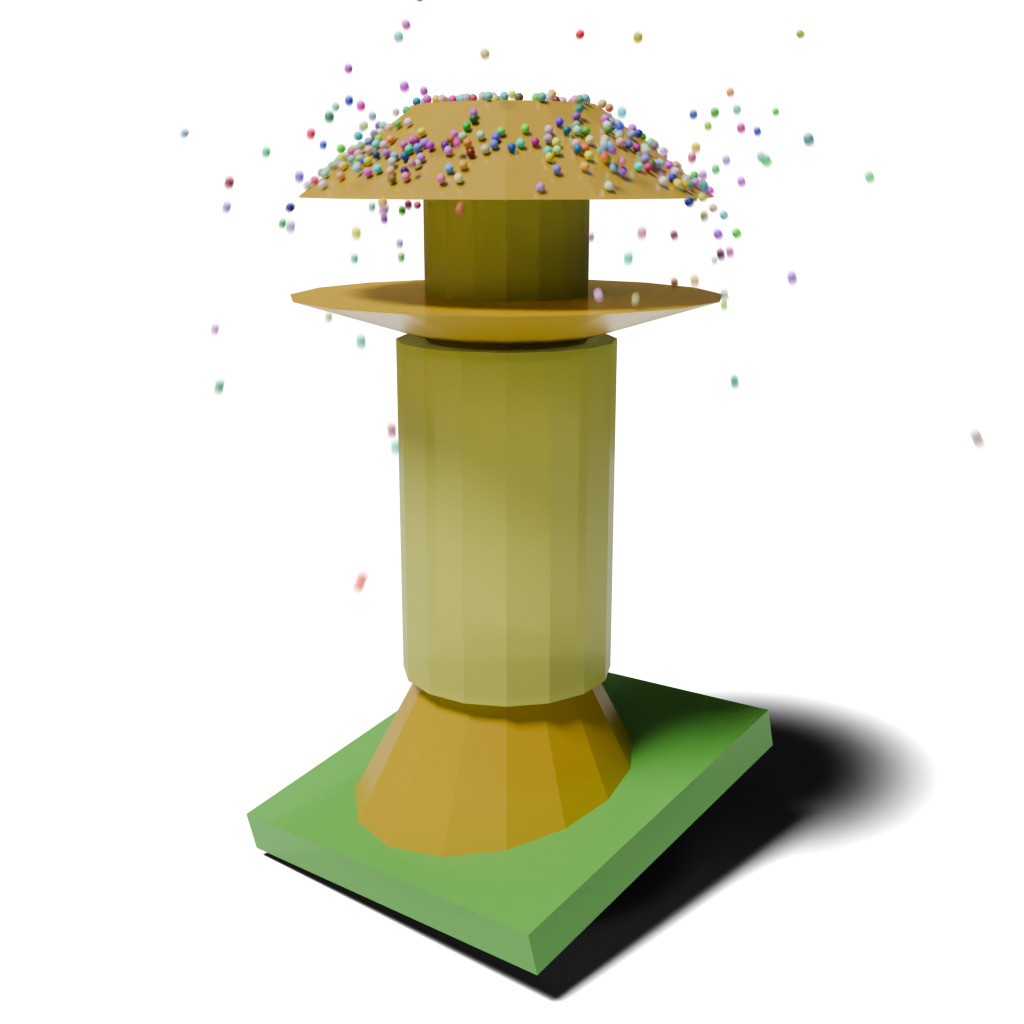} &
        \includegraphics[width=0.25\linewidth]{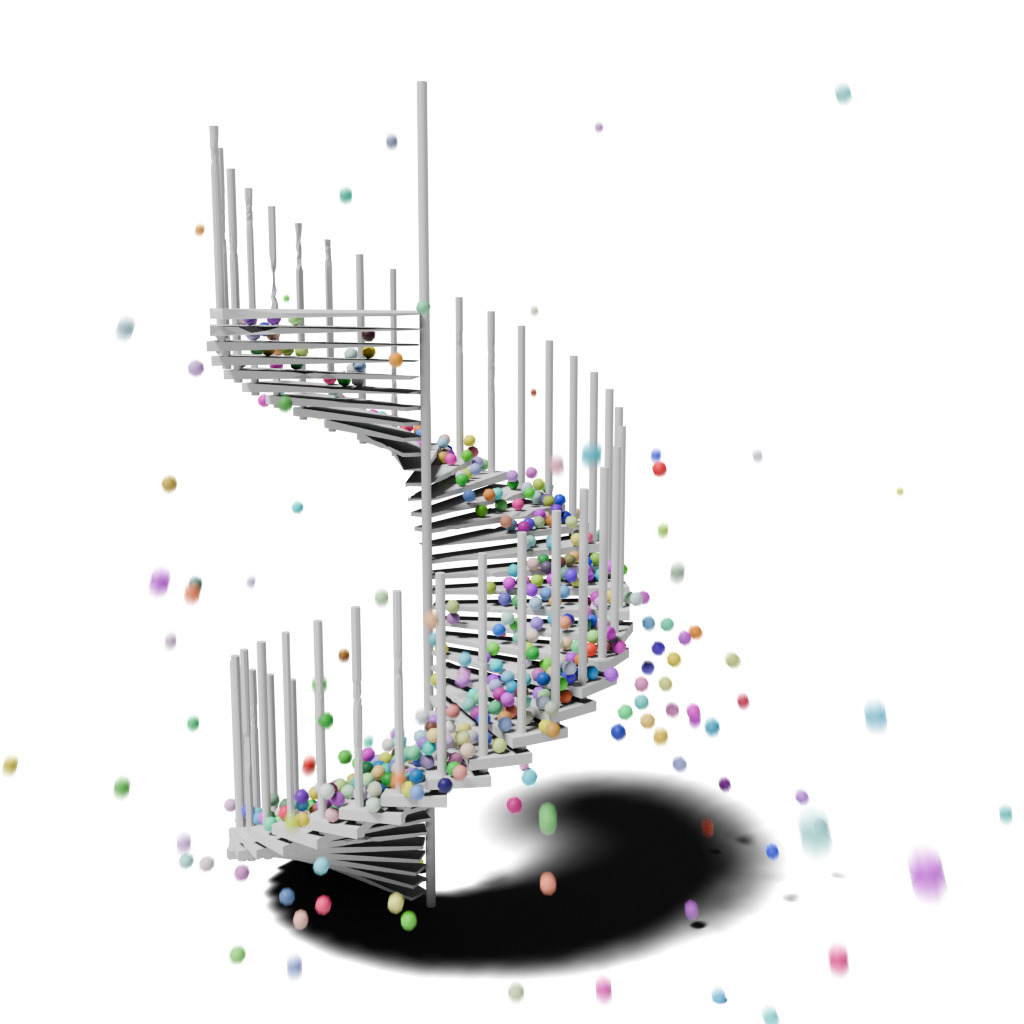} & \includegraphics[width=0.25\linewidth]{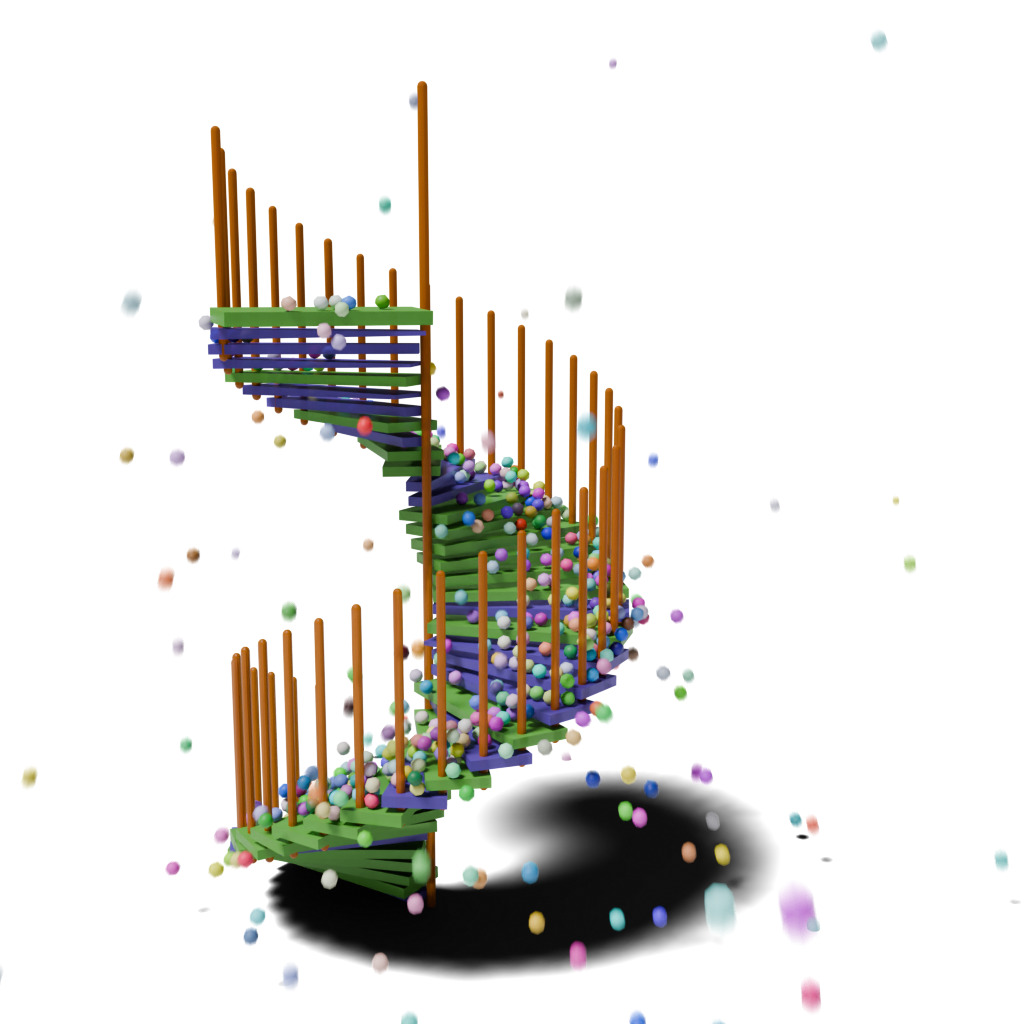}
        \\
        $|F| = 3404$ & 1 Box, 2 Cylinders, 3 Frustums & $|F| = 6358$ & 24 boxes, 40 Capsules, 16 Prisms \\

        \includegraphics[width=0.25\linewidth]{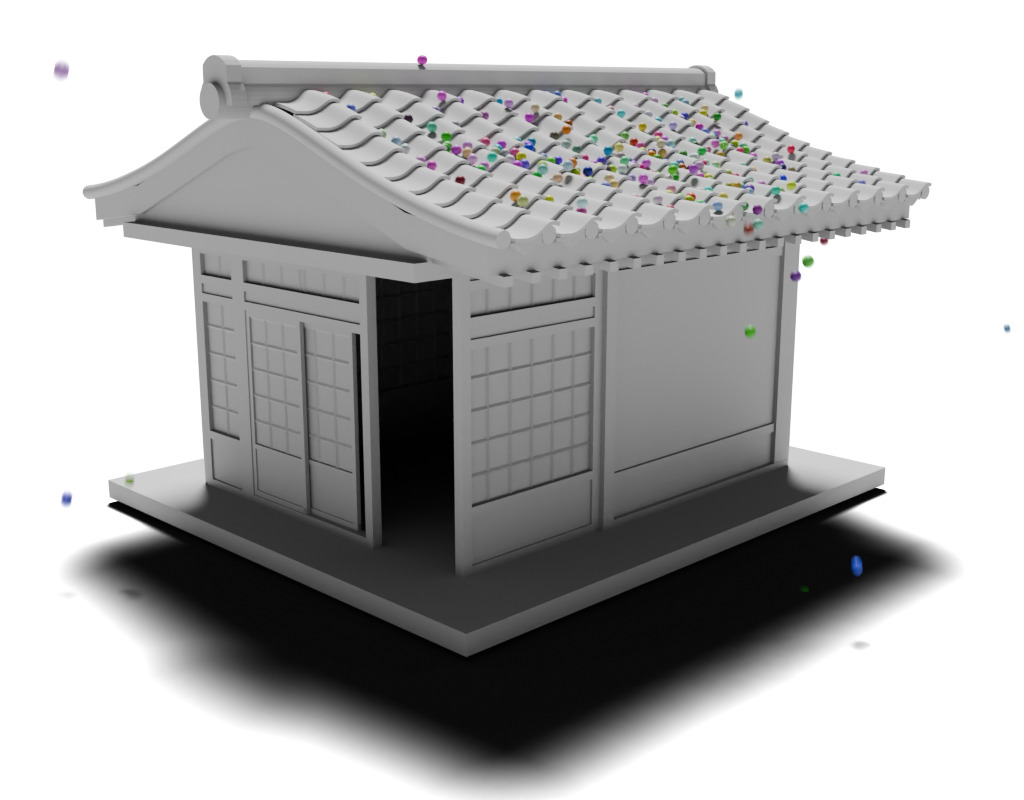} &
        \includegraphics[width=0.25\linewidth]{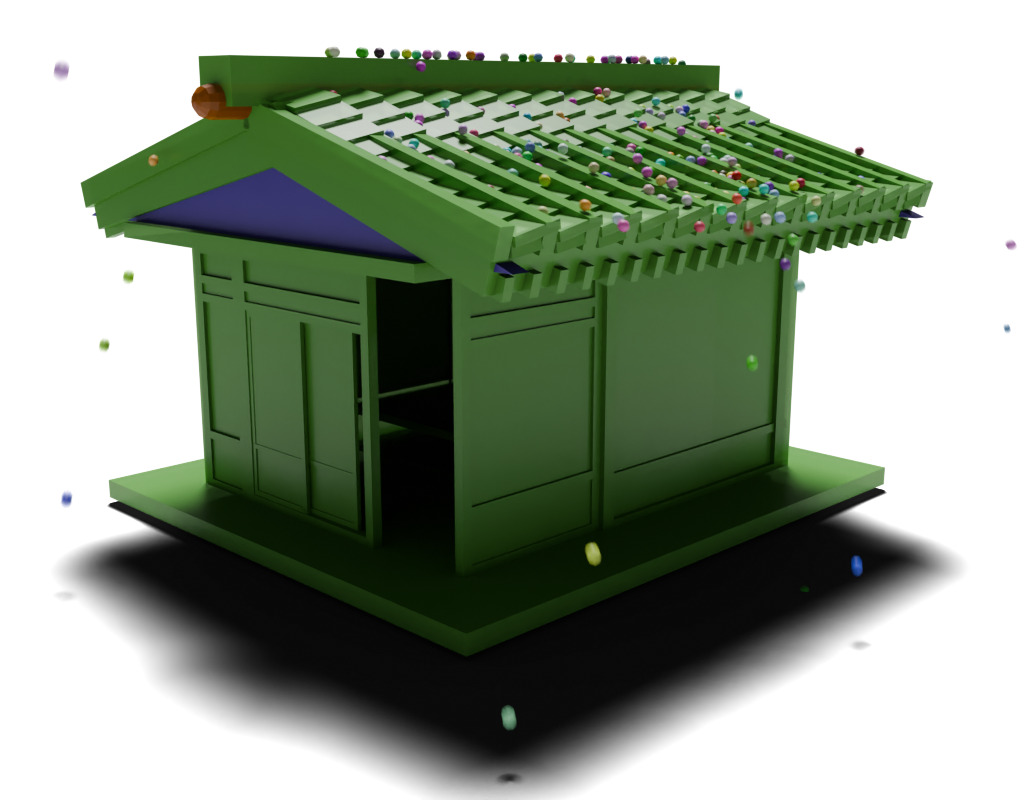} &
        \includegraphics[width=0.25\linewidth]{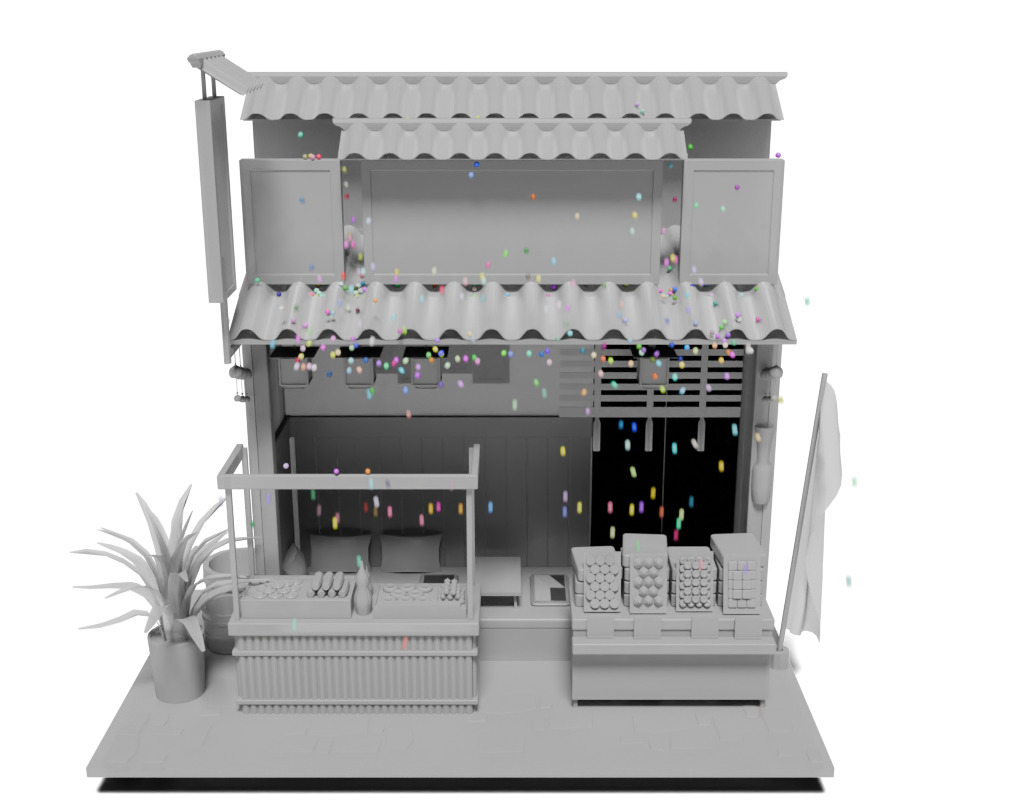} &
        \includegraphics[width=0.25\linewidth]{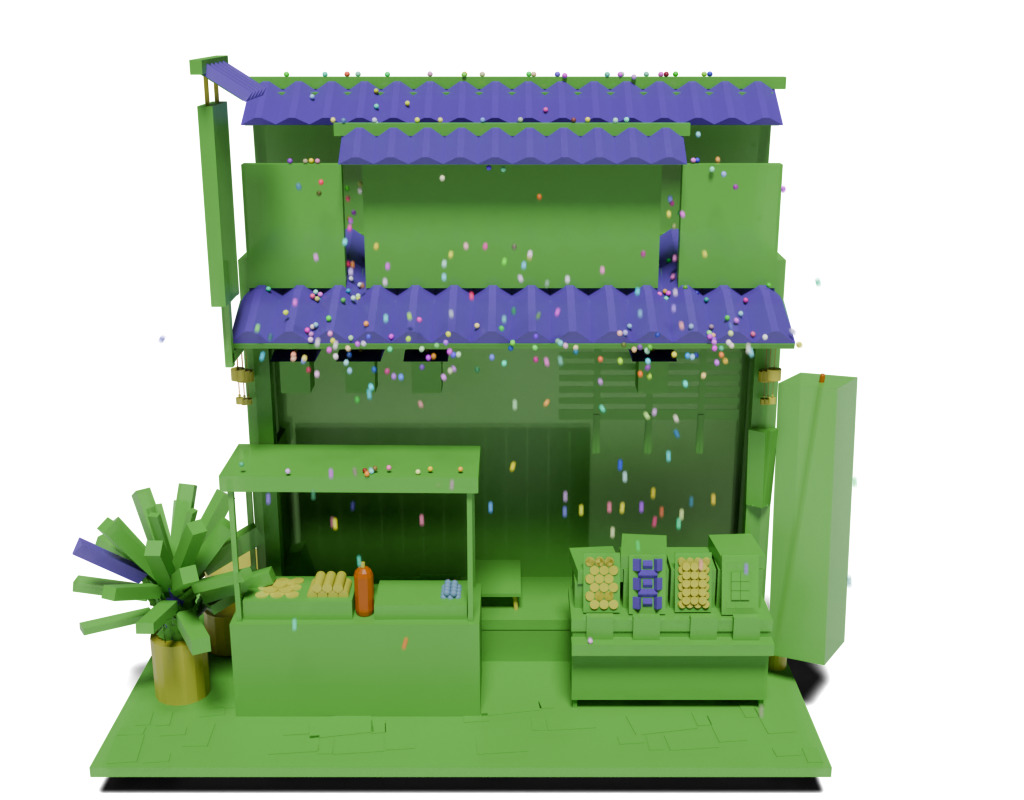} \\
        $|F| = 233965$ & 409 Boxes, 1 Cap, 2 Prism & $|F| = 119639$ & \tiny 645 Boxes, 22 Cap, 145 Cyl, 118 Prisms, 22 Sph \\
        
        \includegraphics[width=0.25\linewidth]{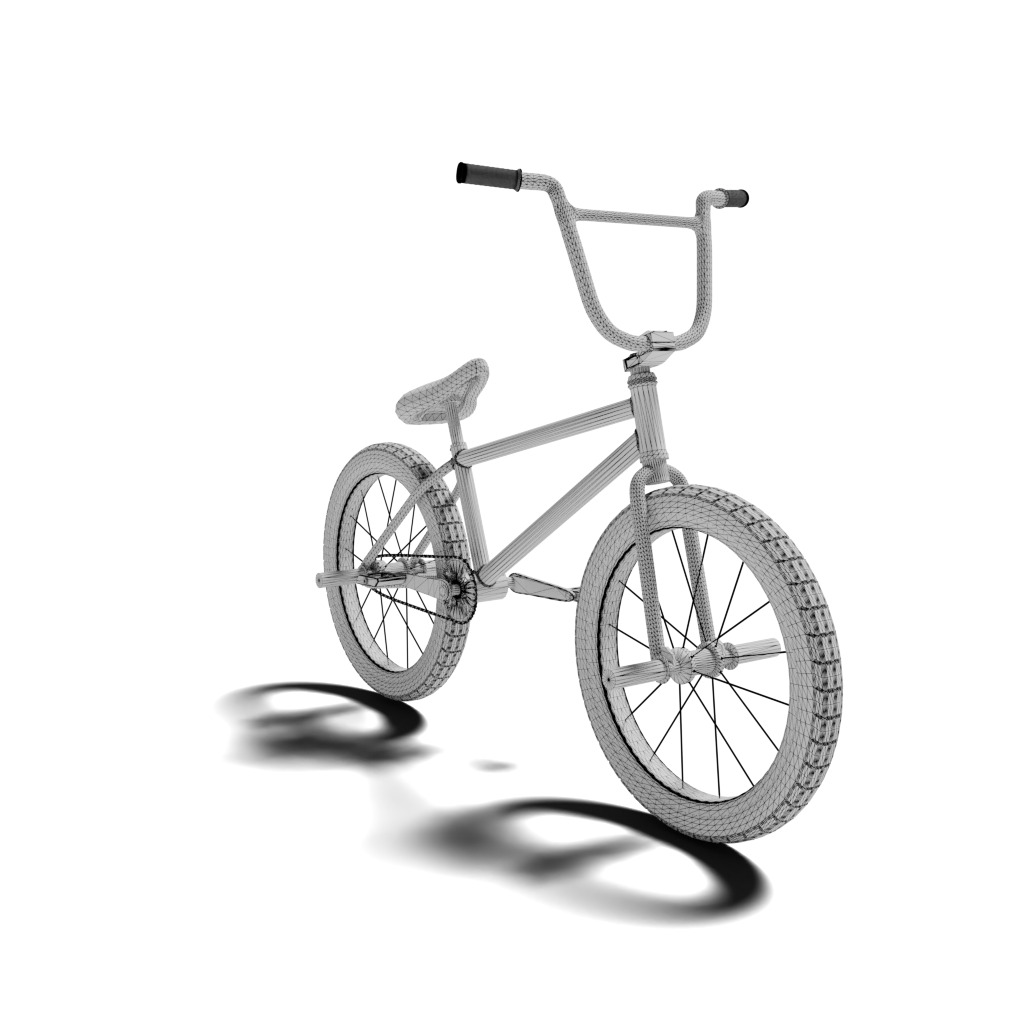} &
        \includegraphics[width=0.25\linewidth]{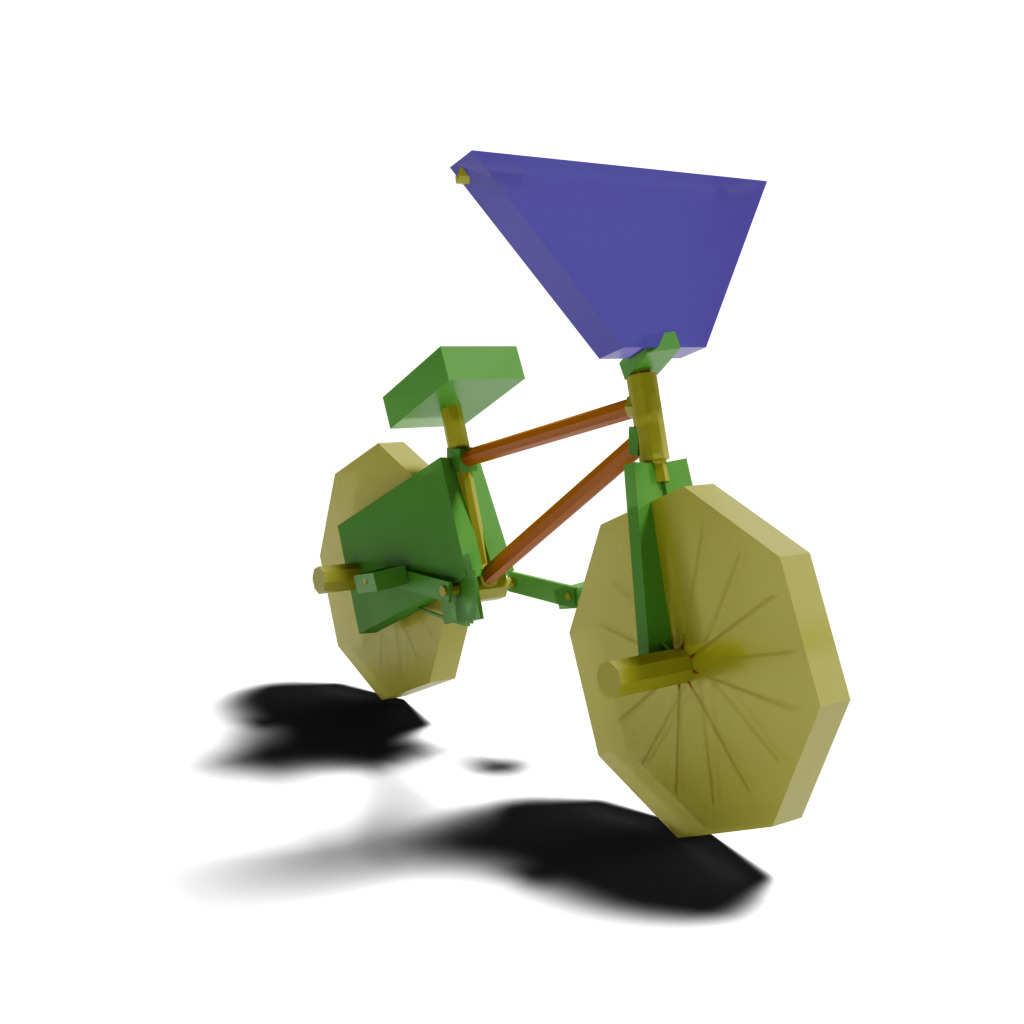} &
        
        \includegraphics[width=0.25\linewidth]{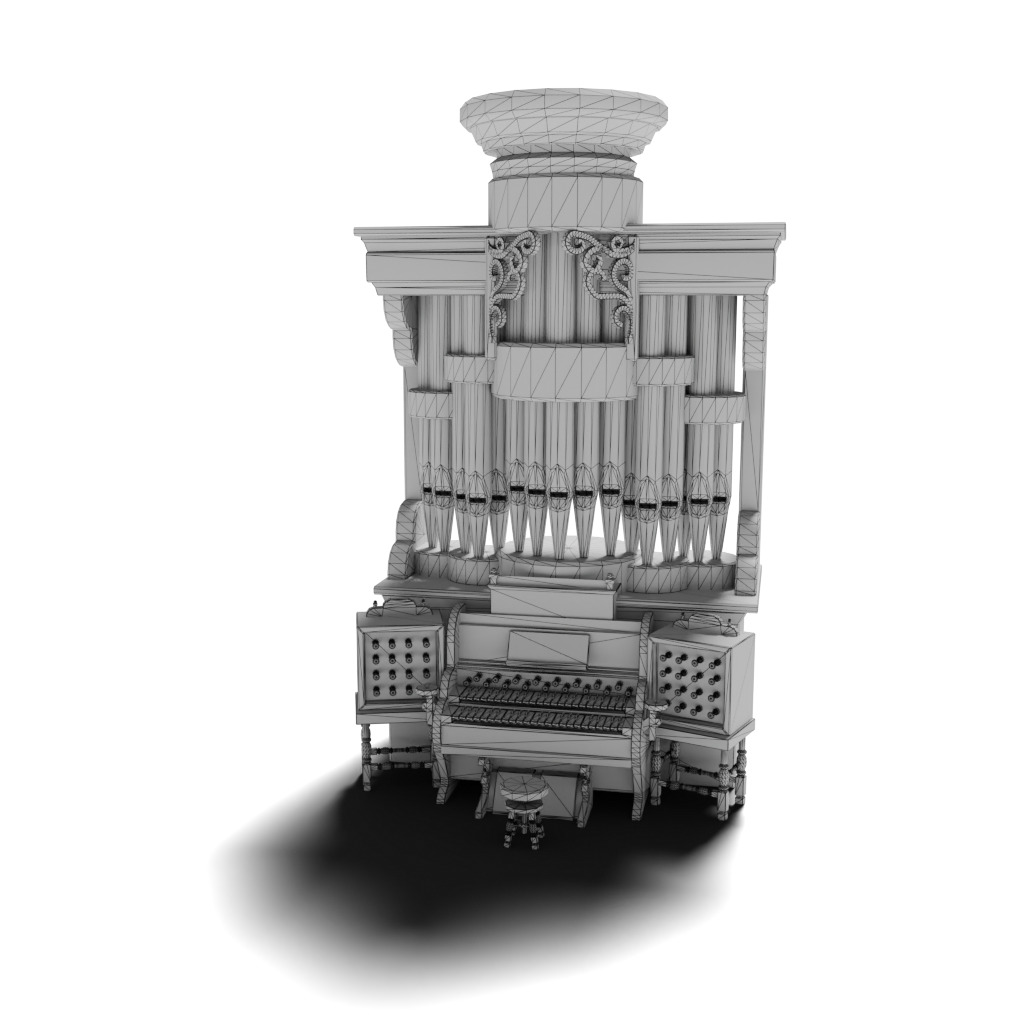} &
        \includegraphics[width=0.25\linewidth]{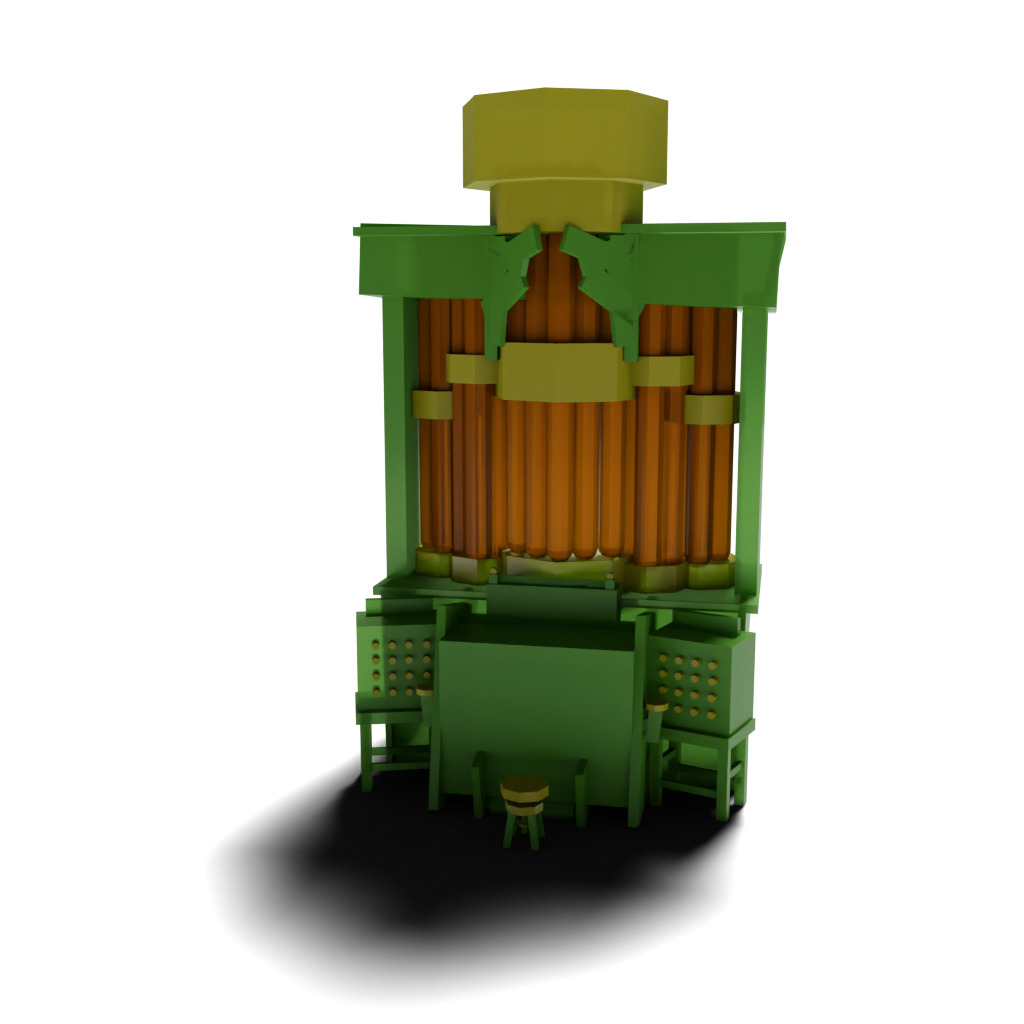}
        \\
        $|F| = 54995$ & 50 Boxes, 34 Cap, 22 Cyl, 2 Prism & $|F| = 20875$ &
        \footnotesize 94 Boxes, 17 Cap, 67 Cyl, 2 Sph \\

        \includegraphics[width=0.25\linewidth]{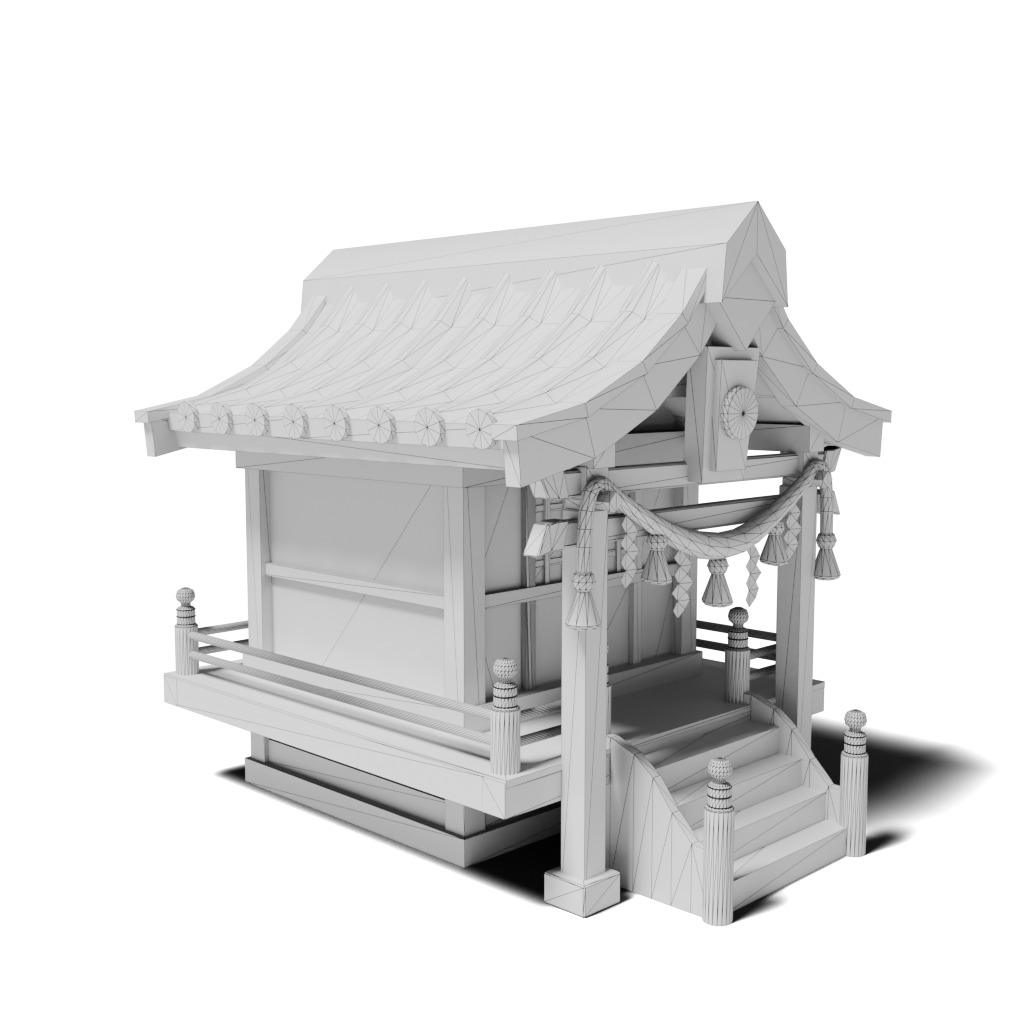} & \includegraphics[width=0.25\linewidth]{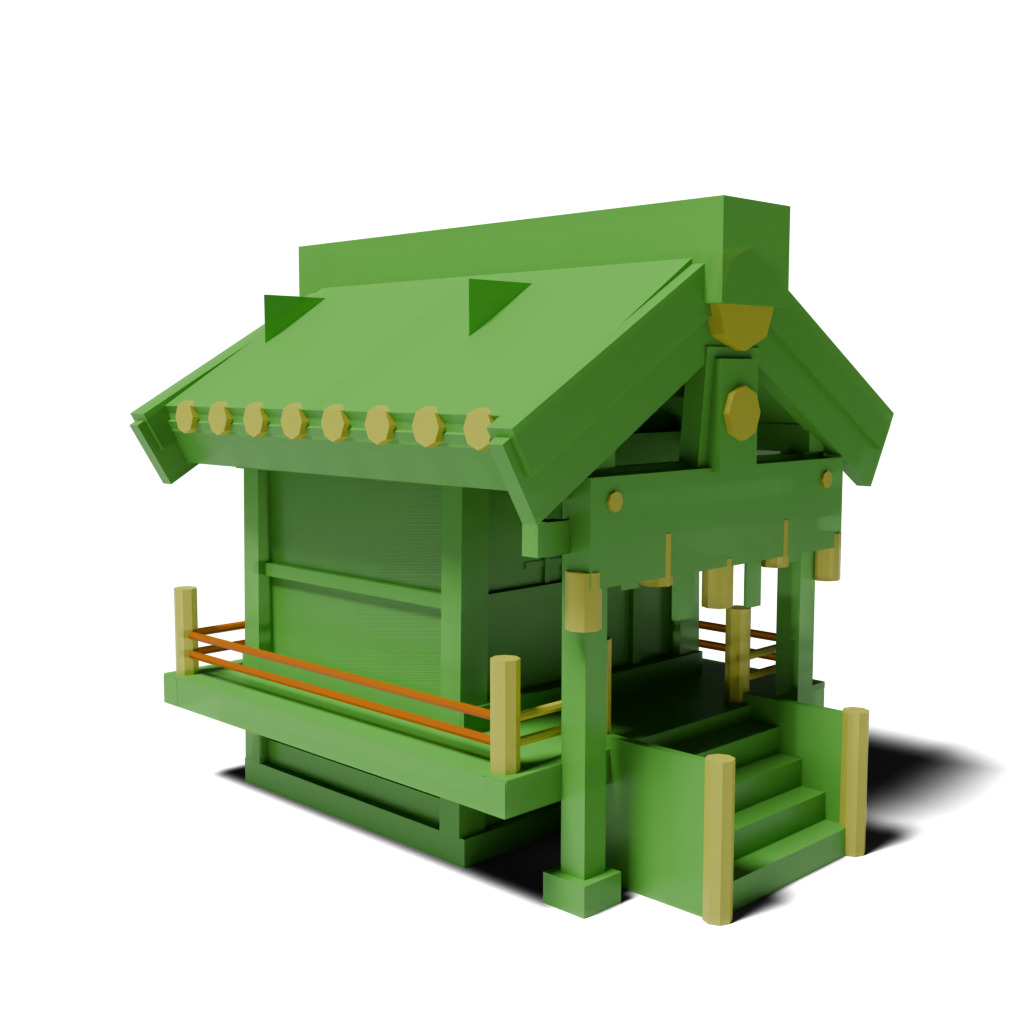} &
        \includegraphics[width=0.25\linewidth]{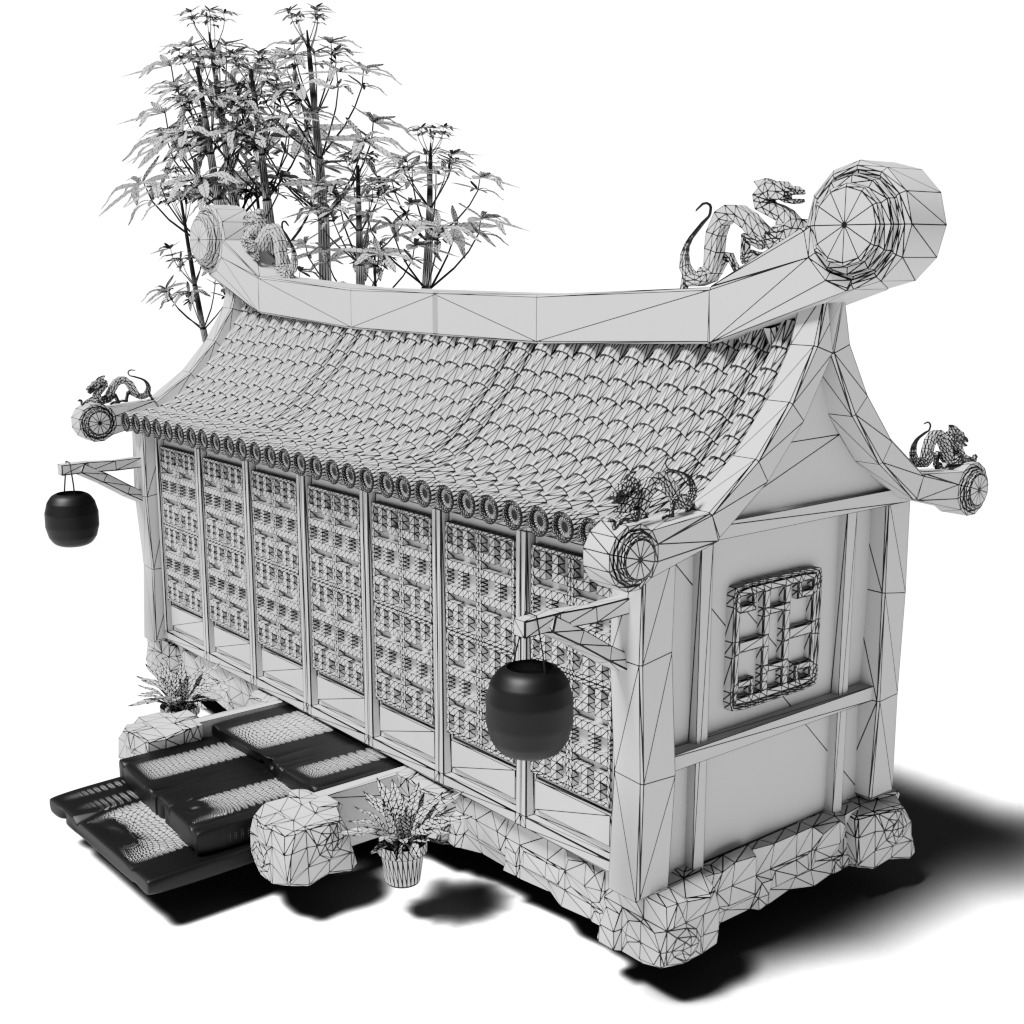} & \includegraphics[width=0.25\linewidth]{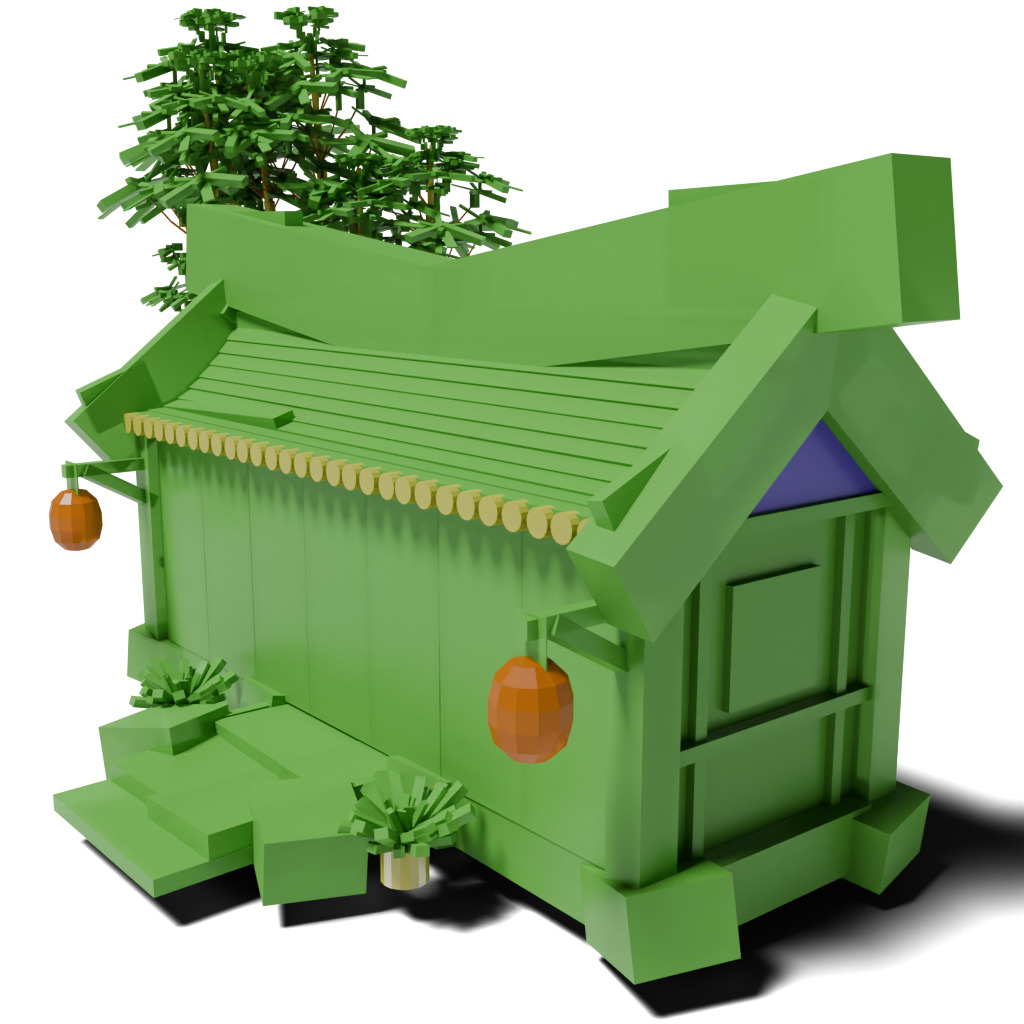} \\
        $|F| = 6280$ & 77 Boxes, 6 Capsules, 40 Cylinders & $|F| = 267267$ & \scriptsize 3675 Boxes, 223 Cap, 131 Cyl, 2 Prisms \\\\
    \end{tabular}
    \caption{Additional results from running our approach on a variety of input meshes. Green indicates bounding boxes, yellow for cylinders, blue for trapezoidal prisms, and red for capsules, light blue for spherres. Our approach greatly cuts down on the number of input elements. \ccby BlenderFace, \ccby RubaQewar, \ccby neos\_nelia, \ccby Deerex, \ccby Hoody468, \ccby Nathan Sioui, \ccby neih.D, \ccby Scritta.}
    \label{fig:additional_qualitative_results}
    % \Description{
    % Our approach on a number of additional outputs:
    % From left-to-right, top-to-bottom:
    % 1. A bicycle
    % 2. A spiral staircase
    % 3. A chimney pipe
    % 4. A church organ
    % 5. A small japanese house
    % 6. An asian storefront
    % 7. A shinto shrine
    % 8. An asian style hut with bamboo behind it.
    % }
\end{figure*}
}

{
\setlength{\tabcolsep}{0em}
\begin{figure*}
    \renewcommand{\arraystretch}{0.9}
    \centering
    \begin{tabular}{c c c c}
        \multicolumn{4}{c}{Additional Comparisons} \\
        Input Mesh & Convex Primitive Decomposition & CoACD & V-HACD \\
        \includegraphics[width=0.25\linewidth]{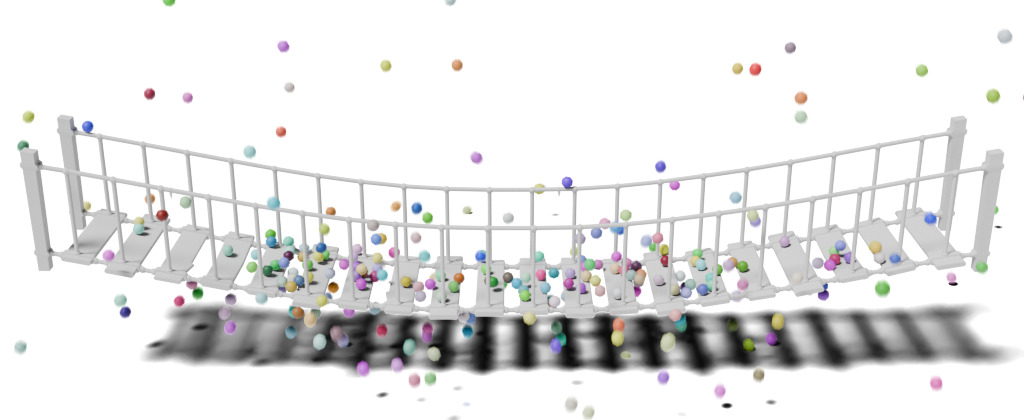} &
        \includegraphics[width=0.25\linewidth]{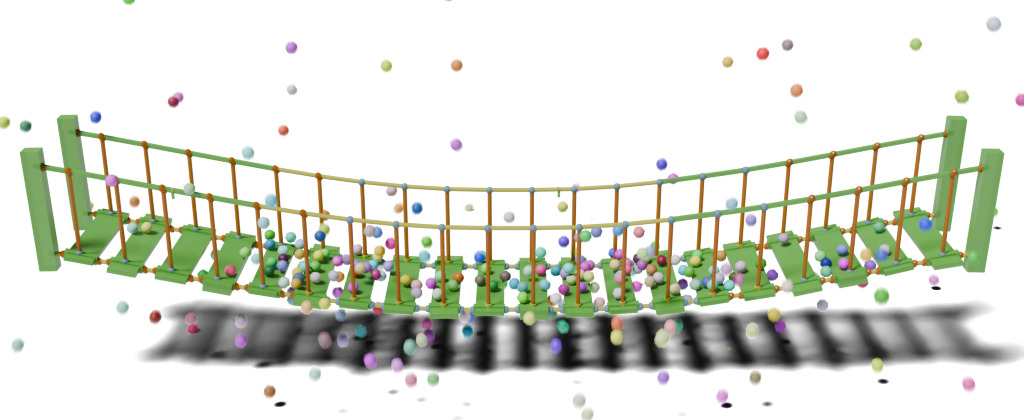} &
        \includegraphics[width=0.25\linewidth]{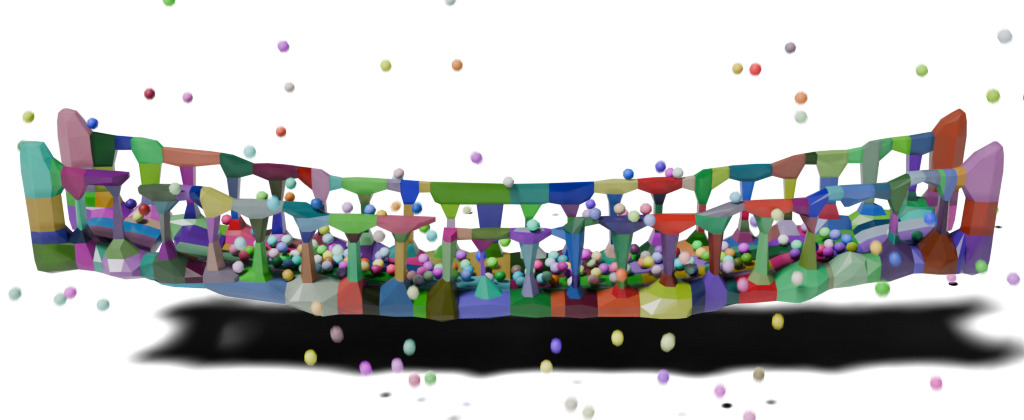} &
        \includegraphics[width=0.25\linewidth]{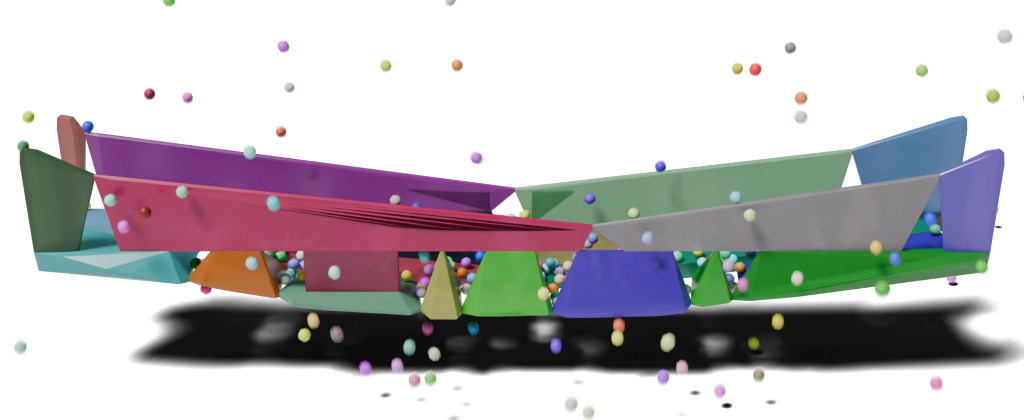} \\
        
        |F| = 7256 & 132 Boxes, 40 Cap, 170 Cyl, 16 Sph & 279 Hulls (|F| = 12246) & 27 Hulls (|F| = 2376) \\
        \includegraphics[width=0.15\linewidth]{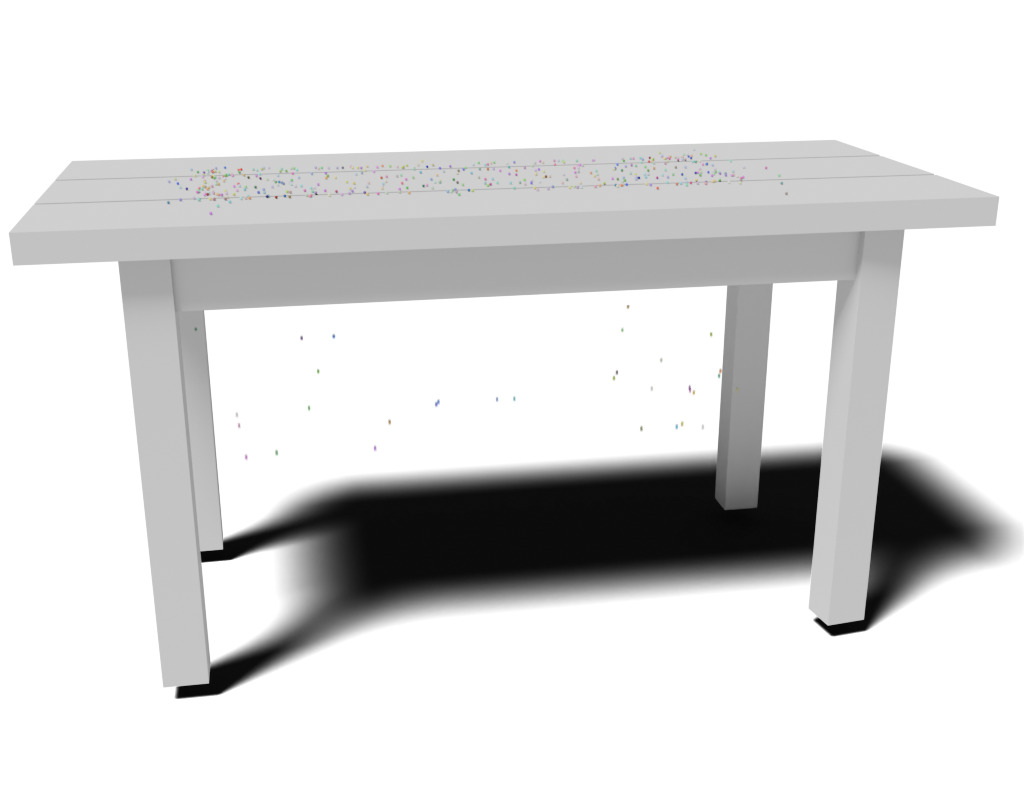} &
        \includegraphics[width=0.15\linewidth]{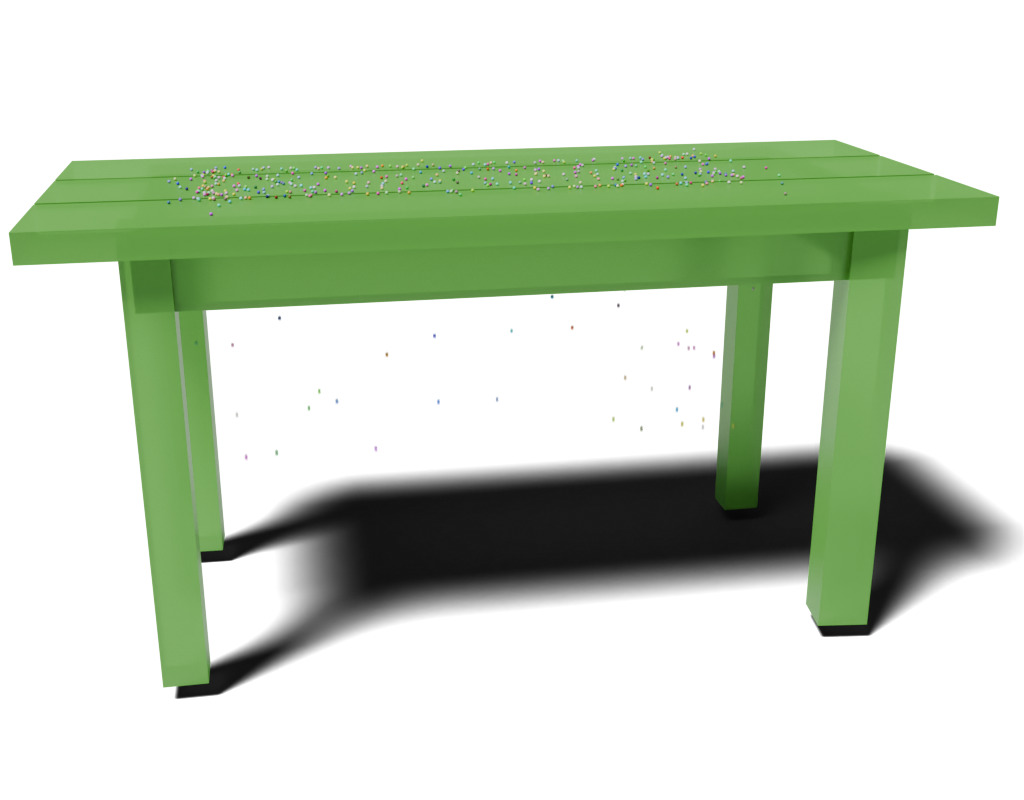} &
        \includegraphics[width=0.15\linewidth]{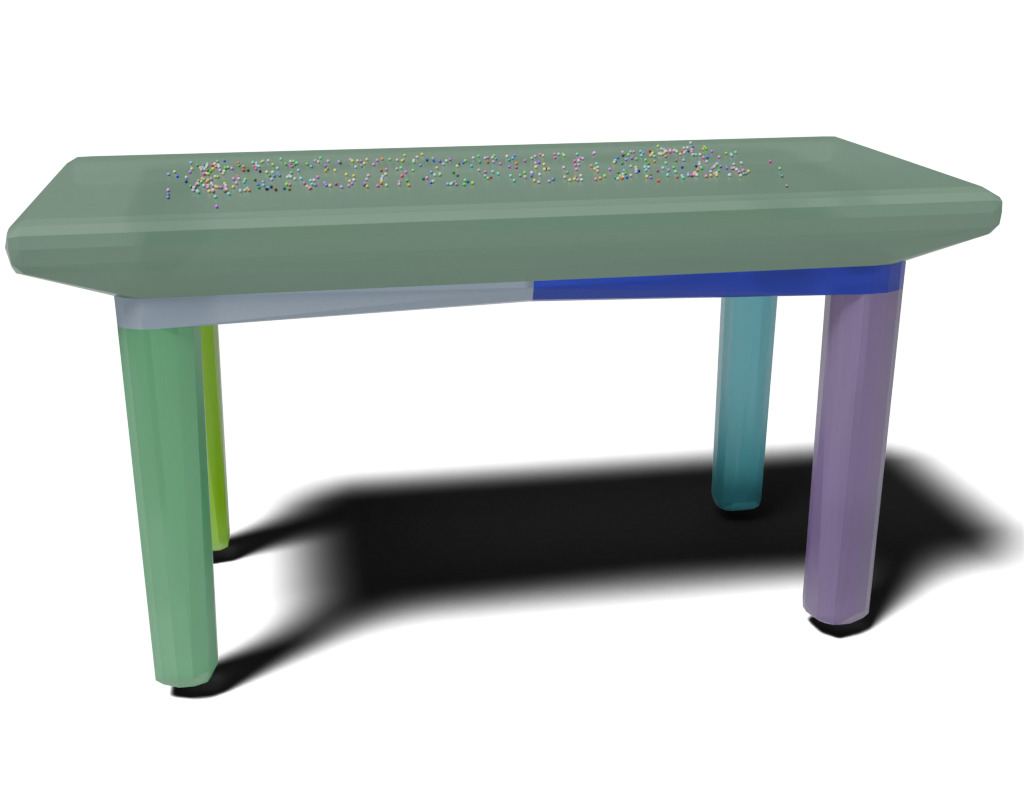} &
        \includegraphics[width=0.15\linewidth]{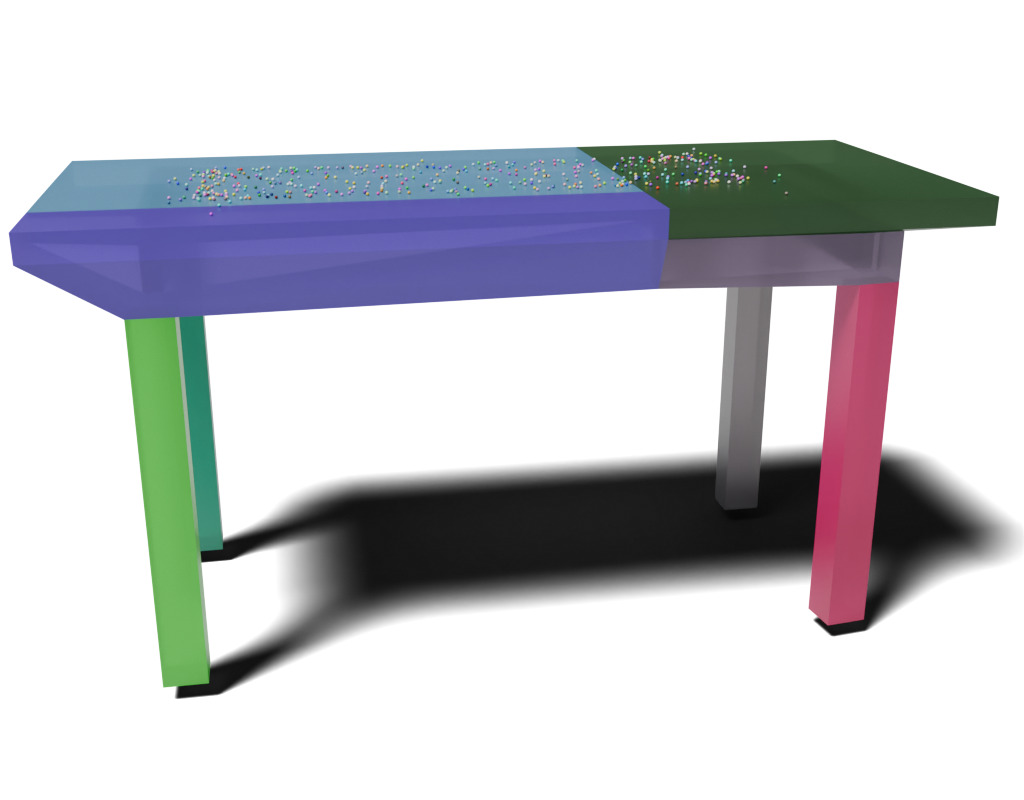} \\
        |F| = 66 & 11 Boxes & 11 Hulls (|F| = 1210) & 12 Hulls (|F| = 266) \\

        \includegraphics[width=0.2\linewidth]{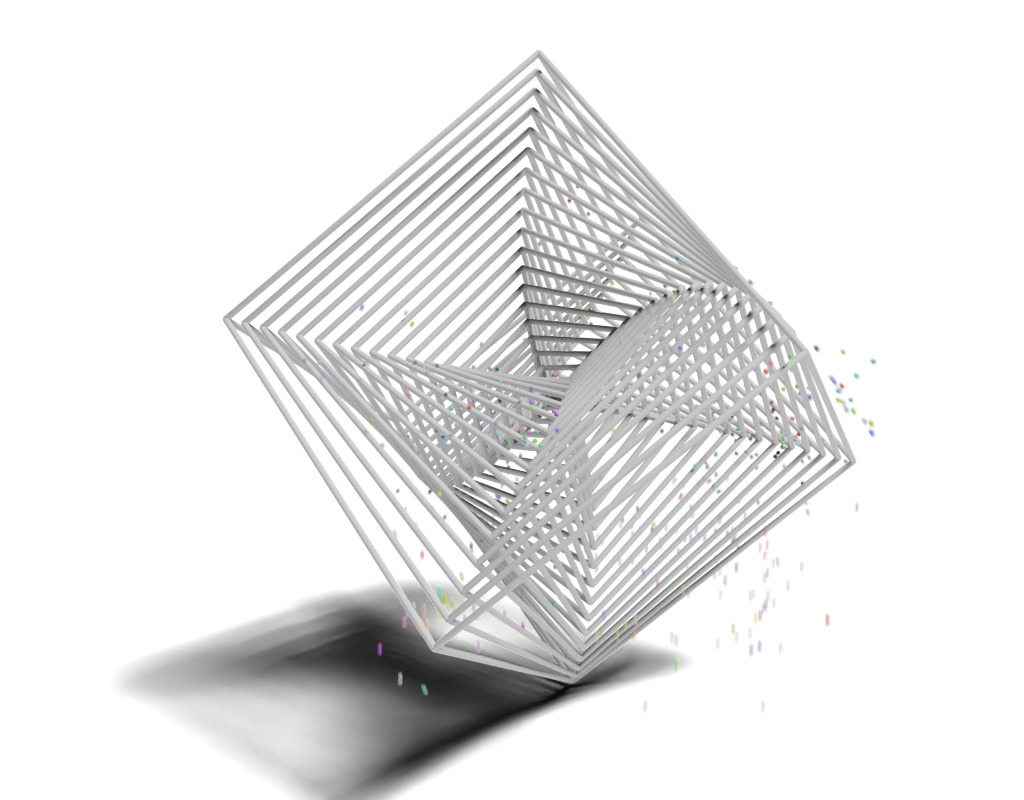} &
        \includegraphics[width=0.2\linewidth]{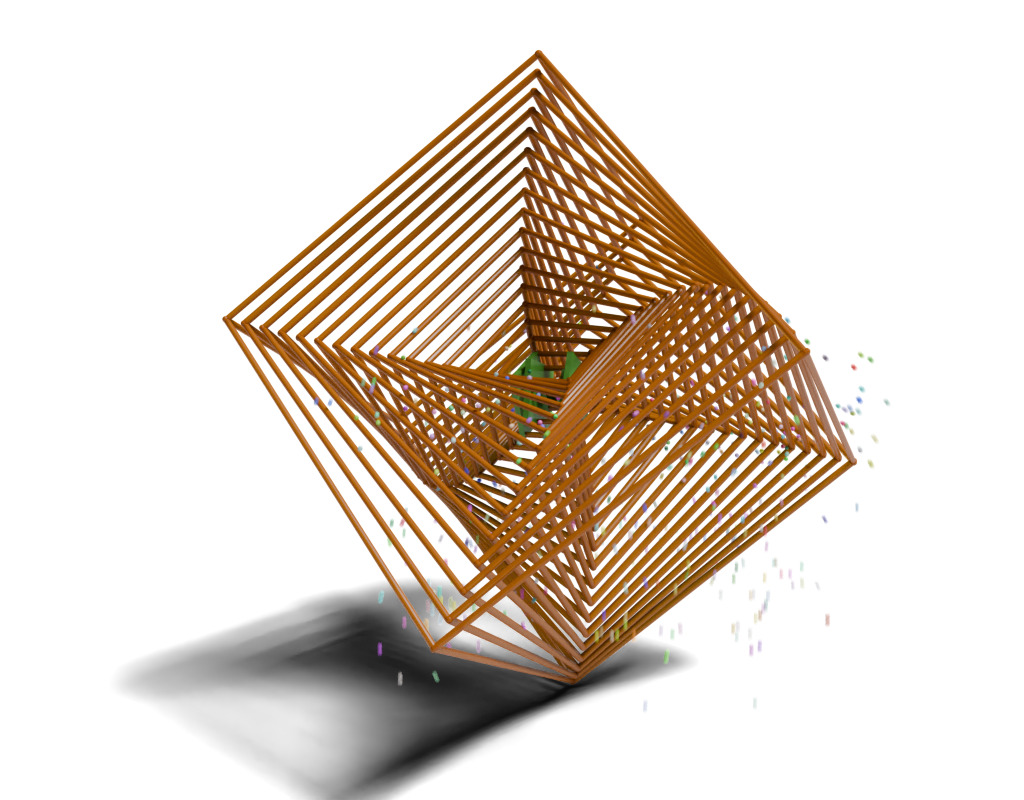} &
        \includegraphics[width=0.2\linewidth]{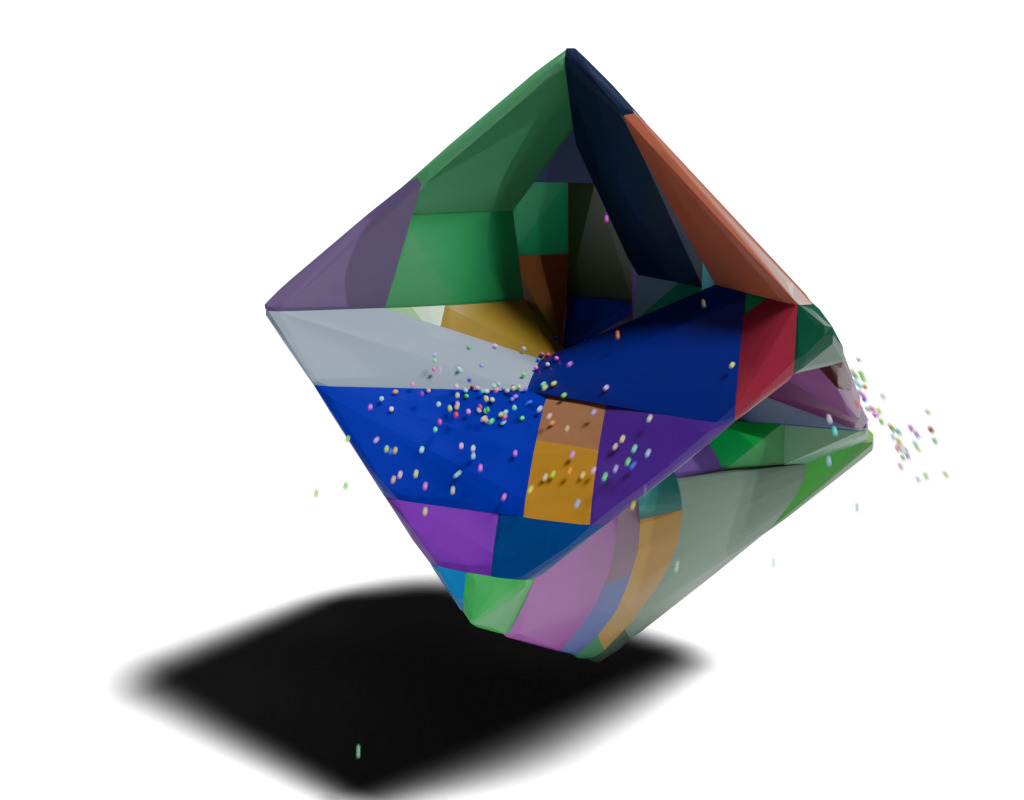} &
        \includegraphics[width=0.2\linewidth]{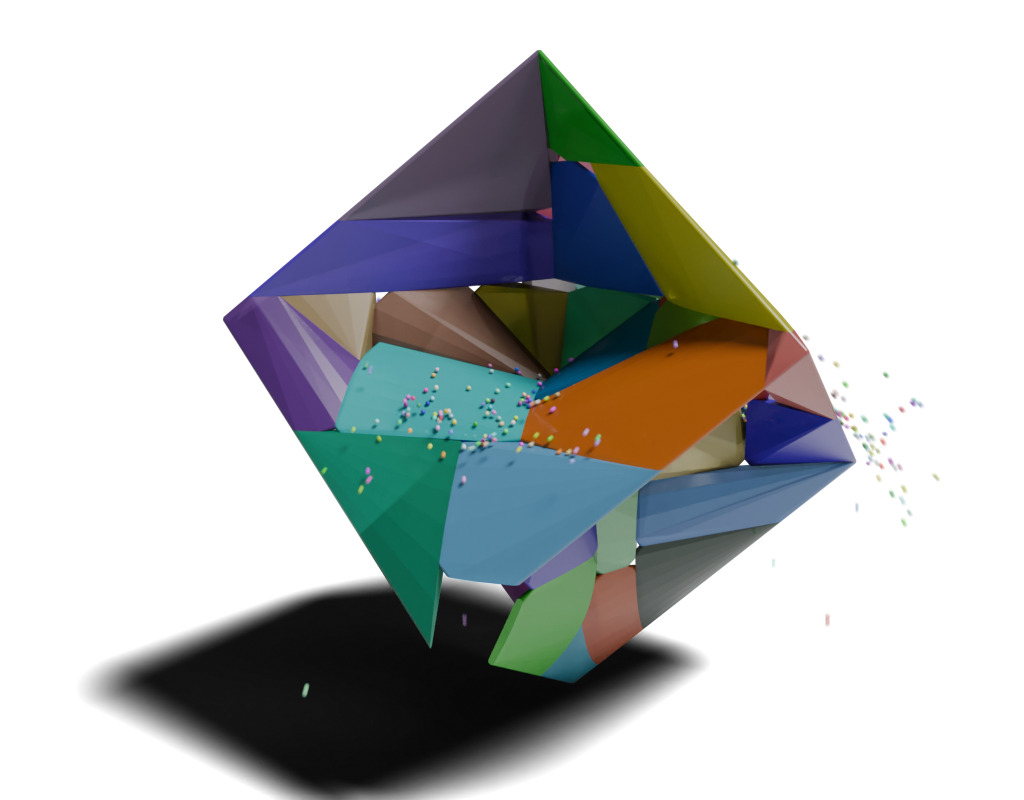} \\
        |F| = 5756 & 4 Boxes, 191 Cap, 21 Cyl & 86 Hulls (|F| = 7294) & 44 Hulls (|F| = 7459)  \\
        
        \includegraphics[width=0.14\linewidth]{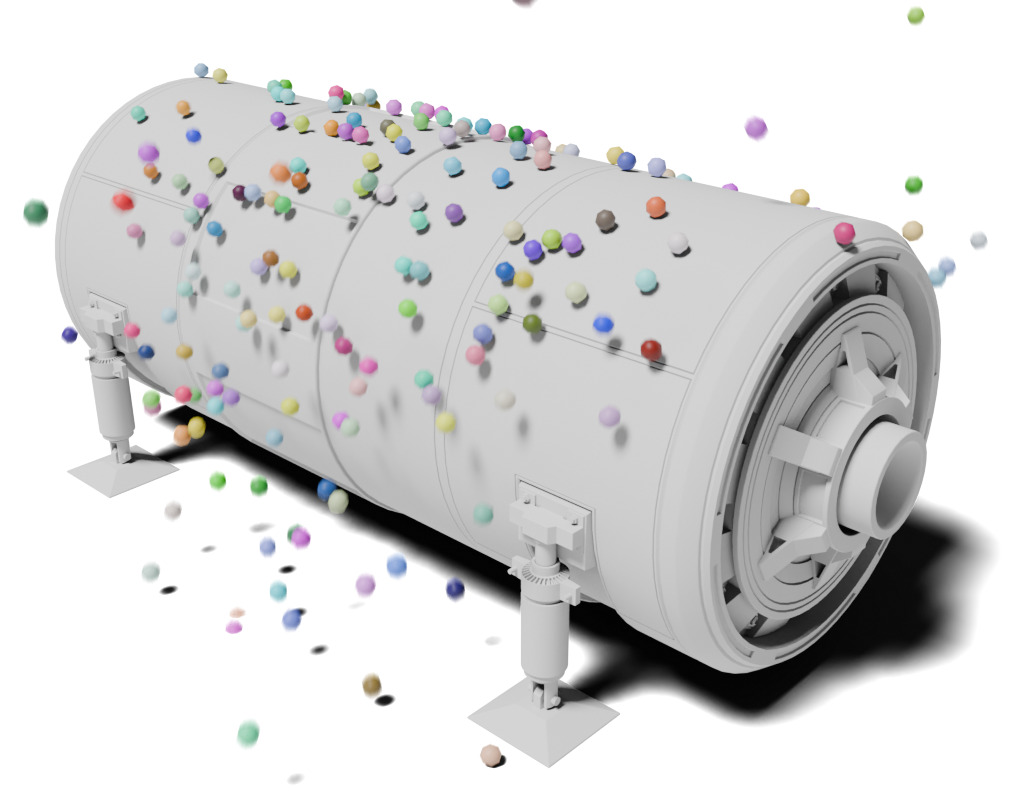} &
        \includegraphics[width=0.14\linewidth]{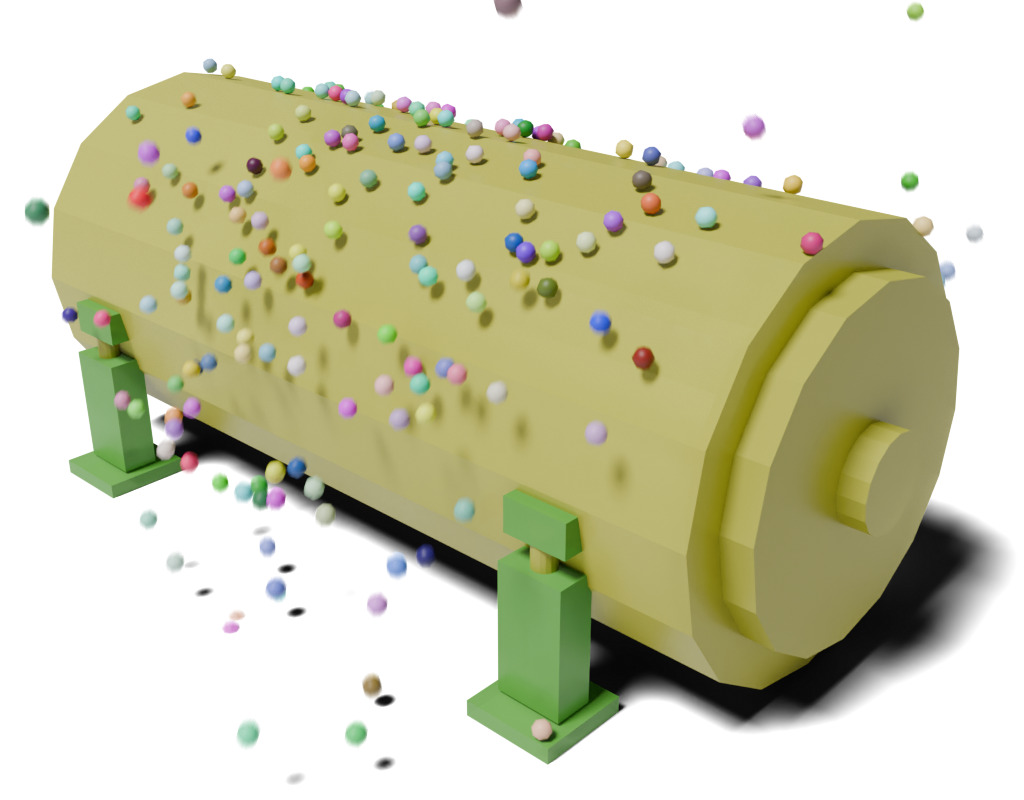} &
        \includegraphics[width=0.14\linewidth]{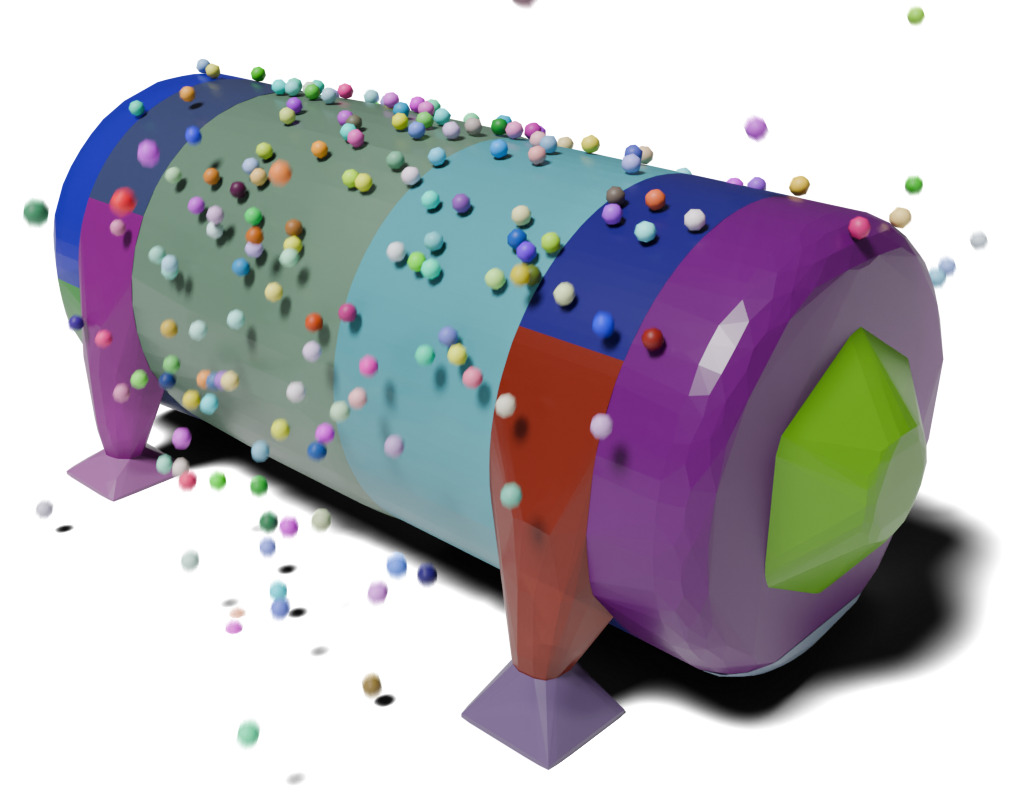} &
        \includegraphics[width=0.14\linewidth]{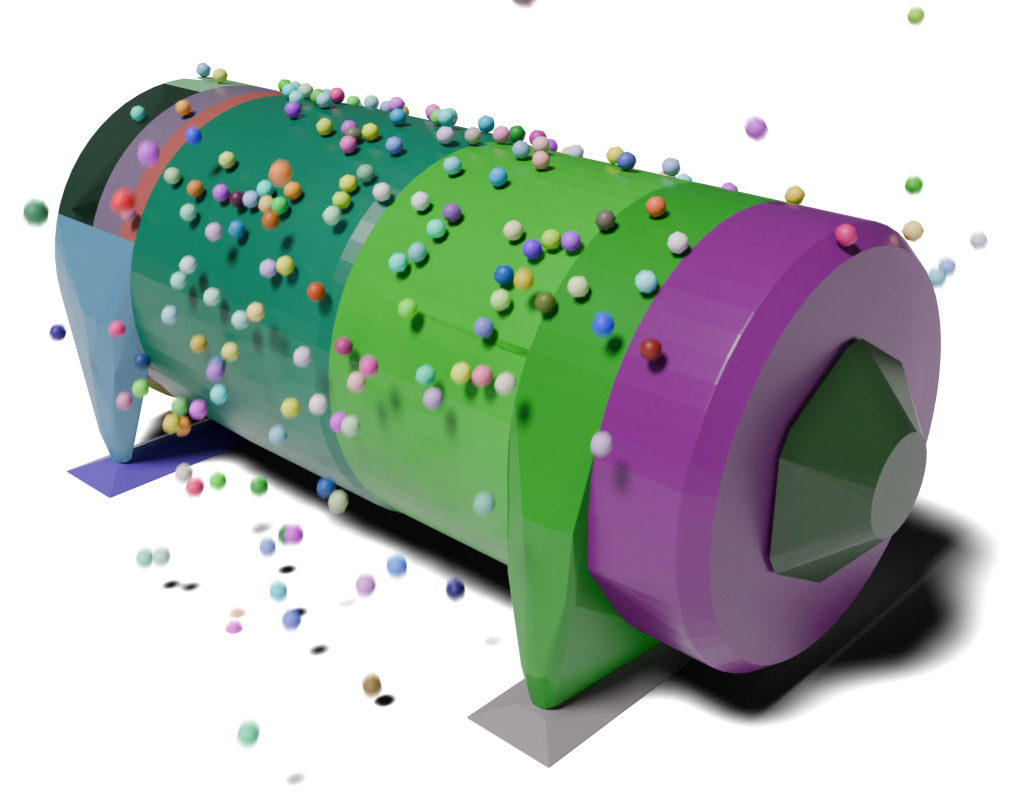} \\
        |F| = 15306 & 16 Boxes, 7 Cylinders & 19 Hulls (|F| = 4784) & 17 Hulls (|F| = 3672) \\

        \includegraphics[width=0.08\linewidth]{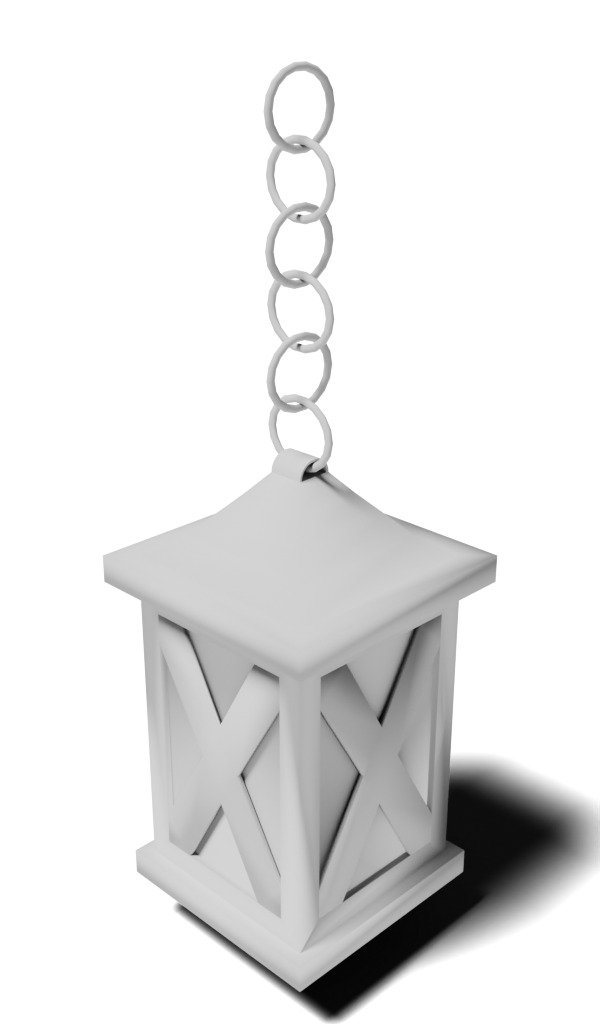} &
        \includegraphics[width=0.08\linewidth]{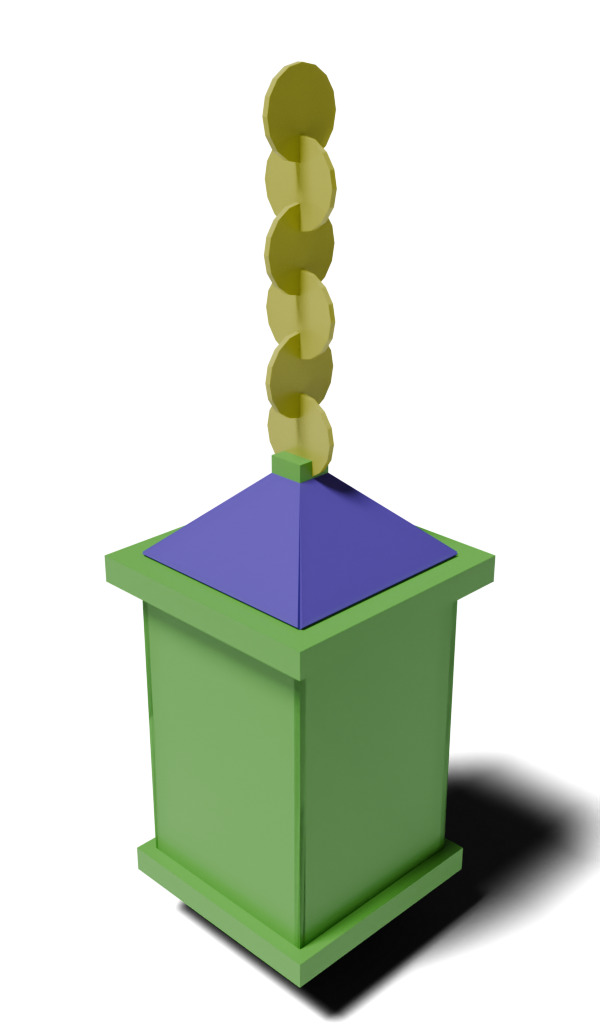} &
        \includegraphics[width=0.08\linewidth]{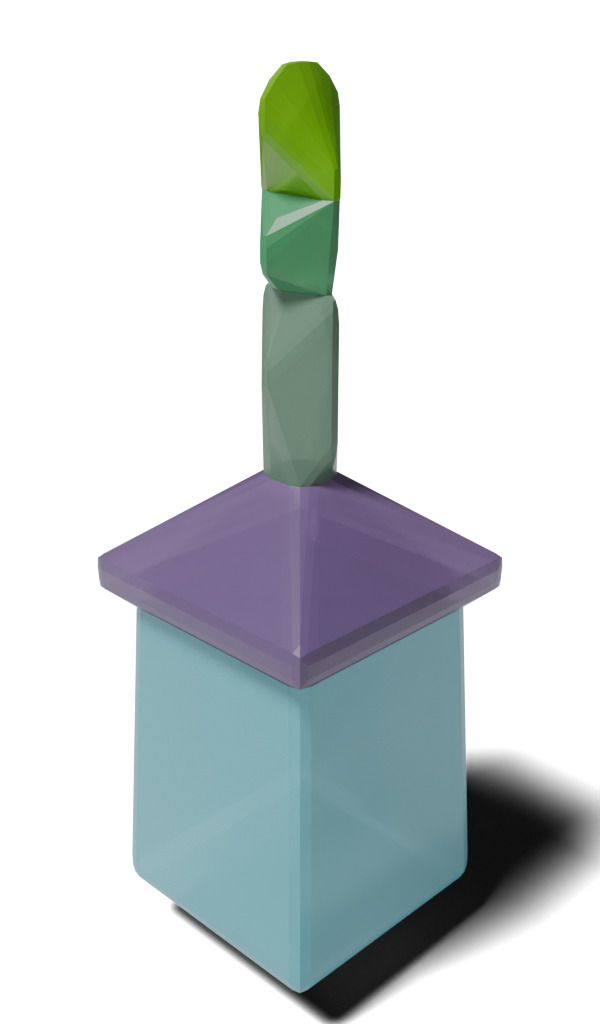} &
        \includegraphics[width=0.08\linewidth]{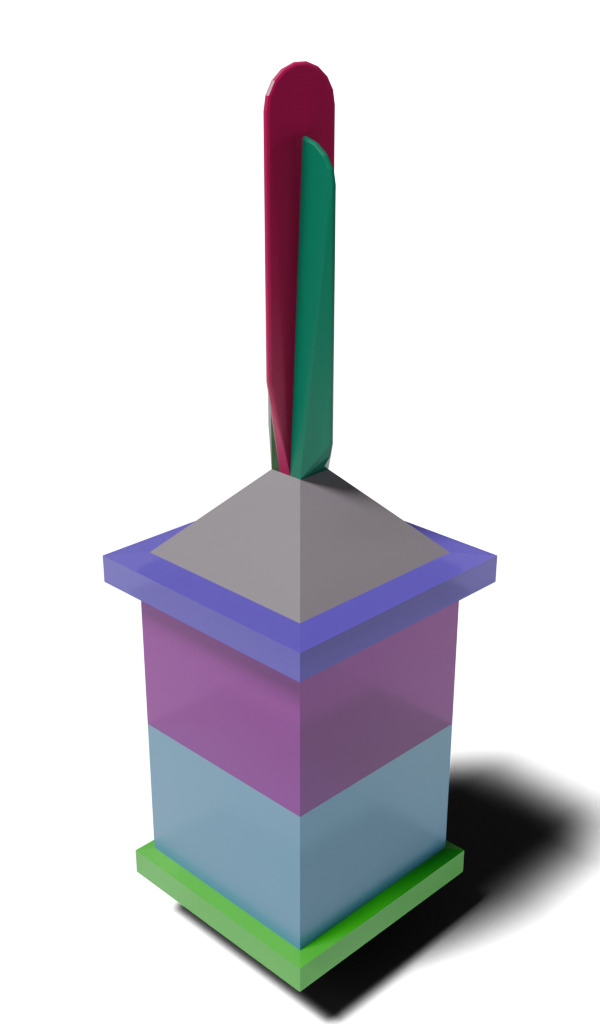} \\
        |F| = 1360 & 17 Boxes, 6 Cylinders, 4 Prisms & 5 Hulls (|F| = 716) & 8 Hulls (|F| = 394) \\

        \includegraphics[width=0.2\linewidth]{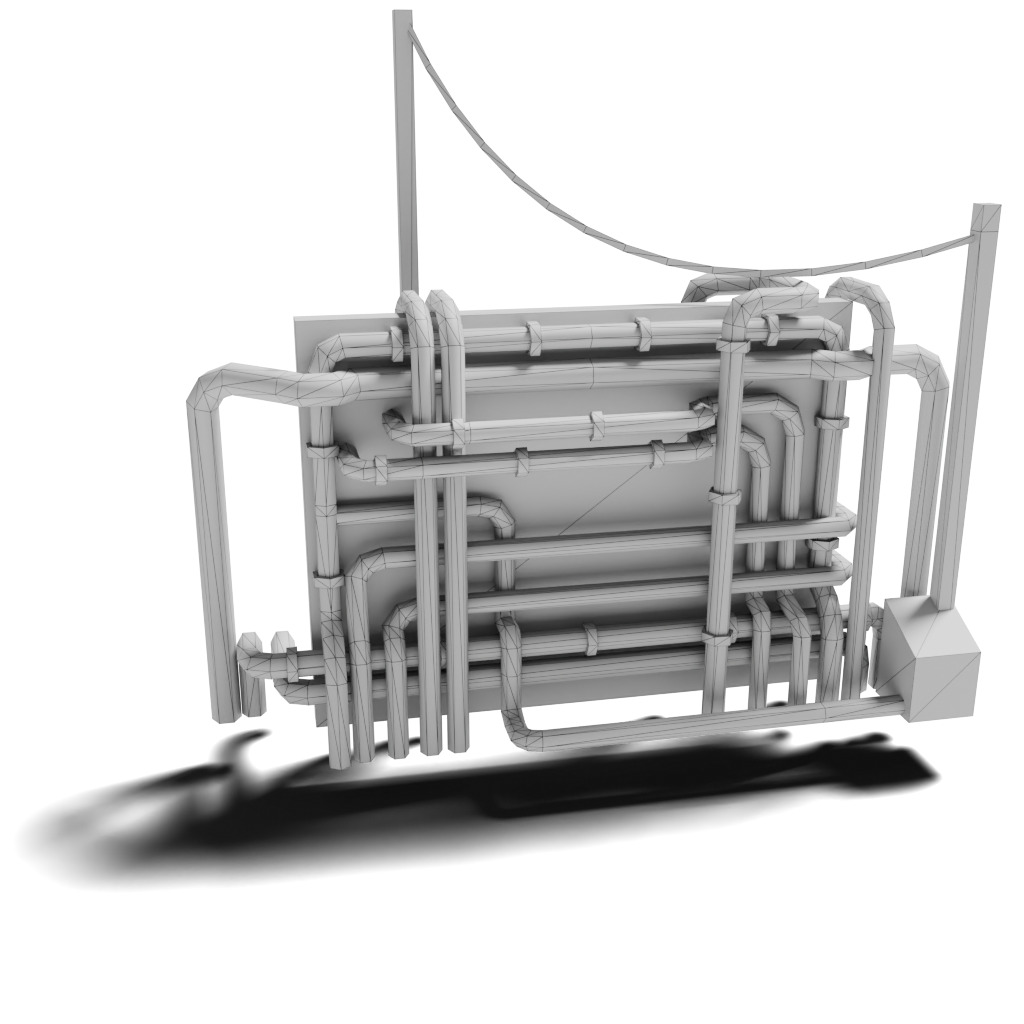} &
        \includegraphics[width=0.2\linewidth]{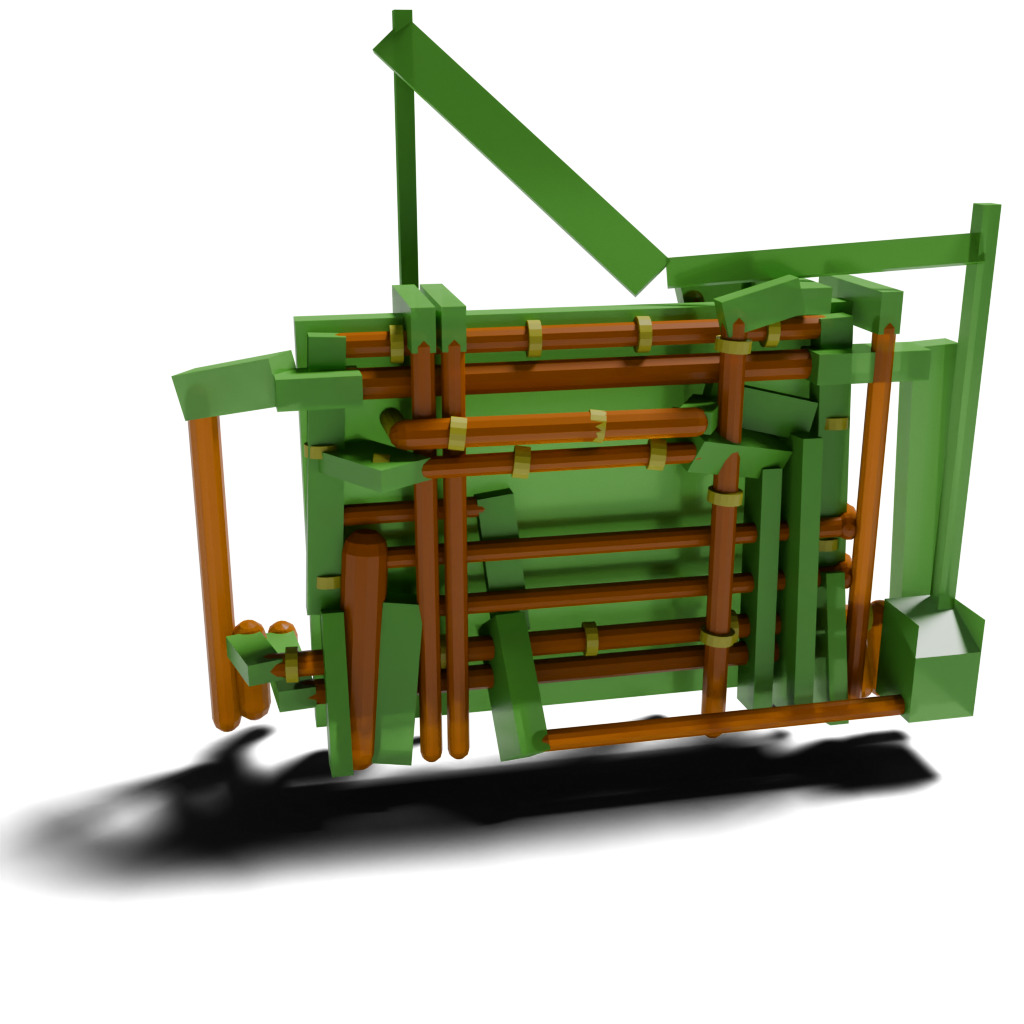} &
        \includegraphics[width=0.2\linewidth]{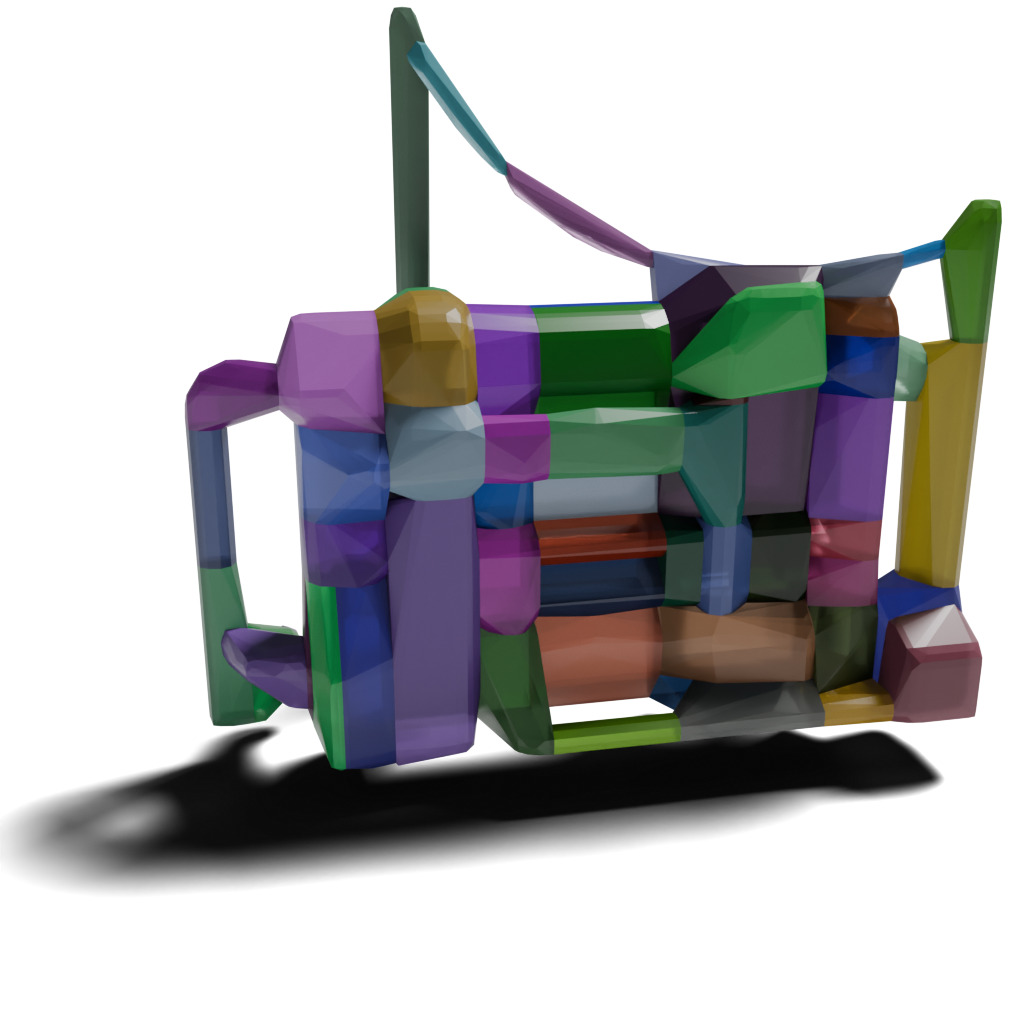} &
        \includegraphics[width=0.2\linewidth]{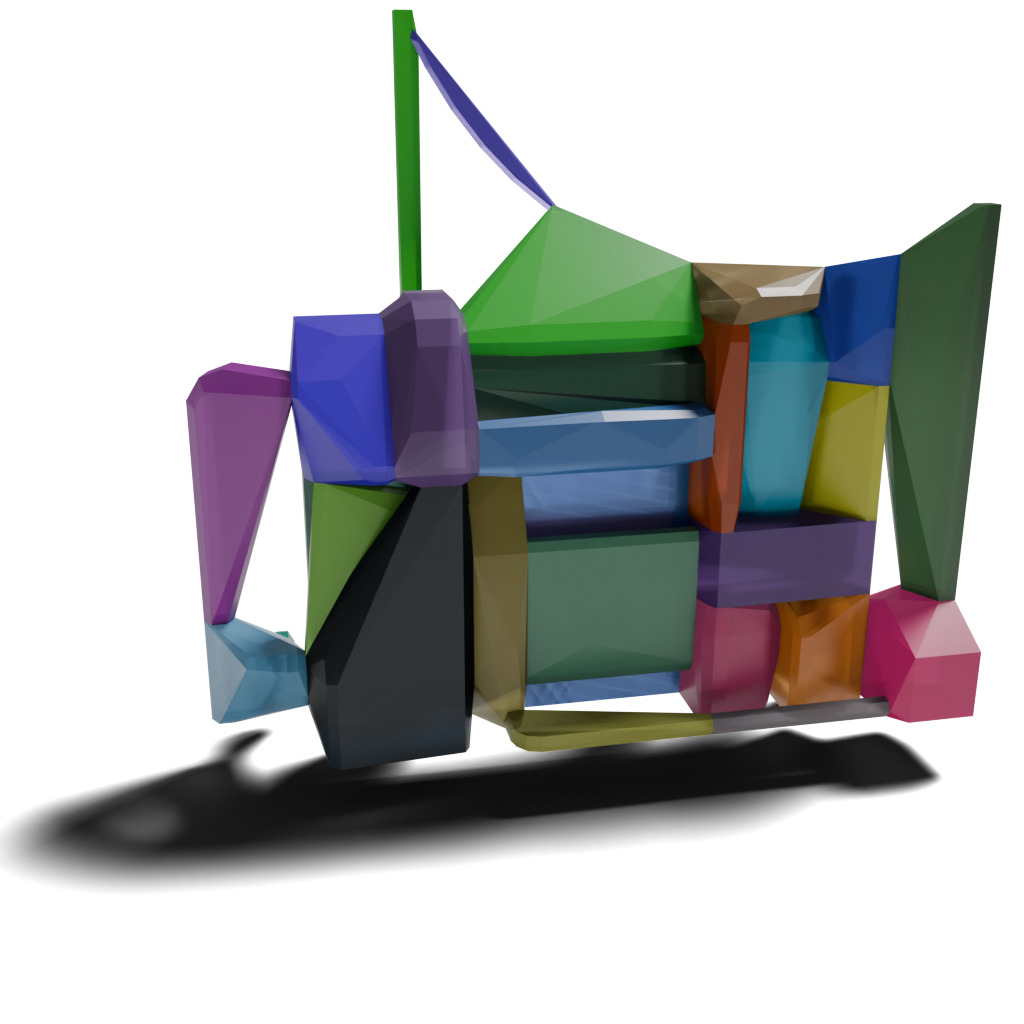} \\
        |F| = 1545 & 44 Boxes, 31 Cap., 25 Cyl. & 70 Hulls (|F| = 7564) & 35 Hulls (|F| = 1840) \\

        \includegraphics[width=0.25\linewidth]{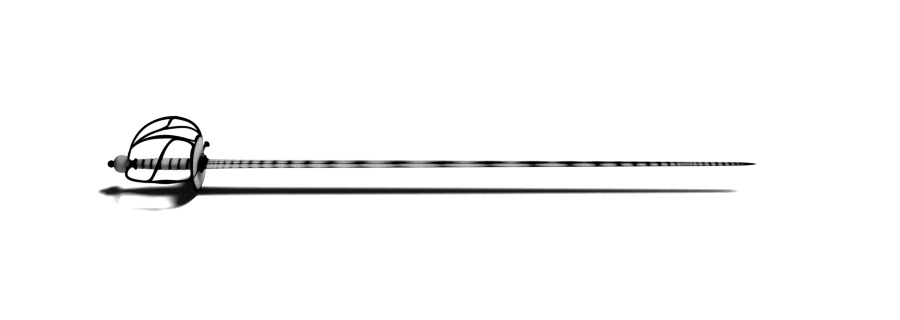} &
        \includegraphics[width=0.25\linewidth]{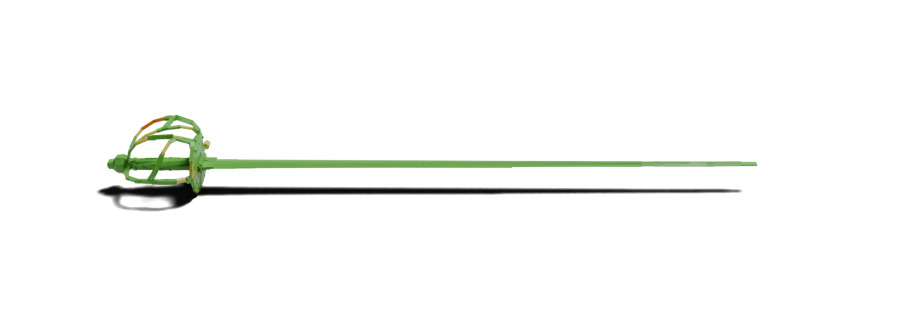} &
        \includegraphics[width=0.25\linewidth]{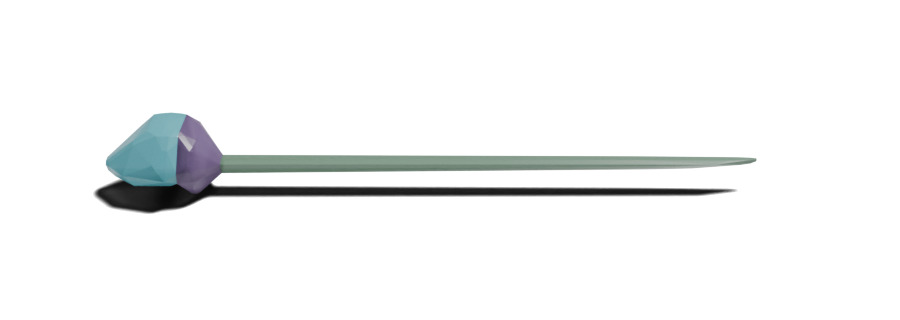} & \includegraphics[width=0.25\linewidth]{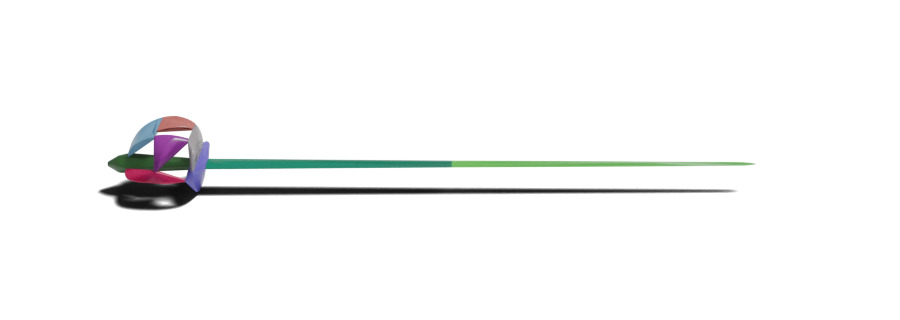} \\
        |F| = 35288 & 273 Boxes, 5 Cap, 18 Cyl, 8 Prisms & 3 Hulls (|F| = 596) & 9 Hulls (|F| = 10182) \\
    \end{tabular}
    \caption{Additional comparison from running our approach on a variety of input meshes. Green indicates bounding boxes, yellow cylinders, blue trapezoidal prisms, red capsules, light blue spheres, and light orange frustums. Our approach more closely adheres to the input mesh while containing fewer elements than approximate convex decomposition approaches. Our approach also uses the fewest bytes on all models shown. \ccby lucq, \ccby toAflame, \ccby maxpsr, \ccby Blenderolokos, \ccby Aerial\_Knight, \ccby RitorDP.}
    \label{fig:additional_comparisons}
    % \Description{
    % Our approach compared to CoACD and V-HACD on a number of additional outputs:
    % First column is the input, 2nd column is our approach, third column is CoACD, fourth column is V-HACD.
    % From top-to-bottom:
    % 1. A bridge
    % 2. A table
    % 3. A cylindrical container of some sort
    % 4. A lantern hanging by a chain
    % 5. A wall covered by pipes
    % 6. a French sword with an ornate guard
    % }
\end{figure*}
}

% AUTOGENERATED TABLE manually modified
\begin{table*}
\centering
\scriptsize
\begin{tabular}{|c|| c|c|c|c ||c|c|c|c|c|c|c||c|c||c|c|}
\hline
& \multicolumn{4}{c||}{Input} & \multicolumn{7}{c||}{Ours} & \multicolumn{2}{c||}{CoACD} & \multicolumn{2}{c|}{V-HACD} \\\hline 
\hline
Name & \#F & \#V  & Mani. & WT & \#Box & \#Cap & \#Sph & \#Cyl & \#Fru & \#Pri & Time (s) & \#F & \#Hull & \#F & \#Hull \\\hline
Armadillo & 99976 & 49990 & \greencheck & \greencheck & 195 & & & 4 & & 1 & 12.303 & 9632 & 24 & 11275 & 15 \\\hline
Armored Charizard & 4649 & 3273 & \greencheck & \redX & 144 & 2 & & 10 & & 32 & 0.417 & 7170 & 41 & 2402 & 26 \\\hline
Bell & 10686 & 10578 & \greencheck & \redX & 47 & 4 & & 2 & & & 1.418 & 11546 & 63 & 6198 & 22 \\\hline
Bicycle & 54995 & 102354 & \greencheck & \redX & 50 & 34 & & 22 & & 2 & 9.373 & 4904 & 30 & 9106 & 23 \\\hline
Canon At1 & 9624 & 13094 & \greencheck & \redX & 8 & & & 4 & & & 1.531 & 7484 & 28 & 3174 & 15 \\\hline
Cardboard Boxes & 119 & 202 & \redX & \redX & 34 & & & & & 10 & 0.006 & 5568 & 65 & 496 & 40 \\\hline
Casio Keyboard & 2804 & 3730 & \greencheck & \redX & 16 & & & & & & 0.511 & 440 & 1 & 658 & 6 \\\hline
Chimney Pipe & 3404 & 4864 & \greencheck & \redX & 1 & & & 2 & 3 & & 0.641 & 5354 & 27 & 2334 & 30 \\\hline
Chitinous Knight & 87258 & 89239 & \greencheck & \redX & 47 & 1 & & 1 & & & 16.084 & 4316 & 16 & 13019 & 14 \\\hline
Chuo House & 173163 & 209450 & \redX & \redX & 1023 & 10 & & 22 & & 18 & 32.894 & 3198 & 29 & 5200 & 27 \\\hline
Church Organ & 20875 & 26902 & \greencheck & \redX & 94 & 17 & 2 & 67 & & & 2.392 & 5550 & 54 & 3122 & 39 \\\hline
Cinema Scan & 423713 & 351719 & \greencheck & \redX & 475 & 1 & & 17 & & 5 & 62.922 & 8384 & 87 & 6030 & 37 \\\hline
Cube & 5756 & 7401 & \greencheck & \redX & 4 & 191 & & 24 & & & 0.516 & 7286 & 87 & 7076 & 44 \\\hline
Cyberpunk Atm & 6566 & 10447 & \greencheck & \redX & 63 & & & 24 & & & 0.659 & 3912 & 25 & 1458 & 16 \\\hline
Cyberpunk Bike & 31916 & 37040 & \greencheck & \redX & 34 & 3 & 4 & 16 & & 1 & 4.379 & 4614 & 15 & 9626 & 42 \\\hline
Dojo & 4569 & 4133 & \greencheck & \redX & 470 & 1 & & 8 & & 4 & 0.208 & 9866 & 149 & 2354 & 50 \\\hline
Dungeon (Precise) & 17986 & 59852 & \greencheck & \redX & 532 & & & 9 & & & 1.680 & 6984 & 90 & 1298 & 17 \\\hline
Espresso Machine & 19738 & 29059 & \greencheck & \redX & 60 & 16 & 1 & 101 & & & 2.500 & 10878 & 93 & 4668 & 35 \\\hline
Fantasy Asian House & 267627 & 289886 & \greencheck & \redX & 3675 & 223 & & 131 & & 2 & 39.965 & 3548 & 28 & 10940 & 35 \\\hline
Fps-Hands & 4076 & 4600 & \greencheck & \redX & 145 & & & 5 & & & 0.533 & 2472 & 12 & 2440 & 14 \\\hline
Fractal & 8184 & 7816 & \redX & \redX & 5434 & & & & & & 0.168 & 53148 & 1126 & 3010 & 52 \\\hline
French Halfbasket & 35288 & 41255 & \greencheck & \redX & 273 & 5 & & 18 & & & 5.383 & 596 & 3 & 10182 & 9 \\\hline
Greek Vase & 12545 & 14464 & \redX & \redX & 379 & & & & & & 0.713 & 18624 & 62 & 12790 & 51 \\\hline
Gun & 10256 & 8994 & \greencheck & \redX & 112 & & & 13 & & 8 & 1.203 & 3354 & 17 & 2308 & 20 \\\hline
Hover Bike & 9327 & 7025 & \greencheck & \redX & 66 & & & 6 & & & 1.011 & 3924 & 16 & 2850 & 19 \\\hline
Jpn Corridor & 200 & 308 & \greencheck & \redX & 33 & & & & & & 0.011 & 4754 & 34 & 542 & 20 \\\hline
Jpn House & 119620 & 141206 & \greencheck & \redX & 645 & 22 & 22 & 145 & & 118 & 17.247 & 6332 & 75 & 4010 & 34 \\\hline
Jpn House 2 & 233965 & 235330 & \redX & \redX & 409 & 1 & & & & 2 & 50.214 & 3876 & 39 & 2184 & 24 \\\hline
Jpn House 3 & 41462 & 42444 & \greencheck & \greencheck & 645 & & & 1 & & 3 & 4.282 & 1970 & 13 & 1404 & 13 \\\hline
Jpn Paddle Crab & 567566 & 550825 & \greencheck & \redX & 1146 & & & 14 & & 3 & 134.896 & 11484 & 64 & 16579 & 22 \\\hline
Shinto Shrine & 6280 & 6638 & \greencheck & \redX & 77 & 6 & & 40 & & & 0.780 & 7116 & 69 & 2270 & 26 \\\hline
Kagiya Edo Tokyo & 22819 & 18820 & \redX & \redX & 668 & 1 & & 207 & & 19 & 2.400 & 1088 & 7 & 652 & 8 \\\hline
Lantern & 1360 & 1862 & \greencheck & \redX & 17 & & & 6 & & 4 & 0.148 & 716 & 5 & 392 & 8 \\\hline
Lekythos & 1075 & 1148 & \greencheck & \redX & 110 & & & 3 & & 7 & 0.106 & 9056 & 40 & 3498 & 42 \\\hline
Lethe & 4782 & 8219 & \greencheck & \redX & 112 & 1 & 13 & 62 & & 10 & 0.490 & 4172 & 25 & 2140 & 18 \\\hline
Marble Track & 32558 & 47029 & \greencheck & \redX & 106 & 1 & 4 & 117 & & 10 & 5.460 & 7594 & 50 & 3946 & 18 \\\hline
Maze & 37191 & 40663 & \greencheck & \redX & 7000 & & & & & & 1.450 & 70674 & 684 & 3628 & 105 \\\hline
Mech & 8732 & 15771 & \greencheck & \redX & 42 & 2 & & 5 & & 4 & 1.130 & 5618 & 29 & 2632 & 22 \\\hline
Melon Pallet & 45016 & 45342 & \greencheck & \redX & 38 & & & & & & 6.015 & 9790 & 54 & 4900 & 23 \\\hline
Militech Robot & 52373 & 29422 & \greencheck & \redX & 116 & & & 68 & & & 5.925 & 13050 & 66 & 5784 & 25 \\\hline
Modular Fence Corner & 4642 & 3257 & \greencheck & \redX & 3 & & & 5 & & & 0.592 & 5240 & 13 & 1602 & 10 \\\hline
Old Tree & 4226 & 4921 & \greencheck & \redX & 98 & 1 & & & & 1 & 0.476 & 6204 & 38 & 3079 & 15 \\\hline
Pipe Wall & 1545 & 1791 & \greencheck & \redX & 44 & 31 & & 25 & & & 0.131 & 7564 & 70 & 1840 & 35 \\\hline
Pipes & 1200 & 2154 & \greencheck & \redX & 21 & & & 22 & & 7 & 0.127 & 5666 & 27 & 3032 & 42 \\\hline
Potted Plant & 4861 & 6830 & \greencheck & \redX & 57 & 2 & & 8 & & 5 & 0.535 & 10560 & 70 & 4600 & 46 \\\hline
Raspberry Plant & 487445 & 427594 & \greencheck & \redX & 245 & 3 & & 5 & & 1 & 135.427 & 8724 & 76 & 9640 & 34 \\\hline
Road & 306 & 522 & \greencheck & \redX & 5 & & & & & & 0.044 & 132 & 1 & 66 & 6 \\\hline
Rope Bridge & 7256 & 7624 & \greencheck & \redX & 132 & 40 & 16 & 170 & & & 0.515 & 12254 & 279 & 2376 & 27 \\\hline
Shinto Watchtower & 131367 & 134770 & \redX & \redX & 1387 & 1 & 80 & 1 & & 11 & 16.909 & 12968 & 122 & 12408 & 75 \\\hline
Shirakawago House & 15965 & 21329 & \greencheck & \redX & 1299 & 11 & 2 & 80 & & & 1.487 & 11424 & 100 & 3009 & 41 \\\hline
Snowflake Orb & 77418 & 38395 & \greencheck & \greencheck & 1273 & & & 156 & & 66 & 7.822 & 22768 & 78 & 39433 & 39 \\\hline
Sony Pc & 15512 & 24533 & \greencheck & \redX & 171 & 6 & & 42 & & 5 & 2.341 & 3440 & 30 & 3864 & 18 \\\hline
Space Fighter & 34965 & 55060 & \greencheck & \redX & 158 & & 4 & 23 & & & 8.472 & 4512 & 16 & 5593 & 39 \\\hline
Spacestation V3 & 35488 & 89668 & \greencheck & \redX & 1667 & 8 & 91 & 181 & & & 3.808 & 6170 & 165 & 2132 & 28 \\\hline
Speeder Bike & 11926 & 15152 & \greencheck & \redX & 239 & 1 & 3 & 20 & & 20 & 1.305 & 3924 & 20 & 6096 & 42 \\\hline
Spider Tank & 106464 & 204581 & \greencheck & \redX & 244 & 11 & 3 & 125 & & 2 & 15.844 & 7560 & 41 & 4104 & 19 \\\hline
Spiral Staircase & 6358 & 7437 & \redX & \redX & 24 & 40 & & & & 16 & 0.861 & 3556 & 35 & 4502 & 30 \\\hline
Guard Rail & 14744 & 14559 & \greencheck & \redX & 16 & 13 & & 11 & & & 1.985 & 2428 & 29 & 2836 & 14 \\\hline
Table & 66 & 154 & \greencheck & \redX & 11 & & & & & & 0.005 & 1210 & 11 & 266 & 12 \\\hline
Tank & 15306 & 21338 & \greencheck & \redX & 16 & & & 7 & & & 2.636 & 4784 & 19 & 3672 & 17 \\\hline
Teacup & 1465 & 2360 & \greencheck & \redX & 117 & & & 3 & & & 0.076 & 10670 & 37 & 3554 & 39 \\\hline
Training Dummy & 20816 & 17483 & \greencheck & \redX & 8 & & & 19 & & & 2.981 & 4586 & 14 & 4108 & 17 \\\hline
Turtle Castle & 378627 & 405918 & \greencheck & \redX & 384 & 1 & & 56 & & 6 & 64.410 & 7852 & 49 & 7942 & 24 \\\hline
Robot Vera & 47790 & 66243 & \greencheck & \redX & 138 & & 12 & 146 & & & 5.829 & 9774 & 46 & 7874 & 24 \\\hline
War Tank & 18477 & 26601 & \redX & \redX & 57 & 3 & & 5 & & & 2.474 & 7992 & 98 & 2236 & 29 \\\hline
Wat Benchamabophit & 999956 & 1115136 & \greencheck & \redX & 1621 & 17 & 6 & 54 & & 8 & 178.146 & 6376 & 56 & 6096 & 44 \\\hline
White Headed Vulture & 115998 & 65304 & \greencheck & \redX & 198 & & & & & 2 & 16.866 & 4006 & 15 & 6488 & 18 \\\hline
Yeahright & 377344 & 377084 & \greencheck & \greencheck & 761 & 13 & 4 & 22 & & & 70.787 & 4860 & 47 & 38432 & 30 \\\hline\end{tabular}
\caption{Primitive counts of our approach on our dataset from Sketchfab, compared to CoACD~\cite{coacd} and V-HACD~\cite{vhacd}. Empty cells indicate 0. Non-manifoldness indicates if there are any non-manifold edges, and (WT) watertightness indicates if there are no boundary edges. Convex primitive decomposition's output varies heavily with the structure of the input mesh, sometimes favoring one kind of primitive over others. \label{tab:complete_results}}
\end{table*}

% AUTOGENERATED TABLE
\begin{table*}
\centering
\scriptsize
\begin{tabular}{|c||c|c||c|c||c|c|}
\hline
1-way (New $\rightarrow$ Input) Dist.$^\downarrow$ & \multicolumn{2}{c||}{Ours} & \multicolumn{2}{c||}{CoACD} & \multicolumn{2}{c|}{V-HACD} \\\hline
Name & Hausdorff & Chamfer & Hausdorff & Chamfer & Hausdorff & Chamfer \\\hline\hline
Armadillo & 0.10065 & 0.01314 & 0.08252 & 0.01008 & \textbf{0.07888} & \textbf{0.00848} \\\hline
Armored Charizard & \textbf{0.01523} & \textbf{0.00071} & 0.07099 & 0.01056 & 0.07472 & 0.01009 \\\hline
Bell & 0.06347 & 0.01062 & \textbf{0.03188} & \textbf{0.00663} & 0.09992 & 0.01308 \\\hline
Bicycle & \textbf{0.01196} & \textbf{0.00221} & 0.03536 & 0.00935 & 0.04943 & 0.00718 \\\hline
Canon At1 & 0.12554 & 0.02809 & \textbf{0.12344} & 0.01619 & 0.12567 & \textbf{0.01438} \\\hline
Cardboard Boxes & \textbf{0.01531} & \textbf{0.00236} & 0.03449 & 0.00758 & 0.05353 & 0.00563 \\\hline
Casio Keyboard & \textbf{0.01388} & \textbf{0.00296} & 0.02015 & 0.00997 & 0.04523 & 0.00490 \\\hline
Chimney Pipe & 0.13351 & 0.02427 & \textbf{0.08515} & 0.01136 & 0.08925 & \textbf{0.00826} \\\hline
Chitinous Knight & 0.07392 & 0.01502 & \textbf{0.06393} & 0.01201 & 0.06716 & \textbf{0.00719} \\\hline
Chuo House & 0.10229 & \textbf{0.00340} & \textbf{0.08021} & 0.01410 & 0.08105 & 0.01114 \\\hline
Church Organ & \textbf{0.03531} & 0.00746 & 0.03630 & 0.00946 & 0.03745 & \textbf{0.00493} \\\hline
Cinema Scan & 0.07168 & \textbf{0.00214} & \textbf{0.03566} & 0.00906 & 0.05282 & 0.00793 \\\hline
Cube & \textbf{0.00260} & \textbf{0.00025} & 0.02967 & 0.00810 & 0.04041 & 0.00707 \\\hline
Cyberpunk Atm & \textbf{0.04043} & \textbf{0.00746} & 0.07036 & 0.01175 & 0.05880 & 0.00847 \\\hline
Cyberpunk Bike & 0.07587 & 0.01264 & \textbf{0.03430} & 0.00912 & 0.04883 & \textbf{0.00599} \\\hline
Dojo & 0.11107 & 0.01811 & \textbf{0.03127} & 0.00792 & 0.09150 & \textbf{0.00762} \\\hline
Dungeon Level (Precise) & \textbf{0.00570} & \textbf{0.00294} & 0.01758 & 0.00548 & 0.05483 & 0.00483 \\\hline
Espresso Machine & \textbf{0.02850} & \textbf{0.00691} & 0.07415 & 0.01195 & 0.06583 & 0.01048 \\\hline
Fantasy Asian House & 0.06878 & \textbf{0.00547} & 0.05872 & 0.01137 & \textbf{0.05344} & 0.00748 \\\hline
Fps-Hands & \textbf{0.01239} & \textbf{0.00383} & 0.03822 & 0.00944 & 0.04366 & 0.00614 \\\hline
Fractal & \textbf{0.00003} & \textbf{0.00003} & 0.02123 & 0.00506 & 0.03494 & 0.00619 \\\hline
French Halfbasket & \textbf{0.00596} & \textbf{0.00065} & 0.03245 & 0.00838 & 0.01955 & 0.00204 \\\hline
Greek Vase & \textbf{0.00669} & \textbf{0.00176} & 0.03433 & 0.00998 & 0.11744 & 0.01602 \\\hline
Gun & \textbf{0.03779} & \textbf{0.00673} & 0.05785 & 0.01105 & 0.05903 & 0.00723 \\\hline
Hover Bike & \textbf{0.05884} & \textbf{0.00960} & 0.06923 & 0.01254 & 0.07398 & 0.01110 \\\hline
Jpn Corridor & \textbf{0.00770} & \textbf{0.00003} & 0.02189 & 0.00832 & 0.12483 & 0.00890 \\\hline
Jpn House & 0.09144 & 0.02295 & \textbf{0.03314} & 0.00817 & 0.25382 & \textbf{0.00716} \\\hline
Jpn House 2 & \textbf{0.01592} & \textbf{0.00195} & 0.03046 & 0.00715 & 0.03925 & 0.00452 \\\hline
Jpn House 3 & \textbf{0.01791} & \textbf{0.00021} & 0.12074 & 0.01888 & 0.12718 & 0.01599 \\\hline
Jpn Paddle Crab & \textbf{0.02933} & \textbf{0.00562} & 0.07334 & 0.00910 & 0.08255 & 0.00969 \\\hline
Shinto Shrine & 0.04818 & 0.00908 & \textbf{0.04068} & 0.00880 & 0.05948 & \textbf{0.00786} \\\hline
Kagiya Edo Tokyo & \textbf{0.00642} & \textbf{0.00176} & 0.10794 & 0.02253 & 0.10737 & 0.01558 \\\hline
Lantern & 0.11232 & 0.01439 & \textbf{0.10980} & 0.01612 & 0.11202 & \textbf{0.01331} \\\hline
Lekythos & \textbf{0.04035} & 0.00738 & 0.04216 & 0.01160 & 0.06151 & \textbf{0.00709} \\\hline
Lethe & 0.07164 & \textbf{0.00751} & \textbf{0.04467} & 0.00822 & 0.09024 & 0.00923 \\\hline
Marble Track & \textbf{0.01486} & \textbf{0.00053} & 0.02851 & 0.00630 & 0.10007 & 0.00621 \\\hline
Maze & \textbf{0.00057} & \textbf{0.00043} & 0.00701 & 0.00247 & 0.01625 & 0.00194 \\\hline
Mech & 0.08841 & 0.01407 & 0.06509 & 0.00984 & \textbf{0.05552} & \textbf{0.00670} \\\hline
Melon Pallet & \textbf{0.01604} & \textbf{0.00334} & 0.05276 & 0.00656 & 0.05174 & 0.00426 \\\hline
Militech Robot & \textbf{0.00808} & \textbf{0.00221} & 0.04393 & 0.00707 & 0.05294 & 0.00760 \\\hline
Modular Fence Corner & 0.25632 & 0.02719 & \textbf{0.02813} & 0.00839 & 0.06559 & \textbf{0.00594} \\\hline
Old Tree & 0.08825 & 0.00616 & \textbf{0.03537} & \textbf{0.00410} & 0.06434 & 0.01040 \\\hline
Pipe Wall & \textbf{0.00977} & \textbf{0.00240} & 0.03148 & 0.00733 & 0.05593 & 0.00590 \\\hline
Pipes & \textbf{0.01968} & 0.00718 & 0.03984 & 0.00967 & 0.04104 & \textbf{0.00656} \\\hline
Potted Plant & 0.05526 & 0.01271 & \textbf{0.05503} & \textbf{0.01049} & 0.08425 & 0.01146 \\\hline
Raspberry Plant & 0.06606 & \textbf{0.00817} & \textbf{0.03220} & 0.00833 & 0.04783 & 0.00877 \\\hline
Road & \textbf{0.00153} & 0.00066 & 0.01749 & 0.01075 & 0.00344 & \textbf{0.00006} \\\hline
Rope Bridge & 0.05778 & 0.01795 & \textbf{0.01827} & 0.00678 & 0.03575 & \textbf{0.00648} \\\hline
Shinto Watchtower & \textbf{0.03843} & 0.00958 & 0.03871 & 0.00944 & 0.05066 & \textbf{0.00851} \\\hline
Shirakawago House & \textbf{0.06299} & \textbf{0.00490} & 0.12086 & 0.01269 & 0.12261 & 0.01411 \\\hline
Snowflake Orb & \textbf{0.01533} & \textbf{0.00449} & 0.02962 & 0.00795 & 0.03527 & 0.00569 \\\hline
Sony Pc & 0.05983 & \textbf{0.00839} & \textbf{0.04663} & 0.01307 & 0.11451 & 0.01695 \\\hline
Space Fighter & 0.04089 & 0.00609 & 0.03990 & 0.00673 & \textbf{0.03338} & \textbf{0.00309} \\\hline
Spacestation V3 & \textbf{0.02719} & \textbf{0.00248} & 0.02855 & 0.00641 & 0.11113 & 0.03307 \\\hline
Speeder Bike & \textbf{0.05146} & 0.00897 & 0.05253 & 0.00956 & 0.05333 & \textbf{0.00728} \\\hline
Spider Tank & \textbf{0.00962} & \textbf{0.00192} & 0.03645 & 0.00769 & 0.05948 & 0.00760 \\\hline
Spiral Staircase & 0.13282 & 0.01846 & \textbf{0.03745} & 0.00972 & 0.05840 & \textbf{0.00797} \\\hline
Guard Rail & \textbf{0.00502} & \textbf{0.00173} & 0.03474 & 0.00778 & 0.04421 & 0.00988 \\\hline
Table & \textbf{0.00207} & \textbf{0.00098} & 0.03840 & 0.01053 & 0.03449 & 0.00234 \\\hline
Tank & \textbf{0.02078} & \textbf{0.00533} & 0.16587 & 0.02053 & 0.16880 & 0.02079 \\\hline
Teacup & \textbf{0.01361} & \textbf{0.00415} & 0.03497 & 0.00972 & 0.08308 & 0.00593 \\\hline
Training Dummy & \textbf{0.03855} & 0.01072 & 0.09124 & 0.01129 & 0.05419 & \textbf{0.00929} \\\hline
Turtle Castle & \textbf{0.05321} & \textbf{0.00784} & 0.09978 & 0.01941 & 0.10281 & 0.01814 \\\hline
Robot Vera & \textbf{0.01120} & \textbf{0.00115} & 0.03402 & 0.00710 & 0.04165 & 0.00605 \\\hline
War Tank & 0.04930 & 0.00802 & \textbf{0.04800} & 0.00721 & 0.05458 & \textbf{0.00662} \\\hline
Wat Benchamabophit & \textbf{0.01643} & \textbf{0.00196} & 0.03528 & 0.00801 & 0.08985 & 0.00699 \\\hline
White Headed Vulture & \textbf{0.03381} & \textbf{0.00180} & 0.09739 & 0.01401 & 0.09997 & 0.01310 \\\hline
Yeahright & \textbf{0.00333} & \textbf{0.00092} & 0.03685 & 0.00939 & 0.06662 & 0.01072 \\\hline
\end{tabular}
\caption{\label{tab:distance_comparison} Full set of comparisons of our approach on our dataset. Convex primitive decomposition is on average closer to the input mesh than CoACD~\cite{coacd} and V-HACD~\cite{vhacd} on both the chamfer and hausdorff distance. Distances are normalized as $\frac{\text{Hausdorff\slash Chamfer New $\rightarrow$ Input}}{\lVert\text{Bounding Box Diag}\rVert_2}$.
}
\end{table*}

% AUTOGENERATED TABLE
\begin{table*}
\centering
\scriptsize
\begin{tabular}{|c|c|c|c|c|c|c|c|c|}
\hline
Model Name & \multicolumn{2}{c|}{Ours} & \multicolumn{3}{c|}{CoACD} & \multicolumn{3}{c|}{V-HACD} \\\hline
 & Floats & Total & Floats & Ints & Total & Floats & Ints & Total \\\cline{0-1}\cline{4-5}\cline{7-8}
Bytes Used & 4 & Bytes & 4 & 2 & Bytes & 4 & 2 & Bytes \\\hline\hline
Armadillo & 1989 & \textbf{ 7956 } & 14592 & 28896 & 116160 & 17835 & 33825 & 138990 \\\hline
Armored Charizard & 1876 & \textbf{ 7504 } & 11001 & 21510 & 87024 & 3810 & 7206 & 29652 \\\hline
Bell & 512 & \textbf{ 2048 } & 17697 & 34638 & 140064 & 9726 & 18594 & 76092 \\\hline
Bicycle & 914 & \textbf{ 3656 } & 7536 & 14712 & 59568 & 14268 & 27318 & 111708 \\\hline
Canon At1 & 108 & \textbf{ 432 } & 11394 & 22452 & 90480 & 4959 & 9522 & 38880 \\\hline
Cardboard Boxes & 450 & \textbf{ 1800 } & 8742 & 16704 & 68376 & 987 & 1488 & 6924 \\\hline
Casio Keyboard & 160 & \textbf{ 640 } & 666 & 1320 & 5304 & 1065 & 1974 & 8208 \\\hline
Chimney Pipe & 48 & \textbf{ 192 } & 8193 & 16062 & 64896 & 3813 & 7002 & 29256 \\\hline
Chitinous Knight & 484 & \textbf{ 1936 } & 6570 & 12948 & 52176 & 20706 & 39057 & 160938 \\\hline
Chuo House & 10652 & 42608 & 4971 & 9594 & \textbf{ 39072 } & 8142 & 15600 & 63768 \\\hline
Church Organ & 1536 & \textbf{ 6144 } & 8649 & 16650 & 67896 & 5019 & 9366 & 38808 \\\hline
Cinema Scan & 4931 & \textbf{ 19724 } & 13098 & 25152 & 102696 & 9882 & 18090 & 75708 \\\hline
Cube & 1545 & \textbf{ 6180 } & 11451 & 21858 & 89520 & 11013 & 21228 & 86508 \\\hline
Cyberpunk Atm & 798 & \textbf{ 3192 } & 6018 & 11736 & 47544 & 2346 & 4374 & 18132 \\\hline
Cyberpunk Bike & 500 & \textbf{ 2000 } & 7011 & 13842 & 55728 & 15357 & 28878 & 119184 \\\hline
Dojo & 4807 & \textbf{ 19228 } & 15693 & 29598 & 121968 & 3894 & 7062 & 29700 \\\hline
Dungeon (Precise) & 5383 & 21532 & 11016 & 20952 & 85968 & 2127 & 3894 & \textbf{ 16296 } \\\hline
Espresso Machine & 1423 & \textbf{ 5692 } & 16875 & 32634 & 132768 & 7389 & 14004 & 57564 \\\hline
Fantasy Asian House & 39250 & 157000 & 5490 & 10644 & \textbf{ 43248 } & 16524 & 32820 & 131736 \\\hline
Fps-Hands & 1485 & \textbf{ 5940 } & 3780 & 7416 & 29952 & 3813 & 7320 & 29892 \\\hline
Fractal & 54340 & 217360 & 86478 & 159444 & 664800 & 4845 & 9030 & \textbf{ 37440 } \\\hline
French Halfbasket & 2891 & 11564 & 912 & 1788 & \textbf{ 7224 } & 16737 & 30546 & 128040 \\\hline
Greek Vase & 3790 & \textbf{ 15160 } & 28308 & 55872 & 224976 & 19779 & 38370 & 155856 \\\hline
Gun & 1299 & \textbf{ 5196 } & 5133 & 10062 & 40656 & 3627 & 6924 & 28356 \\\hline
Hover Bike & 702 & \textbf{ 2808 } & 5982 & 11772 & 47472 & 4491 & 8550 & 35064 \\\hline
Jpn Corridor & 330 & \textbf{ 1320 } & 7335 & 14262 & 57864 & 936 & 1626 & 6996 \\\hline
Jpn House & 9005 & \textbf{ 36020 } & 9948 & 18996 & 77784 & 6360 & 12030 & 49500 \\\hline
Jpn House 2 & 4119 & \textbf{ 16476 } & 6048 & 11628 & 47448 & 3552 & 6552 & 27312 \\\hline
Jpn House 3 & 6490 & 25960 & 3033 & 5910 & 23952 & 2316 & 4212 & \textbf{ 17688 } \\\hline
Jpn Paddle Crab & 11591 & \textbf{ 46364 } & 17610 & 34452 & 139344 & 27912 & 49737 & 211122 \\\hline
Shinto Shrine & 1092 & \textbf{ 4368 } & 11088 & 21348 & 87048 & 3660 & 6810 & 28260 \\\hline
Kagiya Edo Tokyo & 8345 & 33380 & 1674 & 3264 & 13224 & 1068 & 1956 & \textbf{ 8184 } \\\hline
Lantern & 256 & \textbf{ 1024 } & 1104 & 2148 & 8712 & 636 & 1176 & 4896 \\\hline
Lekythos & 1198 & \textbf{ 4792 } & 13824 & 27168 & 109632 & 5598 & 10494 & 43380 \\\hline
Lethe & 1723 & \textbf{ 6892 } & 6408 & 12516 & 50664 & 3420 & 6420 & 26520 \\\hline
Marble Track & 2012 & \textbf{ 8048 } & 11691 & 22782 & 92328 & 6438 & 11838 & 49428 \\\hline
Maze & 70000 & 280000 & 330 & 212022 & 425364 & 6105 & 10884 & \textbf{ 46188 } \\\hline
Mech & 513 & \textbf{ 2052 } & 8601 & 16854 & 68112 & 4185 & 7896 & 32532 \\\hline
Melon Pallet & 380 & \textbf{ 1520 } & 15009 & 29370 & 118776 & 8127 & 14700 & 61908 \\\hline
Militech Robot & 1636 & \textbf{ 6544 } & 19971 & 39150 & 158184 & 9144 & 17352 & 71280 \\\hline
Modular Fence Corner & 65 & \textbf{ 260 } & 7938 & 15720 & 63192 & 2472 & 4806 & 19500 \\\hline
Old Tree & 998 & \textbf{ 3992 } & 9534 & 18612 & 75360 & 4878 & 9237 & 37986 \\\hline
Pipes & 441 & \textbf{ 1764 } & 8661 & 16998 & 68640 & 4872 & 9096 & 37680 \\\hline
Pipe Wall & 832 & \textbf{ 3328 } & 11766 & 22692 & 92448 & 3069 & 5520 & 23316 \\\hline
Potted Plant & 695 & \textbf{ 2780 } & 16260 & 31680 & 128400 & 7437 & 13800 & 57348 \\\hline
Raspberry Plant & 2517 & \textbf{ 10068 } & 13542 & 26172 & 106512 & 16023 & 28920 & 121932 \\\hline
Road & 50 & \textbf{ 200 } & 204 & 396 & 1608 & 135 & 198 & 936 \\\hline
Rope Bridge & 2854 & \textbf{ 11416 } & 20055 & 36762 & 153744 & 3891 & 7128 & 29820 \\\hline
Shinto Watchtower & 14325 & \textbf{ 57300 } & 20184 & 38904 & 158544 & 19698 & 37224 & 153240 \\\hline
Shirakawago House & 13635 & 54540 & 17736 & 34272 & 139488 & 4854 & 9027 & \textbf{ 37470 } \\\hline
Snowflake Orb & 14548 & \textbf{ 58192 } & 34620 & 68304 & 275088 & 67422 & 118299 & 506286 \\\hline
Sony Pc & 2101 & \textbf{ 8404 } & 5340 & 10320 & 42000 & 6336 & 11592 & 48528 \\\hline
Spacestation V3 & 18357 & 73428 & 10245 & 18510 & 78000 & 3432 & 6396 & \textbf{ 26520 } \\\hline
Space Fighter & 1757 & \textbf{ 7028 } & 6864 & 13536 & 54528 & 8802 & 16779 & 68766 \\\hline
Speeder Bike & 2769 & \textbf{ 11076 } & 6006 & 11772 & 47568 & 9642 & 18288 & 75144 \\\hline
Spider Tank & 3426 & \textbf{ 13704 } & 11586 & 22680 & 91704 & 6510 & 12312 & 50664 \\\hline
Spiral Staircase & 696 & \textbf{ 2784 } & 5544 & 10668 & 43512 & 7110 & 13506 & 55452 \\\hline
Guard Rail & 328 & \textbf{ 1312 } & 3816 & 7284 & 29832 & 4473 & 8508 & 34908 \\\hline
Table & 110 & \textbf{ 440 } & 1881 & 3630 & 14784 & 471 & 798 & 3480 \\\hline
Tank & 209 & \textbf{ 836 } & 7290 & 14352 & 57864 & 5781 & 11016 & 45156 \\\hline
Teacup & 1191 & \textbf{ 4764 } & 16227 & 32010 & 128928 & 5676 & 10662 & 44028 \\\hline
Training Dummy & 213 & \textbf{ 852 } & 6963 & 13758 & 55368 & 6375 & 12324 & 50148 \\\hline
Turtle Castle & 4305 & \textbf{ 17220 } & 12072 & 23556 & 95400 & 12660 & 23826 & 98292 \\\hline
Robot Vera & 2450 & \textbf{ 9800 } & 14937 & 29322 & 118392 & 12339 & 23622 & 96600 \\\hline
War Tank & 626 & \textbf{ 2504 } & 12576 & 23976 & 98256 & 3564 & 6708 & 27672 \\\hline
Wat Benchamabophit & 16819 & \textbf{ 67276 } & 9900 & 19128 & 77856 & 9624 & 18288 & 75072 \\\hline
White Headed Vulture & 2002 & \textbf{ 8008 } & 6099 & 12018 & 48432 & 10590 & 19464 & 81288 \\\hline
Yeahright & 7871 & \textbf{ 31484 } & 7572 & 14580 & 59448 & 59139 & 115296 & 467148 \\\hline
\end{tabular}
\caption{\label{tab:raw-memory-cost} Comparison of memory cost for each model in our dataset, measured in bytes. Our approach on whole uses less memory than the other approaches. For floats, we treat them as 4 bytes (single-precision \texttt{float} in C), and for integers we treat them as unsigned 2 byte integers (\texttt{uint16\_t} in C), even though in some cases it would overflow.}
\end{table*}

\begin{figure*}
    \centering
    \begin{tabular}{c c}
        \multicolumn{2}{c}{Frame Time$^\downarrow$  During Ball-Dropping Sim. of Ours, CoACD, V-HACD} \\
        Similar Behavior for All Meshes & Ours Has Closer Behavior To Input \\

        {\small Training Dummy (Vertex Merging Fig.~\ref{fig:ablate-vertex-merging})} &
        {\small Shinto Watchtower (Fig.~\ref{fig:teaser})} \\ 
        \includegraphics[width=0.49\linewidth]{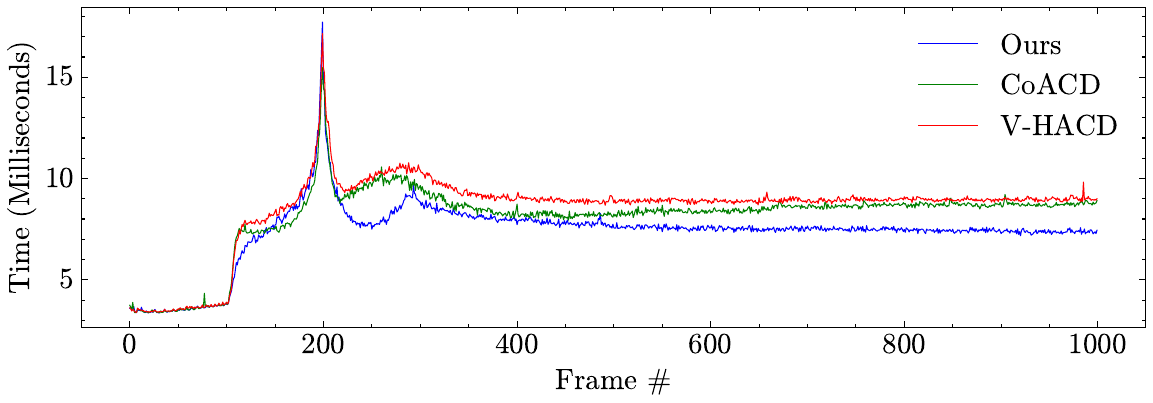} &
        \includegraphics[width=0.49\linewidth]{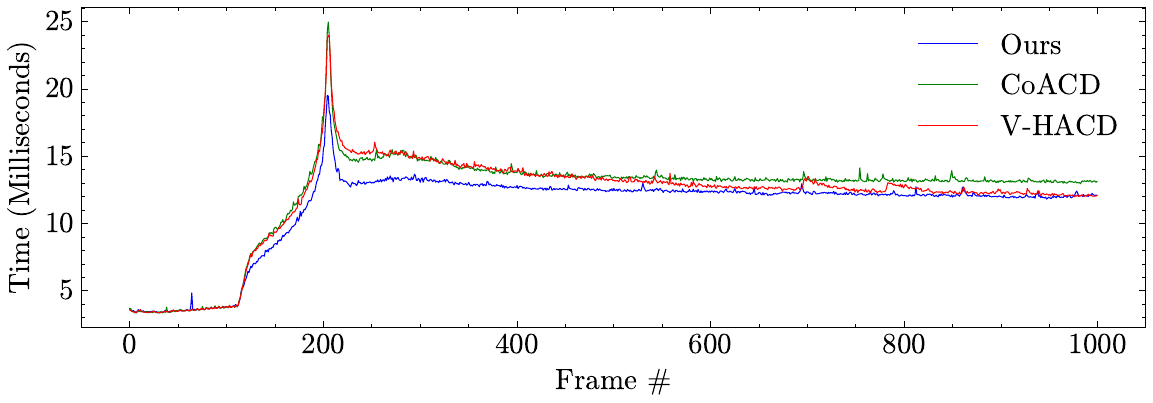} \\
        
        {\small Cylindrical Water Tank (Fig.~\ref{fig:additional_comparisons})} & {\small Rope Bridge (Fig.~\ref{fig:additional_comparisons})} \\
        \includegraphics[width=0.49\linewidth]{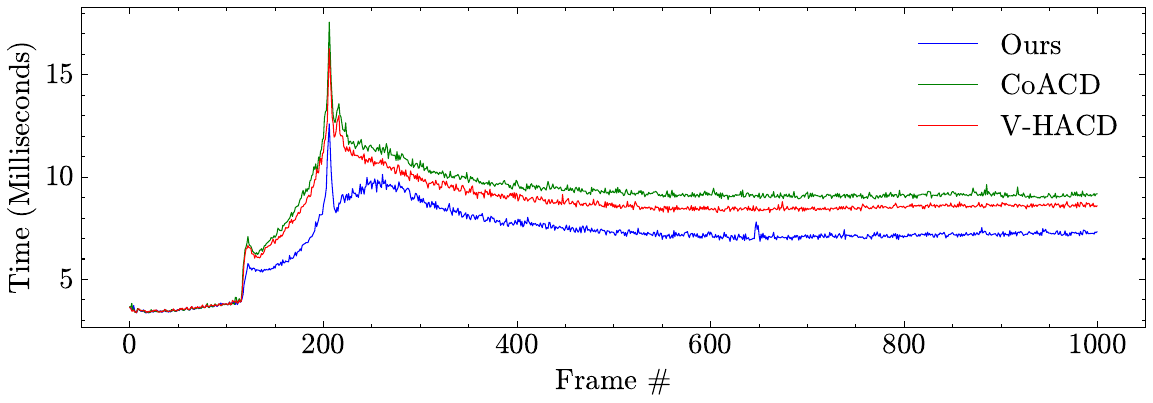} &
        \includegraphics[width=0.49\linewidth]{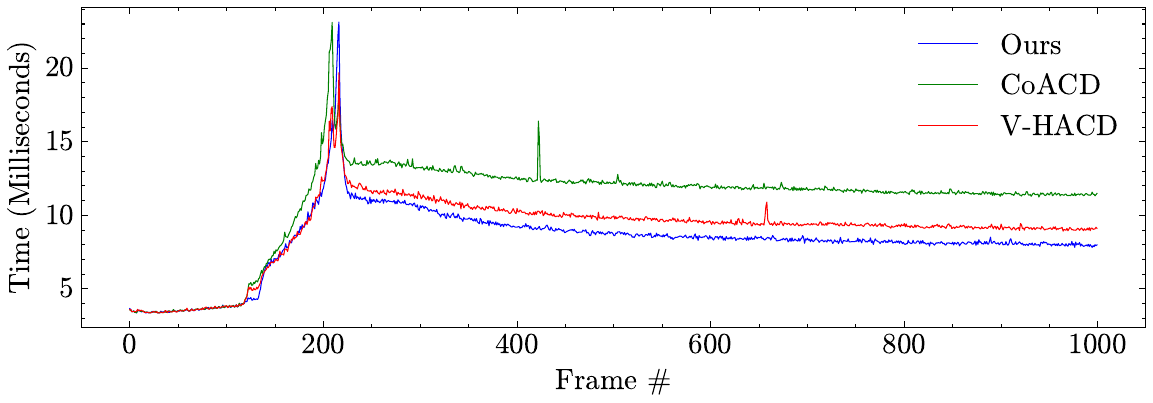} \\
        
        {\small Bell} &
        {\small Table (Fig.~\ref{fig:additional_comparisons})} \\
        \includegraphics[width=0.49\linewidth]{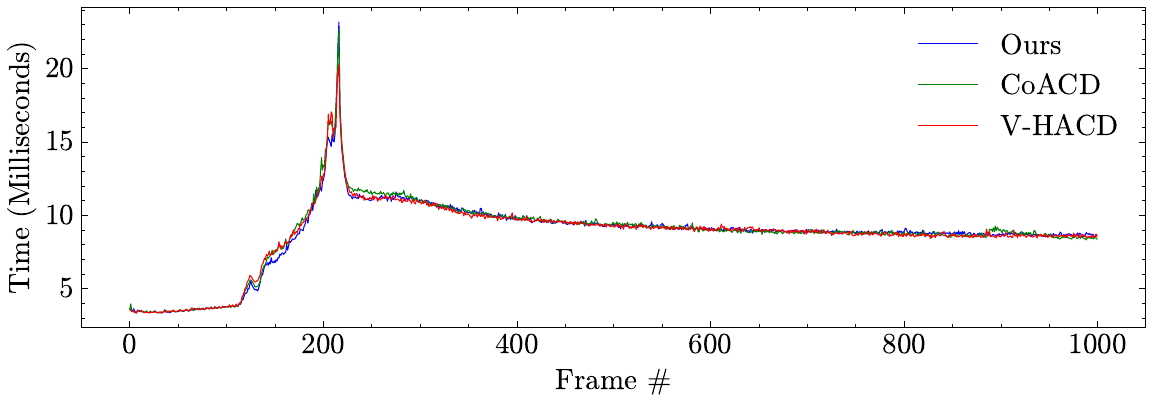} &
        \includegraphics[width=0.49\linewidth]{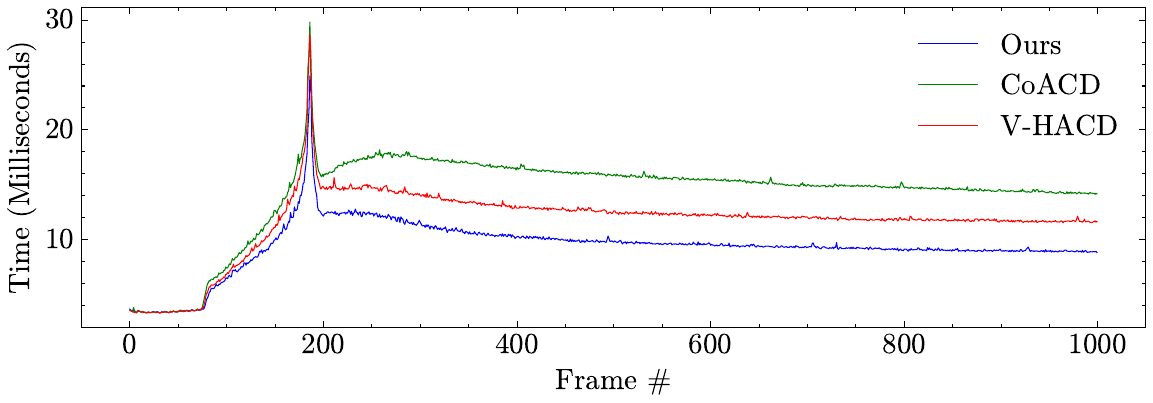} \\
        
        {\small Canon AT1 Camera} &
        {\small Cube (Precise) (Fig.~\ref{fig:cube_failure})} \\
        \includegraphics[width=0.49\linewidth]{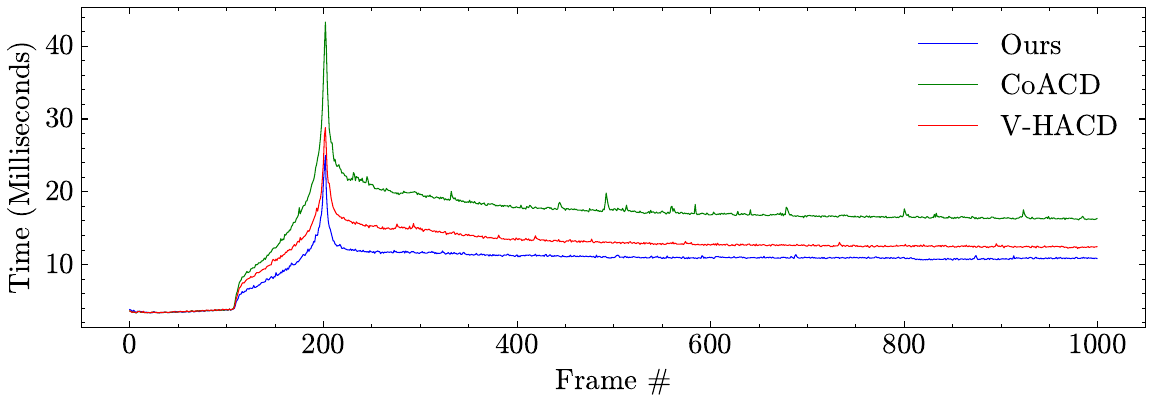} &
        \includegraphics[width=0.49\linewidth]{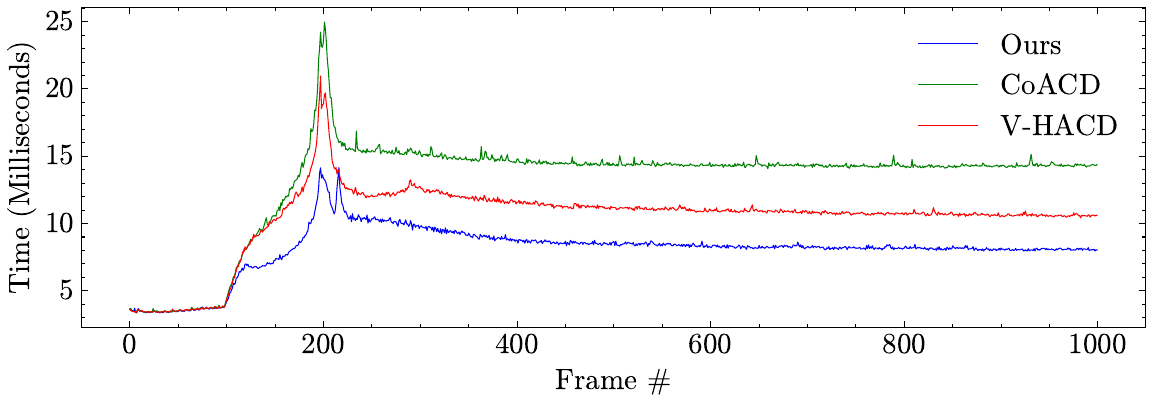} \\
        
        {\small War Tank} & {\small Spiral Staircase (Fig.~\ref{fig:additional_qualitative_results})} \\
        \includegraphics[width=0.49\linewidth]{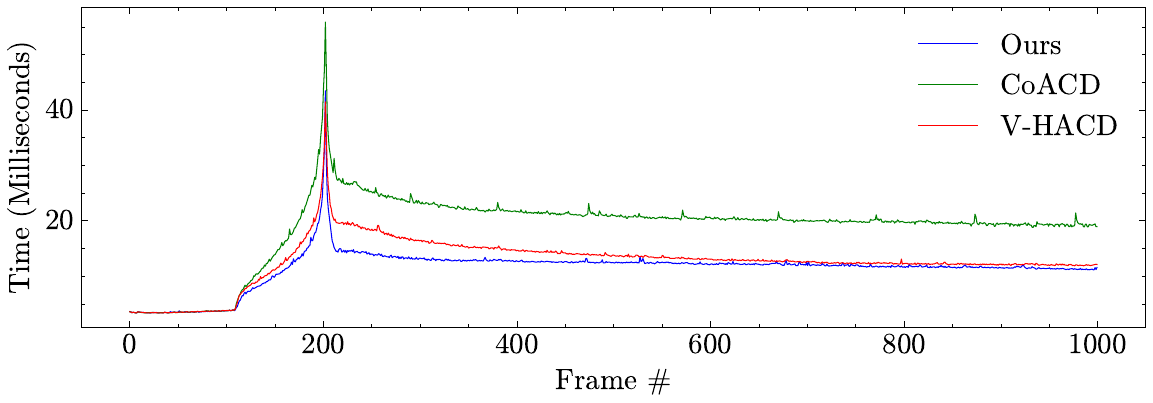} &
        \includegraphics[width=0.49\linewidth]{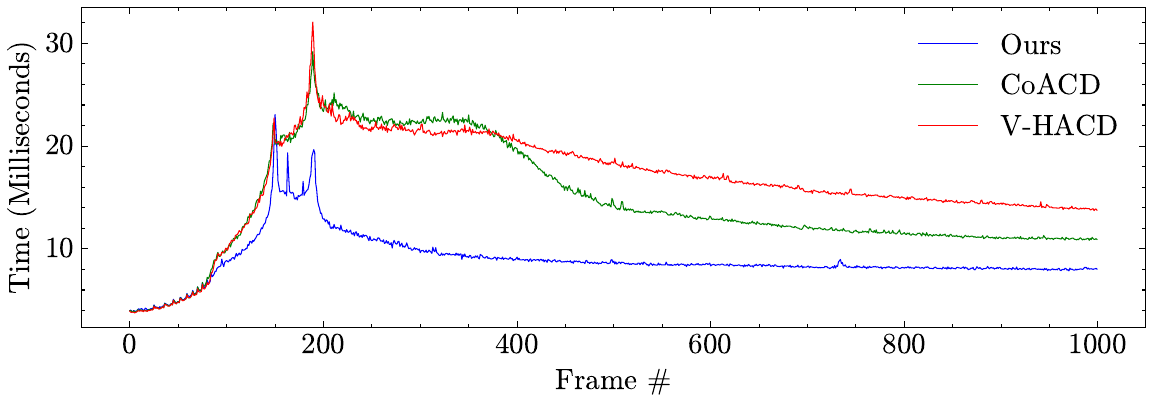} \\
    \end{tabular}

    \caption{Comparison of frame duration for collision detection when dropping 5000 spheres on different meshes, where the left-column is meshes where the behavior of all colliders is relatively similar, and the right column is the performance when our collider has visually more similar performance to the input. In both cases, our approach has comparable or better performance than CoACD and V-HACD.}
    
    \label{fig:additional-sim-times}
    %\Description{A chart comparing the duration of each frame for collision detection, for our approach, CoACD and V-HACD. The three approaches are quite similar, our approach's peak is below CoACD and above V-HACD. The tail of our plot is noticeably lower.}
    %\vspace{-1em}
\end{figure*}
\begin{figure*}
    \centering
    \scriptsize
    \begin{tabular}{c c}
        \multicolumn{2}{c}{Frame Time$^\downarrow$  During Ball-Dropping Sim. of Ours, CoACD, V-HACD (All similar to Input)} \\
        {\small Jpn Corridor, Ours = 35 Primitives, CoACD = 34 Hulls} &
        {\small Espresso Machine, Ours = 178 Primitives, CoACD = 93 Hulls} \\
        \includegraphics[width=0.45\linewidth]{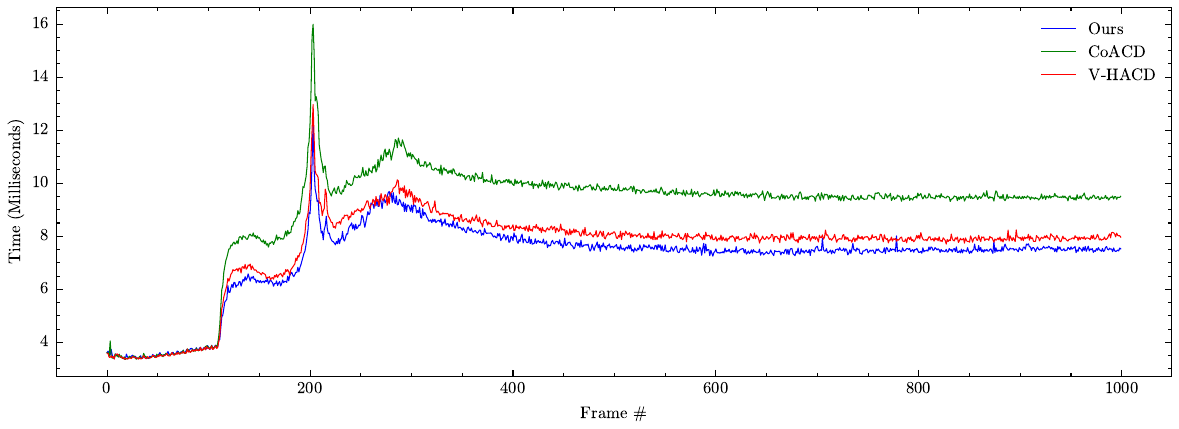} &
        \includegraphics[width=0.45\linewidth]{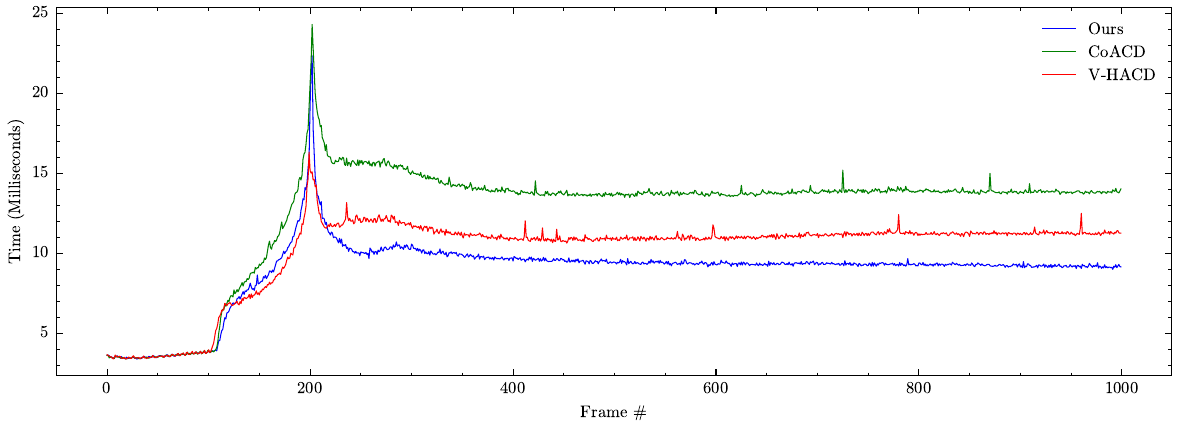} \\
        {\small Melon Pallet, Ours = 38 Primitives, CoACD = 54 Hulls} &
        {\small Jpn Paddle Crab, Fig.~\ref{fig:organic-meshes}, Ours = 1165 Primitives, CoACD = 64 Hulls} \\
        \includegraphics[width=0.45\linewidth]{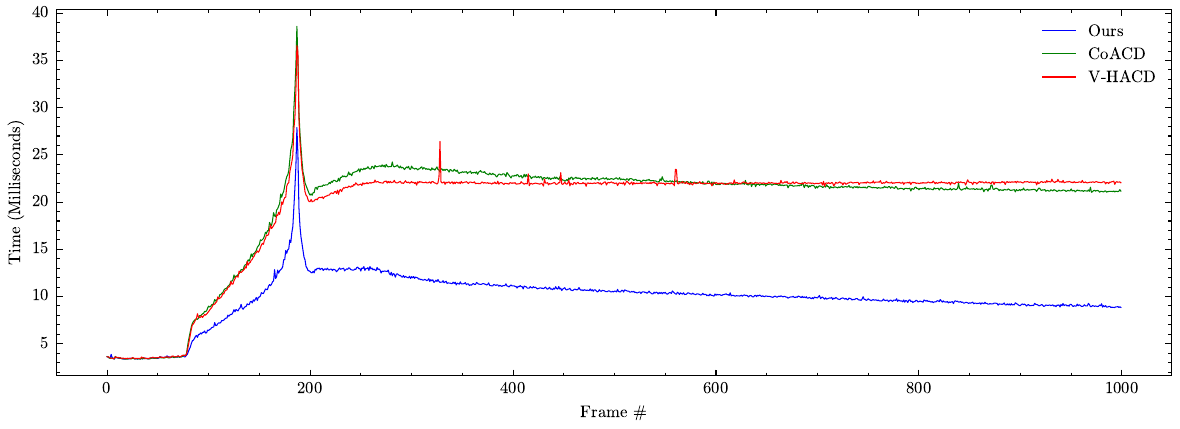} &
        \includegraphics[width=0.45\linewidth]{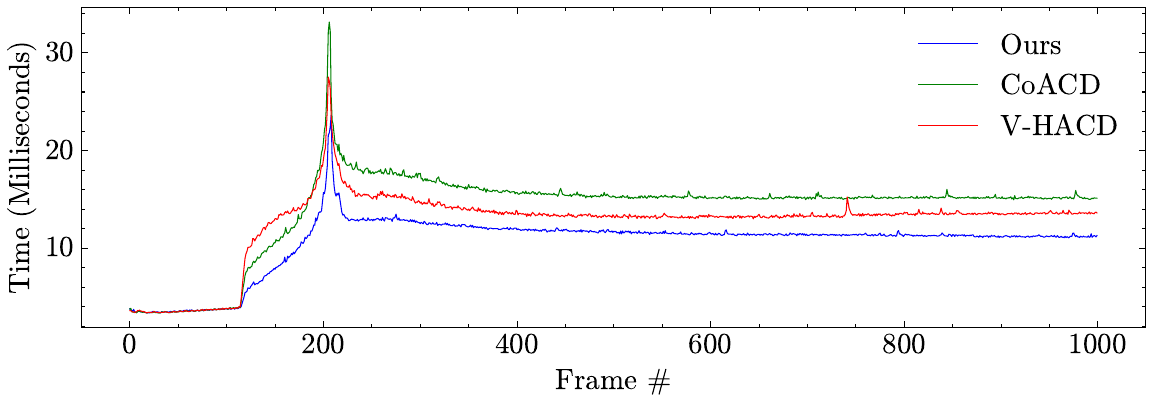} \\
        {\small Church Organ, Fig.~\ref{fig:additional_qualitative_results}, Ours = 180 Primitives, CoACD = 57} & {\small Space Fighter, Ours = 158 Primitives, CoACD = 16 Hulls} \\
        \includegraphics[width=0.45\linewidth]{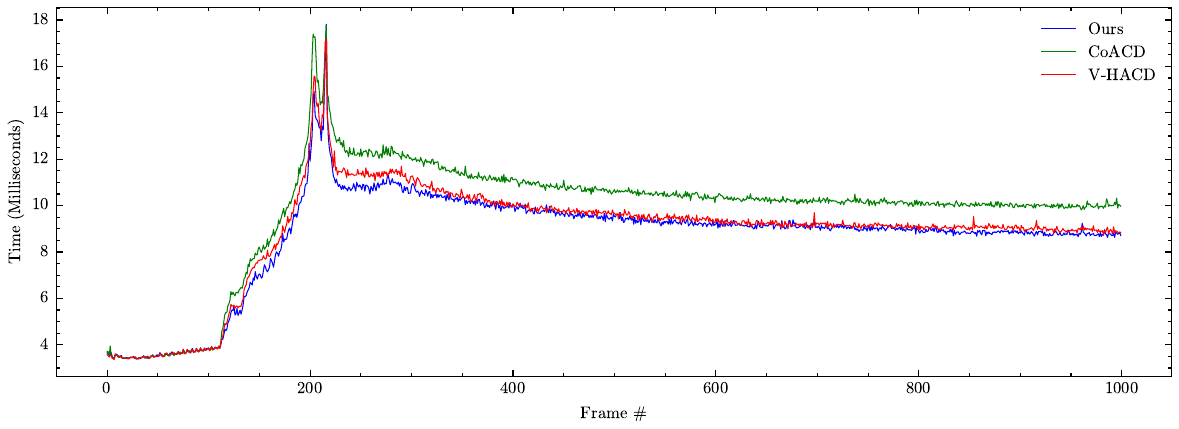} &
        \includegraphics[width=0.45\linewidth]{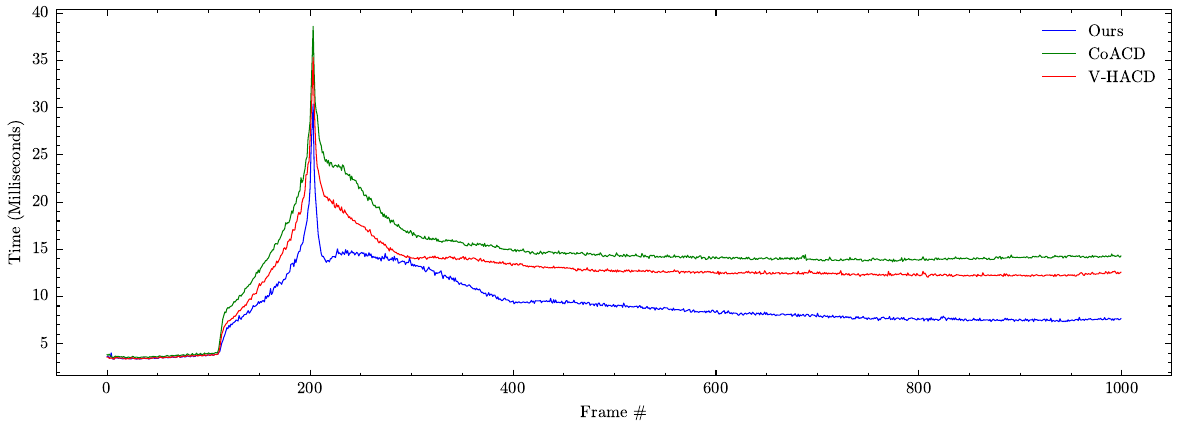} \\
    \end{tabular}

    \caption{Comparison of our approach against CoACD and V-HACD, considering the number of components in our approach versus CoACD. Our approach is always faster or equal to approximate convex decomposition, regardless of the number of components.}
    
    \label{fig:sim-times-component-comparison}
    %\Description{A chart comparing the duration of each frame for collision detection, for our approach, CoACD and V-HACD. The three approaches are quite similar, our approach's peak is below CoACD and above V-HACD. The tail of our plot is noticeably lower.}
\end{figure*}

\begin{figure*}
    \centering
    \setlength{\tabcolsep}{2pt}
    \begin{tabular}{c c c c}
        Input & Ours & Input & Ours \\
        \includegraphics[width=0.2\linewidth]{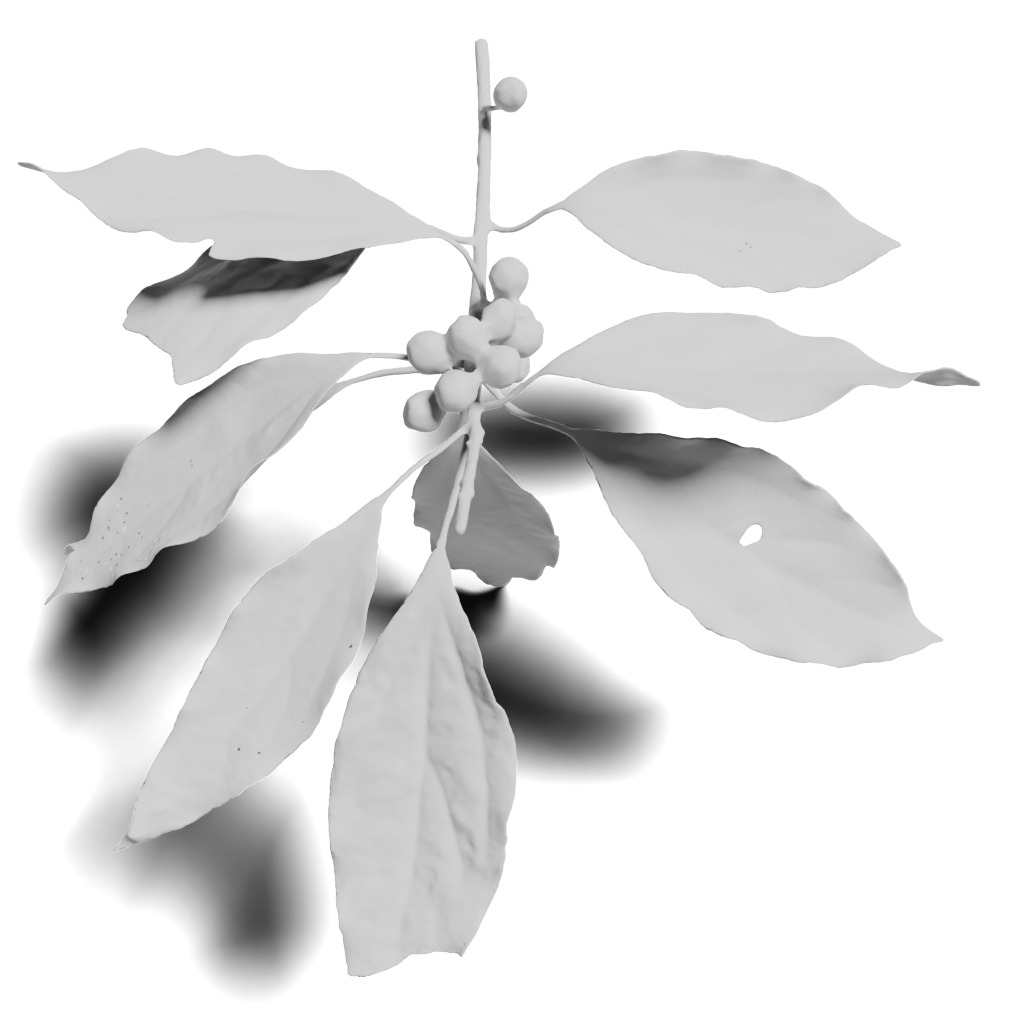} & 
        \includegraphics[width=0.2\linewidth]{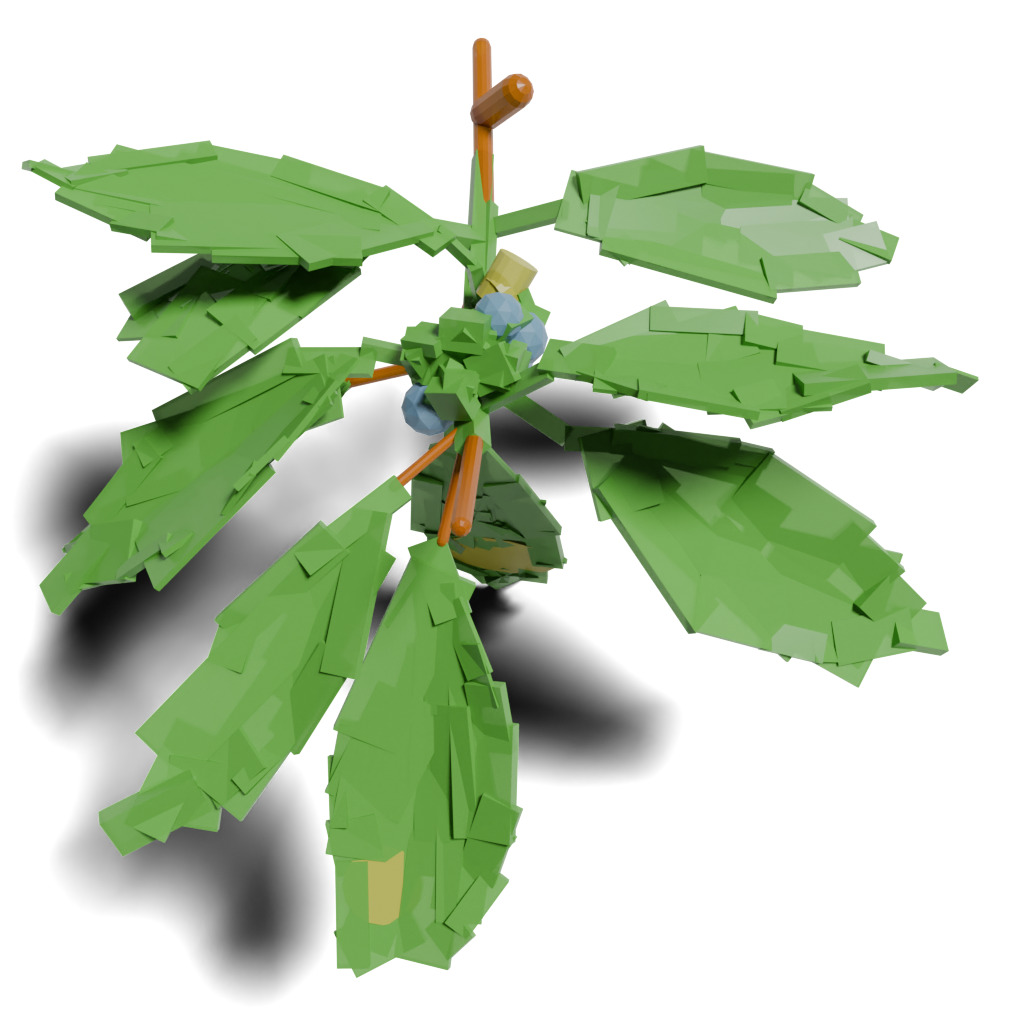} &
        \includegraphics[width=0.2\linewidth]{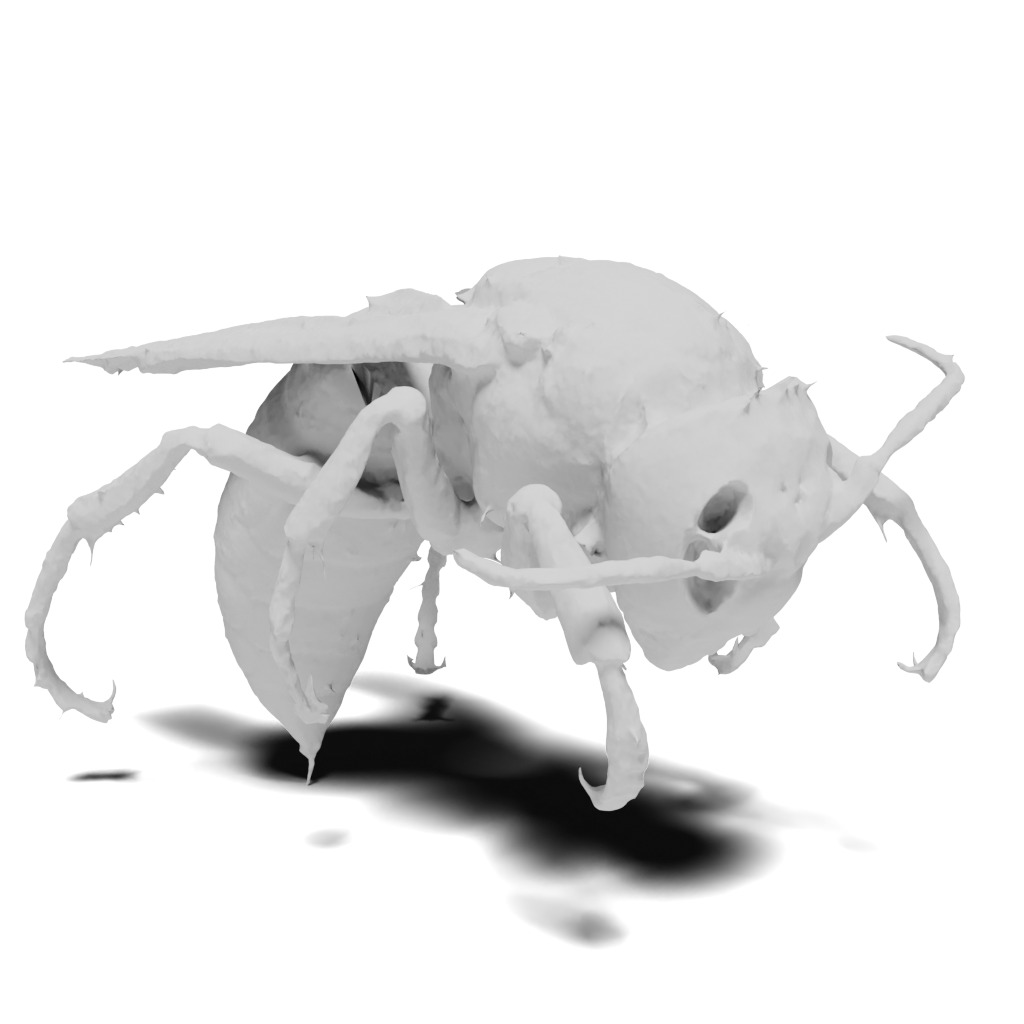} & 
        \includegraphics[width=0.2\linewidth]{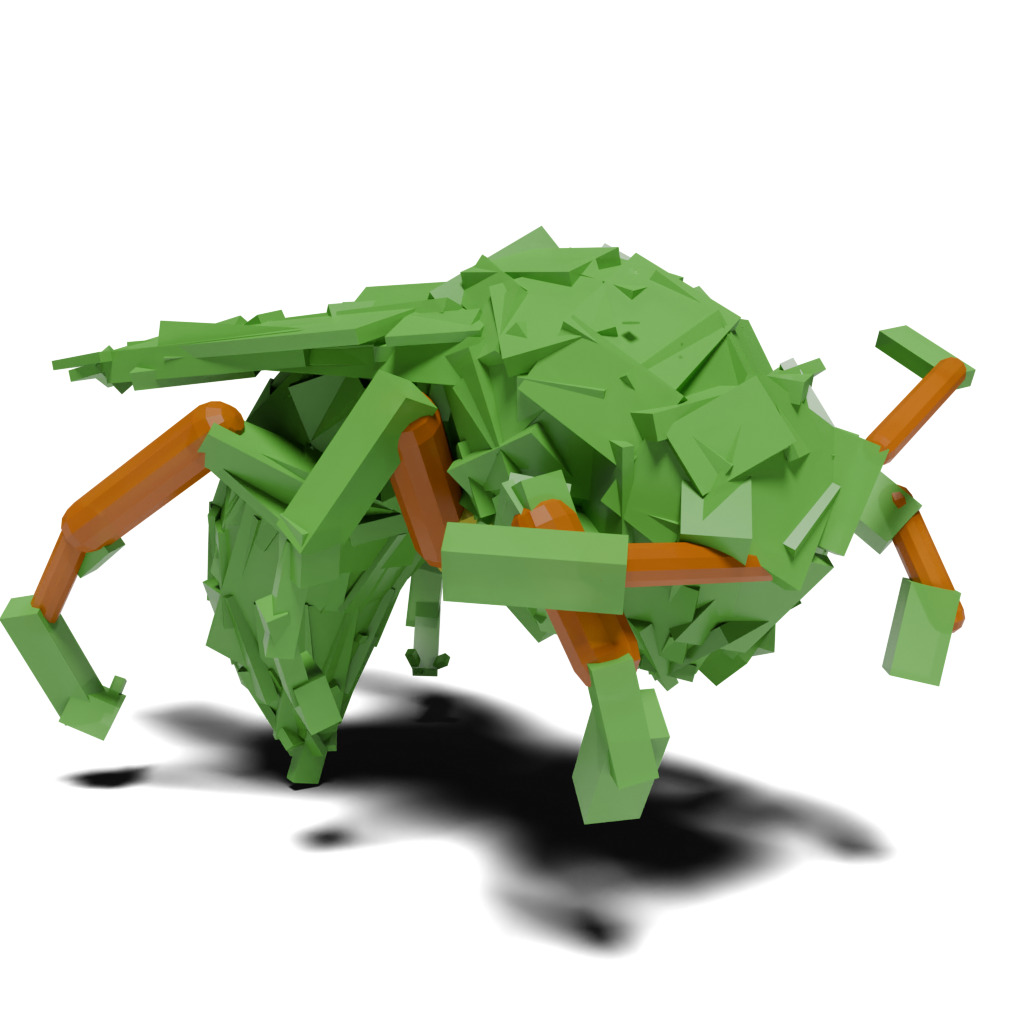} \\
        |F| = 497084 & 288 Boxes, 8 Cap, 4 Cyl, 4 Sph & |F| = 375224 & 311 Boxes, 12 Cap, 5 Cyl, 3 Sph \\
    \end{tabular}
    \caption{We show our approach on organic meshes, which are not represented heavily in our dataset. Our approach is primarily designed for artist created meshes for games, but can capture coarse details on organic meshes. A larger number of primitives are used, as each primitive alone cannot capture curved surfaces well. For these models, we change weights so that spheres are 0.7, capsules are 0.85, and isosceles prisms are disallowed. The current approach for these kinds of mesh is manual creation, convex hulls, or coarse approximations. \cczero ffish.asia / floraZia.com.}
    \label{fig:organic-meshes}
    %\Description{Decomposition of a plant with some berries, and a giant hornet with our approach}
\end{figure*}

\begin{figure*}
    \centering
    \setlength{\tabcolsep}{1pt}
    \begin{tabular}{c c c c c c}
        Input & Ours & V-HACD & Input & Ours & V-HACD \\
 
        \includegraphics[width=0.14\linewidth]{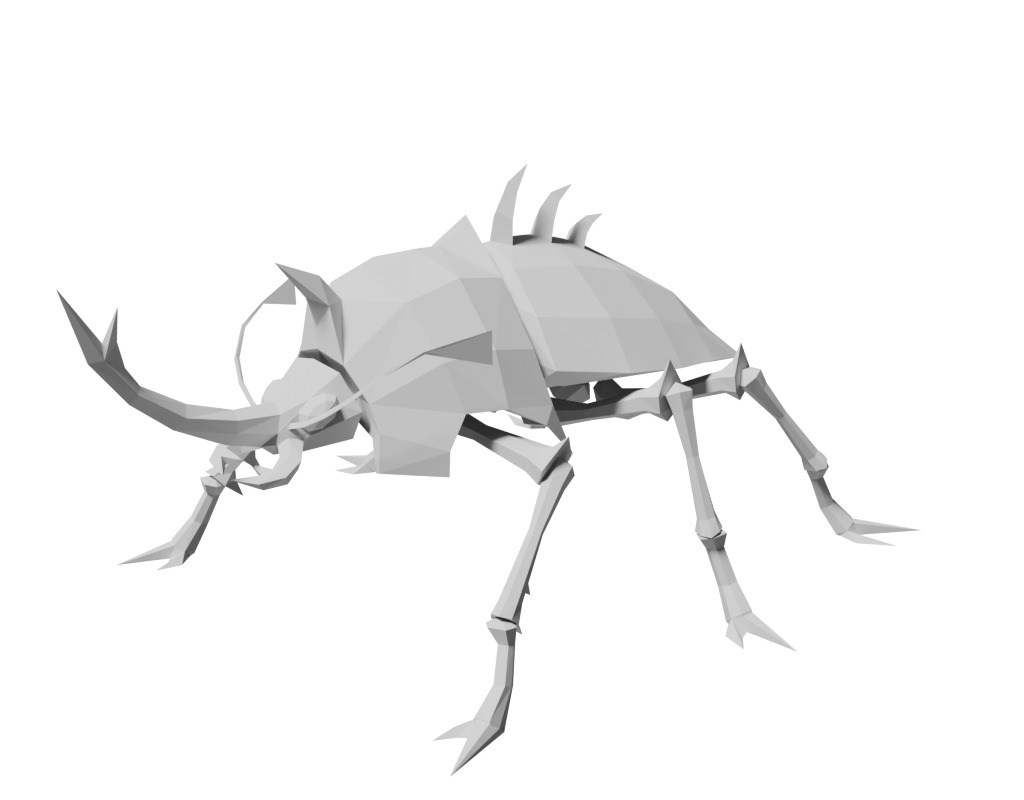} & 
        \includegraphics[width=0.14\linewidth]{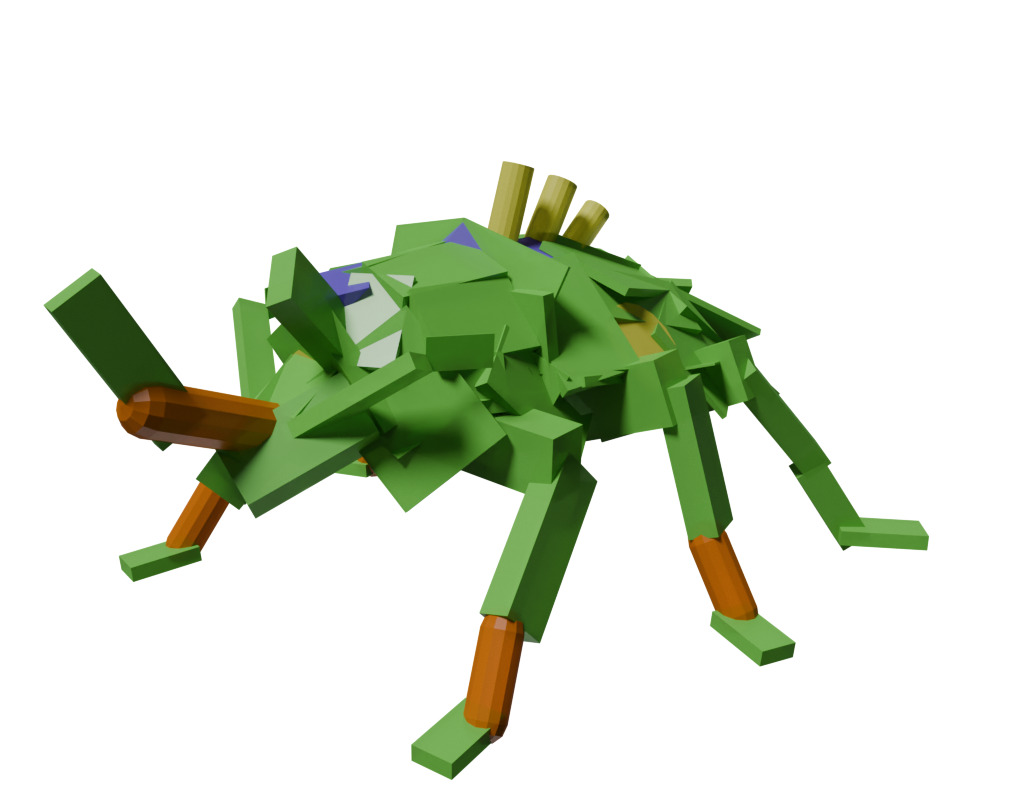} & 
        \includegraphics[width=0.14\linewidth]{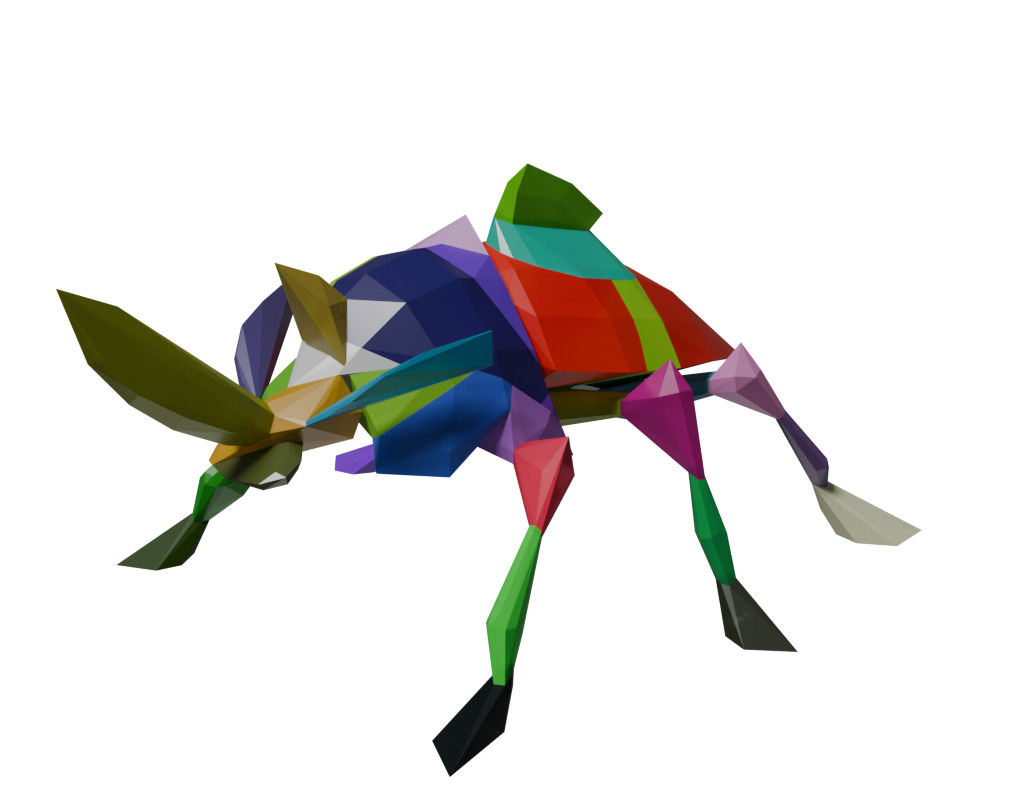} &
        
        \includegraphics[width=0.14\linewidth]{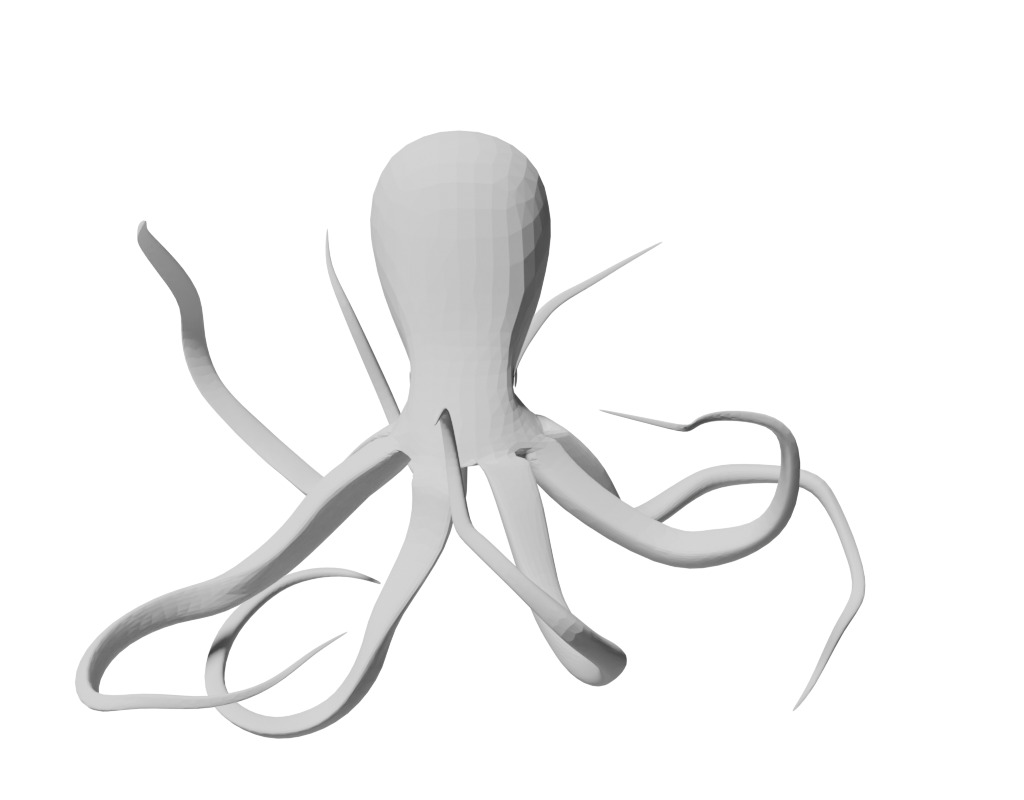} & 
        \includegraphics[width=0.14\linewidth]{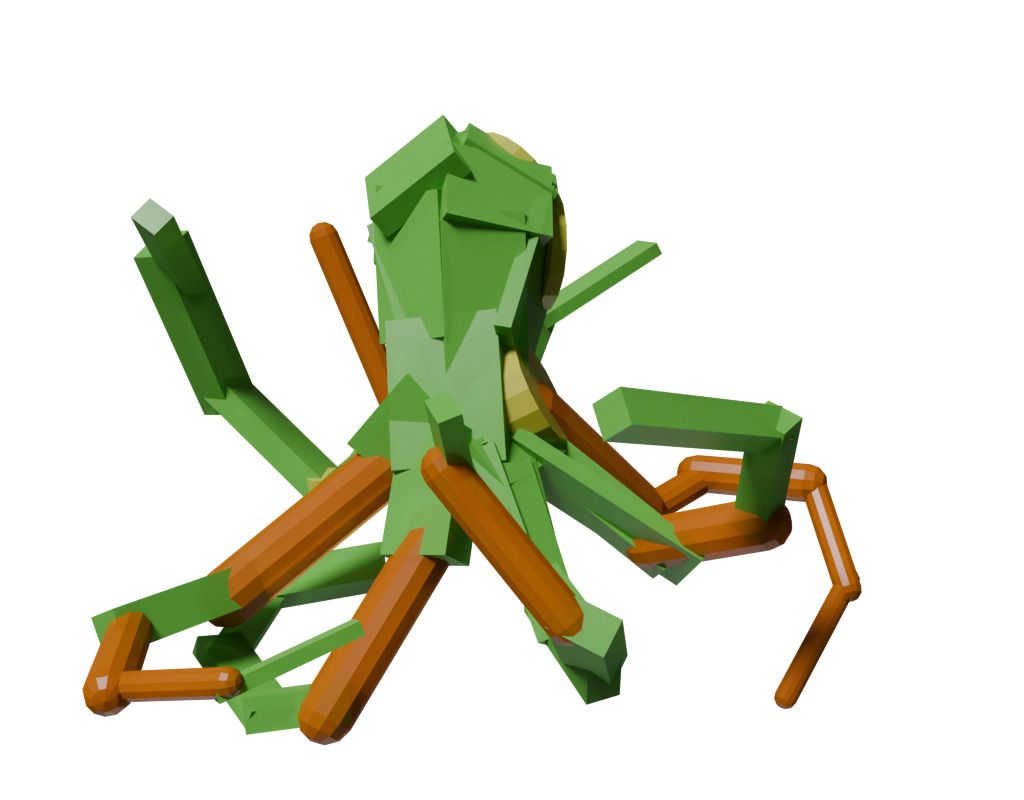} & 
        \includegraphics[width=0.14\linewidth]{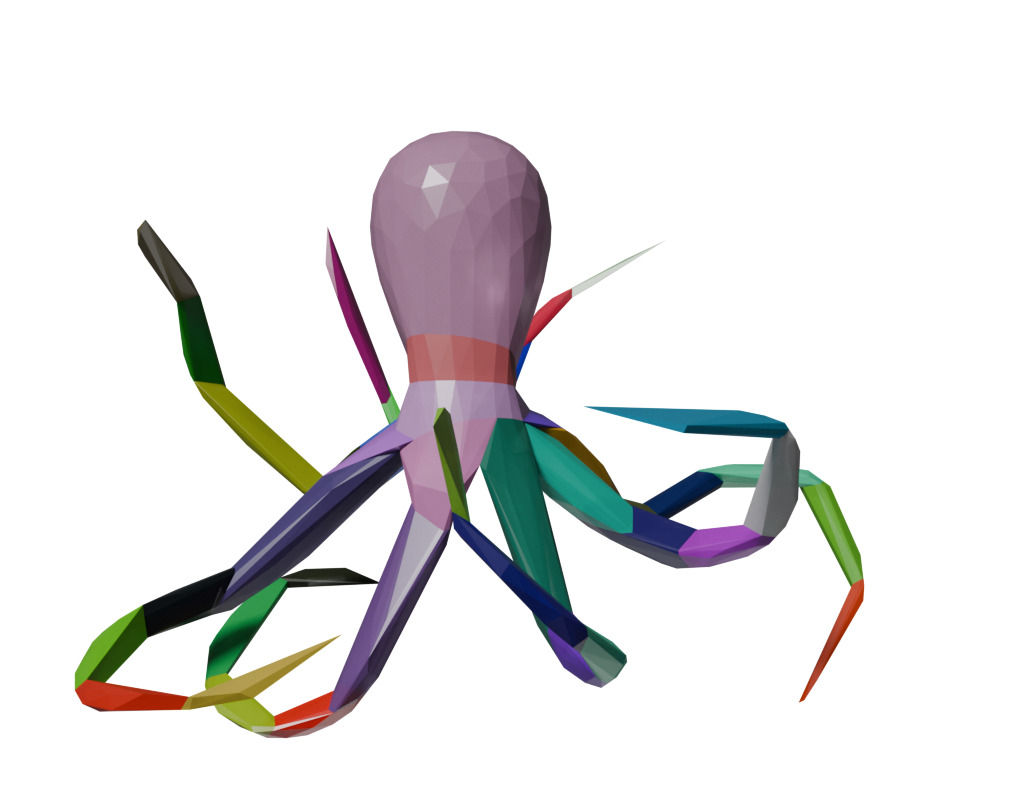} \\

        |F| = 1884 & 69 Boxes, 7 Cap, 6 Cyl, 8 Prism & 50 Hulls (|F| = 4272) & 
        |F| = 33872 & 48 Boxes, 14 Cap, 5 Cyl & 47 Hulls (|F| = 5606) \\
        \multicolumn{3}{c}{\includegraphics[width=0.45\linewidth]{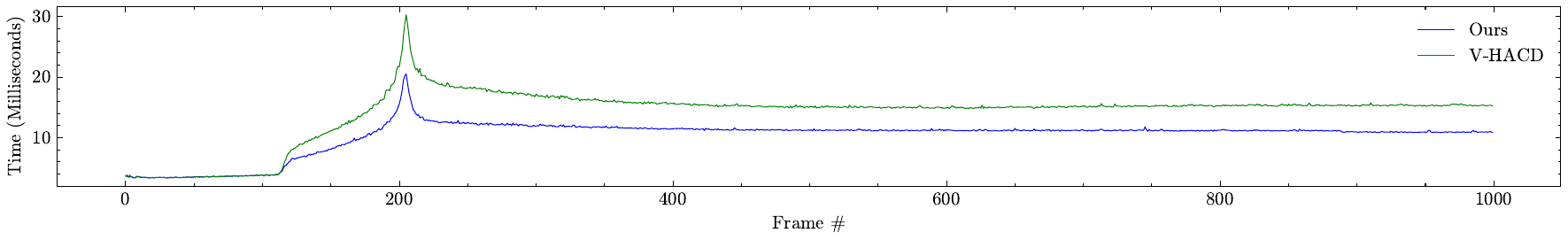}} &
        \multicolumn{3}{c}{\includegraphics[width=0.45\linewidth]{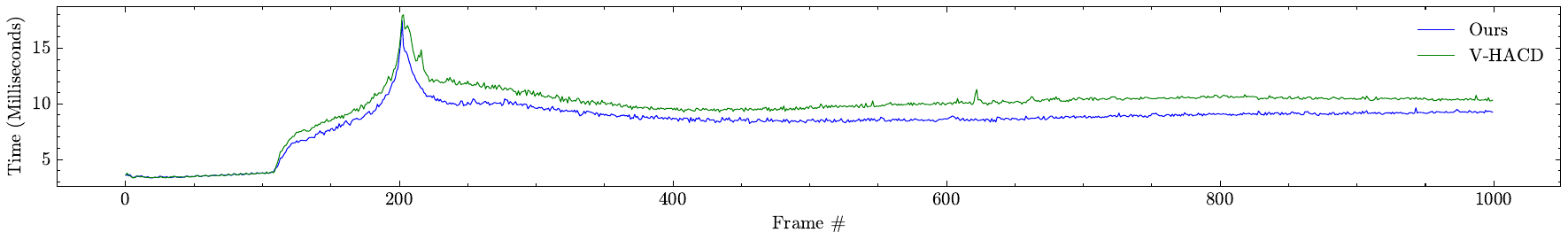}} \\

        \includegraphics[width=0.14\linewidth]{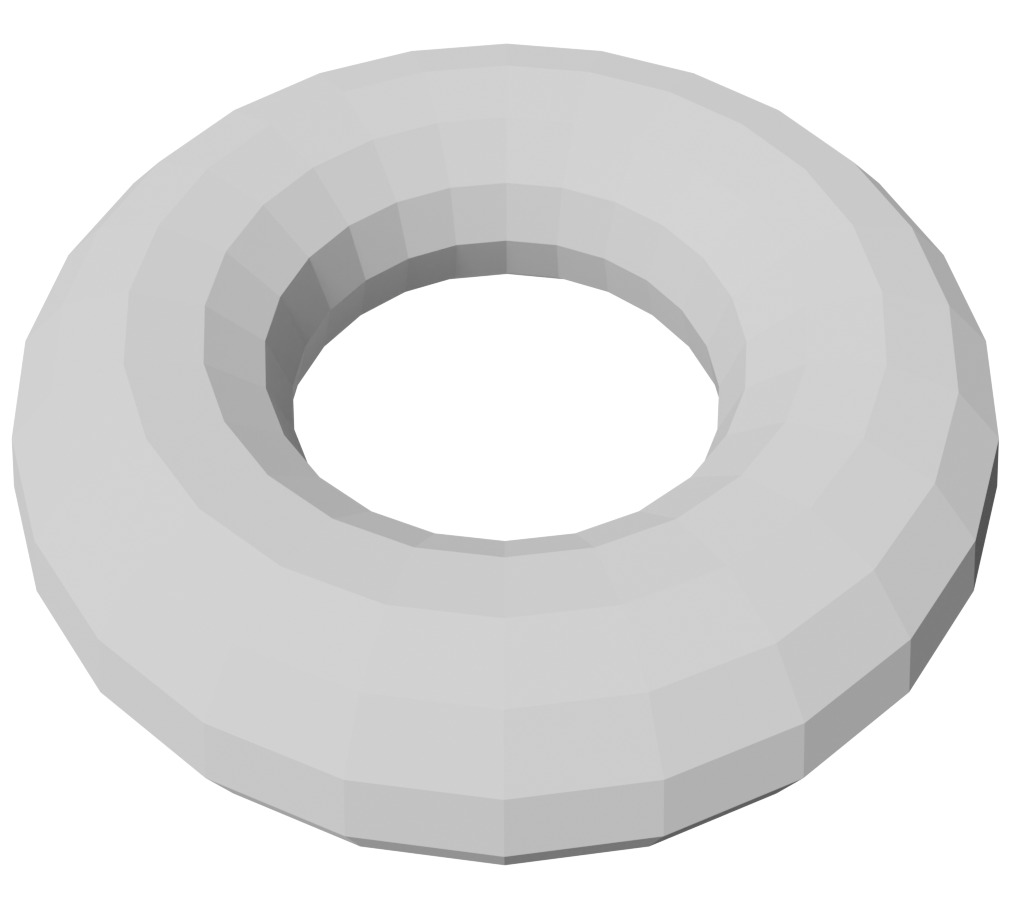} & 
        \includegraphics[width=0.14\linewidth]{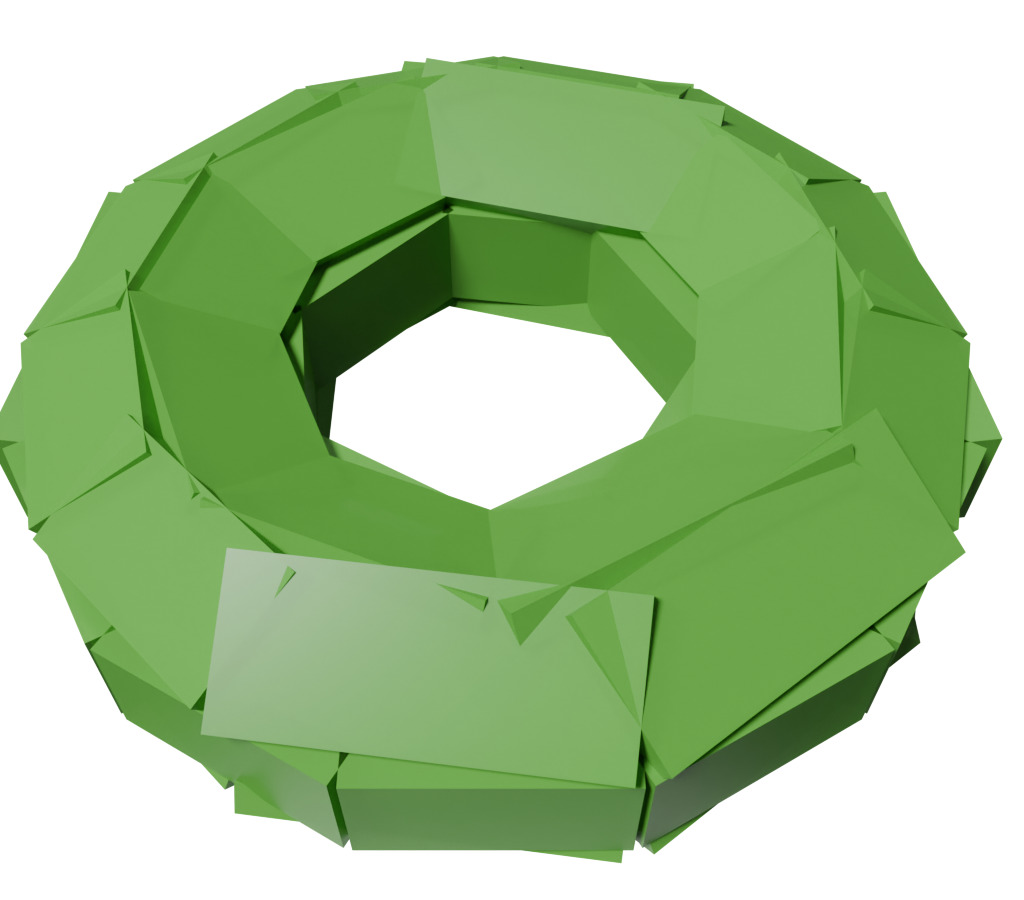} & 
        \includegraphics[width=0.14\linewidth]{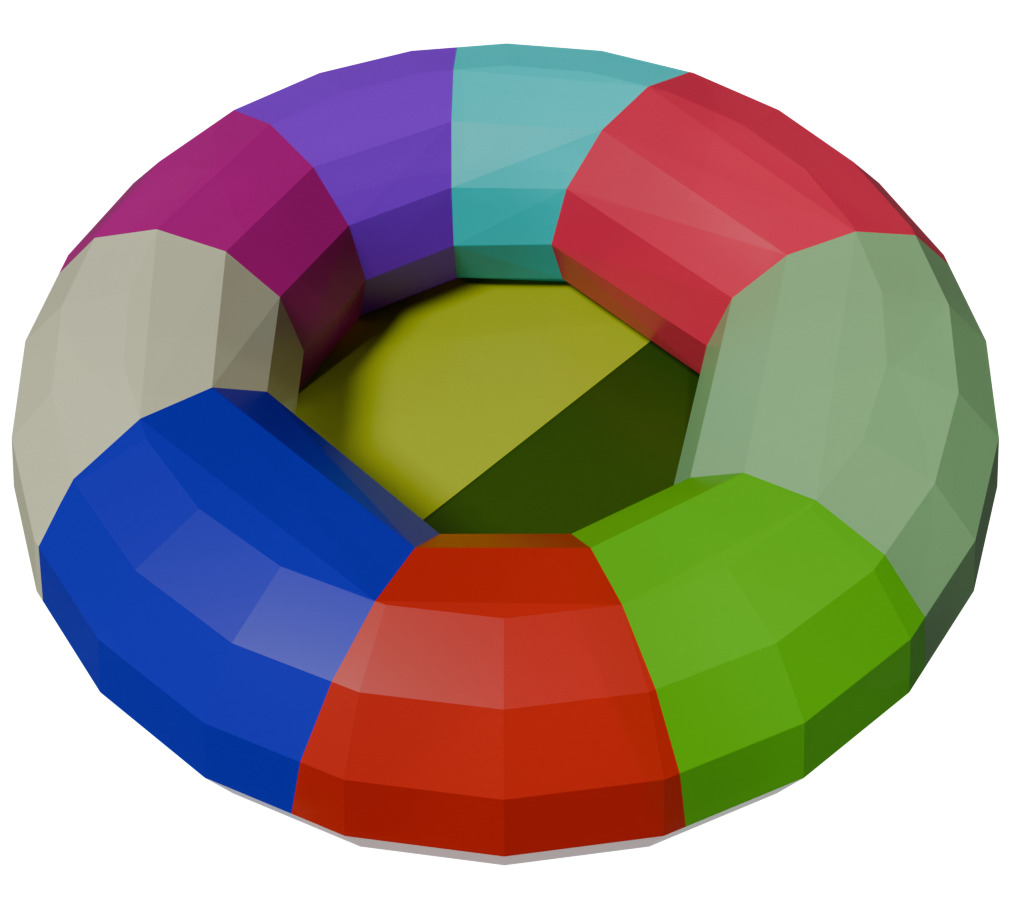} &
                
        \includegraphics[width=0.15\linewidth]{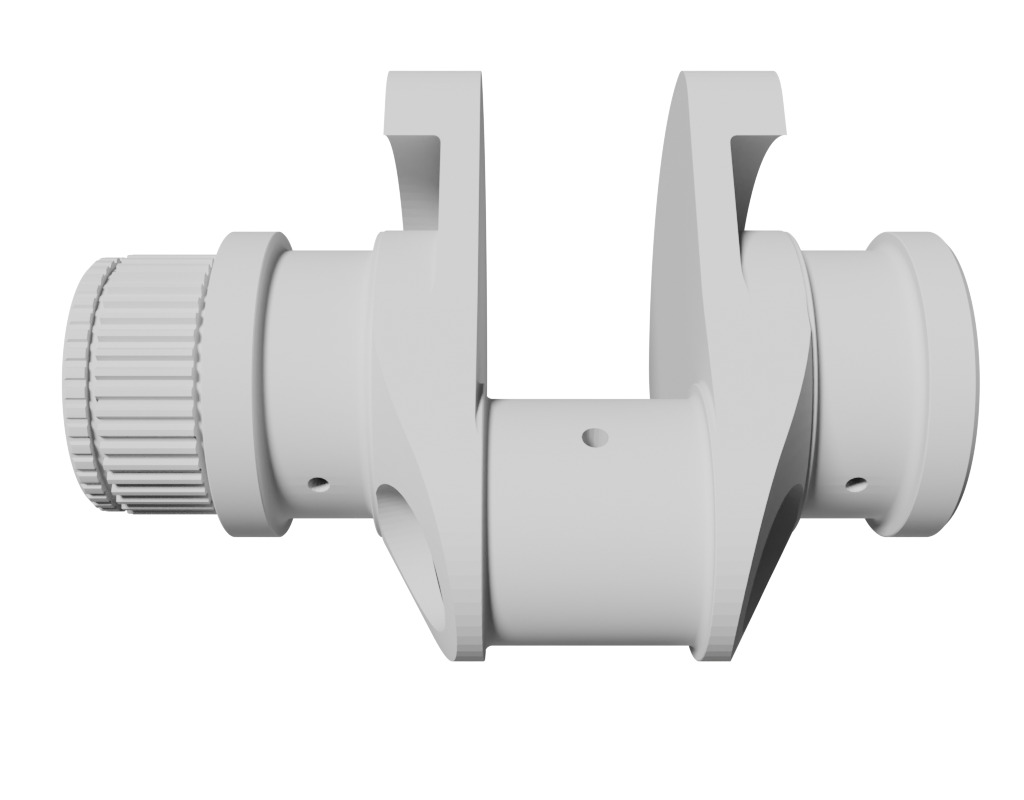} & 
        \includegraphics[width=0.15\linewidth]{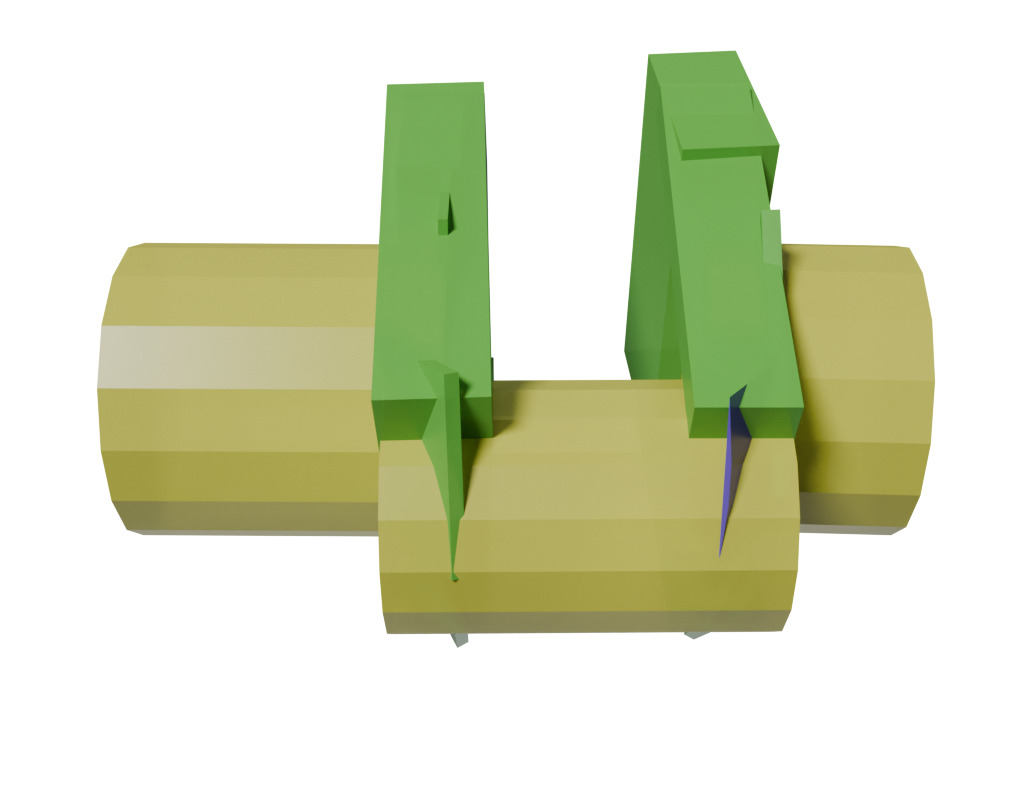} & 
        \includegraphics[width=0.15\linewidth]{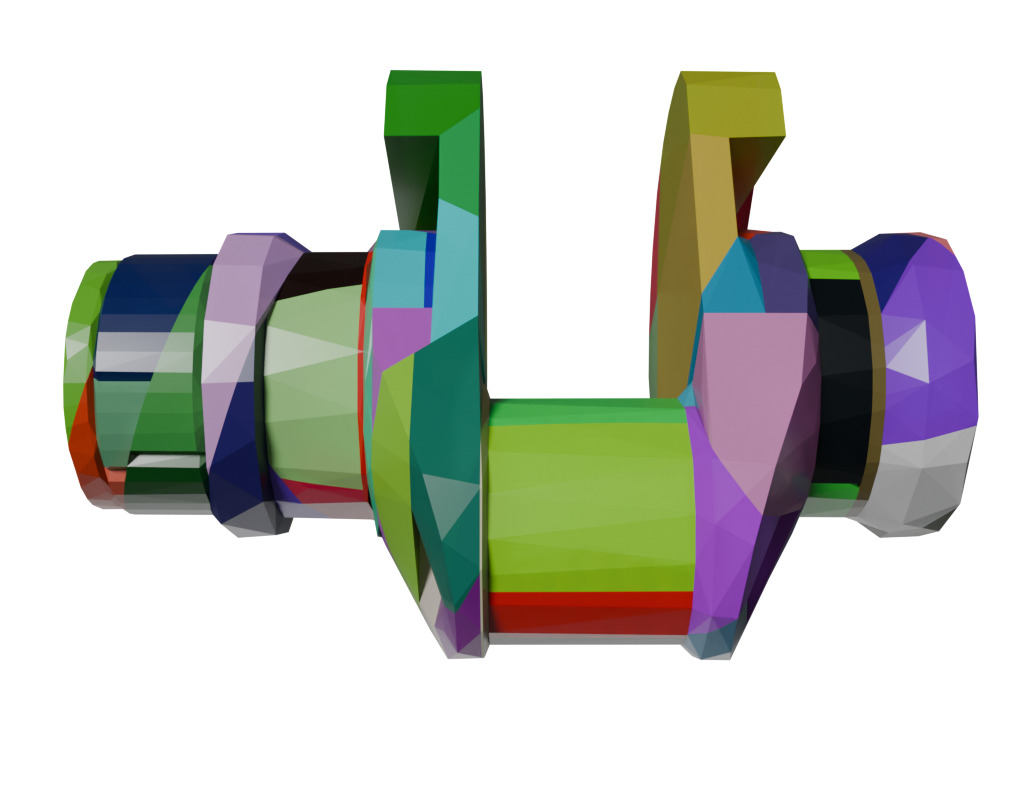} \\
        |F| = 576 & 54 Boxes, 1 Prism & 21 Hulls (|F| = 2340) &
        |F| = 100028 & 11 Boxes, 3 Cyl, 2 Prism & 72 Hulls (|F| = 8812)
        \\
        \multicolumn{3}{c}{\includegraphics[width=0.45\linewidth]{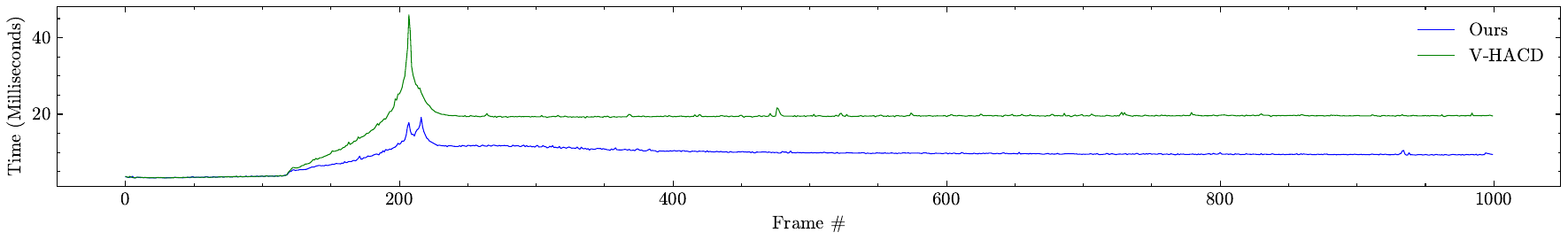}} &
        \multicolumn{3}{c}{\includegraphics[width=0.45\linewidth]{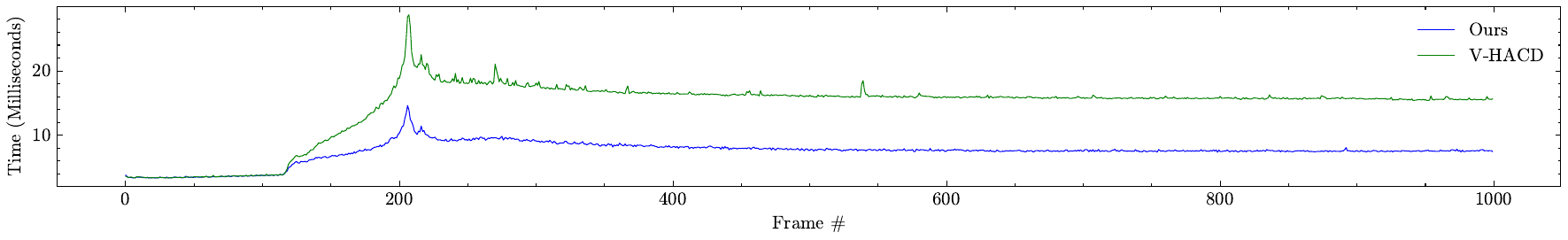}} \\
    \end{tabular}
    \caption{\label{fig:vhacd-data-comp} Comparison of our approach to V-HACD~\cite{vhacd} on a few models from their dataset. Our approach's decomposition often breaks up the input mesh in similar positions to V-HACD, while improving collision performance.}
    %\Description{Top-Left: Torus, Top-Right: Dancing Ballerina, Bottom-Left: Hornbug, Bottom-Right: Octopus. Ours is shown decomposed into mostly green boxes, V-HACD is decomposed into multicolor components.}
\end{figure*}

\end{document}